\documentclass{aa}

\usepackage{graphics}

\usepackage{graphicx}
\usepackage{multicol}
\usepackage{longtable}
\usepackage{supertabular}
\usepackage{txfonts}
\usepackage{color,soul}

\usepackage{natbib,twoopt}
\usepackage{amsmath} 
\usepackage[breaklinks=true]{hyperref} 

\bibpunct{(}{)}{;}{a}{}{,} 
\makeatletter
\newcommandtwoopt{\citeads}[3][][]{\href{http://adsabs.harvard.edu/abs/#3}%
{\def\hyper@linkstart##1##2{}%
\let\hyper@linkend\@empty\citealp[#1][#2]{#3}}}
\newcommandtwoopt{\citepads}[3][][]{\href{http://adsabs.harvard.edu/abs/#3}%
{\def\hyper@linkstart##1##2{}%
\let\hyper@linkend\@empty\citep[#1][#2]{#3}}}
\newcommandtwoopt{\citetads}[3][][]{\href{http://adsabs.harvard.edu/abs/#3}%
{\def\hyper@linkstart##1##2{}%
\let\hyper@linkend\@empty\citet[#1][#2]{#3}}}
\newcommandtwoopt{\citeyearads}[3][][]%
{\href{http://adsabs.harvard.edu/abs/#3}
{\def\hyper@linkstart##1##2{}%
\let\hyper@linkend\@empty\citeyear[#1][#2]{#3}}}
\makeatother

\usepackage{color}
\definecolor{mygreen}{RGB}{0,128,0}

\hypersetup{colorlinks=true,linkcolor=blue,citecolor=blue,urlcolor=blue}

\begin{document}

\title{Stellar and substellar companions from Gaia EDR3}
\subtitle{Proper-motion anomaly and resolved common proper-motion pairs\thanks{Tables \ref{PMa-sample}, \ref{CPMHIP-sample}, and \ref{CPMGaia-sample} are available in electronic form at the CDS via anonymous ftp to \url{cdsarc.u-strasbg.fr} (130.79.128.5) or via \url{http://cdsweb.u-strasbg.fr/cgi-bin/qcat?J/A+A/}}}
\titlerunning{Stellar and substellar companions from Gaia EDR3}
\authorrunning{P. Kervella et al.}
\author{
Pierre~Kervella\inst{1}
\and
Fr\'ed\'eric Arenou\inst{2}
\and
Fr\'ed\'eric~Th\'evenin\inst{3}
}
\institute{
LESIA, Observatoire de Paris, Universit\'e PSL, CNRS, Sorbonne Universit\'e, Universit\'e de Paris, 5 place Jules Janssen, 92195 Meudon, France, \email{pierre.kervella@observatoiredeparis.psl.eu}.
\and
GEPI, Observatoire de Paris, Universit\'e PSL, CNRS, 5 Place Jules Janssen, 92190 Meudon, France.
\and
Universit\'e C\^ote d'Azur, Observatoire de la C\^ote d'Azur, CNRS, Lagrange UMR 7293, CS 34229, 06304, Nice Cedex 4, France.
}
\date{Received ; Accepted}
\abstract
{The multiplicity fraction of stars, down to the substellar regime, is a parameter of fundamental importance for stellar formation, evolution, and planetology. The census of multiple stars in the solar neighborhood is however incomplete.}
{Our study is aimed at detecting companions of Hipparcos catalog stars from the proper motion anomaly (PMa) they induce on their host star, namely, the difference between their long-term Hipparcos-Gaia and short-term Gaia proper motion vectors.
We also aim to detect resolved, gravitationally bound companions of the Hipparcos catalog stars ($117,955$ stars) and of the Gaia EDR3 stars closer than 100\,pc (542,232 stars).}
{Using the Hipparcos and EDR3 data, we revised the PMa catalog for the Hipparcos stars. In order to identify gravitationally bound visual companions of our sample, we searched the Gaia EDR3 catalog for common proper-motion (CPM) candidates.}
{The detection of tangential velocity anomalies with a median accuracy of $\sigma(\Delta v_\mathrm{T})=26$\,cm\,s$^{-1}$ per parsec of distance is demonstrated with the EDR3. This improvement by a factor 2.5 in accuracy, as compared to Gaia DR2, results in PMa detection limits on companions that are well into the planetary mass regime for many targets. We identify 37,515 Hipparcos stars presenting a PMa at significant level (S/N>3), namely, a fraction of 32\% (compared to 30\% for the DR2) and 12,914 (11\%) hosting CPM bound candidate companions. After including the Gaia EDR3 renormalised unit weight error (RUWE>1.4) as an additional indicator, 50,720 stars of the Hipparcos catalog (43\%) exhibit at least one signal of binarity. Among the Gaia EDR3 stars located within 100\,pc, we find CPM bound candidate companions for 39,490 stars (7.3\% of the sample).} 
{The search for companions using a combination of the PMa, CPM, and RUWE indicators significantly improves the exhaustivity of the multiplicity survey. The detection of CPM companions of very bright stars (heavily saturated on the Gaia detectors) that are classical benchmark objects for stellar physics provides a useful proxy for estimating their distance with a higher accuracy than with Hipparcos.}

\maketitle

\section{Introduction}

Binary and multiple stars are essential objects in many fields of astrophysics and the statistics of stellar multiplicity, down to planetary mass companions, is an observable of fundamental importance.
For stellar physics, binaries allow for the precise determination of stellar masses, down to sub-percent accuracy \citepads{2021A&ARv..29....4S}.
Pairs of stars sharing the same age and initial chemical composition but with, for instance, slightly different masses, are valuable and highly constraining test cases for stellar models.
Eclipsing binaries, detected in large numbers by space photometry missions (see, e.g., \citeads{2016AJ....151...68K}) are key targets, both for modeling \citepads{2017A&A...608A..62H} and distance determinations \citepads{Pietrzynski:2019aa}.
The influence of binary stars on the formation of our Galaxy, its evolution, and composition has many facets.
Multiplicity deeply influences the physical mechanisms through which stars form, affecting the stellar initial mass function.
The evolution of binary stars may also diverge considerably from that of single stars, for instance, through mass exchange.
This is particularly common during the final stages of their evolution, resulting in spectacular events such as novae or type Ia supernovae.
The coalescence of the compact products of the evolution of massive binary stars is also a major source of gravitational wave emission \citepads{2017PhRvL.119p1101A}.
Giant and telluric planets with extremely diverse properties are now known in large numbers, mainly from the radial velocity and transit techniques \citepads{2016PASP..128f6001F, 2015ARA&A..53..409W, 2014PASP..126..827H} but also from direct imaging \citepads{2019A&A...631A.155B, 2019AJ....158...13N, 2015ApJ...815..108M}.
Stellar binarity has a major impact on the stability of planetary systems (see, e.g., \citeads{2016AJ....152....8K, 2014ApJ...791..111W}).
High-precision astrometry offers a complementary way to detect and characterize exoplanets through the detection of their influence on the space trajectory of their host stars.

Thanks to the unprecedented accuracy of its astrometric measurements and its sensitivity to faint objects, Gaia provides us with a direct way to constrain the presence of companions, exploring the planetary mass regime for a large number of stars in the solar neighborhood.
The long time baseline of 24.75\,years between the Hipparcos and Gaia Early Data Release 3 (EDR3) position measurement epochs opens the possibility to determine the  long-term proper motion (PM) vectors of the Hipparcos catalog stars with a high level of accuracy.
For a single star, the long-term and short-term PM vectors are identical (apart from the geometrical perspective acceleration), but they diverge in the presence of a secondary orbiting body.
The presence of a faint secondary object results in a shift of the barycenter of the system away from its photocenter (usually located close to the primary star's position). The orbital motion of the pair induces a time-dependent displacement of the photocenter around the center of mass.
The comparison of the long-term PM vector with the Gaia or Hipparcos short-term PM vectors therefore opens the possibility to search for an orbiting companion through its effect on the PM of the primary target.
Historically, this principle was first employed by \citetads{1844MNRAS...6R.136B} to discover the invisible companion of \object{Sirius}, the white dwarf \object{Sirius B}, and it was also applied to various types of stars, for instance, by \citetads{1999A&A...346..675W}, \citetads{2004ASPC..318..141J}, \citetads{2005AJ....129.2420M}, \citetads{2007A&A...464..377F}, and \citetads{2008ApJ...687..566M}.
In the present work, we measured the PM offset as a "proper motion anomaly" (PMa), namely, a difference between the "instantaneous" PM vector from the Hip2 or EDR3 catalogs and the long-term PM vector.

In Sect.~\ref{HG-PMa}, we present a revision of the PMa of Hipparcos stars using new astrometry from the EDR3 \citepads{GaiaEDR3content}. After briefly defining the PMa and describing how it can be interpreted in terms of companion properties, we evaluate the sensitivity, completeness, and accuracy of our PMa catalog. We then introduce, in Sect.~\ref{cpm-section}, the procedure we adopted to identify common proper-motion (CPM) gravitationally bound candidate companions. In Sect.~\ref{discussion}, we discuss the global results of our survey, the possible use of the renormalised unit weight error (RUWE) parameter as an additional indicator for binarity, and the combination of the PMa and CPM techniques. Finally, we present sample analyses of specific targets in Sect.~\ref{examples}, followed by our conclusions in Sect.~\ref{conclusion}.


\section{Hipparcos-Gaia proper motion anomaly \label{HG-PMa}}

\subsection{General principle}

The principle underlying the detection of companions from their influence on the PM of a star relies on the comparison of the long-term and short-term PMs of this star. For a single star, the long-term PM determined from the positions measured at the Hipparcos and EDR3 epochs (24.75 years apart) is identical to the short term PM measured by each mission over a few years.
For a binary star, the short-term PM includes in addition the tangential component of its orbital velocity. As the latter is changing with time over the orbital period of the system, a deviation appears between the short-term and long-term PMs of the star, due to the curvature of its sky trajectory. 
The PMa, namely, the difference between the short-term and long-term PM, is therefore an efficient and sensitive indicator to detect non-single stars, as it is a proxy for the orbital velocity of the photocenter of the system around its center of mass.
Thanks to the long time baseline between the Hipparcos and Gaia epochs, the PMa can now be measured with a very high accuracy, which translates to substellar mass sensitivity for the companion of nearby stars.
Further details on the PMa are available in \citetads{2019A&A...623A..72K}.
Examples of analyses of binary and multiple stars based on Hipparcos and Gaia astrometry can be found for instance in the following studies:
\citetads{2018ApJS..239...31B}, \citetads{2019AJ....158..140B}, \citetads{2019ApJ...871L...4D}, \citetads{2019A&A...623A.116K}, \citetads{2020ApJ...904L..25C}, \citetads{2020MNRAS.496.1922B}, \citetads{2021A&A...645A...7K}, \citetads{2021ApJS..254...42B}, \citetads{2021arXiv210907525B}, and \citetads{2021arXiv210608249K}.

\subsection{Input data, basic corrections, and PMa computation\label{inputdata}}

We adopted the Hipparcos catalog at epoch J1991.25 (\citeads{2007ASSL..350.....V}, hereafter `Hip2',  117\,955 sources) and the Gaia EDR3 catalog \citepads{2016A&A...595A...1G, 2021A&A...650C...3G, GaiaEDR3content, 2020yCat.1350....0G} at epoch J2016.0.
For the collection of most of the data used in the present work, we made extensive use of the \texttt{astroquery} library \citepads{2017ascl.soft08004G} distributed as part of the \texttt{Astropy} library \citepads{2013A&A...558A..33A,2018AJ....156..123A} to access the ViZieR online database \citepads{2000A&AS..143...23O} at the CDS.
For the cross-identification of the Hip2 stars in the EDR3 catalog, we started from the \texttt{gaiaedr3.hipparcos2\_best\_neighbour} list provided as part of the EDR3 \citepads{2017A&A...607A.105M,2019A&A...621A.144M}, which has 99\,525 records (84.4\% of Hip2).
For the missing Hip2 sources, we searched the EDR3 catalog shifting the Hip2 source position to epoch J2016.0 using the Hip2 PM vector.
We then classically employed magnitude, parallax, and angular proximity criteria to select the most probable candidate source in the EDR3 catalog.
A total of 116\,343 sources are present in our PMa catalog (98.6\% of Hip2 catalog), out of which 568 stars (0.5\%) have neither DR2 or EDR3 PMa vectors (due, e.g., to the Gaia PM vector being unavailable) and 1535 stars have no EDR3 PMa vector (1.3\%).
We applied the corrections to the EDR3 parallaxes as prescribed by \citetads{2021A&A...649A...4L}\footnote{\url{https://www.cosmos.esa.int/web/gaia/edr3-code}}, and corrected the PM of bright sources for the spin of the Gaia frame with respect to the ICRS determined by \citetads{2021A&A...649A.124C}. We also inflated the parallax error bars according to \citetads{2021MNRAS.506.2269E}.

For simplicity, we use $\mu_{\mathrm{\alpha}}$ to denote the PM along the right ascension axis, $\mu_{\mathrm{\alpha}} \cos(\delta)$.
We collected ancillary data (photometry, radial velocity, etc.) and estimated the mass and radius of each star following the methodology described by \citetads{2019A&A...623A..72K}.
The long-term Hipparcos-Gaia vector is computed from the difference in coordinates between the Hipparcos and Gaia catalogs, scaled by the time difference between the two catalogs (24.75 years for Gaia EDR3). The PMa vector coordinates are computed by subtracting the Hipparcos-Gaia PM vector from the individual Hipparcos and Gaia vectors, and the associated uncertainties are computed using a simple Monte Carlo approach. This computation is conducted in three dimensions for the stars located within 20\,pc of the Sun, to take properly into account the light time propagation and perspective acceleration. They are particularly important for the nearest stars with a fast PM (\object{Proxima Centauri}, \object{Barnard's star}...). For stars beyond this distance, a two-dimensional computation (in tangential coordinates) was implemented to reduce the computation time, as the perspective acceleration is negligible, but still taking into account the light time propagation.

As a remark, the Hipparcos astrometry of visual pairs with separations of 10 to $20\arcsec$ is often distorted owing to the satellite's measuring system design (Tokovinin, private comm.). As a result, the PMa computed for the components of such physically unrelated pairs can be spurious.

\subsection{Companion properties and sensitivity function\label{sensitivity_pma}}

As discussed by \citetads{2019A&A...623A..72K}, the mass of the companion of a primary star exhibiting a PMa signal can be constrained using the measured tangential velocity anomaly.
It is, however, degenerate with its orbital radius $r$ following the relation:
\begin{equation}\label{m2mass}
\frac{m_2}{\sqrt{r}} = \sqrt{\frac{m_1}{G}}\,v_\mathrm{1} = \sqrt{\frac{m_1}{G}}\,\left( \frac{\Delta \mu [\mathrm{mas\,a}^{-1}] }{\varpi [\mathrm{mas}]} \times 4740.470 \right)
,\end{equation}
where $m_1$ is the mass of the primary star, $m_2$ the mass of the companion, $G$ the universal gravitational constant, $\Delta \vec{\mu}$ the PMa, $v_1$ the tangential orbital velocity of the primary star, and $\varpi$ its parallax.
The sensitivity of the PMa technique in terms of secondary mass therefore decreases linearly with the distance of the target.
In this expression, we assume that the orbit is circular and observed ``face-on,'' and that the photocenter of the system is located close to the primary star (the secondary source is faint compared to the primary).
Also, the practical sensitivity of the PMa technique is limited by the time window smearing of the short-term PM measurements (Hipparcos or Gaia), as well as the limited time baseline between the Hipparcos and Gaia epochs for the estimation of the long-term PM vector (see below).
For a more realistic definition of the expected companion properties, we include the uncertainty on the orbit inclination in a statistical way, following Sect. 3.6 of \citetads{2019A&A...623A..72K}. The influence of the orbital eccentricity is limited (in a statistical sense), as it does not introduce a global bias, but it will affect individual measurements obtained, for instance, near the periastron or apastron for which the orbital velocity takes extreme values.

\begin{figure}
\includegraphics[width=\hsize]{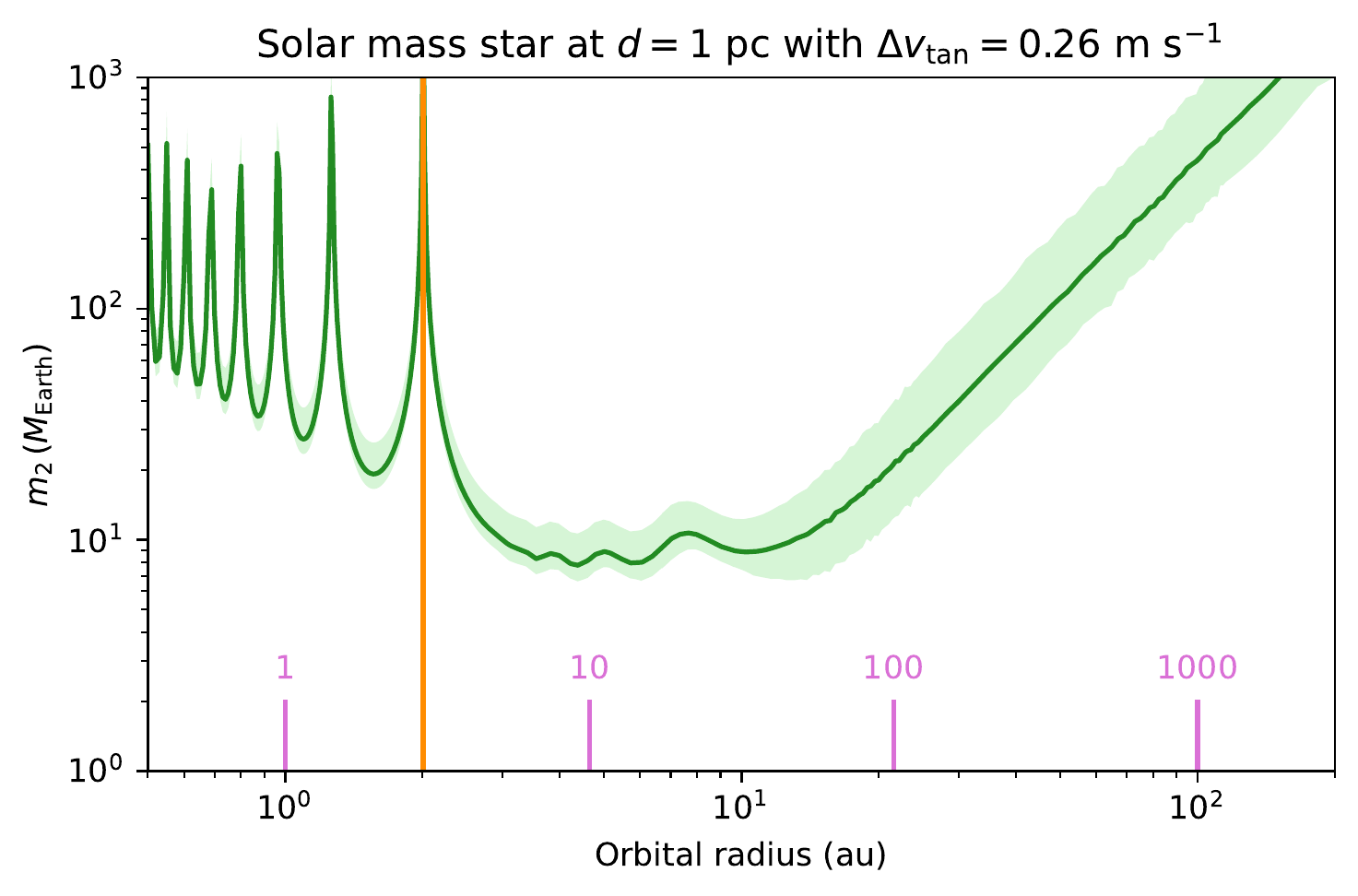}
\caption{Sensitivity function $m_2 = f(r)$ for a solar mass star at a distance of 1\,pc. The pink markings show a selection of orbital periods in years, and the orbital radius corresponding to an orbital period equal to the Gaia EDR3 duration (34 months) is displayed as an orange vertical line.\label{Sensitivity-function-m2r}}
\end{figure}

The sensitivity function $m_2 = f(r)$ is affected by the fact that the Hipparcos and Gaia catalog measurements are smeared over the observing time window of the two missions. The astrometric transits were obtained over a period of $\delta t_\mathrm{H} = 1227$\,d \citepads{1997A&A...323L..49P}, $\delta t_\mathrm{G2} = 668$\,d \citepads{2018A&A...616A...1G}, and $\delta t_\mathrm{G3} = 1038$\,d \citepads{GaiaEDR3content}, respectively, for Hipparcos, Gaia DR2 and Gaia EDR3. This drastically reduces the sensitivity of the survey to companions with orbital periods shorter than these time windows.
The sensitivity also decreases for long period orbits due to the fact that we subtract the long-term $\vec{\mu_\mathrm{HG}}$ PM vector from the short-term Gaia PM vector. For long orbital periods (typically longer than about five$ $\,times the Hipparcos-Gaia time span), the subtraction of $\vec{\mu_\mathrm{HG}}$ removes a significant part of the signature of the orbital motion of the photocenter of the system around the barycenter. This reduces the PMa signal and, therefore, the sensitivity to low-mass objects.
Figure~\ref{Sensitivity-function-m2r} shows the sensitivity function for a solar mass star located at a distance of 1\,pc, with a tangential velocity anomaly of 0.26\,m\,s$^{-1}$ corresponding to the median accuracy of EDR3 PMa measurements. The domain shaded in green shows the geometrical uncertainty due to the unconstrained orbital inclination.
The "spikes" visible in Fig.~\ref{Sensitivity-function-m2r} for orbital radii smaller than that corresponding to the Gaia time window are due to the fact that when the orbital period corresponds to the EDR3 time window is divided by an integer, the PMa signal becomes null. This results in a non-detection of the companion independently of its mass.

\subsection{Properties of the PMa catalog}

An extract of the PMa catalog is presented in Table~\ref{PMa-sample}.

\subsubsection{Completeness of the sample\label{completeness}}

\begin{figure}
\includegraphics[width=\hsize]{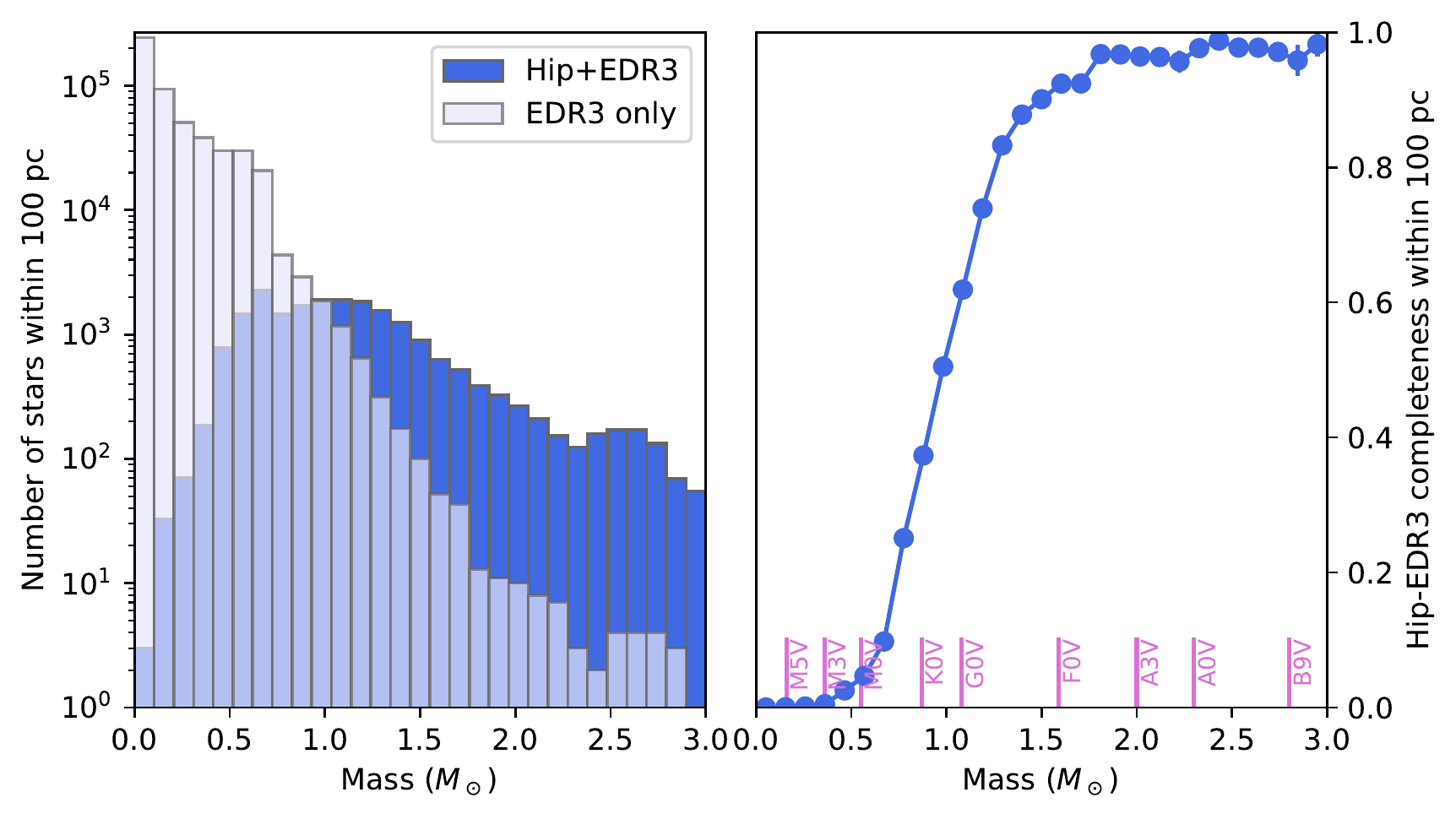}
\caption{Number of stars as a function of mass (left panel) and completeness of the Hipparcos-EDR3 proper motion anomaly sample within 100 pc (right panel).\label{PMa-completeness}}
\end{figure}

We estimate the completeness of the Hipparcos-EDR3 sample within 100 pc for stellar-mass objects using as a basis the full EDR3 catalog within the scope of this distance. As shown by \citetads{2021A&A...649A...6G}, the degree of completeness of the Gaia EDR3 Catalogue of Nearby Stars (GCNS) within 100\,pc is at an excellent level. The deep $G \approx 21$ limiting magnitude of Gaia corresponds to the apparent brightness of the lowest mass stars at 100\,pc (see also Sect.~\ref{sensitivity-combination}). The EDR3 catalog is thus highly complete for stellar mass objects down to the hydrogen-burning limit up to this distance and gives a good fiducial to estimate the Hipparcos completeness. The distribution of the number of stars as a function of mass is shown in Fig.~\ref{PMa-completeness} for the Hipparcos+EDR3 and full EDR3 samples. The completeness of the Hipparcos-EDR3 catalog compared to the EDR3 for low-mass stars below $0.5\,M_\odot$ located within 100\,pc ($\varpi_\mathrm{G3}>10$\,mas) is only $\approx 0.07\%$, whereas it is higher than 80\% for stars more massive than the Sun.

\subsubsection{Accuracy}

\begin{figure}
\includegraphics[width=\hsize]{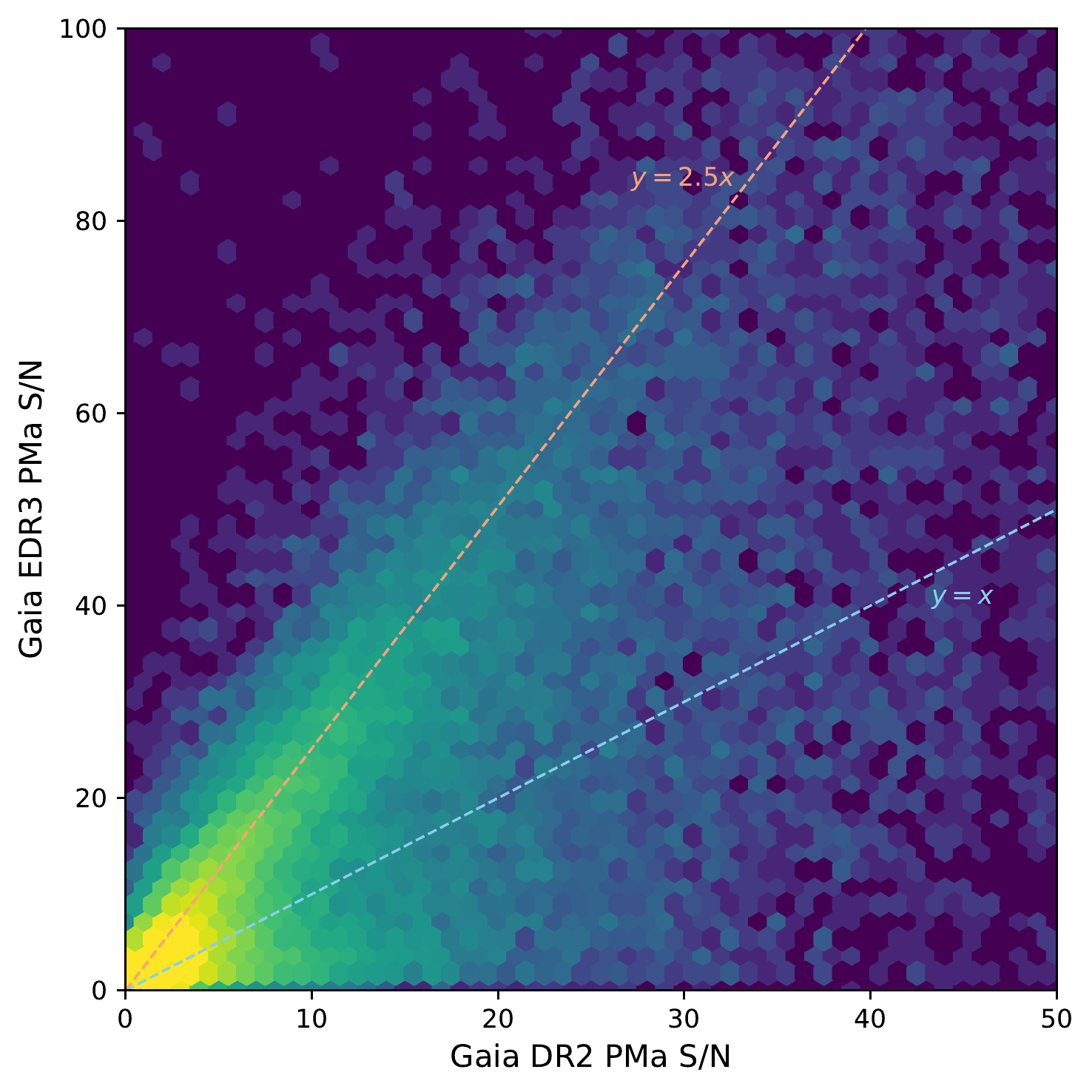}
\caption{2D histogram of the PMa signal-to-noise ratio between Gaia DR2 and Gaia EDR3 analyses.
The increase in S/N of the PMa signal in the EDR3 is typically a factor of 2.5 (pink dashed line).\label{PMa-G2-G3}}
\end{figure}

The median accuracy of the determined $\Delta \mu_\mathrm{G3}$ PMa vectors is $\sigma(\Delta \mu_\mathrm{G3}) = 56\,\mu$as\,a$^{-1}$, corresponding to an accuracy on the tangential velocity anomaly of $\sigma(\Delta v_\mathrm{tan, G3}) = 26$\,cm\,s$^{-1}$\,pc$^{-1}$ (i.e., normalized to a distance of 1\,pc). This corresponds to an improvement of a factor 2.5 compared to the accuracy of the Gaia DR2 PMa values presented by \citetads{2019A&A...623A..72K}. This improvement is also visible in Fig.~\ref{PMa-G2-G3}, which shows a 2D histogram of the measured PMa signal-to-noise (S/N) values from Gaia DR2 and EDR3.

The median uncertainties of the Hip2 catalog positions in RA and Dec are $\sigma(\alpha[\mathrm{Hip2]}) = 0.7$\,mas and $\sigma(\delta[\mathrm{Hip2]}) = 0.6$\,mas, resulting in a median contribution to the uncertainty on the Hip2-EDR3 long-term proper motion of $\sigma(\mu_\mathrm{HG}) = 37\,\mu$as\,a$^{-1}$. On the other hand, the median uncertainty of the Gaia EDR3 PM vector norm for stars brighter than $G=12$ is $\sigma(\mu_\mathrm{G3}) = 27\,\mu$as\,a$^{-1}$ and is expected to decrease in the Gaia DR4 and DR5 to $\sigma(\mu_\mathrm{G4}) \approx 6\,\mu$as\,a$^{-1}$ and $\sigma(\mu_\mathrm{G5}) \approx 2\,\mu$as\,a$^{-1}$ for bright stars\footnote{\texttt{\url{https://www.cosmos.esa.int/web/gaia/science-performance}} retrieved in September 2021}. While the Hipparcos astrometry is already dominant in the error budget of the PMa vector determination, its use in combination with the future DR4 and DR5 epoch astrometry will still be a powerful asset in characterizing companions with long orbital periods (of several centuries). This will help to bridge the gap between the astrometric companion detections (from the Gaia epoch astrometry) and the spatially resolved CPM companions (see Sect.~\ref{sensitivity-combination}).

\subsubsection{Internal and external validation}

\begin{figure}
\includegraphics[width=\hsize]{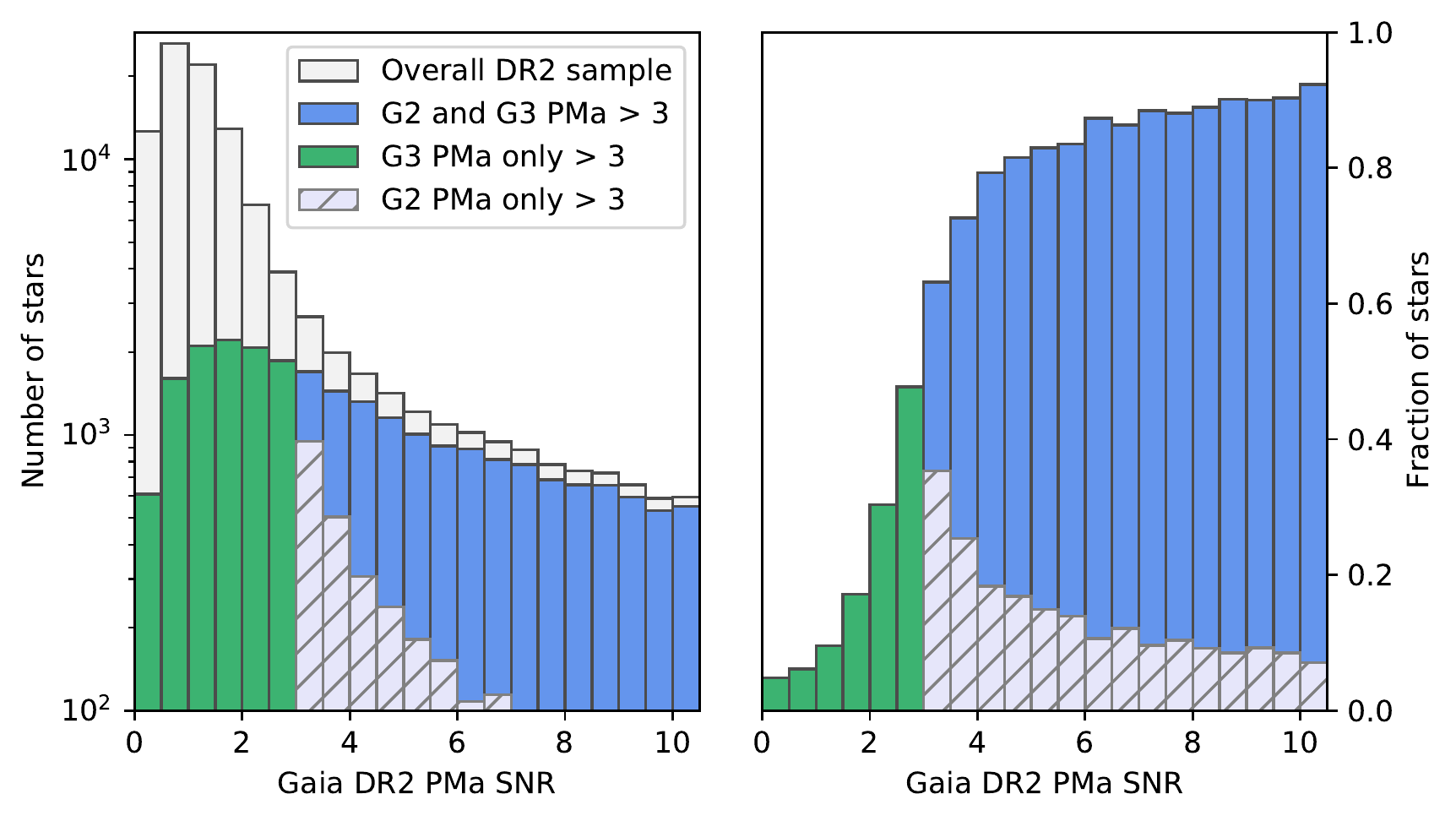}
\caption{Histogram of the S/N of the PMa signal from Gaia DR2 (light grey), the stars presenting a PMa S/N > 3 both in DR2 and EDR3 (medium blue), the stars with a PMa S/N > 3 only in  DR2 (hatched light blue), and the stars with a PMa S/N > 3 only in the EDR3 (green).
The right panel shows the corresponding fraction of the stars in the total sample per S/N bin.\label{Hip2-confirmed23-histo}}
\end{figure}

Among the sample of stars that show a significant PMa detection (S/N>3) in the DR2 \citepads{2019A&A...623A..72K}, 88.5\% are confirmed with an EDR3 PMa S/N larger than 3 (Table~\ref{G2G3-confirmations}; Fig.~\ref{Hip2-confirmed23-histo}). In addition, thanks to the improved accuracy of the EDR3 measurements, 10,423 stars exhibit a PMa S/N greater than 3, while they were below this limit in the DR2. Overall, the EDR3 increases the accuracy and reliability of the PMa detections, removing a significant number of spurious detections and confirming most of the DR2 signals.
For 3\% of the Hipparcos sources, a significant PMa signal (S/N > 3) was found using the DR2, which is not confirmed using the EDR3 (S/N<3). In some cases, this could be caused by companions whose orbital period is close to the EDR3 time window, resulting in a strong smearing and the disappearance of the PMa signal. For bright Hipparcos stars, the EDR3 astrometric reduction appears significantly more robust than in the DR2, reducing the biases on their derived EDR3 PM vectors. This results for single stars in a better agreement of their EDR3 PM vectors with the long-term Hipparcos-Gaia PM and, therefore, the disappearance of the PMa signal.

\begin{table}
 \caption{Proper motion anomaly detections and divergences from Gaia DR2 and EDR3.
 \label{G2G3-confirmations}}
 \centering
  \begin{tabular}{lrr}
  \hline
  \hline
    & Number & Fraction  \\ 
  \hline  \noalign{\smallskip}
Objects with DR2 PMa values & 116343 & 100.0 \% \\
DR2 SNR>3 and EDR3 SNR>3   &  27071 & 23.3 \% \\
DR2 SNR>3 and EDR3 SNR<3 &  3490 & 3.0 \% \\
DR2 SNR<3 and EDR3 SNR>3      &  10461 & 9.0 \% \\
   \hline
\end{tabular}
\end{table}

\begin{figure}
\includegraphics[width=\hsize]{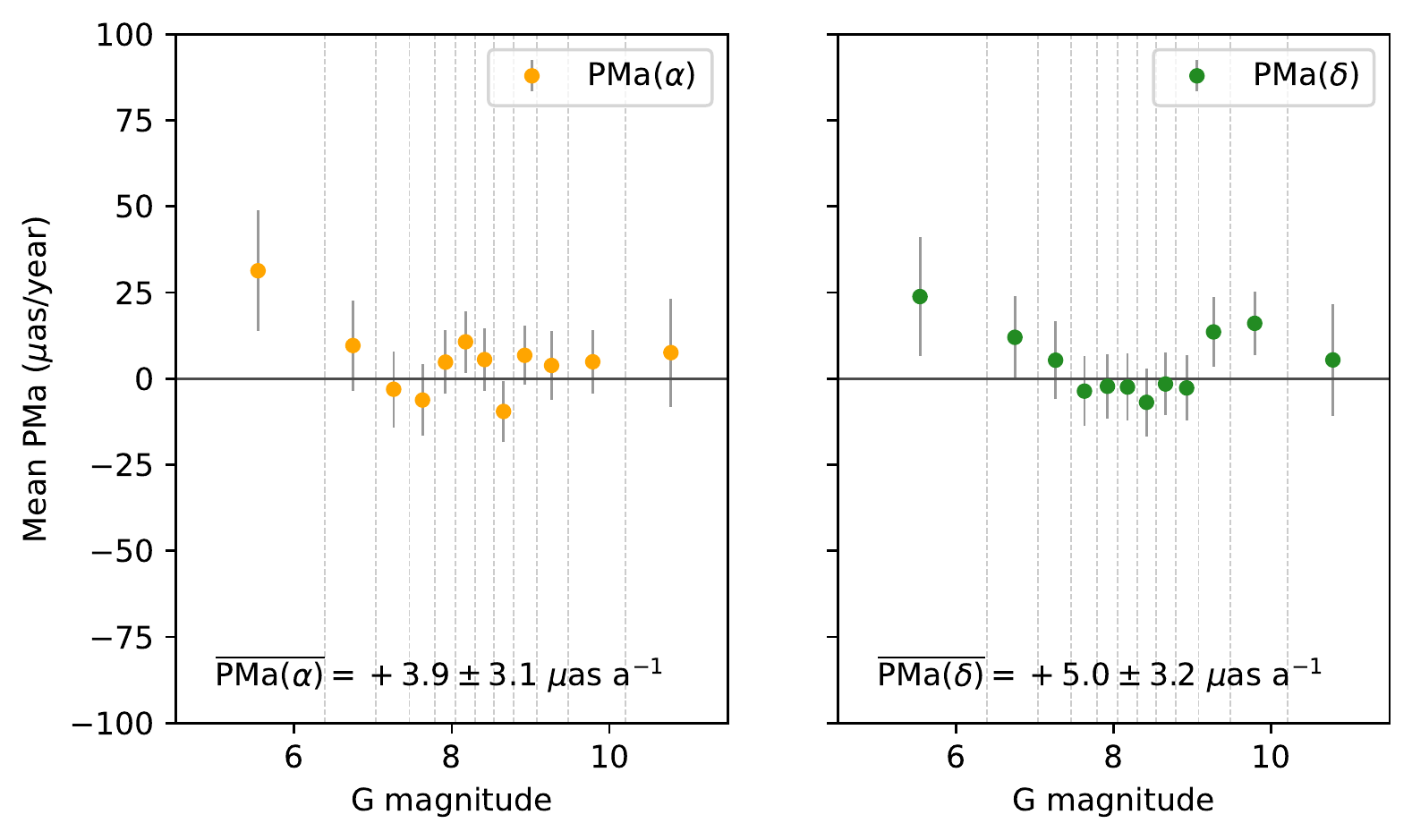}
\includegraphics[width=\hsize]{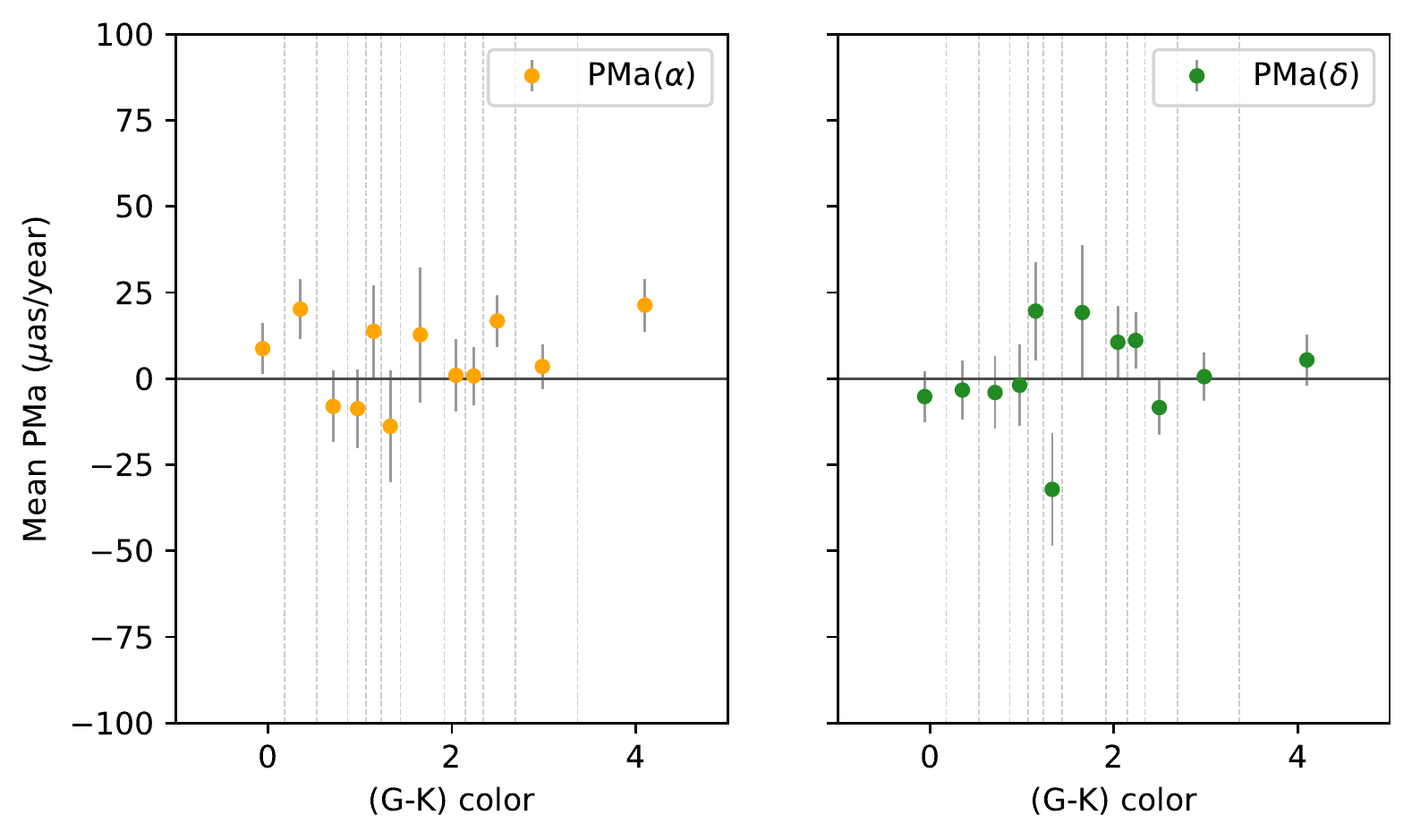}
\caption{Mean value of the proper motion anomaly as a function of the $G$ band magnitude (top panels) and $(G-K)$ color, within bins of 10\,000 stars. The magnitude limits of each bin are shown with dashed lines.
The overall mean value and associated uncertainty is given in each plot.\label{PMa_mean_vs_mag}}
\end{figure}

A mild color dependence of the PM vectors in Gaia EDR3 was found by \citetads{2021A&A...649A.124C} and we applied the recommended correction to the EDR3 catalog values (see also Sect.~\ref{inputdata}). In order to verify that this does not have an effect on the PMa vectors, we computed the mean PMa vectors over bins of 10,000 stars, as a function of their magnitude and visible-infrared color. The result is presented in Fig.~\ref{PMa_mean_vs_mag}. We do not detect any significant bias at a level of $\pm 25\,\mu$as\,a$^{-1}$ ($\pm 12$\ cm\,s$^{-1}$\,pc$^{-1}$). As a significant number of bright Hipparcos stars are close to the Gaia saturation limit, and very diverse in color, this first-order analysis shows that there is no large systematic differential effect due to magnitude or color. However, this test is not intended to demonstrate the absence of a position-dependent effect over the sky (e.g., a sinusoidal bias as a function of right ascension), as the whole sky sample in each bin is averaged to produce the plots presented in Fig.~\ref{PMa_mean_vs_mag}.

\citetads{2018ApJS..239...31B, 2021ApJS..254...42B} recently reported similar analyses of the Hipparcos and Gaia PMs to the present work, respectively, for  Gaia DR2 and EDR3.
No significant systematic difference is present between the present work and \citetads{2021ApJS..254...42B}, with a mean difference in the long-term PM vector of: $\Delta \mu_\alpha = -1.2 \pm 3.7\,\mu\mathrm{as\,a}^{-1}$, $\Delta \mu_\delta = +5.6 \pm 3.9\,\mu\mathrm{as\,a}^{-1}$.
This corresponds to a mean tangential velocity difference of only $\Delta \mu_\alpha = -0.6$\,cm\,s$^{-1}$\,pc$^{-1}$ and $\Delta \mu_\delta = +2.7$\,cm\,s$^{-1}$\,pc$^{-1}$.


\section{Common proper motion companions\label{cpm-section}}

The general principle of our analysis is classically to search for companions of a selection of targets in the EDR3 catalog, based on the proximity of their parallax and PM. As discussed in Sect.~\ref{sample}, we complete our input list of EDR3 catalog targets with the Hipparcos stars that are absent from the Gaia catalog (mainly due to saturation). Comparable works based on the Hipparcos, Gaia DR2 and Gaia EDR3 catalogs can be found in, for instance, \citetads{2011ApJS..192....2S}, \citetads{Jim_nez_Esteban_2019}, \citetads{2019A&A...623A.117K}, \citetads{gonzalez-payo}, \citetads{2019MNRAS.488.4740P}, \citetads{Hartman_2020}, \citetads{2020ARep...64..756S}, \citetads{Zavada_2020}, \citetads{Pearce_2020}, \citetads{2020ARep...64..756S}, \citetads{2021A&A...649A...6G}, and \citetads{2021MNRAS.506.2269E}.

\subsection{Star sample\label{sample}}

We selected two samples of stars for our survey of CPM companion candidates, which are partly overlapping:

\textbf{100\,pc sample}: The EDR3 targets located within 100\,pc ($\varpi_\mathrm{G3}>10$\,mas), supplemented with the missing Hip2 stars located within this distance range. This sample comprises 542,232 individual objects, out of which 21,217 are present both in the Hip2 and EDR3 catalogs, and 262 are present only in the Hip2 catalog (essentially the brightest stars). We chose here the simplified approach of a strict parallax limit for the selection of our sample compared to that of the GCNS \citepads{2021A&A...649A...6G}, as we are not aiming for an exhaustive census of the stars within this distance.

\textbf{Hipparcos stars}: The Hip2 catalog sources that we took into account in the present work comprise 117,628 stars. For these targets, we adopted the Hipparcos-EDR3 long term PM vector (Sect.~\ref{HG-PMa}) and EDR3 parallax when available or, alternatively, the Hip2 PM and parallax. 

A cross-identification of the Hipparcos and DR2 catalogs is presented in \citetads{2019A&A...623A..72K}. The very brightest stars of the Hip2 catalog do not have a counterpart in the EDR3 catalog, as they are heavily saturated (e.g., \object{Sirius}, \object{Betelgeuse}, $\alpha$\,Centauri AB, etc.). However, this is not a limitation for the present CPM companion survey as we adopted the Hip2 parameters (position, parallax, and PM vectors) for these particular targets.

\subsection{Initial search volume}

For each target of our survey, we defined the search range $\delta\varpi$ for the parallax for the candidate companions taking into account: (1) the acceptable difference in distance between the target and its companions and (2) the uncertainties on their respective parallaxes.
To define the search depth in terms of differential distance between the target and the candidate companions, we considered the parallax range $\delta \varpi_A$ defined as:
\begin{equation}
\delta \varpi_A[\arcsec] = \varpi_0[\arcsec] - \frac{1}{\left( 1/ \varpi_0[\arcsec] + dz_\mathrm{max}[\mathrm{pc}] \right)}
\label{nearfar}
,\end{equation}
with $dz_\mathrm{max} = 0.5$\,pc as the maximum acceptable difference in distance and $\varpi_0$ as the parallax of the target. We neglect the difference between the range in parallax corresponding to the far side (with respect to the target) and the near side (larger parallax). We consider the expression of Eq.\,\ref{nearfar} for $\delta \varpi_A[\arcsec]$ symmetrically for the near and far sides. This first term is important for the nearest stars, whose candidate companions may have a significantly different parallax even though they are physically bound (e.g., Proxima and $\alpha$\,Cen~AB).
Secondly, we take into account the uncertainty on the parallax of the main target $\sigma_{\varpi 0}$ via:
\begin{equation}
\delta \varpi_B[\arcsec]  = N\,\sigma_{\varpi0}[\arcsec]
,\end{equation}
where $N=3$ the maximum parallax difference in number of standard deviations.
We therefore queried the Gaia EDR3 catalog with an acceptable parallax range of $[\varpi_0 - \delta \varpi, \varpi_0 + \delta \varpi]$ where:
\begin{align}
\delta \varpi[\arcsec] & = \sqrt{\delta \varpi_A[\arcsec]^2 + \delta \varpi_B[\arcsec]^2},\\
 &= \varpi_0[\arcsec] \sqrt{ \left(\frac{dz_\mathrm{max}[\mathrm{pc}]\, \varpi_0[\arcsec]}{1+dz_\mathrm{max}[\mathrm{pc}]\, \varpi_0[\arcsec]} \right)^2 +\left(N\,\frac{\sigma_{\varpi 0}[\arcsec]}{\varpi_0[\arcsec]}\right)^2}.
\end{align}

The parallax of the primary target ($\varpi_0$) is taken from the EDR3 catalog or, alternatively, from the Hipparcos catalog for the bright stars absent from the EDR3 or those whose parallax is less accurate in the EDR3 than in Hipparcos. In summary, we retrieved from the EDR3 catalog those stars with a parallax within $\pm \delta \varpi$ of the primary target and within 1\,pc in terms of the projected linear separation. We set a minimum search radius of 1\,arcmin and a maximum of $2.5\,\deg$ to avoid overly small (for stars farther than 3.4\,kpc) or large (for stars closer than 23\,pc) search angles. The shape of the resulting search volume in space is a truncated cone with spherical near and far surfaces.

We did not search for candidate companions in the Hipparcos catalog. This means that Hipparcos stars that are absent from the Gaia catalog are not listed as candidate companions of Gaia-only targets (e.g., \object{Sirius A} is not listed as a companion of \object{Sirius B}). However, as we did search the Gaia EDR3 catalog around Hipparcos-only targets, the identified companions are properly listed in the catalog (e.g., Sirius B is listed as a companion of Sirius A). Hipparcos-only companions to Hipparcos-only targets (concerning only a small number of sources) can be found, for instance, in the Hipparcos Catalogue Double and Multiple Systems Annex \citepads{1997A&A...323L..53L}.

\subsection{Photometry, reddening, and physical properties}

We completed the EDR3 record of each star within the search volume with its $K$ band magnitude from the 2MASS catalog \citepads{2006AJ....131.1163S}, the visible $B$, $V,$ and $R$ magnitudes from the NOMAD catalog \citepads{2004AAS...205.4815Z}, and the Hipparcos $H_p$ magnitude (when available).
We added flags for the known binary and multiple systems from the Washington Double Star catalog \citepads{2001AJ....122.3466M} and the double and multiple star annex (DMSA) of the original Hipparcos catalog \citepads{1997ESASP1200.....E}.
The interstellar reddening was neglected for the target stars located within 50\,pc (that is, within the Local Bubble, \citeads{2011ARA&A..49..237F}).
For the more distant objects in our sample, we adopted the color excess $E(B-V)$ predicted by the \texttt{Stilism}\footnote{\url{https://stilism.obspm.fr}} 3D model of the local interstellar medium \citepads{2014A&A...561A..91L, 2017A&A...606A..65C}.
The radial velocities were retrieved from different catalogs as described in \citetads{2019A&A...623A..72K} (\citeads{2002ApJS..141..503N}, \citeads{2018A&A...616A...7S}, \citeads{2007A&A...475..519H}, \citeads{2018A&A...616A...1G}, \citeads{2018A&A...616A...5C}, and \citeads{2012AstL...38..331A}). The stellar masses and radii were estimated from the dereddened photometry following the same procedure as \citetads{2019A&A...623A..72K} (based on \citeads{2000A&AS..141..371G}, \citeads{0004-637X-804-1-64}, \citeads{2016MNRAS.462.2295H}, and \citeads{2004A&A...426..297K}).

\subsection{Selection of common proper motion companions}

Within the field star sample, our selection of the candidate CPM companions is based on a score built from the parallax and PM of the candidate companions located in the search volume, relative to the parameters of the target star.

\subsubsection{Selection on parallax}

The probability that the candidate companion (parallax $\varpi \pm \sigma_\varpi$) and the target (parallax $\varpi_0 \pm \sigma_{\varpi 0}$) are located within $dz_\mathrm{max}=0.5$\,pc of each other along the radial direction to the Sun is given by the probability density function:
\begin{align}
P_\varpi &= PDF(\varpi - \varpi_0; \sigma_\mathrm{tot}) ,\\
       &= \exp \left( - \frac{(\varpi - \varpi_0)^2}{2\,\sigma_\mathrm{tot}^2} \right),
\end{align}
where $\sigma_\mathrm{tot} = \sqrt{\sigma_\varpi^2 + \sigma_{\varpi 0}^2 + (dz_\mathrm{max}\, \varpi_0^2)^2}$.
This quantity gives us the probability that the target and candidate companion are at a compatible distance.

\subsubsection{Selection on relative tangential velocity}

The candidate companions whose differential tangential velocity $\Delta \varv_\mathrm{tan}$ with the target is slower than 5\,km\,s$^{-1}$ are flagged as \texttt{LowV} (low velocity) in the catalog.

To test the possibility that the candidate companion and the target are gravitationally bound, we compare $\Delta \varv_\mathrm{tan}$ with the escape velocity $\varv_\mathrm{esc}$ of the system at the projected linear separation $r = \varpi \Delta \theta$ (with $\Delta \theta$ their angular separation):
$\varv_\mathrm{esc} = \sqrt{2\,G\,(m_1 + m_2)/r}$, where $m_1$ and $m_2$ are the estimates of the masses of the target and candidate companion (when available).
We note that $\varv_\mathrm{esc}$ is an upper limit of the true escape velocity as the actual linear distance between the two stars is larger than $r$.

The probability that the differential velocity $\Delta \varv_\mathrm{tan}$ is lower than $\varv_\mathrm{esc}$ is given by the survival function:
\begin{align}
P_\varv &= 1 - CDF(\Delta \varv_\mathrm{tan}; \varv_\mathrm{esc}; \sigma_{\Delta \varv}) ,\\
        &= 1 - \frac{1}{\sigma_{\Delta \varv} \sqrt{2\pi}}
        \int_{0}^{\Delta \varv_\mathrm{tan}} \exp \left( -\frac{(\varv- M\,\varv_\mathrm{esc})}{2\,\sigma_{\Delta \varv}^2} \right) d\varv,
\end{align}
where $M=2$ is a margin factor, intended to accommodate the unknowns in the determination of the differential tangential velocity, the escape velocity, and the possible presence of perturbing bodies in the considered stellar system.
The tangential velocity $\Delta \varv_\mathrm{tan}$ is the norm of a two-dimensional differential vector, whose coordinates are affected by uncertainties. This induces a systematic positive bias on the estimate of the vector norm (that follows a Rayleigh distribution).
The value of the escape velocity relies on the total mass of the system estimated from photometry, which may be underestimated if additional faint companions are present (e.g., in hierarchical multiple systems). Moreover, in this last configuration, the PM vector of a candidate companion may be affected the additional orbiting body, resulting in a higher tangential velocity.

We reject the candidates whose PM vector has a position angle diverging by more than $\pm 30^\circ$ from the PM vector of the target if it is located within 10,000\,au, and $\pm 10^\circ$ if it is farther from the target. This selection step relies on the hypothesis that the orbital velocity of physical systems is significantly slower than the systemic PM for wide binaries. This criterion rejects only a small fraction of the detected candidates.

In addition to the above velocity criteria, we set a maximum separation of $r=0.5$\,pc for gravitationally bound candidates, that is, $P_\varv$ is set to zero when $r > 0.5$\,pc ($\approx 100$\,kau).

\subsubsection{Score threshold for bound candidates\label{totalscore}}

We define the overall score of each candidate companion as the product of the parallax and velocity compatibility probabilities with the target:
\begin{equation}
P_\mathrm{tot} = P_\varpi\ P_\varv
.\end{equation}

The threshold in total score $P_\mathrm{tot}[\texttt{Bnd}]$ to identify gravitationally bound candidates (flagged as \texttt{Bnd} in the catalog) is an essential parameter to ensure a low degree of contamination of the sample with false positives, while simultaneously preserving valid candidates. To estimate the optimum threshold, we considered two approaches: (1) the overall distribution of the candidate companion scores and (2) the distribution of the linear separations of the companions. For this analysis, we consider the 100\,pc sample including the Hipparcos stars located within this distance.

\begin{figure}
\includegraphics[width=\hsize]{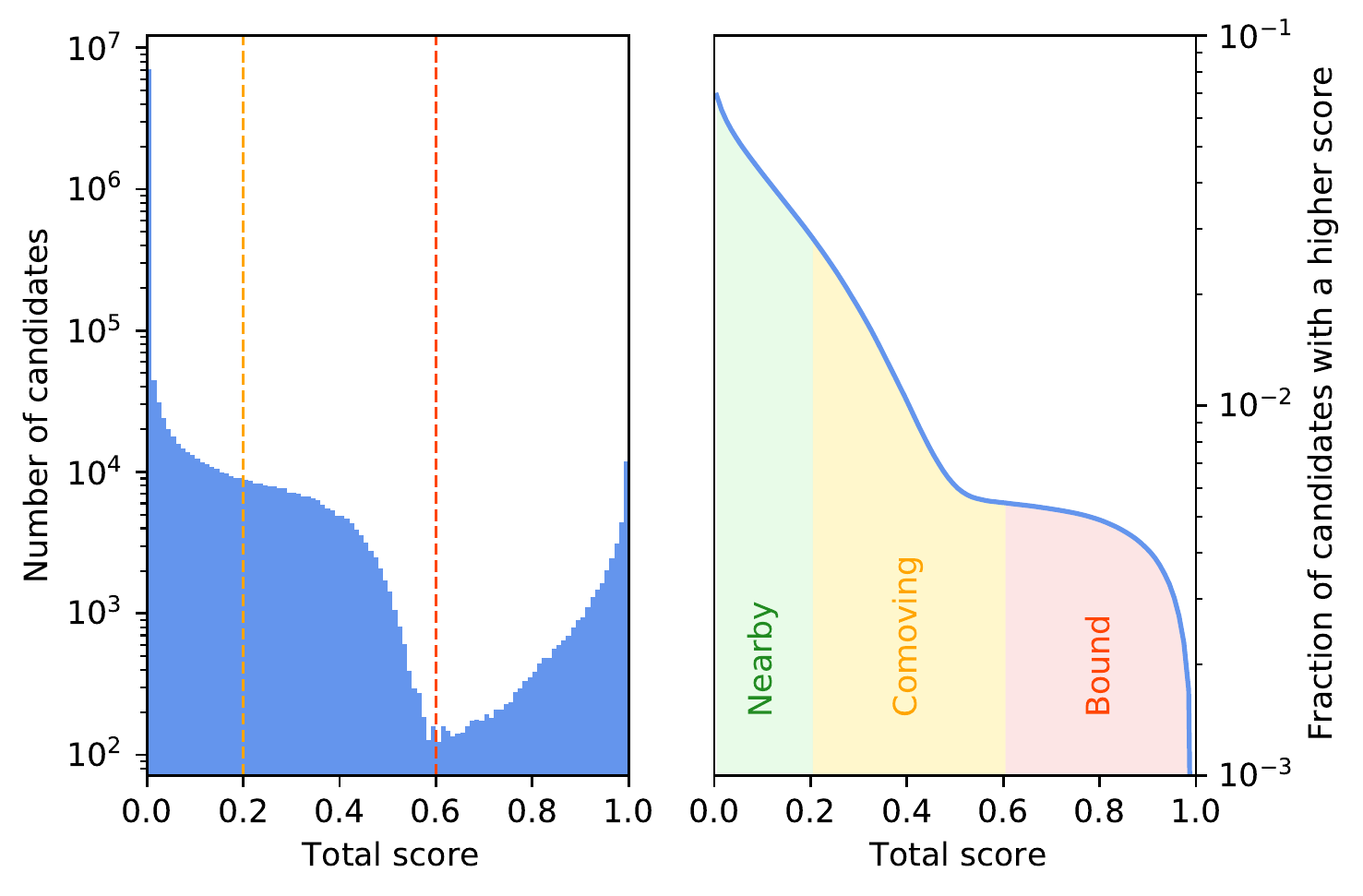}
\caption{Please add brief intro to the caption before describing panels. Left panel: Histogram of the $P_\mathrm{tot}$ total score of candidate companions. Right panel: Fraction of candidate companions with $P_\mathrm{tot}$ higher than a given threshold. \label{GaiaEDR3CPM-Ptot-histo}}
\end{figure}

The overall distribution of the total scores $P_\mathrm{tot}$ of the candidate companions classified as \texttt{LowV} of \texttt{Bnd} is shown in the left panel of Fig.~\ref{GaiaEDR3CPM-Ptot-histo}. Three domains are apparent: the nearby field stars ($P_\mathrm{tot} < 0.2$), the co-moving stars (e.g., within an open cluster,  $0.2 < P_\mathrm{tot} < 0.6$) and the gravitationally bound candidates ($P_\mathrm{tot} > 0.6$). The boundaries between these three samples are visible as the points of inflexion of the histogram, as well as on the fraction of candidate companions above a given threshold (right panel of Fig.~\ref{GaiaEDR3CPM-Ptot-histo}).
The intermediate regime ($0.2 < P_\mathrm{tot} < 0.6$) potentially includes a significant number of bound companions, if the primary target is itself a close binary and its PM vector is affected by the orbital motion.

\begin{figure}
\includegraphics[width=\hsize]{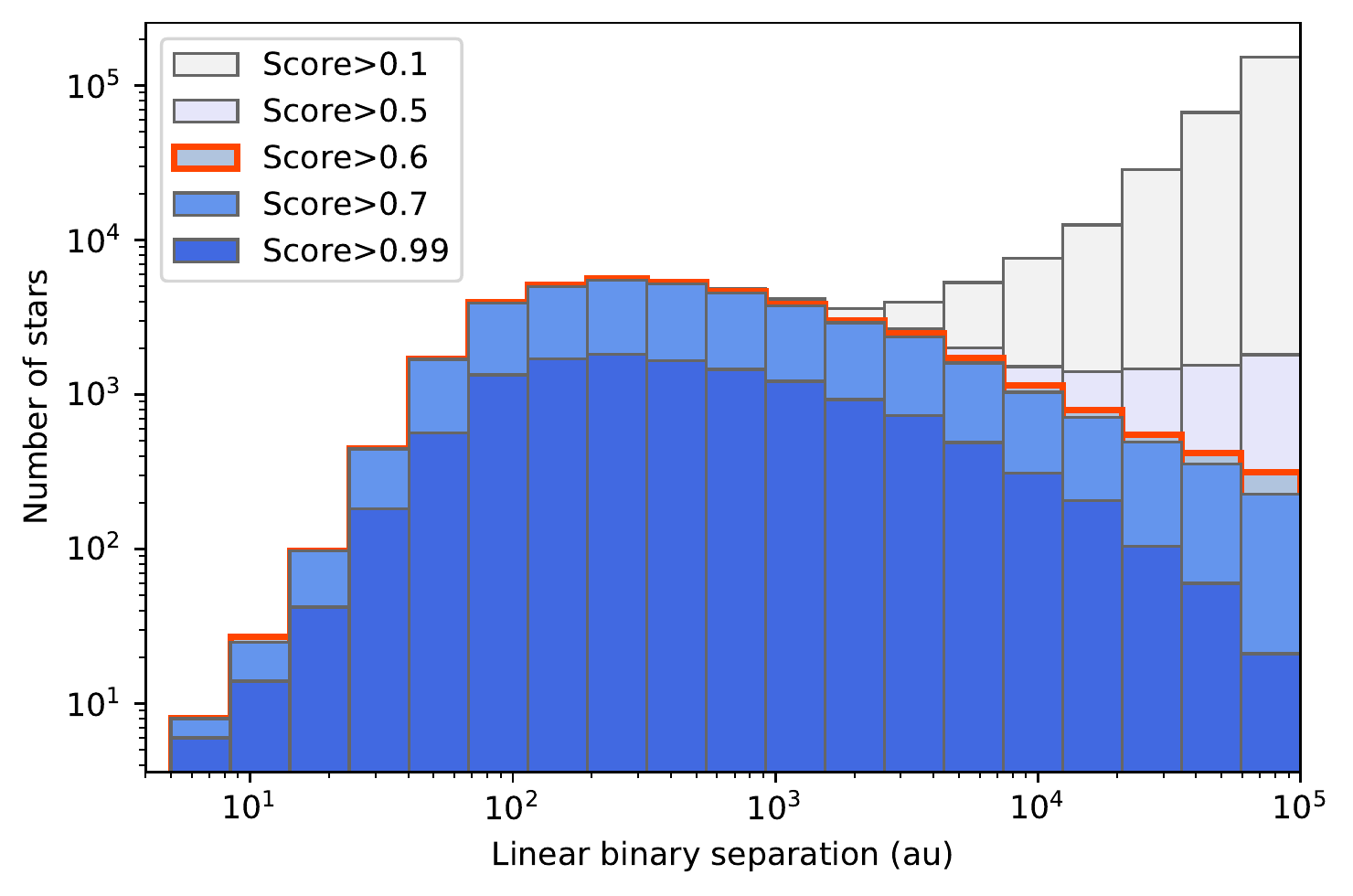}
\caption{Histogram of the linear separation of candidate gravitationally bound companions for different $P_\mathrm{tot}$ total score thresholds.
The histogram for a threshold of 0.6 (corresponding to the optimum) is marked with a red line. \label{GaiaEDR3CPM-LinSep-histo}}
\end{figure}

Another method for determining the $P_\mathrm{tot}[\texttt{Bnd}]$ threshold is to consider the distributions of the number of candidate companions as a function of the linear separation with the primary for different threshold values. These histograms are shown in Fig.~\ref{GaiaEDR3CPM-LinSep-histo}. The histograms for $P_\mathrm{tot}[\texttt{Bnd}] = 0.1$ and 0.5 exhibit a clear divergence in the number of bound candidates for separations above 1000\,au, while it is not present for a threshold of 0.6 and above. This is an indication that a $P_\mathrm{tot}[\texttt{Bnd}] = 0.5$ is too low to prevent the contamination of the candidate bound sample by unbound neighbors. A threshold of $P_\mathrm{tot}[\texttt{Bnd}] = 0.6$ preserves the overall shape of the histogram of the measured separations, compared to the higher 0.7 and 0.99 thresholds, and does not diverge at large separations.

From these two approaches, a probability threshold of $P_\mathrm{tot}[\texttt{Bnd}]=0.6$ for bound candidates appears to be optimal, and we adopt this value to define the gravitationally bound flag (\texttt{Bnd}) in the catalog.
It should be stressed that gravitationally bound companions are present in the catalog below this threshold. For instance, the PM of the components of hierarchical multiple systems are affected by the orbital motion of each star, which results in a potential overestimation of the differential velocities (e.g., between an close binary primary and a third component). In addition, for multiple systems, the presence of an undetected but relatively massive close companion to a primary target (e.g., a main sequence or compact companion to a giant star) potentially results in an underestimation of the total mass. This induces an underestimation of the escape velocity, and potentially bound companions may therefore appear with total scores below the overall bound threshold. For this reason, when searching for bound CPM companions to a given target, the potential companions with $P_\mathrm{tot}$ scores below 0.6 should also be considered as potential candidates.

\subsection{Statistics of the detected CPM companions}

\begin{figure}[ht]
\centering
\includegraphics[width=\hsize]{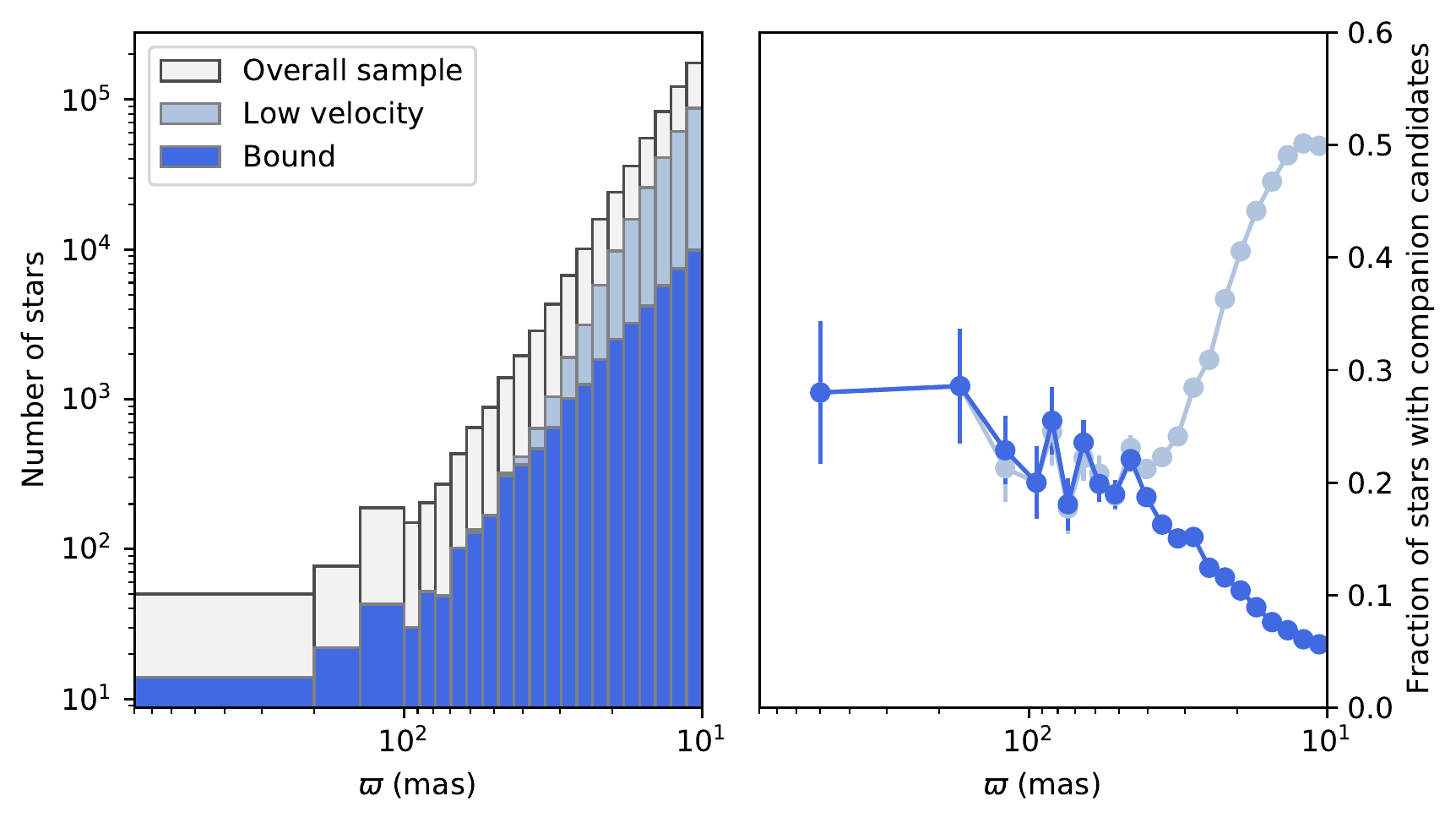}
\caption{Histogram of the detected bound and low velocity candidate companions in the 100\,pc Gaia sample as a function of the parallax of the target (left panel) and binary fraction as a function of the target parallax (right panel).}
\label{binarity_parallax_100pc}
\end{figure}

Extracts of the CPM catalog for the Hipparcos catalog stars and Gaia EDR3 sources within 100\,pc are presented in Tables~\ref{CPMHIP-sample} and \ref{CPMGaia-sample}, respectively.
The histogram of the candidate CPM companions detected in the 100\,pc Gaia sample is presented in Fig.~\ref{binarity_parallax_100pc} as a function of the parallax. For parallaxes $\varpi > 40$\,mas (distance < 25\,pc), the samples of companions flagged as \texttt{LowV} and \texttt{Bnd} are in good agreement, with an overall multiplicity frequency of 20.5\% (Fig.~\ref{binarity_parallax_100pc}, right panel).
Within a distance of 10\,pc, we obtain a multiplicity frequency of 25\%, in good agreement with the 27\% frequency found by \citetads{2021A&A...650A.201R}.
For smaller parallaxes ($\varpi < 40$\,mas), the fraction of \texttt{Bnd} candidates decreases linearly, reaching 10\% at $\varpi = 10$\,mas. Simultaneously, the number of \texttt{LowV} candidates increases rapidly, indicating that the majority of the stars classified in this category are unbound field stars.

\begin{figure}[ht]
\centering
\includegraphics[width=\hsize]{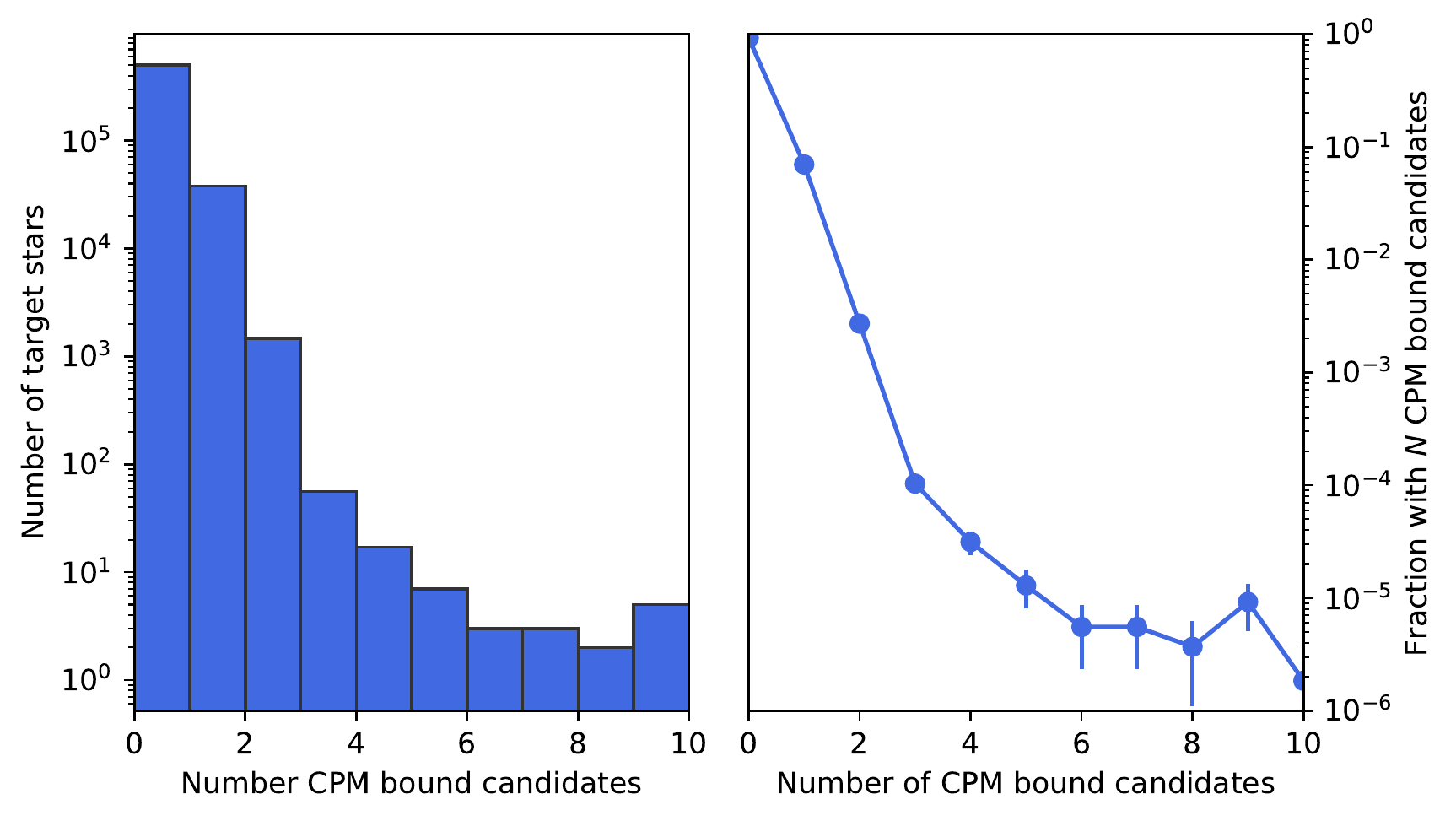}
\caption{Histogram of the number of detected bound candidate companions for stars in the 100\,pc sample (left panel) and fraction of the stars with $N$ candidate bound companions (right panel).}
\label{Nbound-histo_100pc}
\end{figure}

Figure~\ref{Nbound-histo_100pc} shows the histogram of the number of detected gravitationally bound candidates per target stars and the fraction in the 100\,pc sample.
Within the stars with detected bound candidate companions, the large majority has a single companion (96.0\%) or two companions (3.7\%). The sample of 103 stars with more than two bound candidates (0.3\% of all stars with bound candidates) likely contains a significant fraction of stellar groups in clusters that are close to the unbound limit. It should be noted that due to the identical processing for all stars, each member of a system of $N$ stars is counted individually in this total number (i.e., a system of ten candidate bound stars counts for ten stars, each with nine bound candidates). As a consequence, the actual number of high-order multiple systems is therefore very low in our sample.

\section{Discussion\label{discussion}}

\subsection{Binary fraction as a function of primary mass}

\begin{figure}[ht]
\centering
\includegraphics[width=\hsize]{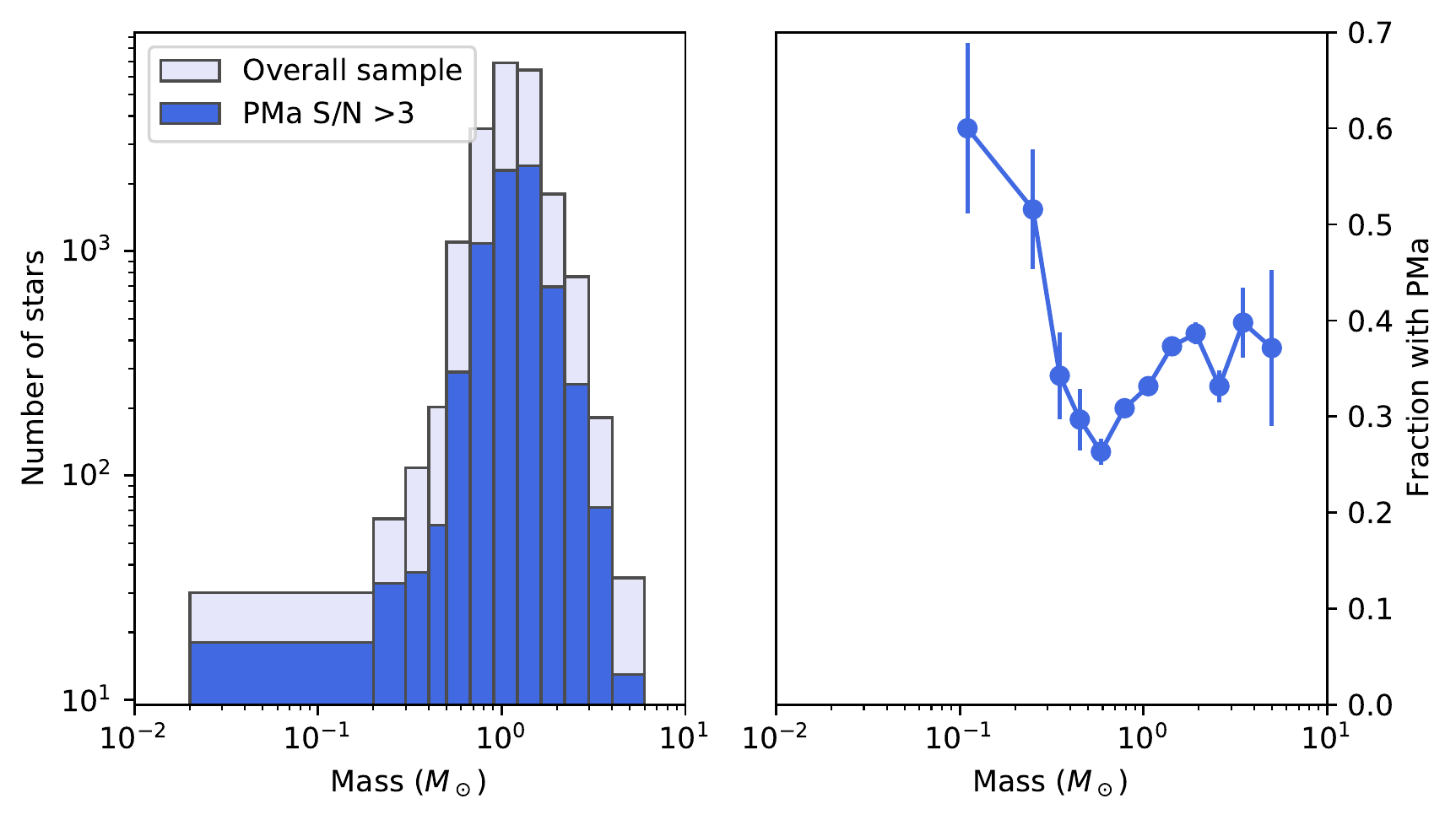}
\includegraphics[width=\hsize]{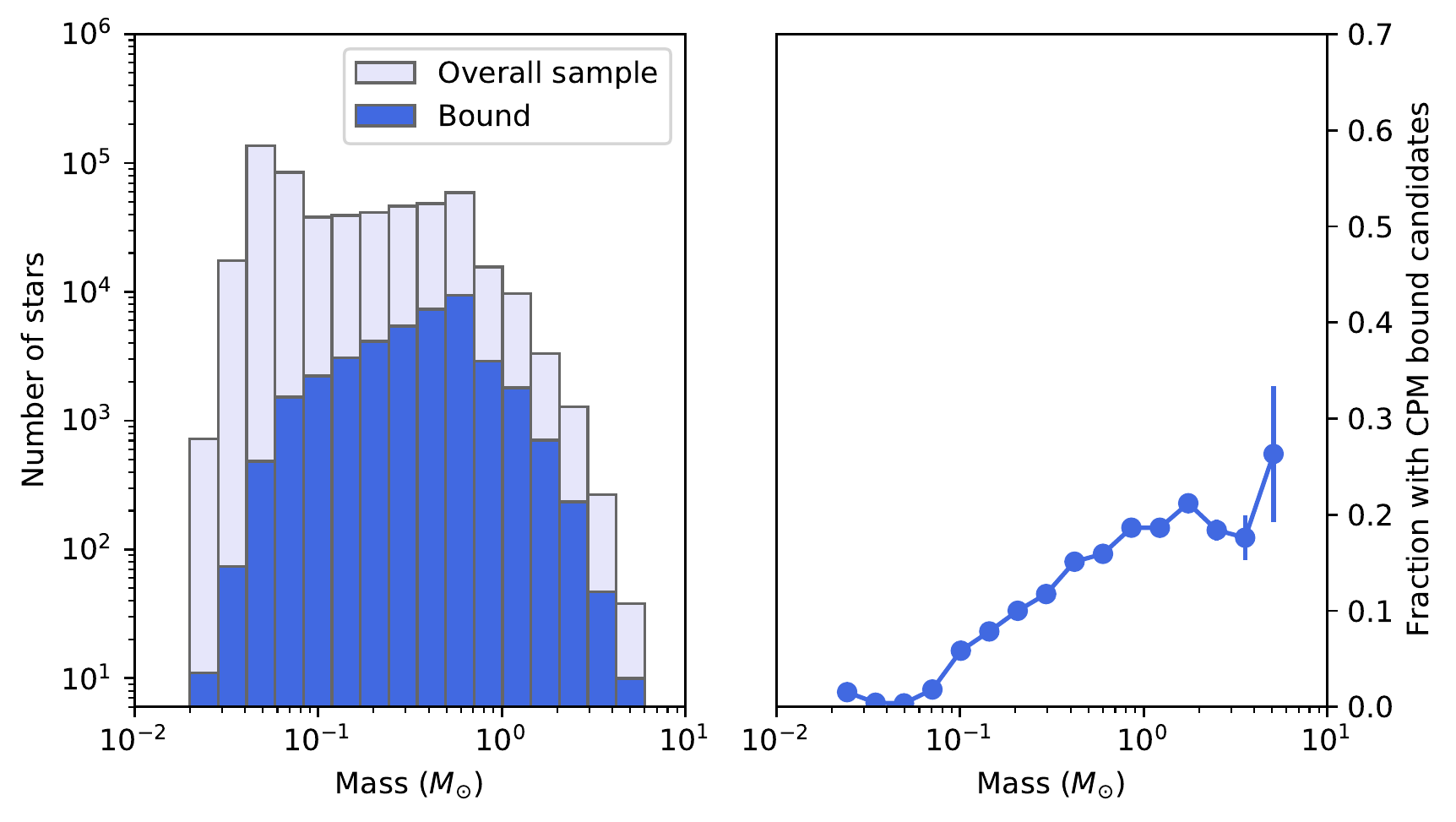}
\caption{Histogram of the Hipparcos stars within 100\,pc exhibiting a PMa S/N larger than 3 (top left panel) and fraction of the overall sample (top right panel), as a function of the primary mass.
The histogram and fraction of the stars with CPM bound candidate companions for the full 100\,pc sample (Gaia+Hipparcos) are shown in the bottom panels.}
\label{binarity_mass_100pc}
\end{figure}

Figure~\ref{binarity_mass_100pc} shows the histograms of the stellar mass of the targets with a significant PMa signal (S/N>3; top panels) and gravitationally bound candidate companions (bottom panels). For those stars more massive than the Sun, the fraction of stars with a PMa signal is $\approx 35$\%, and simultaneously $\approx 20$\% of this sample has bound CPM candidates.

As already reported by \citetads{2019A&A...623A..72K}, the fraction of very low-mass stars of the Hipparcos catalog ($m_1 < 0.3\,M_\odot$) exhibiting a significant PMa signal reaches more than 50\%. This is induced by the very high sensitivity of the PMa companion detection technique (in terms of companion mass) for the nearest very low-mass stars (e.g., \object{Proxima Centauri}, \object{Barnard's star}, \object{Kapteyn's star}...). As a result, we are able to detect the signature of low-mass planetary companions orbiting these objects down to a few tens of Earth masses, which is significantly lower in mass than for the other Hipparcos stars located within 100\,pc. In other words, the PMa signals of the very low-mass Hipparcos stars are likely caused by much lower mass planetary companions than for the rest of the catalog, and the binary fraction consequently appears higher.

\subsection{Gaia RUWE as indicator of binarity}

The Gaia RUWE parameter \citepads{GaiaEDR3astrometric} is generally employed as a statistical quality flag for Gaia data: a value of the RUWE>1.4 indicates that the astrometric parameters of a given source may be degraded.
The majority of high RUWE objects are partially resolved binary stars or tight astrometric binaries with a significant orbit-induced displacement of the photocenter (i.e., those having a low mass ratio between the components).
This is particularly the case when the orbital period is close to 1 year, as it then interferes with the period of the parallactic ellipse measured by Gaia.

The resolving power of Gaia depends on the difference $\Delta G$ in magnitude of the two objects, and is approximately $0.5\arcsec$ for equal magnitude stars (up to $1.2\arcsec$ for $\Delta G = 5$) in the EDR3 as determined by \citetads{2021A&A...649A...6G}.
At a given epoch observation, the pointing of a binary (or double star) by Gaia is more complex than expected \citepads{GaiaEDR3astrometric}. A similar situation already occurred with Hipparcos, and \citetads{1997A&AS..122..571M} coined the word ``\textit{Hippacenter}'' to define the pointing of epoch Hipparcos observations of double stars.
Concerning Gaia, if a double star has a separation well below the angular resolution of the telescope ($\approx 0.1\arcsec$), the ``\textit{Gaiacenter},’’ as we may perhaps designate the epoch pointing for Gaia, is simply the photocenter.
Beyond a $1.2\arcsec$ separation, each component of the pair may be observed individually.
For these two extreme cases, the standard astrometric solution will not be perturbed,beyond the possible effect of the orbital motion of the photocenter (that is, the quantity measured by the PMa observable).
On the contrary, the binaries whose separation lies in the range between $0.1\arcsec-1.2\arcsec$ will have a ``Gaiacenter’’ closer to the primary and that varies with the projected separation along the Gaia transit direction and with the magnitude difference.
With a reference point that is shown not to be be consistent from epoch to epoch, the standard astrometric solution will be perturbed.
In such cases, the derived PMa value should be considered with caution.
While higher values of the RUWE up to 2 or 3 may still provide usable measurements within their stated uncertainties (see \citeads[Sect.~5.3 of]{2021A&A...649A..13M}), there is a higher probability of bias on their astrometry and hence on their PMa.
Further quality parameters provided in the Gaia catalog may be used to test the quality of the astrometry of high RUWE stars.
For instance, applying \texttt{ipd\_frac\_multi\_peak}>3 to the relatively wide binaries ($\approx 1\arcsec$) or \texttt{ipd\_gof\_harmonic\_amplitude}>0.1 to the smaller separations are indications of a photocenter measurement problem (see \citeads[Sect.~3.3 of]{GaiaEDR3validation}).

\begin{figure}
\includegraphics[width=\hsize]{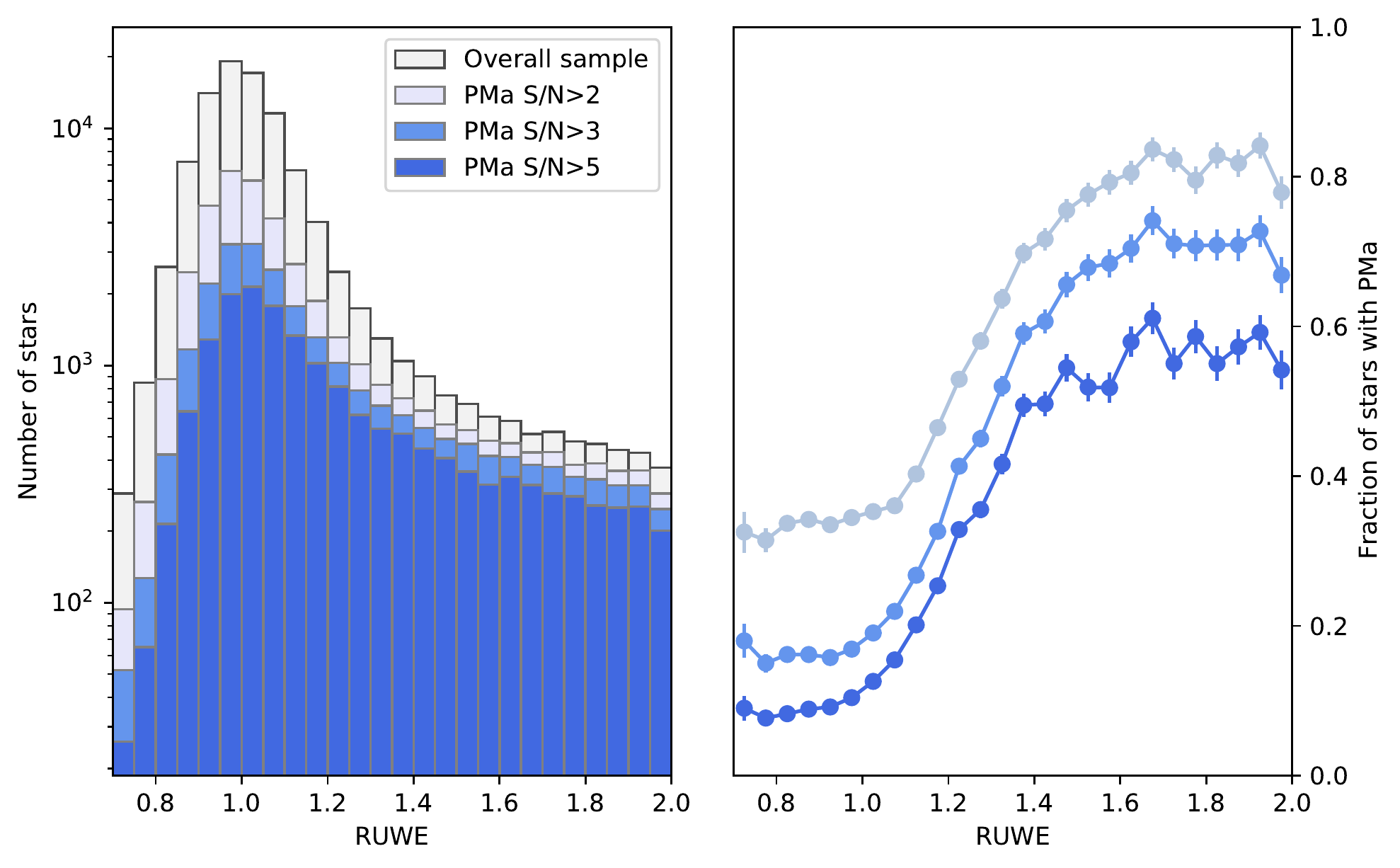}
\caption{Histogram of the PMa S/N values of the Hipparcos catalog stars as a function of their Gaia EDR3 renormalized unit weight error (RUWE).\label{RUWE-PMaSNR-histo}}
\end{figure}

\citetads{2020MNRAS.496.1922B} and \citetads{2021ApJ...907L..33S} demonstrated that the RUWE is actually a reliable indicator of the presence of a close companion.
As shown in Fig.~\ref{RUWE-PMaSNR-histo}, 75\% of the 25,067 Hip2 stars with RUWE>1.4 (representing 21\% of the full Hip2 catalog) exhibit a significant PMa S/N>3. Conversely, 49\% of the 37\,437 Hip2 stars that exhibit a PMa S/N>3 have a RUWE>1.4. We therefore confirm the high correlation between the PMa and RUWE quantities.
As also noted by \citetads{2021ApJ...907L..33S} for eclipsing binaries, there is a smooth transition in the fraction of stars with PMa S/N>3 for RUWE values between 1.0 (20\%) and 1.6 (70\%), which remains stable for higher RUWE values.

The RUWE parameter appears as a valuable indicator of binarity for tight systems (partially resolved or with a large photocenter motion) with angular separations on the order of $1\arcsec$ or below.  It has the important advantage of being available for the full Gaia catalog, whereas the PMa is limited to Hipparcos stars. The RUWE is therefore complementary to the PMa and CPM indicators, with the limitation that the conversion of the RUWE value into constraints on the physical properties of the companion is made difficult given its statistical nature.

\subsection{Combined sensitivity of the PMa and CPM techniques\label{sensitivity-combination}}

\begin{figure*}[t]
\centering
\includegraphics[width=8.5cm]{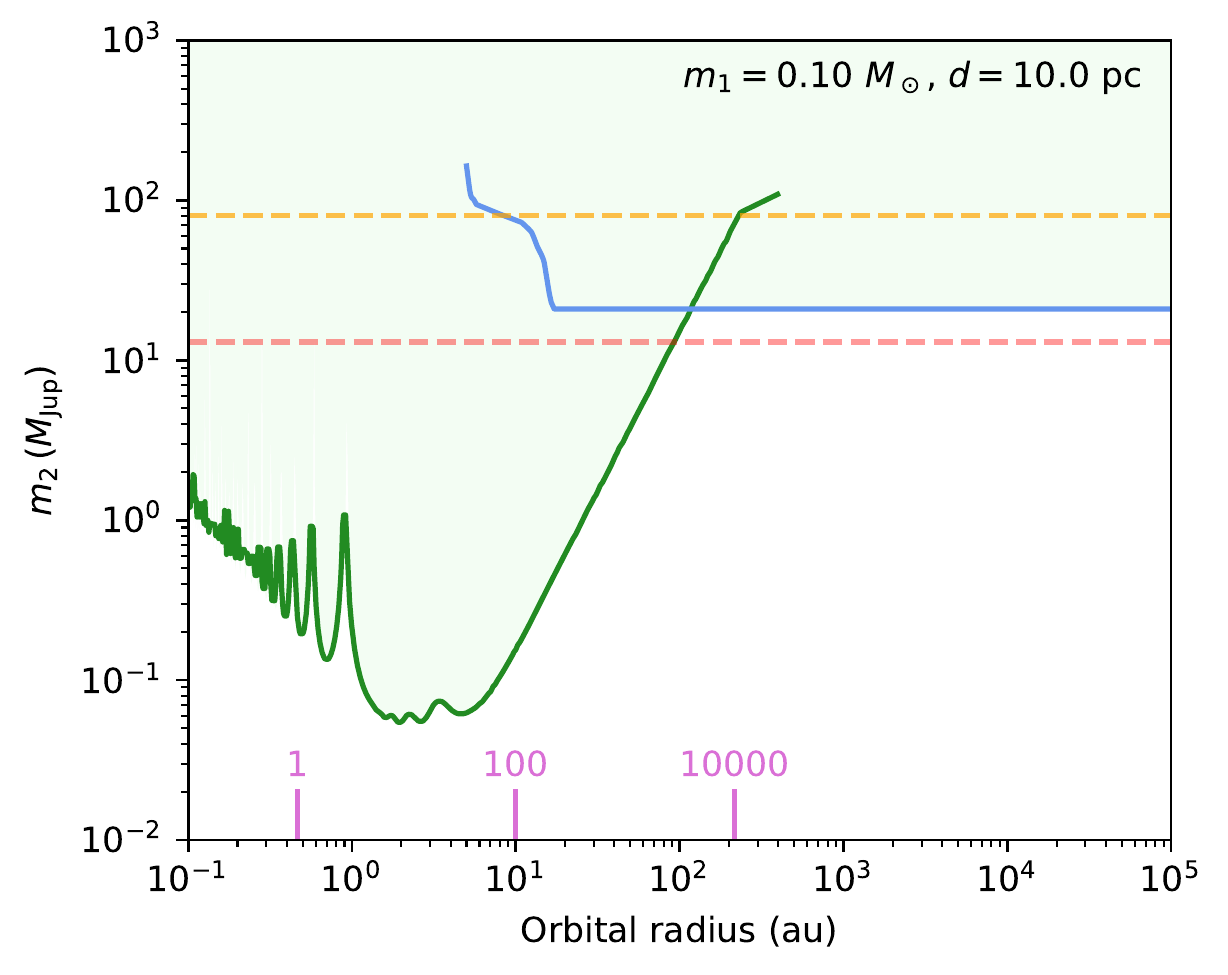}
\includegraphics[width=8.5cm]{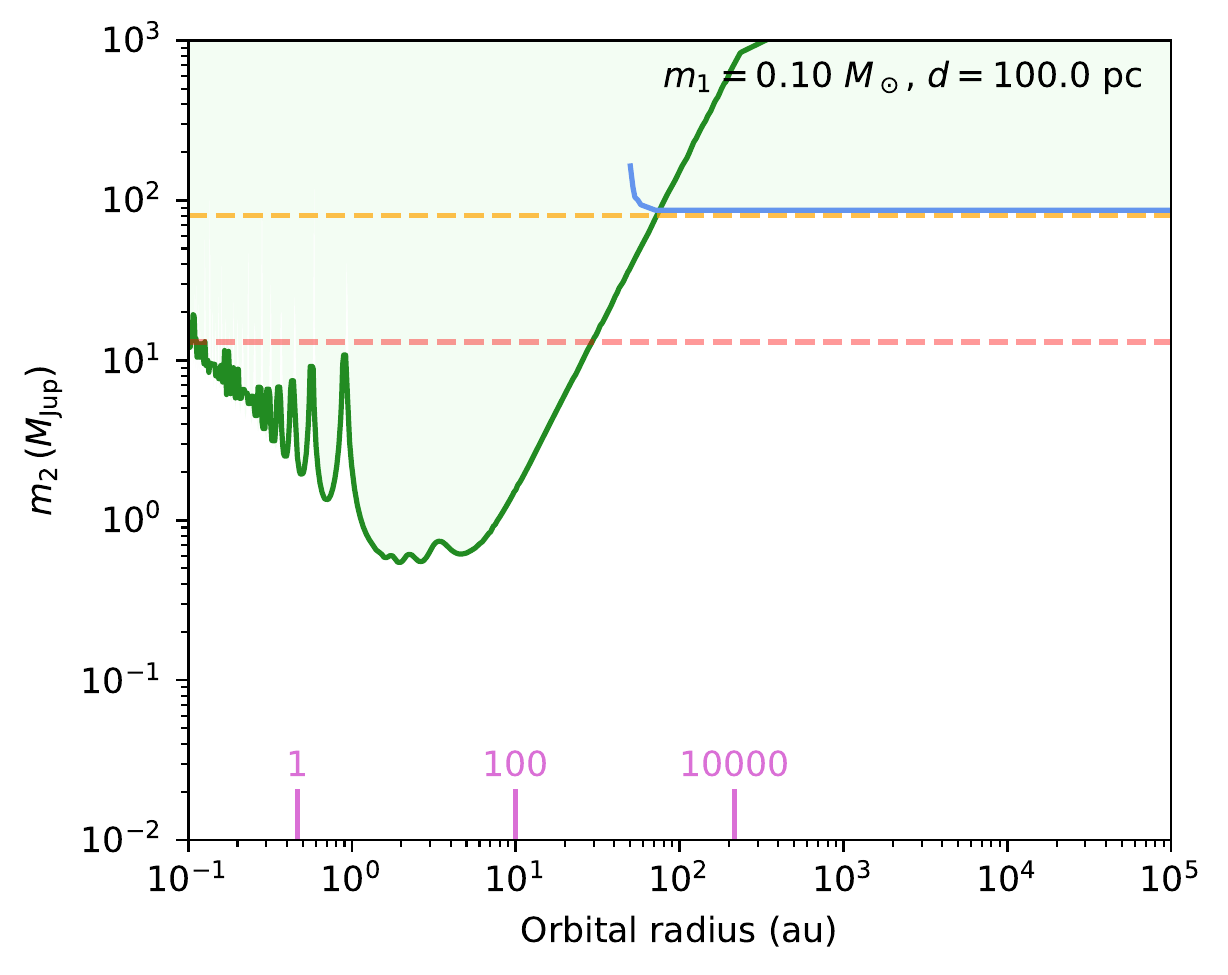}
\includegraphics[width=8.5cm]{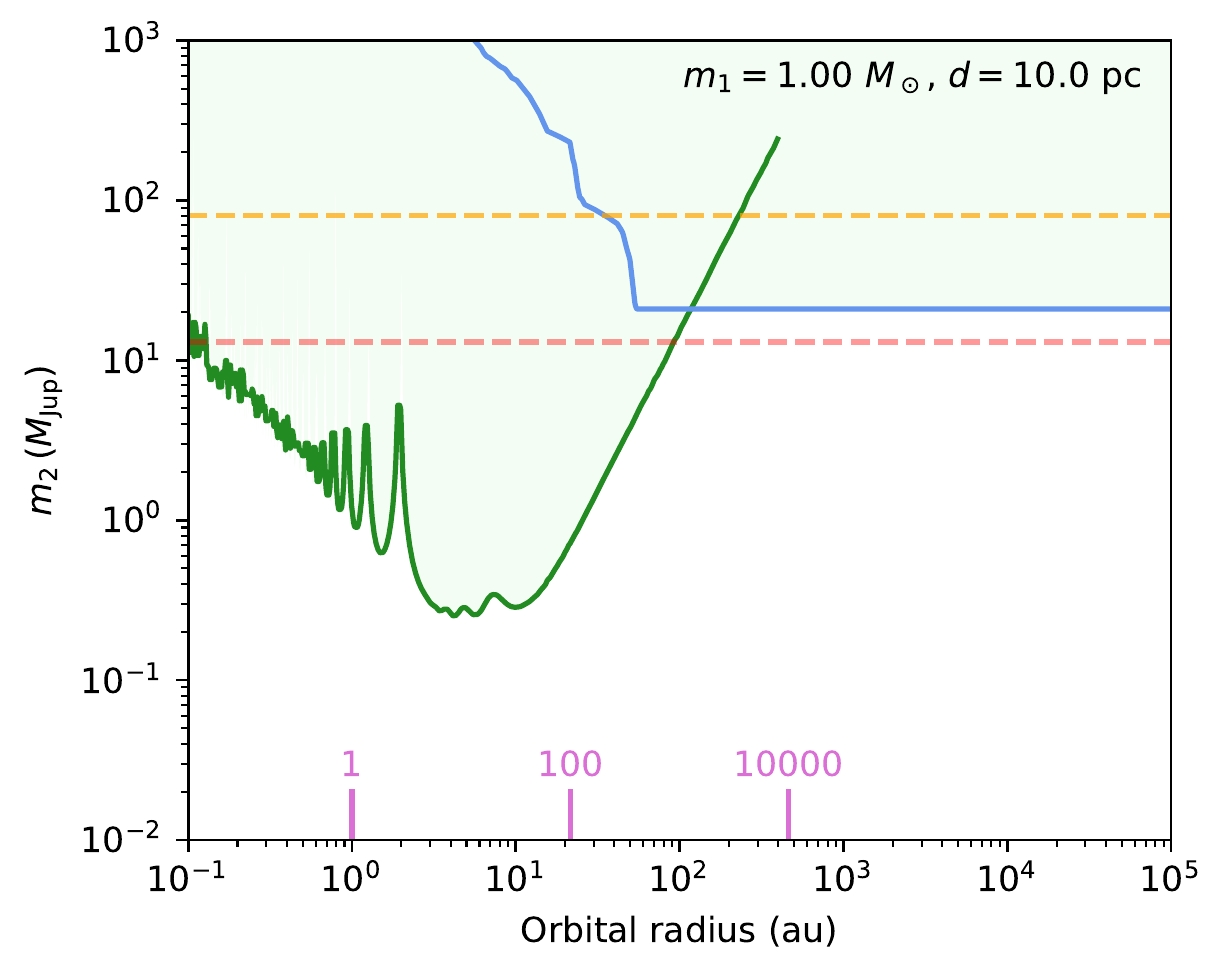}
\includegraphics[width=8.5cm]{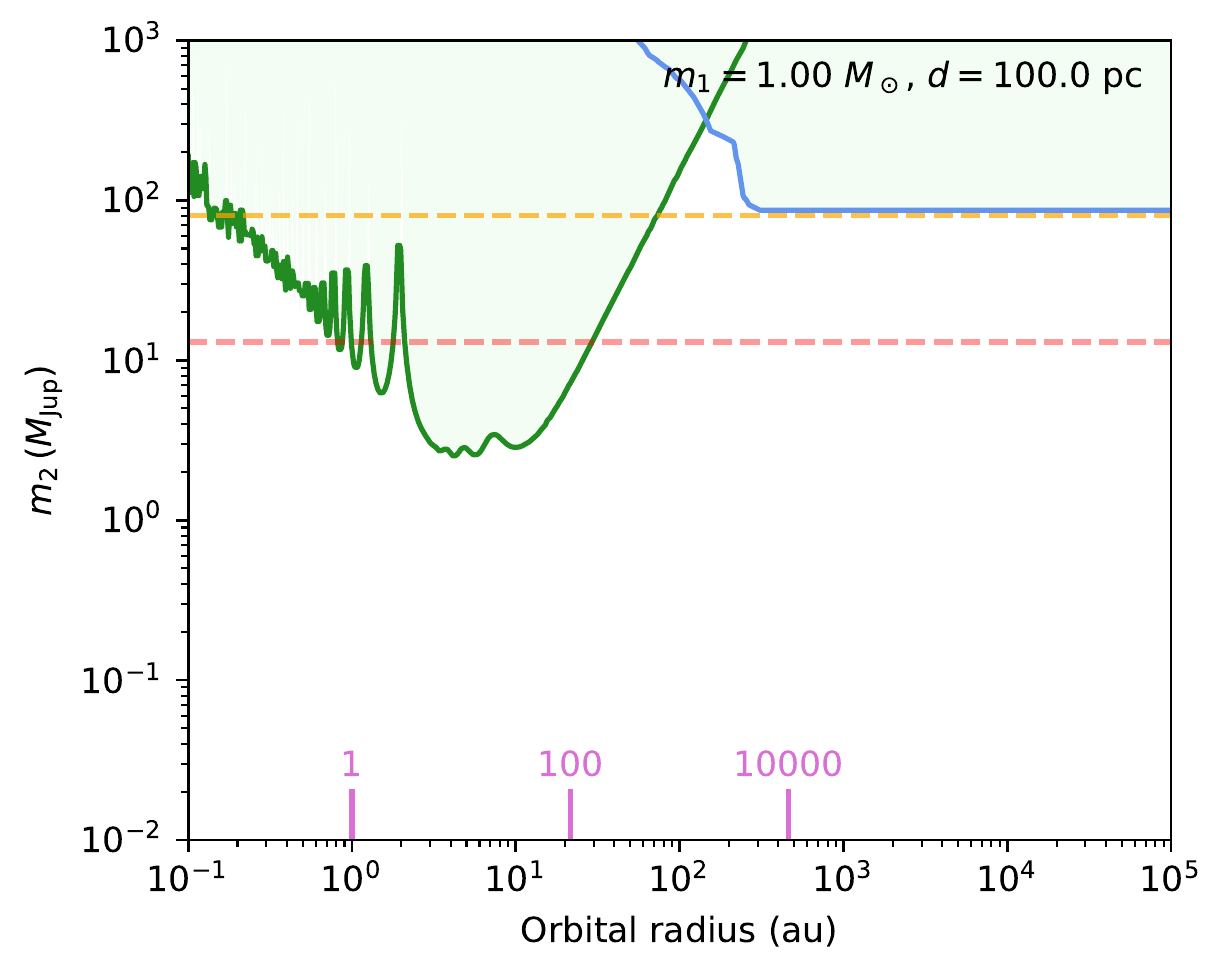}
\includegraphics[width=8.5cm]{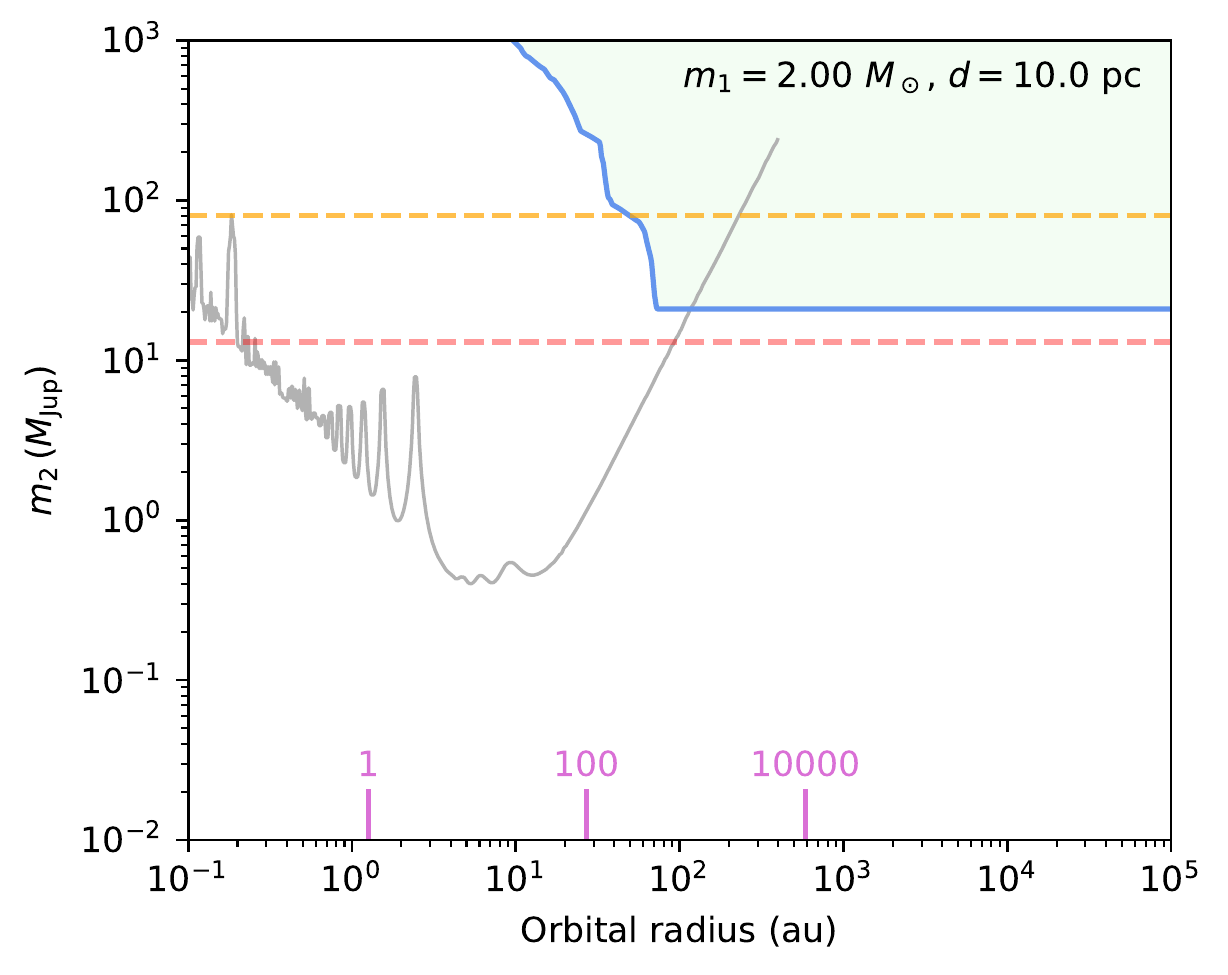}
\includegraphics[width=8.5cm]{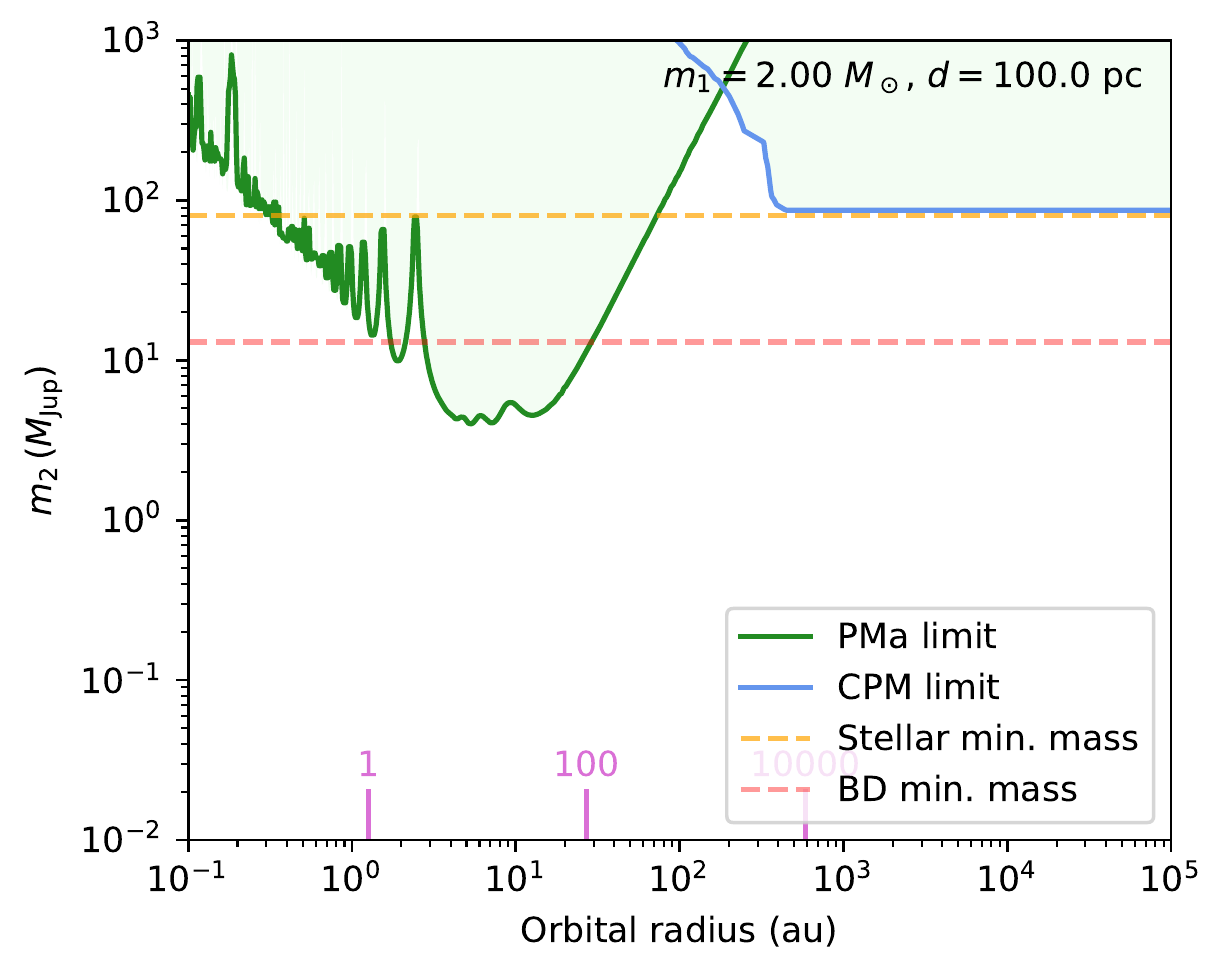}
\caption{Combined sensitivity limits of the PMa and CPM detection techniques for different combinations of target mass and distance.
When the target star is too bright for the application of the PMa technique ($G<3.0$) the sensitivity curve is shown in grey. The substellar ($m_2 = 80\,M_\mathrm{Jup}$) and planetary ($m_2=13\,M_\mathrm{Jup}$) mass limits are shown with dashed lines.}
\label{sensitivity_PMa_CPM}
\end{figure*}

Considering the median accuracy of the PMa vectors from Gaia EDR3 ($\sigma(\Delta \mu_\mathrm{G3}) = 56\,\mu$as\,a$^{-1}$) and Gaia's EDR3 limiting magnitude of $G=20.41$ (from \citeads{2021A&A...649A...6G}, encompassing 85\% of the sources), the domains of companion masses sampled by the combination of the PMa+CPM approaches are shown in Fig.~\ref{sensitivity_PMa_CPM}.
To convert the Gaia limiting magnitude to companion masses (for the CPM technique mass limits), we adopted the $M_G$ magnitude-spectral type relation calibrated by \citetads{2018A&A...619L...8R}. We then approximated the brown dwarf masses from Fig.~8 of \citetads{2018A&A...619L...8R}, for an age of 5\,Ga. The resulting masses should be considered rough approximations, particularly as the brightness of brown dwarfs critically depends on their age. The mass of stars of spectral types earlier than M6V were taken from the tables by \citetads{2012ApJ...746..154P} and \citetads{2013ApJS..208....9P}. We took into account the contrast sensitivity of Gaia as a function of the separation from the target star by inverting the Eq. (2) of \citetads{2021A&A...649A...6G}. 
The diagrams in Fig.~\ref{sensitivity_PMa_CPM} show the complementarity of the PMa and CPM detection techniques. While the PMa technique is sensitive to substellar mass companions ($m_2<80\,M_\mathrm{Jup}$) from $\approx 2$ to a few hundred astronomical units (with a decreasing sensitivity), the CPM technique enables the detection of substellar companions at separations up to tens of thousands of astronomical units. For targets located at 100\,pc, the CPM mass sensitivity limit corresponds to the substellar mass limit ($\approx 80\,M_\mathrm{Jup}$).

Out of the 4063 Hipparcos stars in the 100\,pc sample that have bound candidate companions, 1585 (39\%) also exhibit a PMa S/N>3.
Conversely, out of the 7175 Hipparcos stars within 100\,pc showing a PMa S/N>3, 1585 (22\%) also have bound candidate companions.
The overlap between the PMa and CPM binary samples may have two origins: (1) the observed PMa signal is induced by the wide candidate detected through the CPM technique and (2) triple systems where the close pair is revealed by the PMa method and the wide companion by CPM. As shown in Fig.~\ref{GaiaEDR3CPM-LinSep-histo}, the majority of resolved companions has a linear separation beyond 300\,au. In principle, this corresponds to orbital periods poorly suited for an efficient detection using the PMa technique (Sect.~\ref{sensitivity_pma}), apart from the very nearby stars within $\approx 10$\,pc for which the sensitivity overlap is significant between the two approaches (Fig.~\ref{sensitivity_PMa_CPM}).
For this reason, the targets beyond this distance for which both a PMa signal and a CPM candidate are detected are in most cases (at least) triple systems composed of a close binary (producing the PMa signal) and a wide companion (from CPM).

\subsection{Overall binary fraction of the Hipparcos catalog}

\begin{table}[ht]
 \caption{Number of stars with PMa, CPM and RUWE>1.4 binarity signals in the Hipparcos catalog. \label{Hip-stat}}
 \centering
  \begin{tabular}{lcc}
  \hline
  \hline
Method & Number of stars & Fraction \\
  \hline  \noalign{\smallskip}
Full catalog & 117,955 & 100\% \\
  \hline  \noalign{\smallskip}
PMa S/N>3 & 37,347 & 32\% \\
CPM bound candidates & 12,914 & 11\% \\
RUWE>1.4 & 25,067 & 21\% \\
  \hline  \noalign{\smallskip}
PMa or CPM & 37,347 & 32\%\\
PMa or CPM or RUWE & 50,720 & 43\%\\
   \hline
\end{tabular}
\end{table}

We list in Table~\ref{Hip-stat} the number and fraction of stars of the full Hipparcos catalog presenting a signal of binarity from the PMa, CPM and RUWE>1.4 indicators. Combining these three indicators, we detect a total of 50,720 stars of the Hipparcos catalog that present a signal of binarity in at least one of the three criteria, corresponding to a fraction of 43\% of the full sample. For comparison, the Hipparcos catalog's Double and Multiple Systems Annex (DMSA) \citepads{1997A&A...323L..49P,1997ESASP.402...13L} comprises 17,917 entries, corresponding to a fraction of 15\% of the catalog.
From the analysis of Gaia EDR3, \citetads{2021ApJS..254...42B} found 30\% of the Hipparcos stars exhibiting a significant difference between their short-term EDR3 and long-term Hipparcos-Gaia, which is consistent with our PMa binary fraction.

\section{Example analyses of specific targets\label{examples}}

In this section, we present a selection of brief analyses of a sample of representative targets of different types as examples of possible interpretations of the contents of the PMa and CPM catalogs.

In the CPM finding charts, the markers showing the positions of the stars in the field are represented at the EDR3 epoch (2016.0). The positions were translated to the Gaia EDR3 reference epoch when needed (e.g., for Hipparcos-only targets). The background images were retrieved from the Second Generation Digitized Sky Survey Red (DSS2-Red). As these images were taken at various epochs, this leads to an apparent difference in position with the markers for the fast PM stars. The PM vectors $\vec{\mu_\mathrm{HG}}$, $\vec{\mu_\mathrm{Hip}}$, and $\vec{\mu_\mathrm{G3}}$ are shown separately when available, respectively, in light red, magenta, and blue colors. The bound candidate companions (\texttt{Bnd} flag in the catalog) are marked with a yellow star and a red PM vector, while the low velocity stars (\texttt{LowV} flag) are marked with an orange PM vector. When present, the field stars that have compatible parallaxes are marked with blue symbols.

In the figures showing the PMa sensitivity function, the possible combinations of mass and orbital radius for the companion are shown as green, blue, and cyan curves, respectively for the EDR3, DR2, and Hipparcos epochs. The associated uncertainty domains are shaded in the corresponding color. The pink markings indicate the orbital period corresponding to selected orbital radii.

\subsection{Bright stars}

Among the stars brighter than the Gaia saturation limit, we identified 1080 stars with magnitudes $m < 6$ in the $V$, $H_P$, or $G$ bands with bound candidate companions. A subset of this sample for the stars brighter than $m = 3$ is listed in Table~\ref{superbright}.

\begin{figure*}
\centering
\includegraphics[width=8.3cm,page=2]{Figures/CPMfigures/HIP018543cpm.pdf}
\includegraphics[width=8.3cm,page=1]{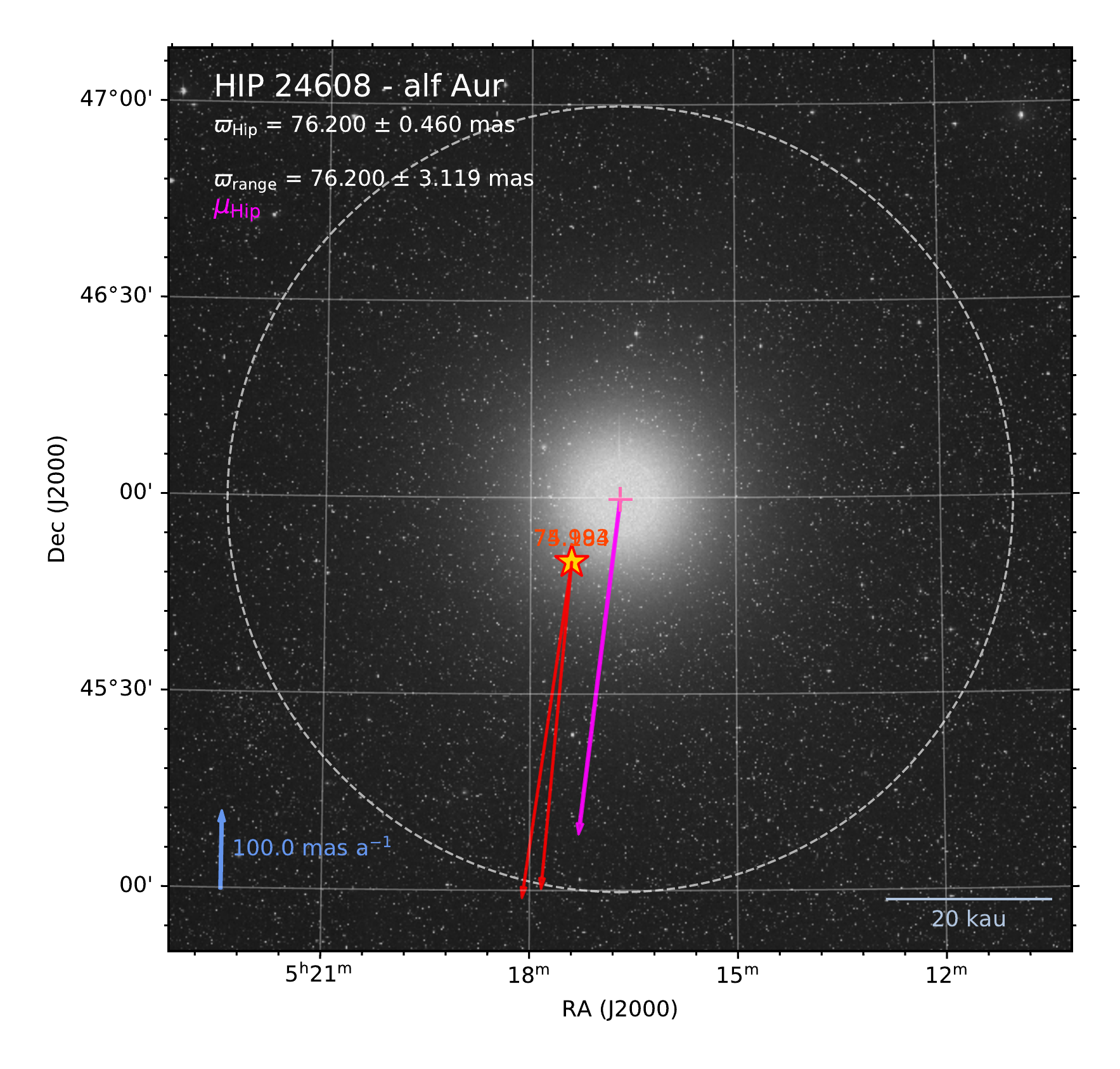}\\
\includegraphics[width=8.3cm,page=2]{Figures/CPMfigures/HIP049669cpm.pdf}
\includegraphics[width=8.3cm,page=2]{Figures/CPMfigures/HIP054061cpm.pdf}\\
\includegraphics[width=8.3cm,page=2]{Figures/CPMfigures/HIP072105cpm.pdf}
\includegraphics[width=8.3cm,page=2]{Figures/CPMfigures/HIP111954cpm.pdf}
\caption{Field charts of the bright stars with bound candidate companions $\gamma$\,Eri, $\alpha$\,Aur, $\alpha$\,Leo, $\alpha$\,UMa, $\epsilon$\,Boo, and $\epsilon$\,PsA. \label{SBright}}
\end{figure*}

\subsubsection{$\gamma$ Eri}

We identified a CPM companion (Fig.~\ref{SBright}), with a very low mass of $\approx 0.1\,M_\odot$, to the nearby red giant star $\gamma$\,Eri\,A (\object{HIP 18543}; spectral type M0III). The projected separation between component B and the primary is 1\,kau, and its Gaia $G$ band magnitude is $G=16.1$.
$\gamma$\,Eri\,A exhibits a moderate PMa in Gaia EDR3 ($S/N=3.2$). This indicates the presence of an additional close-in companion, possibly a  low-mass red dwarf ($M<0.4\,M_\odot$) orbiting within 50\,au of the primary. This PMa signal cannot be explained by the resolved CPM companion, whose mass is insufficient.

\subsubsection{$\alpha$\,Aur (Capella)}

We confirmed the two bound CPM companions \object{GJ 195 AB} of the nearby giant star $\alpha$\,Aur (\object{Capella}, \object{HIP 24608}, \object{HD 34029}; $d=13$\,pc), with estimated masses of 0.53 and $0.57\,M_\odot$. These companions, located at a projected separation of 9.5\,kau from Capella A, were discovered by \citetads{1914AN....197..181F}. As the primary Capella A is itself an equal mass binary \citepads{2011A&A...531A..89W, 2013A&A...560A.113H}, the system is therefore at least a quadruple. 
The very wide unbound CPM companion \object{50 Per} proposed by \citetads{2011ApJS..192....2S} located at a projected separation of 5.4\,pc is outside of the 1\,pc search limit of our survey.

\subsubsection{$\alpha$ Leo (Regulus)}

Next, $\alpha$\,Leo A (\object{HIP 49669}) is known to be a close spectroscopic binary \citepads{2008ApJ...682L.117G} whose companion $\alpha$\,Leo Ab was recently characterized by \citetads{2020ApJ...902...25G} as a 0.3\,M$_\odot$ pre-white dwarf. The main component A is a very-fast-rotating star that is seen almost equator-on \citepads{2005ApJ...628..439M}.
We confirmed that it has two additional bound candidate companions: \object{Gaia EDR3 3880785530720066176} (hereafter $\alpha$\,Leo B) and \object{Gaia EDR3 3880785530720066304} ($\alpha$\,Leo C), which are known to be co-moving with component A since the 19th century \citepads{10.1093/mnras/51.8.460}. They are a pair of relatively low-mass stars that are most likely gravitationally bound together and located at a projected separation of 4,300\,au from $\alpha$\,Leo A (Fig.~\ref{SBright}).

The position angle of $\alpha$\,Leo B with respect to A has slightly evolved from $305.1^\circ$ at epoch 1781.84 (as measured by Herschel) to $307.47^\circ$ at epoch 2016.0. The photometric estimate of the mass of B is around 0.63\,M$_\odot$, corresponding to a K7V spectral type \citepads{2012ApJ...746..154P,2013ApJS..208....9P}. It is only this component that has been identified as bound to Regulus AB, with a very high total score of 0.99. The estimation of the mass of C is complicated as the photometry is scarce, but being 3.5\,magnitudes fainter than component B in the $G$ band, it is likely an M4V red dwarf with a mass around 0.2\,M$_\odot$. This component was not identified by our search algorithm as bound to Regulus AB as its relative velocity of 2.8 km\,s$^{-1}$, caused by the orbital motion of the BC pair, is higher than the escape velocity.
It is possible to take advantage of the Gaia EDR3 parallaxes of components B ($\varpi[B] = 41.310 \pm 0.031$\,mas) and C ($\varpi[C] = 41.242 \pm 0.067$\,mas) to refine the Hipparcos parallax of $\alpha$\,Leo A ($\varpi_\mathrm{Hip}[A] = 41.130 \pm 0.350$\,mas).

\subsubsection{$\alpha$ UMa (Dubhe)}

Then, $\alpha$\,UMa (\object{HIP 54061}) is a very bright ($m_V = 1.8$) spectroscopic binary system. We detect the presence of a very low-mass dwarf companion (\object{Gaia EDR3 862234033499968640}; $m \approx 0.1$\,M$_\odot$) at a projected separation of 550\,au (Fig.~\ref{SBright}). The total score $P_\mathrm{tot}=0.602$ of this star is however close to the limit we adopted for bound candidates (Sect.~\ref{totalscore}). Due to the additional uncertainty on the systemic PM of the primary induced by its binarity, the gravitational link should be considered uncertain.

\subsubsection{$\epsilon$\,Boo}

We identified a candidate brown dwarf CPM companion (\object{Gaia EDR3 1279752168030730496}) to the A0V+K0II-III binary $\epsilon$\,Boo (\object{HIP 72105}; Fig.~\ref{SBright}), at a projected separation of 4.9\,kau (Fig.~\ref{SBright}). An additional CPM companion (\object{Gaia EDR3 1267607615425592448}, \object{2MASS J14454000+2615167}) with a very low relative tangential velocity of $\Delta \varv_\mathrm{tan} = 0.1 \pm 0.2$\,km\,s$^{-1}$ is also identified at a much wider separation of 186\,kau. Thus, $\epsilon$\,Boo may, in fact, be a quadruple system.

\subsubsection{$\epsilon$\,PsA}

\begin{figure}
\includegraphics[width=\hsize]{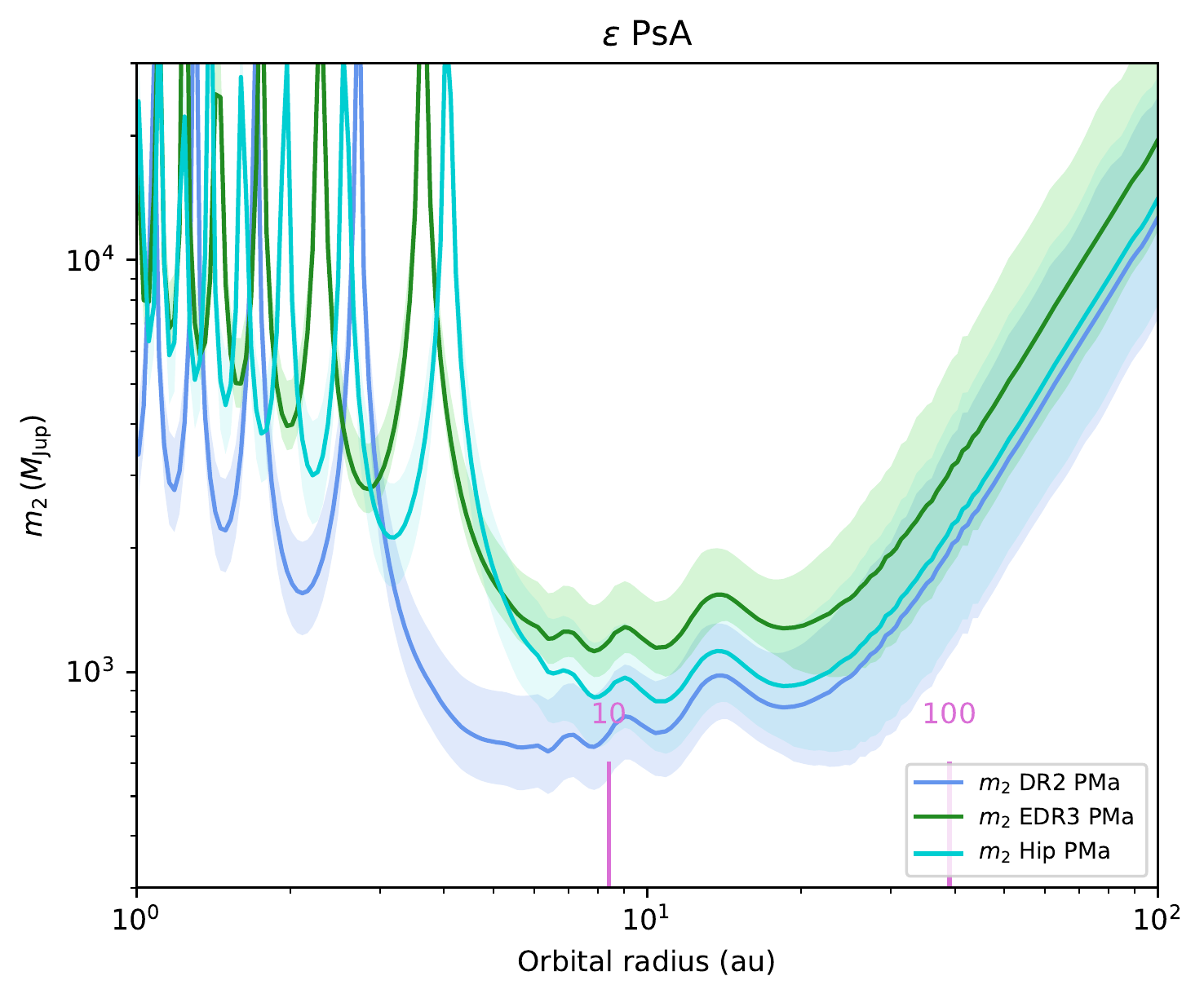}
\caption{PMa sensitivity diagram of the fast rotating Be star $\epsilon$\,PsA. \label{epsPsA-m2r}}
\end{figure}

The emission-line dwarf $\epsilon$\,PsA (\object{HIP 111954}, \object{HD 214748}) of spectral type B8Ve is a fast-rotating star \citepads{2019A&A...621A.123C} that exhibits both a significant PMa signal ($S/N = 12.7$) and a bound CPM candidate companion.
The PMa is visible in Fig.~\ref{SBright} as a difference between the long-term Hipparcos-Gaia PM vector (light green) and the short term Hipparcos and Gaia EDR3 PM vectors. The resolved companion $\epsilon$\,PsA\,B is likely a low-mass red dwarf ($m_B \approx 0.23$\,M$_\odot$), whose tangential velocity difference is only $\Delta \varv_\mathrm{tan} = 0.37 \pm 0.60$\,km\,s$^{-1}$ with respect to $\epsilon$\,PsA\,A. This projected velocity is well below the escape velocity at the projected separation of 11.7\,kau ($v_\mathrm{esc} \approx 0.95$\,km\,s$^{-1}$), considering a mass of $6\,M_\odot$ for the primary.
The observed PMa signal of the main component A cannot be caused by the resolved companion B; rather, the signal indicates the presence of a third component in the system orbiting close to the primary. As shown in Fig.~\ref{epsPsA-m2r}, the companion is possibly a solar mass star orbiting between $\approx 6$ to 30\,au from the primary. Alternatively, it could also be a more massive star orbiting at a larger separation.
The position angle of the Gaia EDR3 tangential velocity anomaly is $PA = 263.8 \pm 2.7 \deg$ for a norm of  $\Delta \varv_\mathrm{tan,G3} = 3.6 \pm 0.3$\,km\,s$^{-1}$ ($S/N=12.7$). The PA coincides modulo $180^\circ$ with the position angle of the gaseous equatorial disk of the Be star, which was found by \citetads{2019A&A...621A.123C} to be $\mathrm{PA}=67^\circ$ (with a high inclination of $i=73^\circ$ on the line of sight). This indicates that the stellar mass close-in companion is possibly orbiting in the same plane as the disk. The PMa is also significant from the Hipparcos catalog ($S/N = 3.9$), with a position angle of $285.9 \pm 9\deg$ and a tangential velocity residual of $\Delta \varv_\mathrm{tan,H} = 2.6 \pm 0.7$\,km\,s$^{-1}$.

\subsubsection{L$_2$ Puppis}

This semi-regular pulsating red giant star L$_2$\,Puppis (\object{HIP 34922}, \object{HD 56096}) exhibits a significant PMa signal in Gaia EDR3 ($S/N = 4.0$) as well as in DR2 ($S/N=3.6$). However, the interpretation of this signal in terms of the presence of a massive companion is not pertinent.
The first reason is that the inhomogeneities present on the surface of giant and supergiant evolved stars (caused by their very large convective cells) affect the position of the photocenter, therefore adding noise to the astrometric measurements \citepads{2011A&A...528A.120C}.
In the case of L$_2$\,Pup, the situation is further complicated by the presence of an inhomogeneous circumstellar dust disk \citepads{2002MNRAS.337...79B, 2014A&A...564A..88K, 2015A&A...576A..46L, 2015A&A...581C...2L, 2020ApJ...901..144N} that is observed almost edge-on ($i=82^\circ$; \citeads{2015A&A...578A..77K, 2016A&A...596A..92K, 2017A&A...601A...5H}).
This disk partially hides the stellar disk and shifts the position of its photocenter in a time-variable way as the star pulsates with a period of $P\approx141$\,days.
The position angle of the EDR3 PMa vector is $PA = 180^\circ$, namely, it is perpendicular to the disk plane. This is consistent with the expected shift of the photocenter as the partially occulted photosphere emerges more or less in a north-south direction above the disk edge.

\begin{figure}
\includegraphics[width=\hsize,page=2]{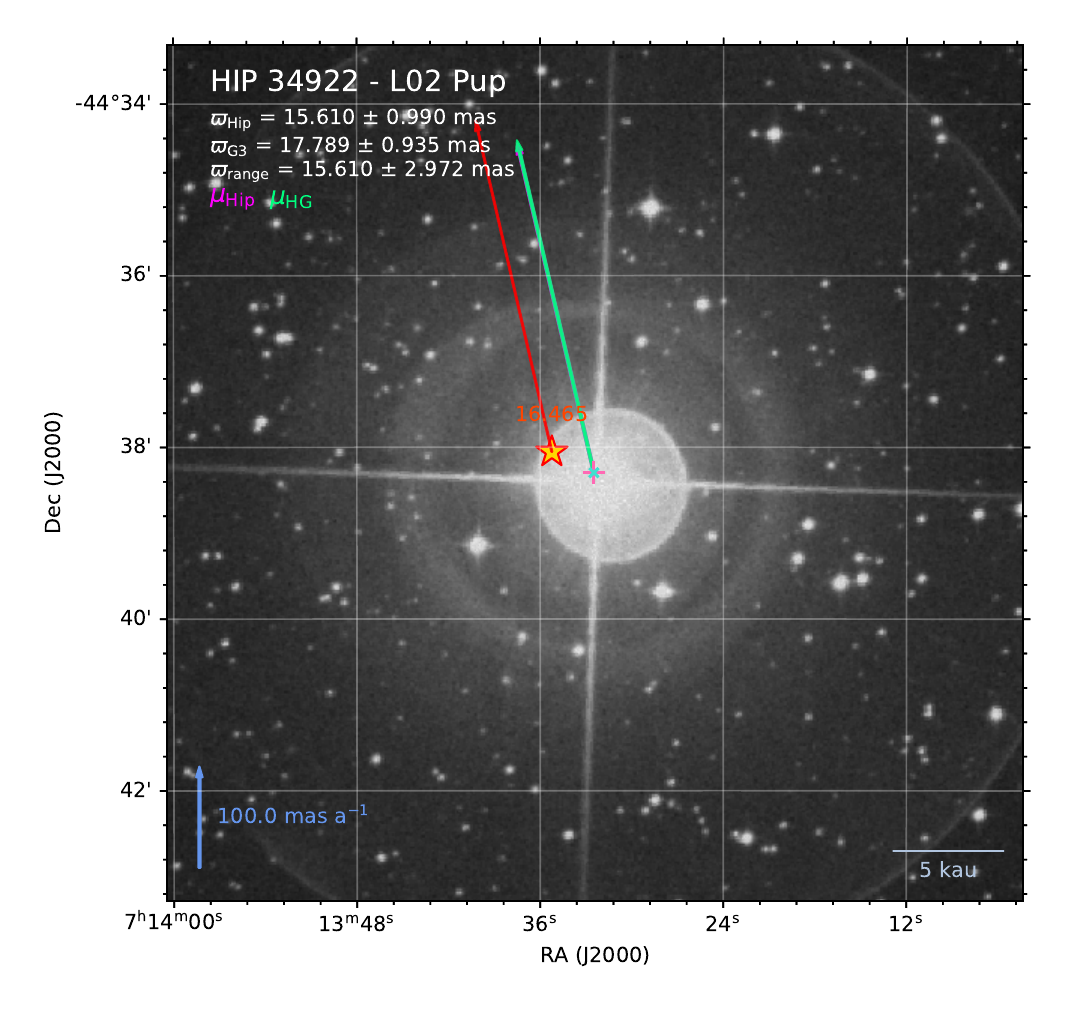}
\caption{Field chart of L$_2$\,Puppis showing its red dwarf companion.\label{HIP034922cpm}}
\end{figure}

We identified a bound candidate CPM companion to L$_2$\,Pup (\object{Gaia EDR3 5559704601965623680}) located at a projected separation of 2100\,au (Fig.~\ref{HIP034922cpm}). L$_2$\,Pup\,B is a faint red dwarf with an estimated mass of $m_B = 0.15\,M_\odot$. Its parallax of $\varpi_\mathrm{G3}[B] = 16.465 \pm 0.028$\,mas) is much more accurate than the parallax of L$_2$\,Pup\,A, both from Hipparcos ($\varpi_\mathrm{H}[A] = 15.61 \pm 0.99$\,mas; \citeads{2007ASSL..350.....V}) and the EDR3 ($\varpi_\mathrm{G3}[A] = 17.79 \pm 0.94$\,mas, RUWE = 8.8). This makes of L$_2$\,Pup\,B a valuable proxy for evaluating the distance of the primary. As a remark, the Gaia DR2 parallax of L$_2$\,Pup\,A was incorrect by a factor two ($\varpi_\mathrm{G2}[A] = 7.36 \pm 0.61$\,mas), likely biased by the variability of the photocenter of the star.

\subsection{Resolved binary stars}

\subsubsection{GJ 65 AB \label{GJ65}}

Gliese 65 is a pair of very low-mass red dwarfs with late M5.5Ve and M6Ve spectral types (\object{GJ65 AB}, \object{Luyten 726-8}, \object{BL Cet}+\object{UV Cet}), which are relatively fast rotators \citepads{2017MNRAS.471..811B}. The two components are both present in the EDR3 catalog (Table~\ref{GJ65-data}). The close proximity of this system ($d=2.7$\,pc) allowed \citetads{2016A&A...593A.127K} to measure their radii using optical interferometry ($R(A)=0.165 \pm 0.006\,R_\odot$, $R(B)=0.159 \pm 0.006\,R_\odot$) and determine their masses ($m(A)=0.1225 \pm 0.0043\,M_\odot$; $m(B) = 0.1195 \pm 0.0043 M_\odot$) from their orbital motion.
These accurate physical parameters make them particularly attractive benchmarks for models of very low-mass stars. The barycentric parallax $\varpi = 373.7 \pm 2.7$\,mas) obtained by \citetads{2016A&A...593A.127K} is in good agreement with the mean EDR3 parallax of the two stars ($\varpi_\mathrm{G3} = 371.92 \pm 0.42$\,mas), although the RUWE is high for the two stars (Table~\ref{GJ65-data}). The mean parallax from the Gaia DR2 catalog ($\varpi_\mathrm{G2} = 371.03 \pm 0.21$\,mas) is within $2.1 \sigma$ of the EDR3 value and it also consistent with the orbital parallax determined by \citetads{2016A&A...593A.127K}.

\begin{table*}
 \caption{Astrometry of the components of the red dwarf binary GJ65 AB from Gaia DR2 and EDR3 and their barycenter, adopting the fractional mass $m_B/(m_A+m_B) = 0.4938 \pm 0.0031$ from \citetads{2016A&A...593A.127K}.
 \label{GJ65-data}}
 \centering
\renewcommand{\arraystretch}{1.2}
 \tiny
  \begin{tabular}{lccccccc}
  \hline
  \hline
 Star & Number & RUWE & RA & Dec & $\mu_\alpha$ & $\mu_\delta$ & $\varpi$  \\ 
 & & & & & (mas\,a$^{-1}$) & (mas\,a$^{-1}$) & (mas) \\
 \hline  \noalign{\smallskip}
 & Gaia DR2 \\
GJ65 A & 5140693571158739840 & 6.5 & 01h39m05.05425s & $-$17d56m54.1548s & $+3385.90 \pm 0.53$ & $+531.97 \pm 0.41$ & $369.96 \pm 0.29$ \\
GJ65 B & 5140693571158739712 & 6.9 & 01h39m05.09051s & $-$17d56m51.9462s & $+3182.81 \pm 0.60$ & $+592.04 \pm 0.46$ & $372.19 \pm 0.30$ \\
GJ65 AB & & & 01h39m05.0722s & $-$17d56m53.0642s  &  &\\
  \hline  \noalign{\smallskip}
& Gaia EDR3 \\
GJ65 A & 5140693571158739840 & 12.4 & 01h39m05.17303s & $-$17d56m53.8796s & $+3385.30 \pm 0.67$  &  $+544.42 \pm 0.38$ & $367.76 \pm 0.83$ \\
GJ65 B & 5140693571158946048 & 10.5 & 01h39m05.20181s & $-$17d56m51.6583s & $+3178.68 \pm 0.43$ & $+584.10 \pm 0.30$ & $373.84 \pm 0.56$ \\
GJ65 AB & & & 01h39m05.1872s & $-$17d56m52.7827s &   & \\
   \hline
\end{tabular}
\end{table*}

\begin{table}
 \caption{Proper motion of the GJ65 AB barycenter from the weighted mean of the Gaia DR2 and EDR3 proper motion vectors of components A and B (first two lines) and from the difference in position between DR2 and EDR3 (last line).
 \label{GJ65-bary}}
 \centering
  \begin{tabular}{lcc}
  \hline
  \hline
Method & $\mu_\alpha$ & $\mu_\delta$ \\
& (mas\,a$^{-1}$) & (mas\,a$^{-1}$) \\
  \hline  \noalign{\smallskip}
Gaia DR2 $\vec{\mu}$ avg. & $+3285.61 \pm 0.40$ &  $+561.63 \pm 0.31$ \\
Gaia EDR3 $\vec{\mu}$ avg. & $+3283.29 \pm 0.40$ & $+563.98 \pm 0.24$ \\
  \hline  \noalign{\smallskip}
DR2-EDR3 pos. & $+3284.66 \pm 0.28$ &  $+562.96 \pm 0.24$ \\
   \hline
\end{tabular}
\end{table}

\begin{figure}
\includegraphics[width=\hsize]{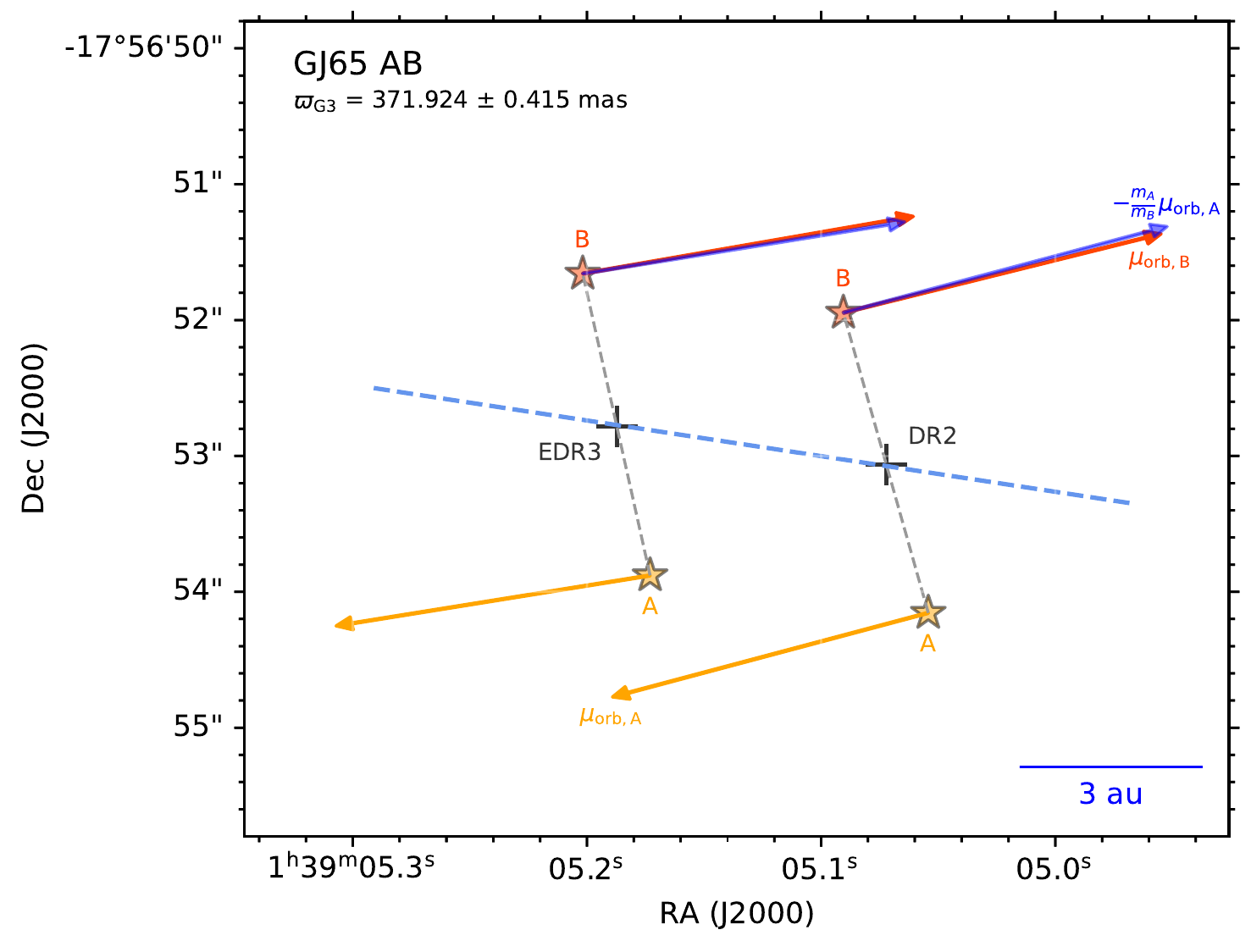}
\caption{Tangential orbital velocity vectors of GJ65 A and B from Gaia DR2 and EDR3.\label{GJ65-orbPM}}
\end{figure}

From the binary orbit, \citetads{2016A&A...593A.127K} estimated the fractional mass $m(B)/m_\mathrm{tot} = m(B)/\left[m(A)+m(B)\right] = 0.4938 \pm 0.0031$ ($\pm 0.6\%$), making it possible to determine the position of their barycenter from the positions of the two stars.
We can estimate the PM vector $\vec{\mu_\mathrm{AB}}$ of the barycenter using two different approaches: from the mean of the PM vectors of the two components (weighted by the inverse of their mass), and from the difference in position of the barycenter between the Gaia DR2 and EDR3 epochs (Table~\ref{GJ65-data}).  Table~\ref{GJ65-bary} gives the resulting measurements of the barycentric PM vector using these two techniques. A difference at a level of $5\sigma$ is present between the DR2 and EDR3 values, which bracket the vector from the DR2 and EDR3 positions. This difference may indicate that the motion of one of the two stars is perturbed by the presence of a third body (details on a related caveat later in this paper).
We computed the PM vector of the barycenter from the difference between its positions at the DR2 and EDR3 epochs. It was then possible to derive the orbital velocity vector of each star A and B by subtracting from the DR2 and EDR3 PM vectors $\vec{\mu}(A)$ and $\vec{\mu}(B)$ the PM of the barycenter $\vec{\mu_\mathrm{AB}}$ through $\vec{\mu_\mathrm{orb}}(A/B) = \vec{\mu}(A/B) - \vec{\mu_\mathrm{AB}}$. A diagram of the resulting PM vectors is presented in Fig.~\ref{GJ65-orbPM}.

For a simple two-star system, the orbital velocity vectors of the two stars are colinear, with opposite directions and their norms are inversely proportional to each star's mass. As discussed by \citetads{2019A&A...623A..72K}, it is therefore possible to search for the signature of an additional massive body orbiting one of the two stars from the orbital velocity anomaly, $\Delta \mu_\mathrm{orb}$, defined as the quantity:
\begin{equation}
\vec{\Delta \mu_\mathrm{orb}}(B) = \vec{\mu_\mathrm{orb}}(B) - \frac{m(A)}{m(B)}\, \vec{\mu_\mathrm{orb}}(A)
.\end{equation}
We obtain the following orbital velocity anomaly vectors from the DR2 and EDR3 data, expressed angularly:
\begin{align}
\vec{\Delta \mu_\mathrm{orb}}(B)[\mathrm{DR2}] & = (+1.93 \pm 0.91 , -2.69 \pm 0.72)\,\mathrm{mas\,a}^{-1} ,\\
\vec{\Delta \mu_\mathrm{orb}}(B)[\mathrm{EDR3}] & = (-2.79 \pm 0.91 , +2.06 \pm 0.61)\,\mathrm{mas\,a}^{-1},
\end{align}
which we can also express in tangential velocity, knowing the parallax of the system:
\begin{align}
\vec{\Delta \varv_\mathrm{orb}}(B)[\mathrm{DR2}] & = (+24.6 \pm 11.5 , -34.3 \pm 9.2)\,\mathrm{m\,s}^{-1} ,\\
 \vec{\Delta \varv_\mathrm{orb}}(B)[\mathrm{EDR3}] & = (-35.5 \pm 11.5 , +26.2 \pm 7.8)\,\mathrm{m\,s}^{-1}.
\end{align}
These residuals are significant at a level of $\approx 3\sigma$ both for the DR2 and EDR3 epochs, which may, in principle, indicate the presence of a third body in orbit around one of the components.

However, these results should be considered as a demonstration of principle and not a detection, due to the large RUWE of the Gaia measurements of the two stars. This high RUWE is possibly caused by the orbital curvature of the trajectories of the two stars, which is not taken into account in the EDR3 astrometric reduction and may result in a bias on the determined PM vectors. Alternatively, the angular proximity of the two stars (EDR3 separation $\approx 2.26\arcsec$) and their apparent brightness ($G(A) = 10.5$ and $G(B) = 10.8$) may induce a mutual contamination of the two stars on the Gaia detectors, depending on the position angle of each observed transit. Due to the relatively short orbital period ($P_\mathrm{orb} = 26.3$\,years), this effect also evolves significantly over time.

When available, the analysis of the epoch astrometry of Gaia will enable a thorough search for low-mass companions from a combined fit of the barycentric PM and parallactic wobble, together with the orbital motion of the components.
As a remark, the differential astrometry of GJ65 AB is monitored using the GRAVITY instrument \citepads[see Sect. 3.6 of]{2017A&A...602A..94G}, with an accuracy on the order of $50\,\mu$as. The objective of this project is to search for the signature of low-mass planets orbiting one of the two stars as a deviation of the differential astrometry of A and B from a two-body orbit.

\subsubsection{61 Cyg AB \label{61Cyg}}

The binary star \object{61 Cyg} AB comprises a K5V primary (\object{ADS 14636A}, \object{GJ 820A}, \object{HD 201091}, \object{HIP 104214}) and a K7V secondary (\object{ADS 14636B}, \object{GJ 820B}, \object{HD 201092}, \object{HIP 104217}). This is the nearest star in the northern hemisphere ($d=3.5$\,pc) and it is thanks to this proximity that \citetads{2008A&A...488..667K} and \citetads{2009ApJ...694.1085V} were able to measure the angular diameters of the two components using optical interferometry. The eccentric orbit of the system ($e \approx 0.4$) and very long orbital period (around 7 centuries, \citeads{2012A&A...546A..69M}) make the dynamical determination of the masses relatively difficult. Existing estimates range from 0.67 to $0.79\ M_\odot$ for A and 0.52 to $0.63\ M_\odot$ for B \citepads{1995Icar..116..359W, 2008A&A...488..667K, 2009ApJ...694.1085V, 2012ApJ...757..112B, 2018RAA....18...94S}.
Following \citetads{2019A&A...623A..72K}, we adopted the masses determined from the photometric mass-luminosity relation by \citetads{2015ApJ...804...64M}: $m(A)= 0.708 \pm 0.053\ M_\odot$ and $m(B)= 0.657 \pm 0.057\ M_\odot$, close to the best-fit values of $m(A) = 0.69\ M_\odot$ and $m(B) = 0.61\ M_\odot$ obtained by \citetads{2008A&A...488..667K} from evolutionary modeling with the CESAM2k code \citepads{1997A&AS..124..597M, 2008Ap&SS.316...61M, 2010ascl.soft10059M}.
The photometric masses correspond to a mass ratio of $m(B)/m(A) = 0.93 \pm 0.11$. From an astrometric determination of the radial velocity of 61 Cyg A and B using Hipparcos and Gaia EDR3, \citetads{2021A&A...652A..45L} obtained a mass ratio $m(B)/m(A) = 0.76 \pm 0.05$, which is $1.6\sigma$ smaller than our adopted value.
\citetads{2019A&A...623A..72K} presented an analysis of the PM of 61 Cyg AB using Hipparcos and Gaia DR2. We hereby extend this analysis using Gaia EDR3 astrometry.

\begin{table}
 \caption{Proper motion of the 61 Cyg AB barycenter from the weighted mean of the Gaia DR2 and EDR3 proper motion vectors of components A and B (first two lines), and from the difference in position between the Hipparcos and the DR2/EDR3 epochs (last two lines).
 \label{61Cyg-bary}}
 \centering
  \begin{tabular}{lcc}
  \hline
  \hline
Method & $\mu_\alpha$ & $\mu_\delta$ \\
& (mas\,a$^{-1}$) & (mas\,a$^{-1}$) \\
  \hline  \noalign{\smallskip}
Gaia DR2 $\vec{\mu}$ avg. & $+4136.10 \pm 0.12$ & $+3204.47 \pm 0.15$ \\
Gaia EDR3 $\vec{\mu}$ avg. & $+4136.17 \pm 0.03$ & $+3204.55 \pm 0.03$ \\
  \hline  \noalign{\smallskip}
Hip-DR2 pos. & $+4133.66 \pm 0.81$ & $+3203.81 \pm 0.17$ \\
Hip-EDR3 pos. & $+4133.71 \pm 0.79$ & $+3203.81 \pm 0.17$ \\
   \hline
\end{tabular}
\end{table}

Following the approach of Sect.~\ref{GJ65}, we first estimated the PM of the barycenter of the system both from the weighted mean of the components' PM vectors and from the difference between the DR2/EDR3 position and the Hipparcos position.
The results are presented in Table~\ref{61Cyg-bary}. Although there is an excellent agreement between the determinations obtained using each of the two methods considered individually, there is a difference $\Delta \mu_\mathrm{AB}$ at a $3\sigma$ level between the two methods. Contrary to GJ65, for which the high RUWE casts doubts on the reliability of Gaia astrometry, the Gaia EDR3 measurements of both components exhibit a satisfactory RUWE level below 1.4 (1.0 and 1.2 for A and B, respectively). This quantity is a difference between the long-term (Hipparcos-Gaia) and short-term (Gaia average) estimates of the barycentric PM vector, therefore equivalent to the PMa defined for individual stars (Sect.~\ref{HG-PMa}).
This observed PMa is robust against a change in the mass ratio of the AB pair. Adopting a lower mass ratio $m(B)/m(A)=0.76$ \citepads{2021A&A...652A..45L} or a higher value of 1.0 (equal mass) for the computation modifies the barycentric PM vectors, but the observed PMa remains significant at a $\approx 3\sigma$ level.

\begin{figure}
\includegraphics[width=\hsize]{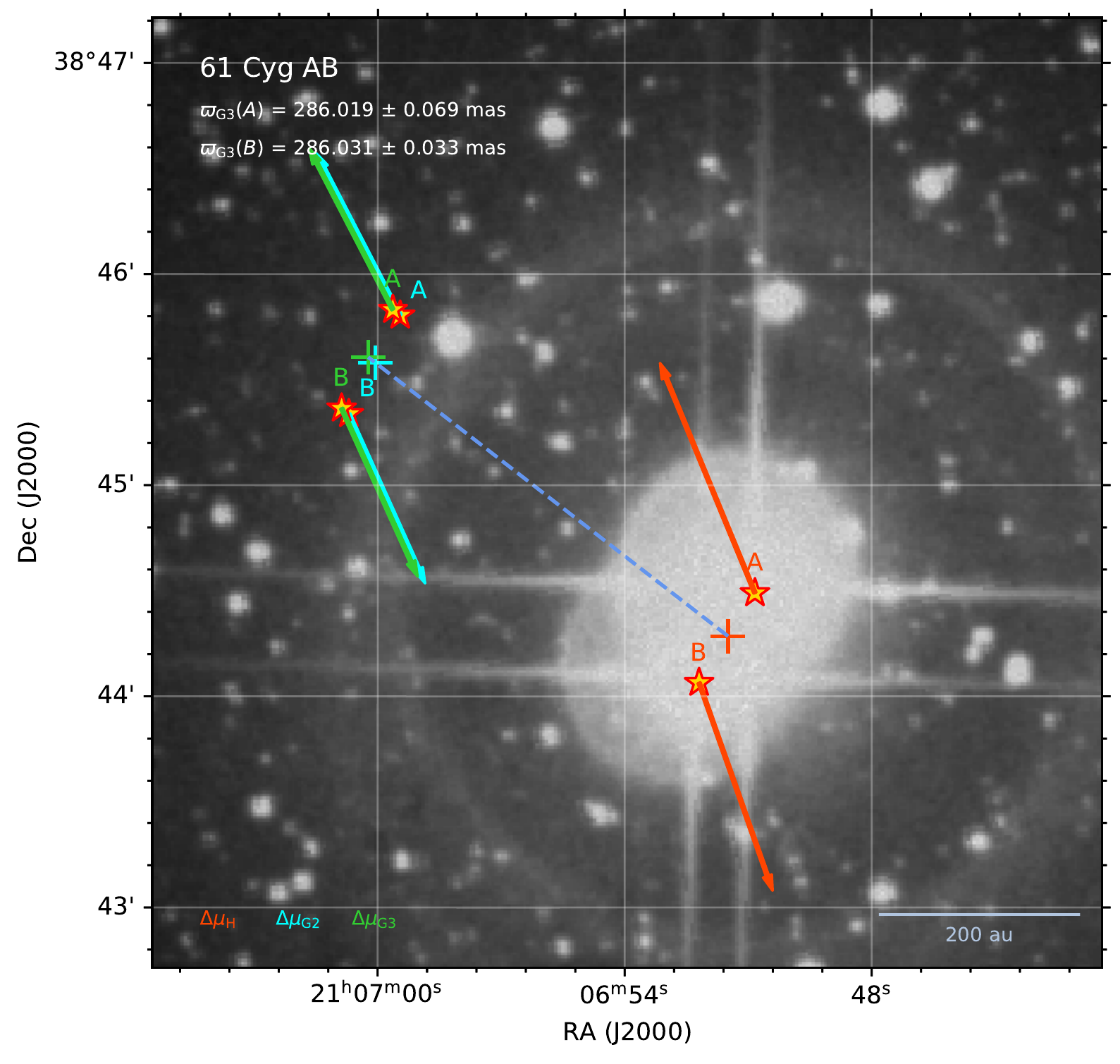}
\includegraphics[width=\hsize]{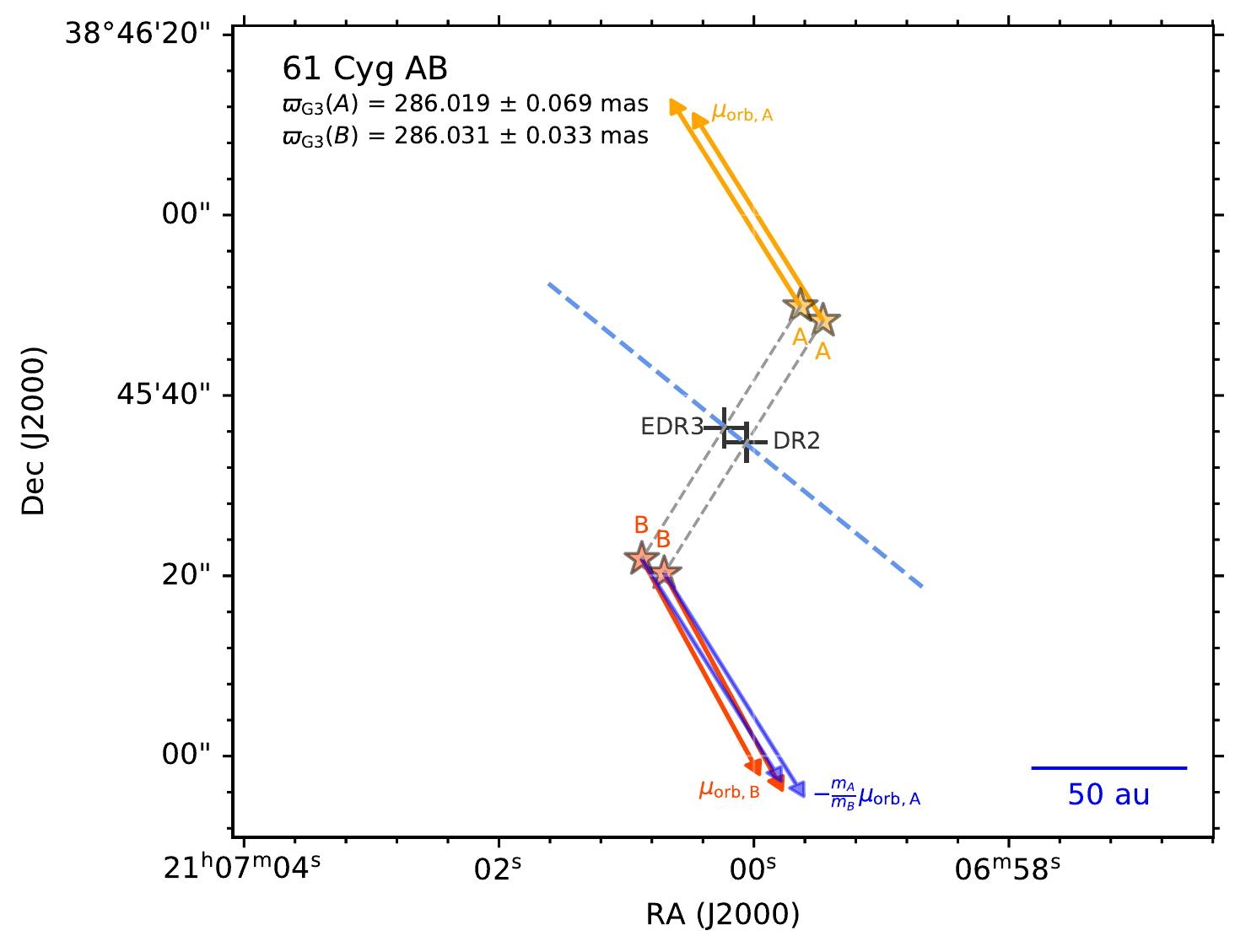}
\caption{Proper motion and tangential orbital velocity of 61 Cyg A and B from Hipparcos, Gaia DR2 and EDR3 (top panel) and enlargement of Gaia DR2 and EDR3 data showing the divergence between the orbital velocity vectors of components A and B (bottom panel). The position of the barycenter is marked with a `+' symbol.\label{61Cyg-orbPM}}
\end{figure}

This significant barycentric PMa indicates the probable presence of a third body orbiting either (1) one of the two components A or B (S-type companion) or (2) the AB pair (circumbinary, P-type companion). In hypothesis (1), the gravitational pull of the putative companion ``drags'' the PM of one of the two components, therefore biasing the short-term barycenter PM computed from the mean of the two component PM vectors. In situation (2), the presence of a very wide companion in circumbinary orbit would shift the PM vectors of both components A and B in the same way. This second hypothesis is, however, unlikely to be correct, as the period of a circumbinary companion would be extremely long (millenial scale). This would induce an undetectable shift on the short-term PM of the pair. The presence of a companion orbiting one of the two stars is therefore the most likely explanation to the observed anomaly on the barycenter PM of 61 Cyg AB.
To further test this hypothesis, we now examine the PM of each component. We first derived the orbital velocity vectors (as in Sect.~\ref{GJ65}) by subtracting the Hipparcos-Gaia barycentric PM from the Gaia PM vector of each star. The resulting vectors are presented in Table~\ref{61Cyg-orbitPM}, together with differential quantities. We observed a divergence in the position angle $\theta$ of the tangential velocity vectors of the two stars, which is also visible in Fig.~\ref{61Cyg-orbPM}. This difference reaches $\Delta \theta_{AB} = 3.2 \pm 1.0\,\deg$ and is consistent between the DR2 and EDR3 epochs. The orbital velocity offset of component B relative to A is significant at a $4.4\sigma$ level at $\Delta \varv_\mathrm{orb} =88 \pm 20$\,m\,s$^{-1}$ at a position angle of $\theta = 74 \pm 6\,\deg$, with consistent values from the DR2 and EDR3 data.

\begin{table*}
 \caption{Orbital velocity vectors of the 61 Cyg components.
 \label{61Cyg-orbitPM}}
 \centering
  \begin{tabular}{lcc}
  \hline
  \hline  \noalign{\smallskip}
 & 61 Cyg A & 61 Cyg B \\
  \hline  \noalign{\smallskip}
\textbf{Gaia DR2} \\
$\vec{\mu_\mathrm{orb}}$ (mas\,a$^{-1}$) & $(+31.54 \pm 0.81,  +47.95 \pm 0.32)$ & $(-26.85 \pm 0.79, -46.28 \pm 0.20)$ \\
$\vec{\mu_\mathrm{orb}}$ position angle $\theta$ & $33.33 \pm 0.69\ \deg$ & $210.12 \pm 0.75\ \deg$ \\
Diff. position angle $\Delta \theta_{AB} = \theta(A) - \theta(B) + 180^\circ$ & \multicolumn{2}{c}{$3.21 \pm 1.01\ \deg$} \\
$\vec{\Delta \mu_\mathrm{orb}} = \vec{\mu_\mathrm{orb}}(B) + (m_A/m_B)\, \vec{\mu_\mathrm{orb}}(A)$ & \multicolumn{2}{c}{$(+4.69 \pm1.13, +1.67 \pm 0.38)$ mas\,a$^{-1}$} \\
$\vec{\Delta \varv_\mathrm{orb}} = \vec{\varv_\mathrm{orb}}(B) + (m_A/m_B)\, \vec{\varv_\mathrm{orb}}(A)$ & \multicolumn{2}{c}{$(+77.7 \pm 18.7, +27.7 \pm 6.3)$ m\,s$^{-1}$} \\
$\vec{\Delta \varv_\mathrm{orb}}$ norm, PA & \multicolumn{2}{c}{$87.1 \pm 21.2$ m\,s$^{-1}$, $+74.6 \pm 6.3\,\deg$} \\
  \hline  \noalign{\smallskip}
\textbf{Gaia EDR3} \\
$\vec{\mu_\mathrm{orb}}$ (mas\,a$^{-1}$) & $(+31.49 \pm 0.77, +47.74 \pm 0.17)$ & $(-26.74 \pm 0.76,  -45.93 \pm 0.17)$ \\
$\vec{\mu_\mathrm{orb}}$ position angle $\theta$ & $33.41 \pm 0.66\ \deg$ & $210.21 \pm 0.70\ \deg$ \\
Diff. position angle  $\Delta \theta_{AB} = \theta(A) - \theta(B) + 180^\circ$ & \multicolumn{2}{c}{$3.20 \pm 0.96\ \deg$} \\
$\vec{\Delta \mu_\mathrm{orb}} = \vec{\mu_\mathrm{orb}}(B) + (m_A/m_B)\, \vec{\mu_\mathrm{orb}}(A)$ & \multicolumn{2}{c}{$(+4.75 \pm 1.08, +1.81 \pm 0.24)$ mas\,a$^{-1}$} \\
$\vec{\Delta \varv_\mathrm{orb}} = \vec{\varv_\mathrm{orb}}(B) +(m_A/m_B)\, \vec{\varv_\mathrm{orb}}(A)$ & \multicolumn{2}{c}{$(+78.7 \pm 17.9, +30.0 \pm 4.0)$ m\,s$^{-1}$} \\
$\vec{\Delta \varv_\mathrm{orb}}$ norm, PA & \multicolumn{2}{c}{$88.5 \pm 19.8$ m\,s$^{-1}$, $+73.5 \pm 5.4\,\deg$} \\
   \hline
\end{tabular}
\end{table*}

\begin{figure}
\includegraphics[width=\hsize]{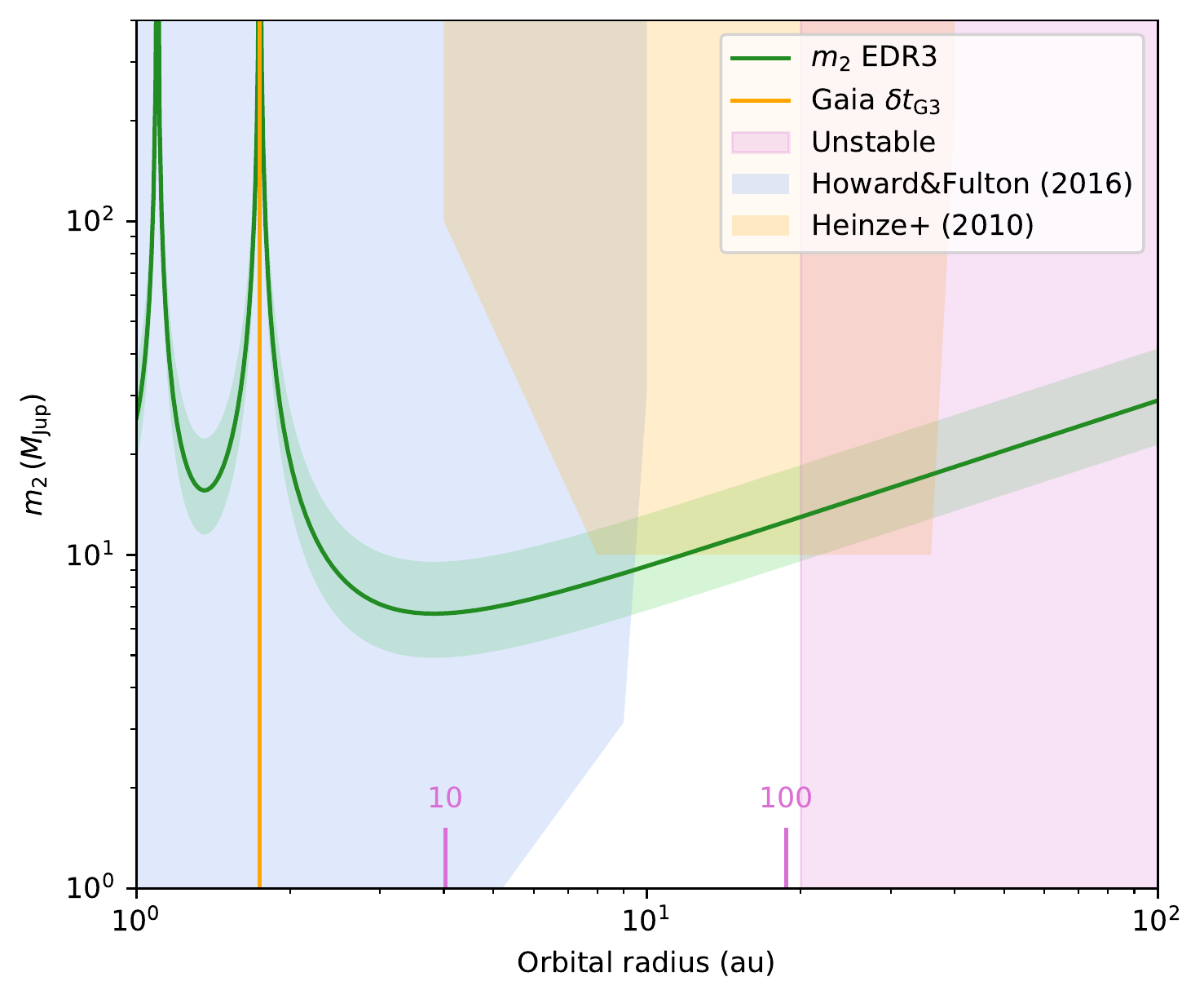}
\caption{Companion properties explaining the observed orbital velocity anomaly of 61\,Cyg AB. The regions excluded from radial velocity data by \citetads{2016PASP..128k4401H} (stars A and B) and from imaging by \citetads{2010ApJ...714.1551H} (star B only) are shown in shaded blue and orange, respectively. The unstable domain from interactions with the other component of 61\,Cyg \citepads{2005A&A...434..355M} is shown in light magenta.  \label{61Cyg-orbMass}}
\end{figure}

The measured orbital velocity anomaly is differential in nature between A and B and it is, in principle, not possible to determine around which of the two stars the companion is orbiting. Qualitatively, the orbital reflex motion due to the companion could result either in an increase or a decrease of the tangential orbital velocity of its host star, depending on the orbital phase.
In principle, the interpretation of the orbital velocity anomaly in terms of companion mass is similar to that of the PMa presented in Sect.~\ref{sensitivity_pma}. As Gaia PMs are average values over the measurement periods, we have a smearing of the velocity signature as in the case of the classical Hipparcos-Gaia PMa. However, as the orbital velocity anomaly is a differential quantity between two ``instantaneous'' velocities (of stars A and B), there is no decrease in sensitivity for very long orbital periods. 
The green domain in Fig.~\ref{61Cyg-orbMass} shows the range of possible combinations of companion mass and orbital radius that would explain the observed orbital velocity anomaly. The plot is drawn for the adopted mass of 61\,Cyg B ($0.657 \pm 0.057\ M_\odot$), but the figure is almost the same for component A. 

According to \citetads{2005A&A...434..355M}, stable orbits of S-type planets are expected for equal-mass binaries up to a star-planet separation of 0.22 times the stellar separation. With a semi-major axis of $a = 24.5\arcsec$ corresponding to $\approx 85$\,au (\citeads{2012A&A...546A..69M}; \citeads{2001AJ....122.3472H}), stable orbits are therefore expected within $\approx 20$\,au of each star. The shaded region in Fig.~\ref{61Cyg-orbMass} shows the domain of unstable orbits at larger separations. The constant velocity anomaly between the DR2 and EDR3 makes a short-period planet unlikely.
In 1943, \citetads{1943PASP...55...29S,1957AJ.....62Q..35S} announced the detection of a massive planet (or brown dwarf) orbiting around one of the components of 61\,Cyg with a period around 5\,years. The presence of a massive companion on such a short period orbit was later disproved by \citetads{1995Icar..116..359W} and \citetads{2008PASP..120..531C}. \citetads{2021AJ....161..134H} identified a low-amplitude radial velocity signal with $K=2.8$\,m\,s$^{-1}$ on 61\,Cyg A with a period of 2\,600\,days ($\approx 7$\,years), which they attributed to stellar activity (see also \citeads{2017ApJ...845...79B}) and classified as a false positive. \citetads{2017AJ....153..208B} found no significant RV signal on both the A or B components. Based on a 10\,000\,days time series of radial velocity measurements, Figs.\,83 and 84 of \citetads{2016PASP..128k4401H} show a non-excluded domain for a high-mass planetary companion of 61\,Cyg~A or B at a separation of 10\,au and above. A radial velocity signal at a level of several 10\,m\,s$^{-1}$ would likely have been detected by recent radial velocity surveys, possibly indicating a high inclination of the planetary orbit and a low radial velocity amplitude. From adaptive optics imaging in the infrared,  \citetads{2010ApJ...714.1551H} obtained detection limits of 8 to10\,$M_\mathrm{Jup}$ between 10 and 30\,au from 61\,Cyg~B (their Fig. 8). However, their assumed age of 2\,Ga for the system appears underestimated (\citeads{2008A&A...488..667K} obtain 6\,Ga), and this older age would result in increased mass detection limits.

 Combining the observed velocity anomaly with these observational constraints, the most probable properties of the exoplanet (or low-mass brown dwarf) present in the 61\,Cyg system are, therefore, a mass of $m_2 \approx 10\,M_\mathrm{Jup}$ and an orbital radius between $\approx 10$ and 20\,au (Fig.~\ref{61Cyg-orbMass}). Shorter orbital periods are in principle also possible in the case of high inclination orbits (see, e.g., \citeads{2021A&A...645A...7K}).
Assuming the same direction on sky as 61\,Cyg AB's orbit for the planetary companion's orbit, the companion would currently be located to the southeast of star B at an angular separation of 3 to $6\arcsec$ or, alternatively, to the northwest of star A within a similar separation range.

\subsection{Exoplanet host stars}

\subsubsection{Proxima Centauri}

The nearest star to the Sun, \object{Proxima Centauri} (\object{GJ 551}, \object{HIP 70890}) is a very low-mass M5.5Ve red dwarf that is a member of the $\alpha$\,Centauri triple system \citepads{2017A&A...598L...7K}. It orbits the main pair $\alpha$\,Cen AB \citepads{2016kervella, 2021A&A...646A...7S} with a very long period of more than 500\,000 years \citepads{2021AJ....162...14A}.
Although the $\alpha$\,Cen AB pair only has one unconfirmed candidate planet \citepads{2021NatCo..12.2651W, 2021NatCo..12..922W}, Proxima Cen hosts one confirmed terrestrial mass planet orbiting in its habitable zone, \object{Proxima b} \citepads{2016Natur.536..437A, 2017A&A...599A.126D, 2020A&A...639A..77S}. With an orbital period of only $P=11.2$\,d and a semi-major axis of $a=0.05$\,au, Proxima b is undetectable astrometrically from the Gaia DR2 or EDR3 catalog data, as these are, respectively, averaged over periods of approximately 2 and 3\,years. As it induces an expected astrometric wobble of less than $3\,\mu$as on its host star, the planet Proxima b will likely remain undetectable even from the individual epoch astrometry collected over the full Gaia mission.
Another candidate planet, Proxima c, has been detected by \citetads{2020SciA....6.7467D} using the radial velocity technique. With an estimated semi-major axis of $a_c = 1.5$\,au, corresponding to an orbital period of $P = 5.2 \pm 0.3$\,a and a radial velocity of $K_c = 1.2 \pm 0.4$\,m\,s$^{-1}$, its minimum mass is estimated to be $m_c \sin i = 5.7 \pm 1.9\,M_\oplus$. Thanks to its longer orbital period and larger expected astrometric signature, Proxima c is in principle detectable using Gaia astrometry. Taking advantage of the marginal Gaia DR2 PMa signal present at a $1.8\sigma$ level in Proxima Cen:

\begin{align}
\vec{\Delta \mu_\mathrm{G2}} & = (+0.218 \pm 0.112, +0.384 \pm 0.215)\ \mathrm{mas\,a}^{-1} ,\\
\vec{\Delta \varv_\mathrm{tan,G2}} & =  (+1.34 \pm 0.69, +2.37 \pm 1.33)\ \mathrm{m\,s}^{-1},
\end{align}

\citetads{2020A&A...635L..14K} determined the orbital inclination and a deprojected mass of $m_c = 12^{+12}_{-5}\,M_\oplus$. From HST-FGS astrometry of Proxima Cen, \citetads{2020RNAAS...4...46B} obtained a comparable mass of $m_c =  18 \pm 5\,M_\oplus$.

The Gaia EDR3 PMa signal is significantly lower than in the DR2  ($S/N = 0.9$):

\begin{align}
\vec{\Delta \mu_\mathrm{G3}} & = (-0.022 \pm 0.046, -0.069 \pm 0.069)\ \mathrm{mas\,a}^{-1} ,\\
\vec{\Delta \varv_\mathrm{tan,G3}} & =  (-0.14 \pm 0.28, -0.42 \pm 0.42)\ \mathrm{m\,s}^{-1}.
\end{align}

\begin{figure}
\includegraphics[width=\hsize]{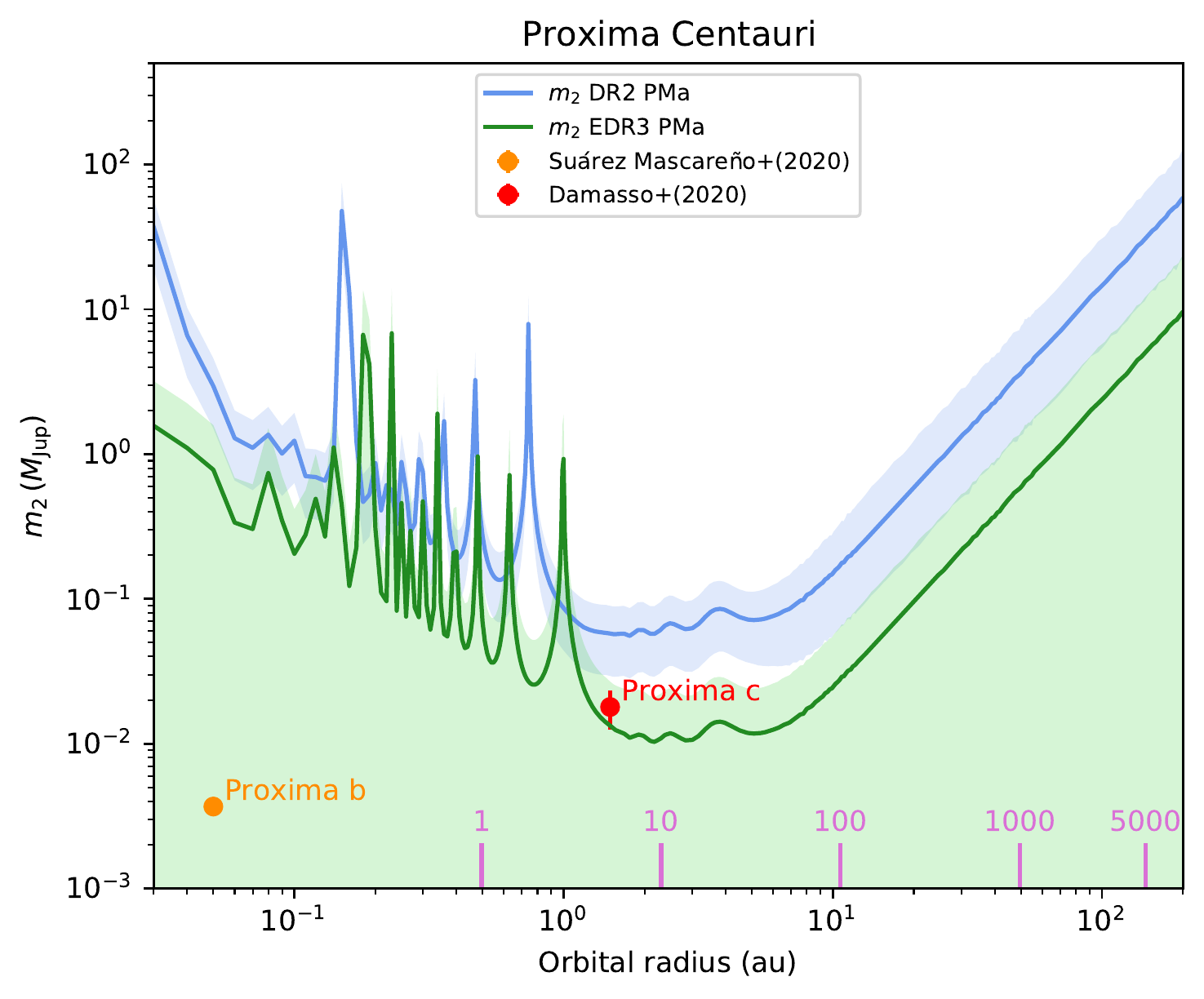}
\caption{PMa sensitivity diagram of Proxima for Gaia DR2 (blue) and EDR3 (green) measurements. The minimum masses of the planets Proxima b \citepads{2020A&A...639A..77S} and c \citetads{2020SciA....6.7467D} are represented with orange and red symbols, respectively. \label{ProximaCentauri-m2r}}
\end{figure}

As shown in Fig.~\ref{ProximaCentauri-m2r}, this PMa level is compatible with a deprojected mass for Proxima c closer to the minimum mass determined by \citetads{2020SciA....6.7467D} than the values estimated by \citetads{2020A&A...635L..14K} and \citetads{2020RNAAS...4...46B}. However, the orbital period of Proxima c ($P = 5.2$\,a) is only $1.8\times$ longer than the integration window of Gaia EDR3 ($\delta t_\mathrm{G3} = 2.8$\,a). Due to the associated smearing effect, this results in a lower sensitivity in the PMa signal, visible as a peak at 1\,au in the EDR3 curve of Fig.~\ref{ProximaCentauri-m2r}. As this decreased sensitivity peak will be further shifted toward larger orbital radii for a longer Gaia integration window, the astrometric signature of Proxima c will likely be detectable only in the epoch astrometry of Gaia (expected with the final Gaia data release).

\subsubsection{$\epsilon$ Eridani}

\begin{figure}
\includegraphics[width=\hsize]{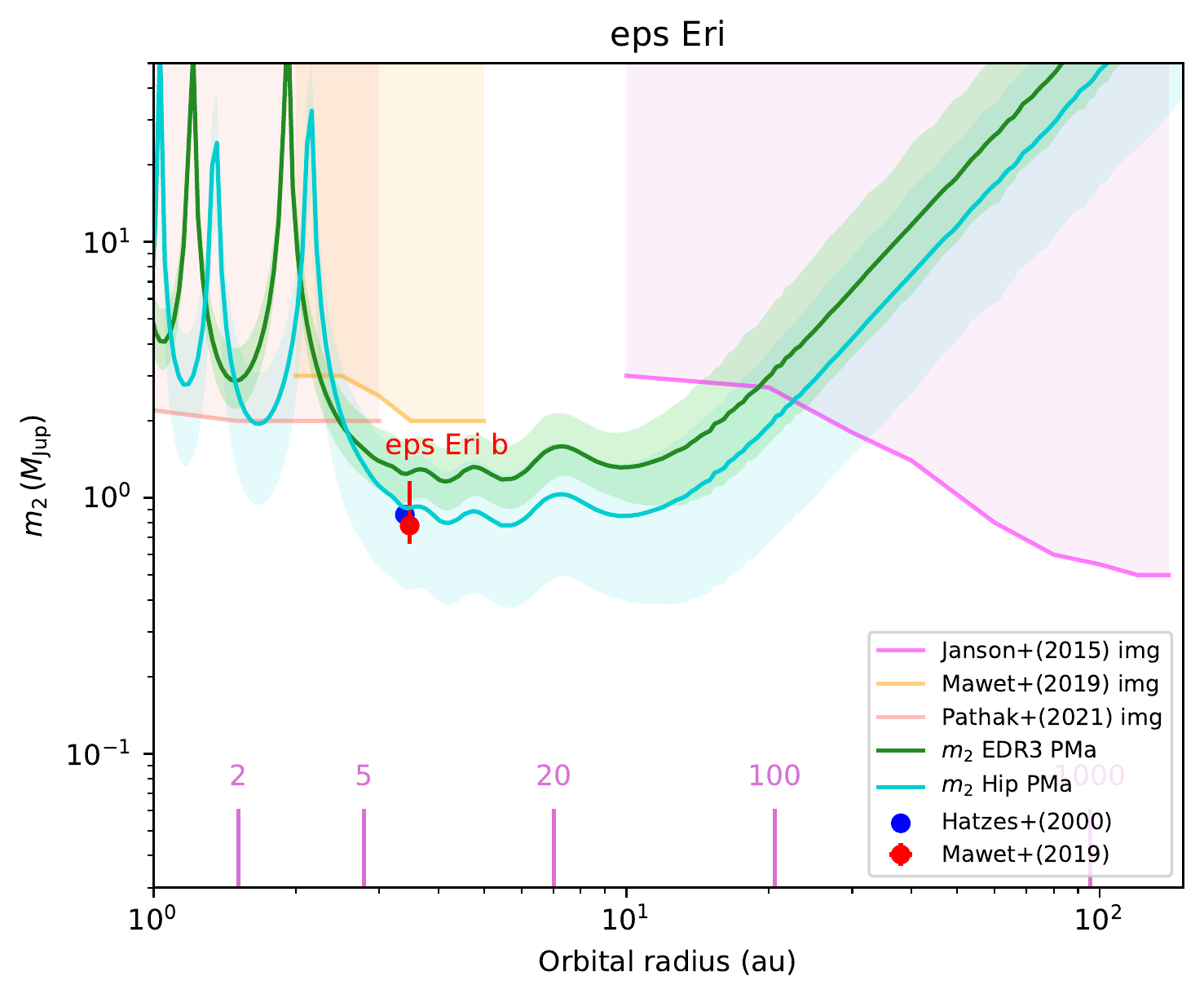}
\caption{PMa sensitivity diagram of $\epsilon$\,Eri for the Hipparcos (cyan) and EDR3 (green) measurements. The shaded regions represent the planet properties excluded by direct imaging searches. \label{epsEri-star-m2r}}
\end{figure}

The young K2V dwarf $\epsilon$\,Eri (\object{GJ 144}, \object{HIP 16357}, \object{HD 22049}) is located at a distance of only $d=3.2$\,pc. The presence of a massive planet orbiting this star was first proposed by \citetads{2000ApJ...544L.145H} from radial velocity data. The presence of this planet was confirmed by \citetads{1538-3881-157-1-33}, who also established its physical properties using the radial velocity technique ($m_b = 0.78^{+0.38}_{-0.12} M_\mathrm{Jup}$, $P_\mathrm{orb} = 7.37 \pm 0.07$\,a, $a = 3.48 \pm 0.02$\,au). However, direct imaging searches for exoplanets around $\epsilon$\,Eri (e.g., \citeads{pathak2021, 1538-3881-157-1-33, 2015A&A...574A.120J}) did not produce any detections.
\citetads{2021arXiv210701090M} analyzed the PM of $\epsilon$\,Eri based on astrometry with URAT telescope \citepads{2015AJ....150..101Z}, as well as Hipparcos and Gaia DR2 and EDR3, and obtained a tangential velocity anomaly of $\vec{\Delta \varv_\mathrm{tan}} = (+6, +13)$\,m\,s$^{-1}$ from the long-term Hipparcos+URAT and the Gaia EDR3 short-term PM, in good agreement with the value we obtain from Hipparcos and EDR3 $\vec{\Delta \varv_\mathrm{tan}}[EDR3] = (+4.7 \pm 2.4, +12.6 \pm 1.8)$\,m\,s$^{-1}$.
The PMa sensitivity diagram (Fig.~\ref{epsEri-star-m2r}) shows the good agreement of the Hipparcos and EDR3 PMa with the properties of $\epsilon$\,Eri b. The Gaia DR2 measurement is not represented as the accuracy of the PM vector is low (three times lower than Hipparcos) and, therefore, it does not set adequate constraints. The planetary properties excluded by the direct imaging searches by \citetads{pathak2021}, \citetads{1538-3881-157-1-33} and \citetads{2015A&A...574A.120J} are represented as shaded areas in Fig.~\ref{epsEri-star-m2r}. This diagram shows the very good complementarity of the astrometric, radial velocity, and direct imaging approaches to characterize planetary systems. We do not identify any CPM companion of $\epsilon$\,Eri in the Gaia EDR3 catalog.

\subsubsection{Kapteyn's star}

\begin{figure}
\includegraphics[width=\hsize]{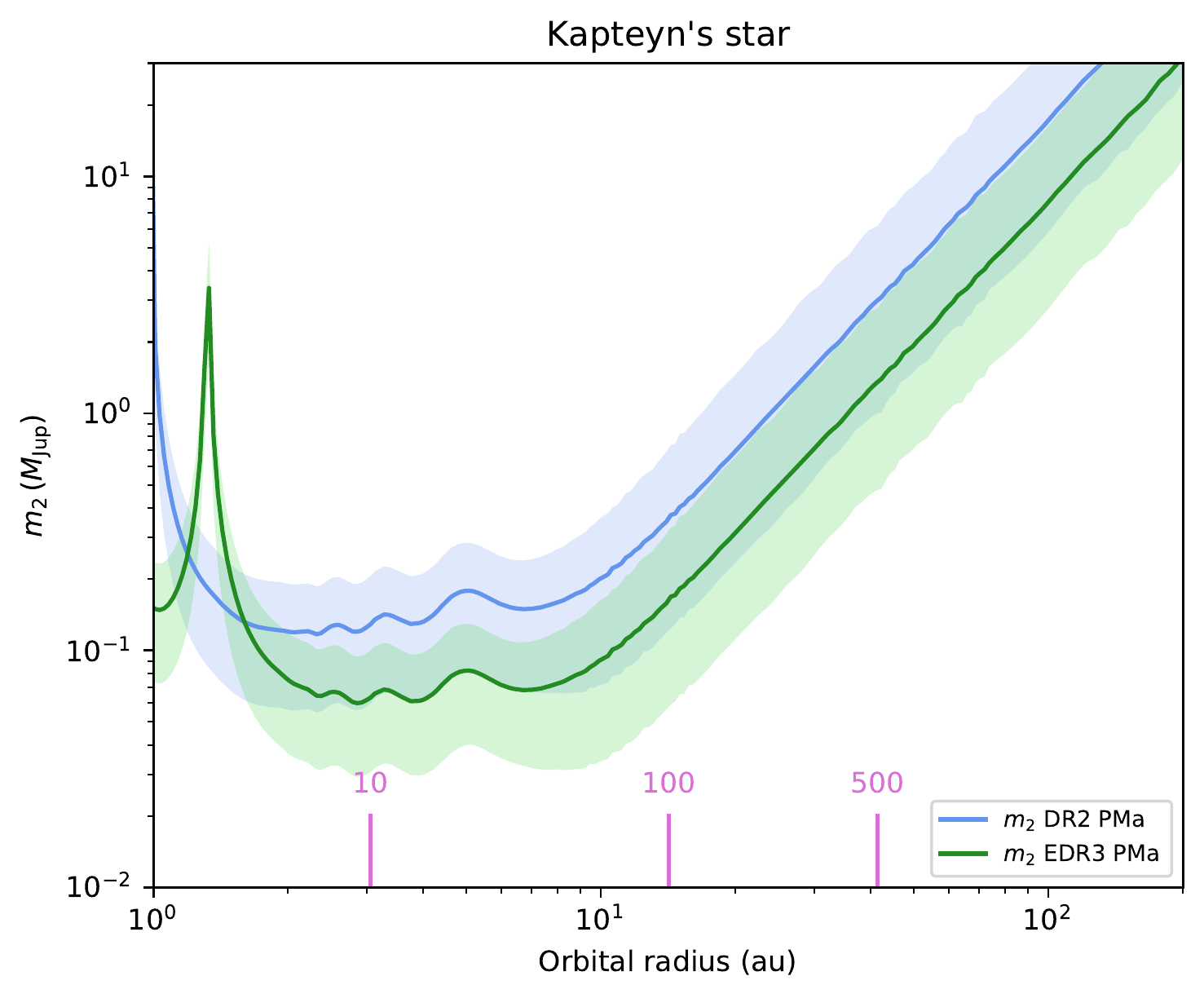}
\caption{PMa sensitivity diagram of Kapteyn's star for Gaia DR2 (blue) and EDR3 (green) proper motion measurements.  \label{Kapteyn-star-m2r}}
\end{figure}

We do not detect any significant PMa signal on the very low-mass red dwarf \object{Kapteyn's star} (\object{HIP 24186}, \object{GJ 191}, \object{HD 33793}) either from the DR2 or EDR3 measurements. The EDR3 residual tangential velocity anomaly is only $\Delta \varv_\mathrm{tan} = 1.46 \pm 0.84\,m\,s^{-1}$, that is, $S/N = 1.7$. This level of agreement between the Hipparcos-Gaia long-term PM vector and the short-term Gaia PM vector is remarkable when compared to the total space velocity of the star of more than 290\,km\,s$^{-1}$. As shown in Fig.~\ref{Kapteyn-star-m2r}, this corresponds to an upper limit of $0.1\,M_\mathrm{Jup}$ on the mass of a companion orbiting between 2 and 10\,au. This negative result is consistent with the non-detection of planetary companions of Kapteyn's star by \citetads{2021AJ....161..230B} from radial velocities.  We do not identify any CPM companion of Kapteyn's star in the Gaia EDR3 catalog.

\subsubsection{$\epsilon$ Indi}

\begin{figure}
\includegraphics[width=\hsize]{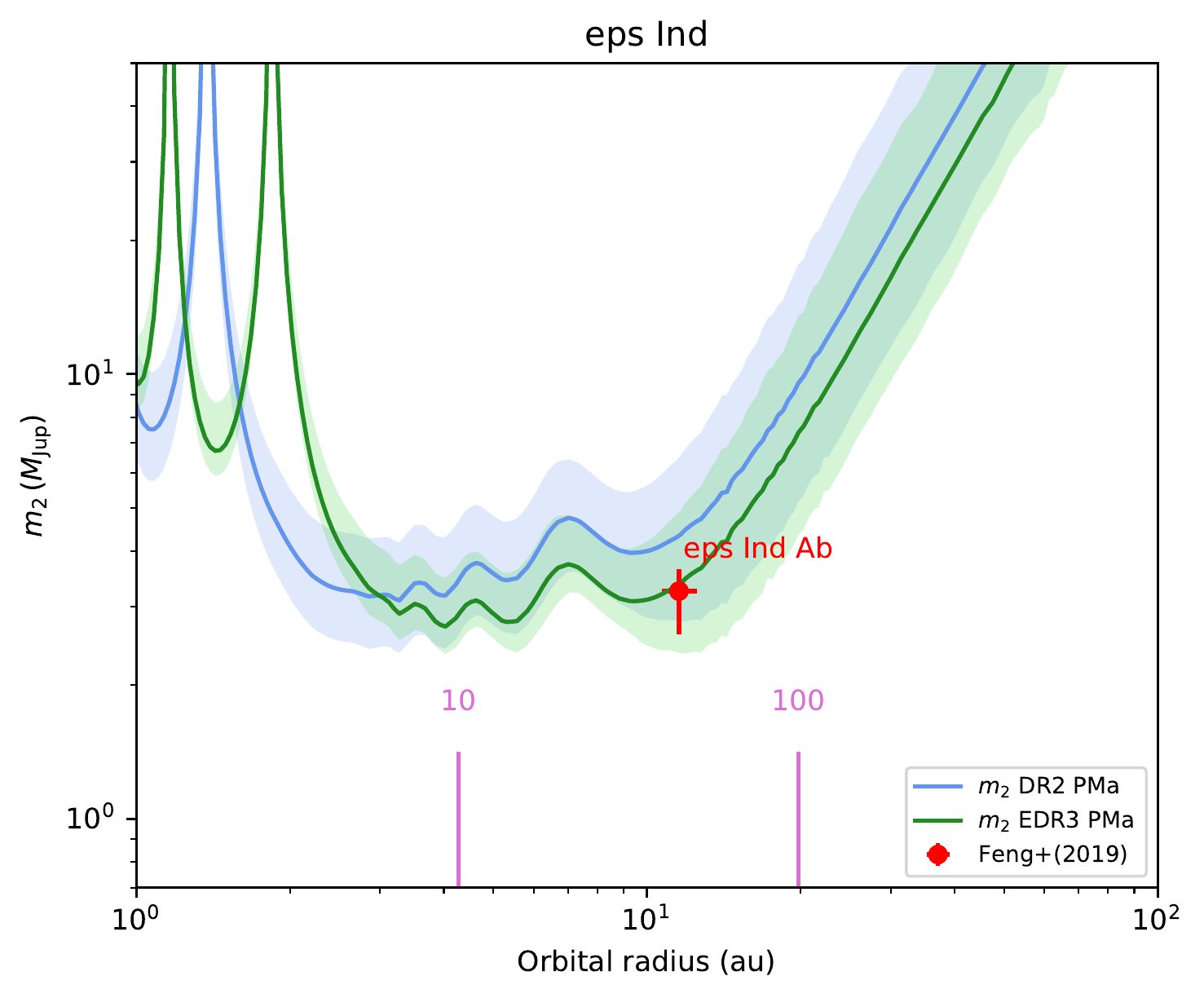}
\caption{PMa sensitivity diagram of $\epsilon$\,Ind A for Gaia DR2 (blue) and EDR3 (green) proper motion measurements. The properties of its massive planet $\epsilon$\,Ind Ab determined by  \citetads{2019MNRAS.490.5002F} are represented with a red point.\label{epsInd-m2r}}
\end{figure}

\begin{figure}
\includegraphics[width=\hsize,page=1]{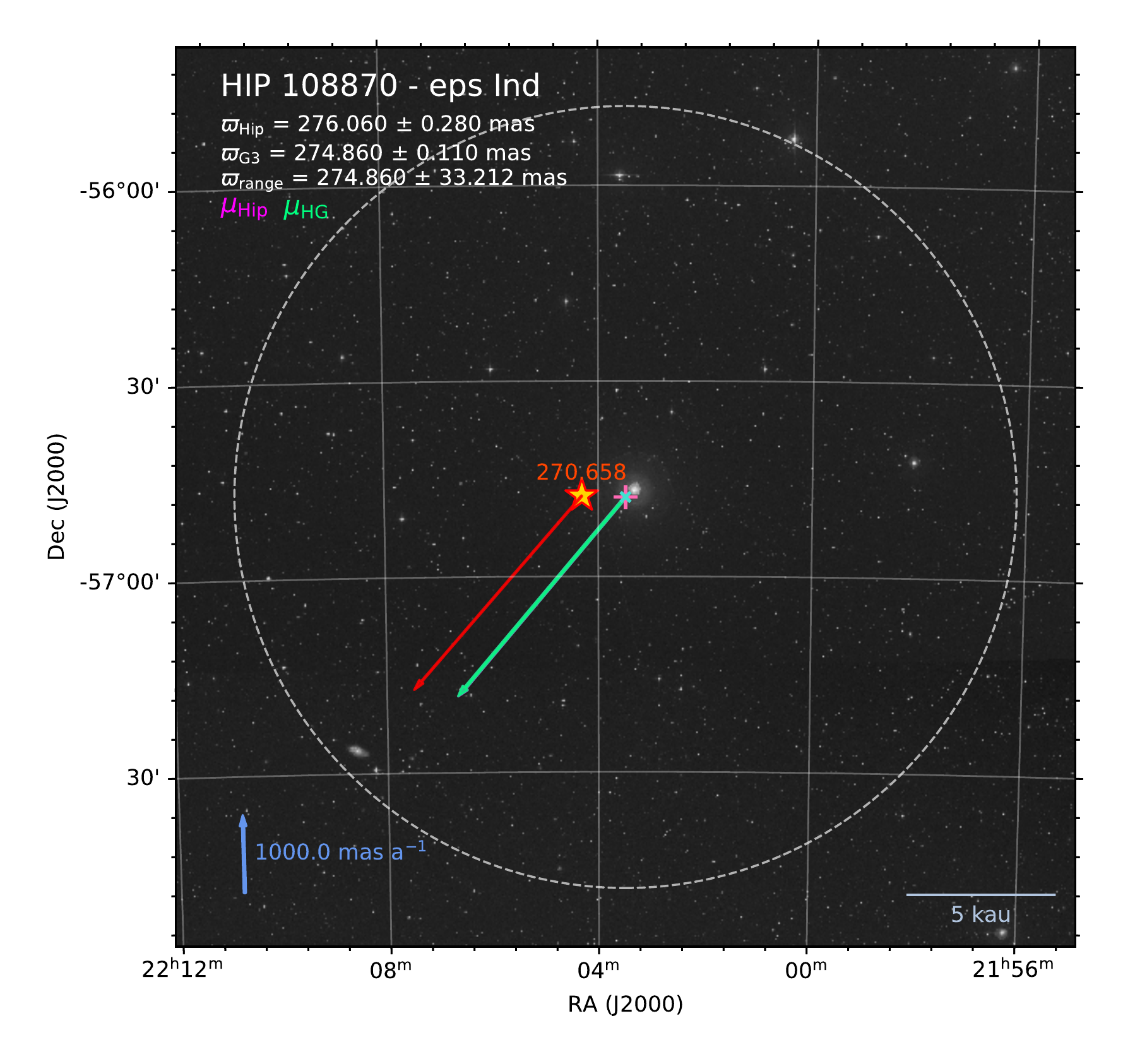}
\caption{Field chart of $\epsilon$\,Ind A with the binary brown dwarf companion $\epsilon$\,Ind~B.\label{HIP108870cpm}}
\end{figure}

The K5V primary star $\epsilon$\,Ind A (\object{GJ 845} A, \object{HIP 108870}) of the triple system $\epsilon$\,Ind hosts a massive exoplanet $\epsilon$\,Ind Ab. It was recently characterized by \citetads{2019MNRAS.490.5002F} as a cold and massive Jupiter analog ($m = 3\,M_\mathrm{Jup}$, $P_\mathrm{orb} = 45$\,a), based on a combination of radial velocity and astrometry from Hipparcos and Gaia. Recent attempts to directly image the planet $\epsilon$\,Ind~Ab in the thermal infrared domain by \citetads{pathak2021} and \citetads{2021A&A...651A..89V} were unsuccessful. We clearly detected the astrometric signature of this planet in the DR2 and EDR3 data, as shown in Fig.~\ref{epsInd-m2r}, with properties compatible with the determination by \citetads{2019MNRAS.490.5002F}.
The secondary $\epsilon$\,Ind B \citepads{2003A&A...398L..29S} is a binary brown dwarf system whose main component $\epsilon$\,Ind Ba (\object{Gaia EDR3 6412596012146801152}) is identified as a bound companion at a linear projected separation of 1.5\,kau and a relative tangential velocity $\Delta \varv_\mathrm{tan}= 1.25 \pm 0.01$\,km\,s$^{-1}$ (Fig.~\ref{HIP108870cpm}).

\subsubsection{$\pi$ Mensae}

\begin{figure}
\includegraphics[width=\hsize]{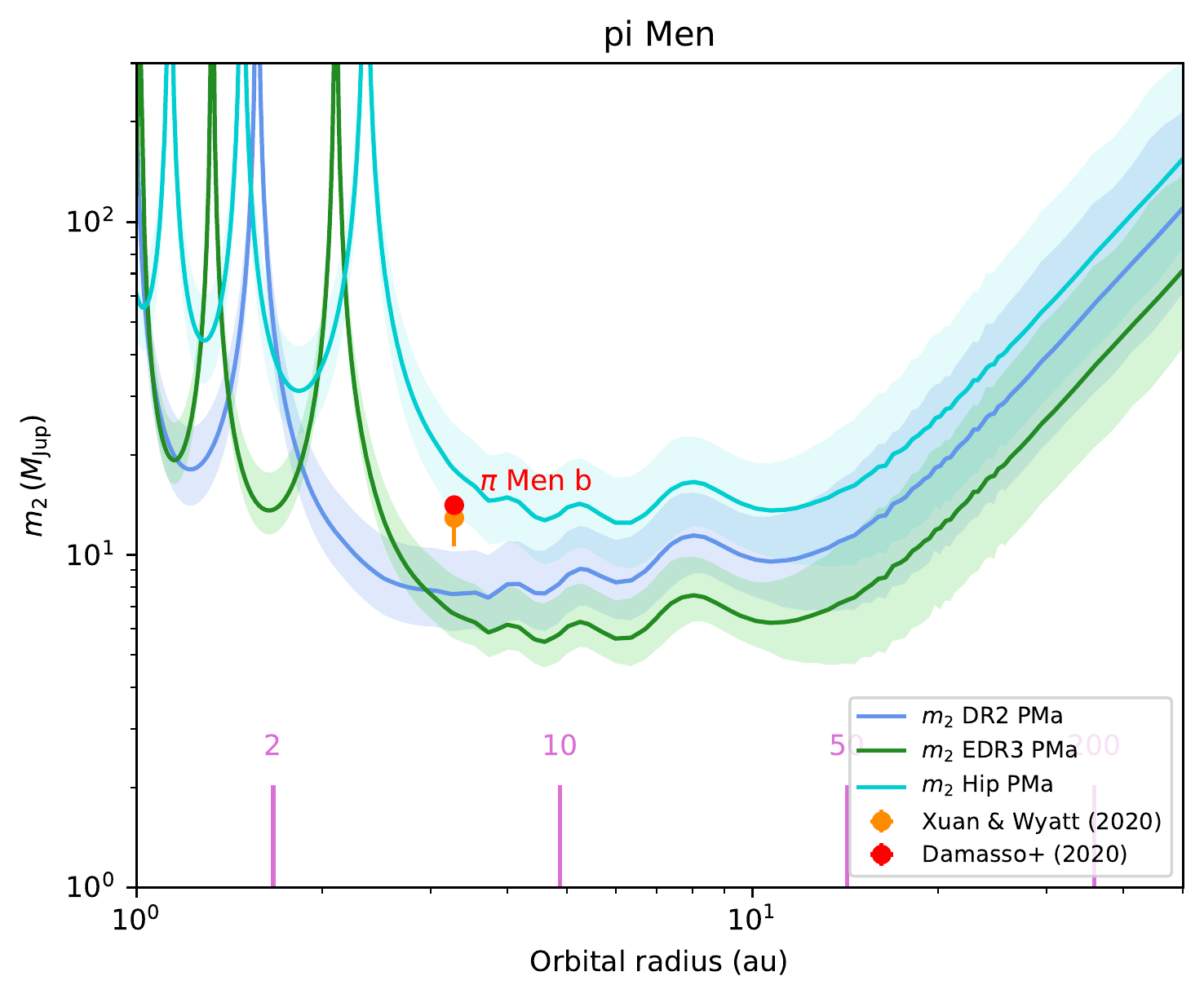}
\caption{PMa sensitivity diagram of $\pi$\,Men for the Hipparcos (cyan), DR2 (blue), and EDR3 (green) proper motion measurements. The properties of its massive planet $\pi$\,Men b determined by \citetads{2020MNRAS.497.2096X} and \citetads{2020A&A...642A..31D} are represented with a red and an orange point, respectively.\label{pi Men-m2r}}
\end{figure}

Based on radial velocity measurements, \citetads{2002MNRAS.333..871J} identified  a massive planet ($\pi$\,Men\,b) orbiting the nearby (18.3\,pc) high-velocity G0V dwarf  $\pi$\,Men (\object{HIP 26394}, \object{HD 39091}), with a period of 5.6\,years. Based on Hipparcos astrometry, \citetads{2011A&A...527A.140R} reported that this companion has a likely mass below $30\,M_\mathrm{Jup}$, and \citetads{2017ApJ...836..139F} classified this star as binary. The discovery of a transiting super-Earth ($\pi$\,Men\,c) with a mass around $5\,M_\oplus$ and an orbital period of 6\,days by \citetads{2018ApJ...868L..39H} and \citetads{2018A&A...619L..10G} considerably renewed the interest in the $\pi$\,Men system.
Using a combination of data sets, including Hipparcos and Gaia astrometry, the mutual inclination of the two planets was found to be remarkably high \citepads{2020MNRAS.497.2096X,2020A&A...640A..73D,2020A&A...642A..31D}.
Additionally, \citetads{2021MNRAS.502.2893K} found from transit spectroscopy that the rotation axis of the star is misaligned by $\approx 24\deg$ with the orbit of the inner super-Earth $\pi$\,Men\,c.
While the latter is beyond reach of a detection from the PMa technique with Gaia, planet b is well within its sensitivity range. We present the mass-orbital radius sensitivity diagram of $\pi$\,Men in Fig.~\ref{pi Men-m2r}. While the predicted mass-orbital radius domains are qualitatively in good agreement between the three catalogs, the PMa signal detected with the DR2 and EDR3 corresponds to a lower mass for planet b than the measured value (by 1 to $2\,\sigma$), while the Hipparcos PMa is slightly higher (by $1\,\sigma$).
These differences are due to the fact that the eccentricity of the orbit of $\pi$\,Men\,b is high at $e_b = 0.642$ \citepads{2020A&A...642A..31D}. A periastron passage of b occurred in J1990.1, within the measurement window of Hipparcos. Recent periastron passages of planet b occurred in J2013.0 and J2018.7, bracketing the measurement windows of Gaia DR2 ($J2014.6-J2016.4$) and EDR3 ($J2014.6-J2017.4$). This means that  Gaia observations essentially cover  the apastron of planet $\pi$\,Men\,b, and therefore give a slower tangential velocity anomaly for the star $\pi$\,Men.
This effect illustrates the limitation of the PMa analysis technique, which assumes a circular orbit for the companion, and accounts for the uncertainty in the inclination in a statistical manner. We did not find any resolved CPM companion of $\pi$\,Men in the Gaia EDR3 catalog.

\subsection{White dwarfs}

\begin{figure}
\includegraphics[width=\hsize]{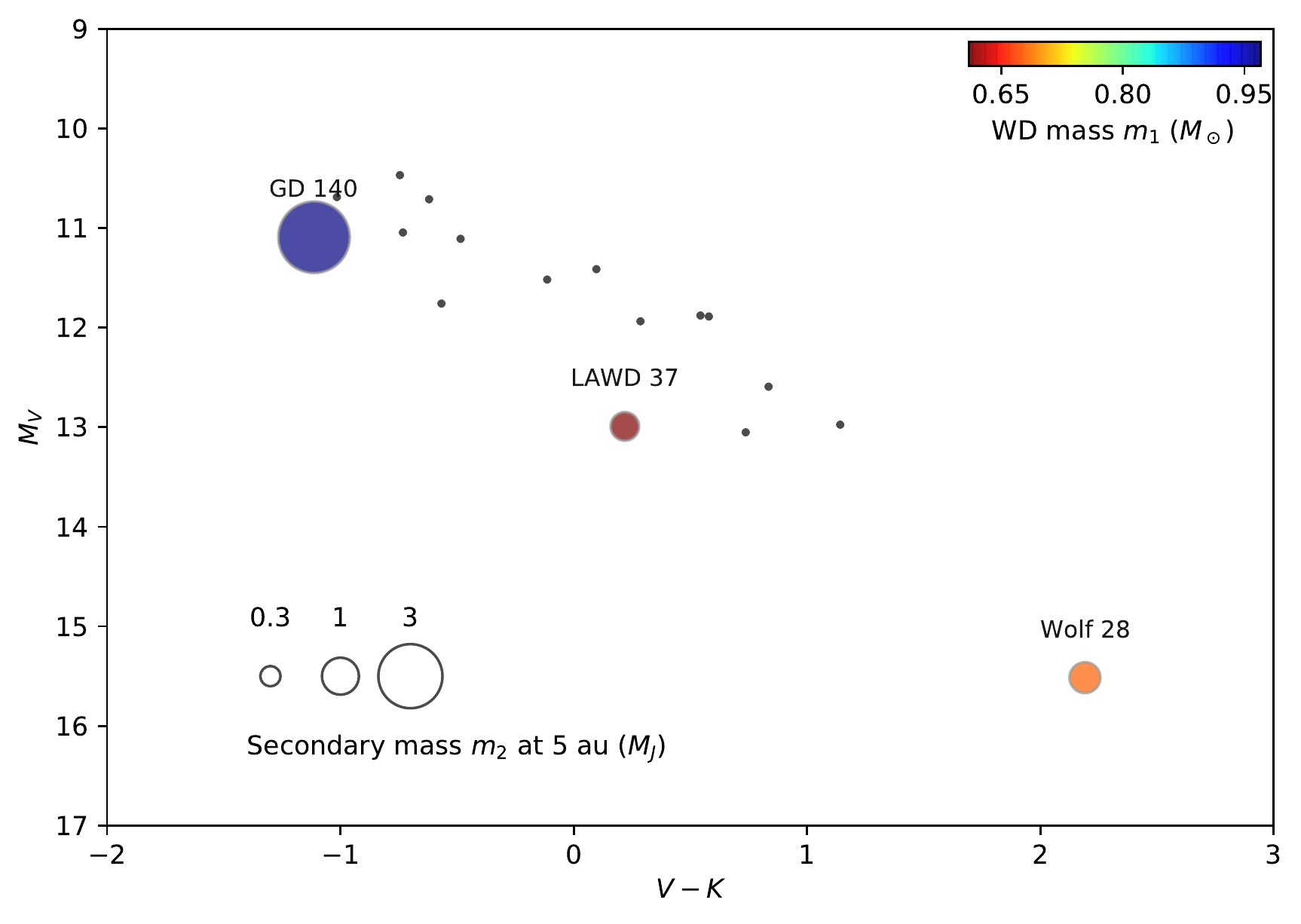}
\caption{Hertzsprung-Russell diagram of the Hipparcos catalog white dwarfs showing the detected EDR3 PMa signals.\label{Hip2-HR-WD}}
\end{figure}

We confirm the significant PMa signal detected by \citetads{2019A&A...623A..72K} in two of the 17 white dwarfs of the Hipparcos catalog (Table~\ref{wd-PMa}; Fig.~\ref{Hip2-HR-WD}): \object{GJ 140} and \object{LAWD 37}. As with the DR2 analysis, \object{Wolf 28} shows an indication of binarity at a $2\,\sigma$ level. The other white dwarfs do not show significant PMa signals, excluding the presence of Jupiter mass companions orbiting within a few astronomical units.
Our PMa sample is limited to the Hipparcos stars, but the Gaia white dwarf sample is naturally much larger (e.g., \citeads{2021arXiv210607669G,2021A&A...649A...6G}). Within 100\,pc, \citetads{2021MNRAS.506.5201R} identified a sample of 112 nearby white dwarf-main sequence binaries based on multi-band photometry.
The determination of the radial velocity of white dwarfs is complicated by the strong gravitational broadening of their spectral lines. As a result, their space velocity vector is affected by a larger uncertainty than normal stars and the PMa is more difficult to measure. Taking advantage of the perspective acceleration for nearby stars, the PMa may also be used to determine astrometrically the radial velocity of white dwarfs and other nearby stars \citepads{2021A&A...652A..45L, 1999A&A...348.1040D}.

\section{Conclusion\label{conclusion}}

Up to a distance of 100\,pc, the combined use of the PMa and CPM techniques enabled us to detect companions down to substellar or even planetary mass using the Gaia EDR3 catalog.
The brightest stars in the sky heavily saturate the Gaia detectors, and the PMa technique is therefore not directly applicable to these targets. We  identified, however, CPM companions based on their Hip2 PM and the EDR3 catalog of surrounding sources. We presented an updated version of the \citetads{2019A&A...623A..72K} catalog of PMa vectors for most of the Hip2 catalog stars, using the EDR3 positions and PM vectors.
We confirm the binary fraction obtained by \citetads{2021A&A...649A...6G}.
From a comparison with the results of our PMa survey, the Gaia RUWE appears as a valuable additional indicator for the presence of companions located within $\approx 1\arcsec$. Combining the PMa, CPM, and RUWE>1.4 indicators of binarity for the Hipparcos catalog stars results in a fraction of 43\% of the targets presenting a significant signal of binarity.
 We presented, as example applications of the PMa and CPM catalogs, analyses of bright star resolved companions, resolved binary stars with individual Gaia PMs, exoplanet host stars, and white dwarfs. We confirm the presence of a significant orbital motion anomaly in the nearby K dwarf binary 61\,Cyg~AB, which we attribute to a low-mass brown dwarf (or high-mass planet) orbiting one of the components. We also recover the perturbation induced by the massive planets orbiting $\epsilon$\,Eri, $\epsilon$\,Ind, and $\pi$\,Men on the PM of their parent stars.
 
The Gaia DR3 catalog will include solutions for unresolved binaries \citepads{2019MmSAI..90..318P} that will enable more refined determinations of the PMa vectors for the Hipparcos stars.
The remarkable complementarity of the PMa and CPM approaches opens up the possibility for testing the binarity of a large sample of objects in the solar neighborhood down to orbital periods of $\approx 3$\,years from the PMa approach, and up to separations of tens of thousands of astronomical units from the CPM approach.
The future availability (in Gaia DR4) of epoch astrometry will eventually waive the present time smearing limitation, and open up the possibility for directly searching for anomalies in the sky trajectory of all Gaia stars.
In synergy with the astrometry, the time series of Gaia photometric and spectroscopic measurements will expand the detection space toward companions with shorter orbital periods, through the transit and radial velocity techniques.
The expected extension of the duration of the Gaia mission up to 2025 will permit the detection of companions with longer orbital periods. As demonstrated in recent works (e.g., \citeads{2018NatAs...2..883S, 2019AJ....158..140B, 2019A&A...632L...9K, 2020A&A...635L..14K, 2021A&A...645A...7K, 2021arXiv210907525B}), the combination of Gaia astrometry with radial velocity and photometric transit measurements will result in highly accurate calibrations of the masses of a large number of planets and brown dwarfs.
Follow-up observations by narrow-angle astrometry using, for instance, adaptive optics \citepads{2010Natur.468.1080M, 2019A&A...621L...8L, 2020A&A...642A..18L}, GRAVITY interferometry \citepads{2019A&A...623L..11G, 2020A&A...633A.110G, 2020A&A...642L...2N, 2021A&A...652A..57K}, or ALMA imaging astrometry \citepads{2021AJ....162...14A, 2021ApJ...916L...2B} will further build up on the Hipparcos and Gaia astrometric measurements, potentially detecting second order astrometric perturbations.
The potential of an infrared astrometric space mission successor to Gaia for the detection and characterization of telluric mass planets is also outstanding \citepads{2019arXiv190712535H}, particularly for planets orbiting low-mass stars and brown dwarfs.

\begin{acknowledgements}
The authors gratefully thank the referee, Dr Andrei A. Tokovinin, for valuable comments and suggestions that led to significant improvements of this paper.
This work has made use of data from the European Space Agency (ESA) mission {\it Gaia} (\url{http://www.cosmos.esa.int/gaia}), processed by the {\it Gaia} Data Processing and Analysis Consortium (DPAC, \url{http://www.cosmos.esa.int/web/gaia/dpac/consortium}).
Funding for the DPAC has been provided by national institutions, in particular the institutions participating in the {\it Gaia} Multilateral Agreement.
The authors acknowledge the support of the French Agence Nationale de la Recherche (ANR), under grant ANR-15-CE31-0012-01 (project UnlockCepheids).
The research leading to these results  has received funding from the European Research Council (ERC) under the European Union's Horizon 2020 research and innovation program (project CepBin, grant agreement No 695099).
This research made use of Astropy\footnote{Available at \url{http://www.astropy.org/}}, a community-developed core Python package for Astronomy \citepads{2013A&A...558A..33A,2018AJ....156..123A}, of the Numpy library \citepads{Harris20} and of the Matplotlib graphics environment \citepads{Hunter:2007}.
This research has made use of the Washington Double Star Catalog maintained at the U.S. Naval Observatory.
We used the SIMBAD and VizieR databases and catalogue access tool at the CDS, Strasbourg (France), and NASA's Astrophysics Data System Bibliographic Services.
The original description of the VizieR service was published in \citetads{2000A&AS..143...23O}.
The Digitized Sky Surveys were produced at the Space Telescope Science Institute under U.S. Government grant NAG W-2166. The images of these surveys are based on photographic data obtained using the Oschin Schmidt Telescope on Palomar Mountain and the UK Schmidt Telescope. 
The UK Schmidt Telescope was operated by the Royal Observatory Edinburgh, with funding from the UK Science and Engineering Research Council, until 1988 June, and thereafter by the Anglo-Australian Observatory. Original plate material is copyright (c) of the Royal Observatory Edinburgh and the Anglo-Australian Observatory. The plates were processed into the present compressed digital form with the permission of these institutions.
This publication makes use of data products from the Two Micron All Sky Survey, which is a joint project of the University of Massachusetts and the Infrared Processing and Analysis Center/California Institute of Technology, funded by the National Aeronautics and Space Administration and the National Science Foundation.
\end{acknowledgements}

\bibliographystyle{aa} 
\bibliography{../Bibliography-Kervella}

\begin{thebibliography}{172}
\expandafter\ifx\csname natexlab\endcsname\relax\def\natexlab#1{#1}\fi

\bibitem[{{Abbott} {et~al.}(2017){Abbott}, {Abbott}, {Abbott}, {Acernese},
  {Ackley}, {Adams}, {Adams}, {Addesso}, {Adhikari}, {Adya}, {Affeldt},
  {Afrough}, {Agarwal}, {Agathos}, {Agatsuma}, {Aggarwal}, {Aguiar}, {Aiello},
  {Ain}, {Ajith}, {Allen}, {Allen}, {Allocca}, {Altin}, {Amato}, {Ananyeva},
  {Anderson}, {Anderson}, {Angelova}, {Antier}, {Appert}, {Arai}, {Araya},
  {Areeda}, {Arnaud}, {Arun}, {Ascenzi}, {Ashton}, {Ast}, {Aston}, {Astone},
  {Atallah}, {Aufmuth}, {Aulbert}, {AultONeal}, {Austin}, {Avila-Alvarez},
  {Babak}, {Bacon}, {Bader}, {Bae}, {Bailes}, {Baker}, {Baldaccini},
  {Ballardin}, {Ballmer}, {Banagiri}, {Barayoga}, {Barclay}, {Barish},
  {Barker}, {Barkett}, {Barone}, {Barr}, {Barsotti}, {Barsuglia}, {Barta},
  {Barthelmy}, {Bartlett}, {Bartos}, {Bassiri}, {Basti}, {Batch}, {Bawaj},
  {Bayley}, {Bazzan}, {B{\'e}csy}, {Beer}, {Bejger}, {Belahcene}, {Bell},
  {Berger}, {Bergmann}, {Bernuzzi}, {Bero}, {Berry}, {Bersanetti}, {Bertolini},
  {Betzwieser}, {Bhagwat}, {Bhandare}, {Bilenko}, {Billingsley}, {Billman},
  {Birch}, {Birney}, {Birnholtz}, {Biscans}, {Biscoveanu}, {Bisht}, {Bitossi},
  {Biwer}, {Bizouard}, {Blackburn}, {Blackman}, {Blair}, {Blair}, {Blair},
  {Bloemen}, {Bock}, {Bode}, {Boer}, {Bogaert}, {Bohe}, {Bondu}, {Bonilla},
  {Bonnand}, {Boom}, {Bork}, {Boschi}, {Bose}, {Bossie}, {Bouffanais}, {Bozzi},
  {Bradaschia}, {Brady}, {Branchesi}, {Brau}, {Briant}, {Brillet}, {Brinkmann},
  {Brisson}, {Brockill}, {Broida}, {Brooks}, {Brown}, {Brown}, {Brunett},
  {Buchanan}, {Buikema}, {Bulik}, {Bulten}, {Buonanno}, {Buskulic}, {Buy},
  {Byer}, {Cabero}, {Cadonati}, {Cagnoli}, {Cahillane}, {Calder{\'o}n
  Bustillo}, {Callister}, {Calloni}, {Camp}, {Canepa}, {Canizares}, {Cannon},
  {Cao}, {Cao}, {Capano}, {Capocasa}, {Carbognani}, {Caride}, {Carney},
  {Carullo}, {Casanueva Diaz}, {Casentini}, {Caudill}, {Cavagli{\`a}},
  {Cavalier}, {Cavalieri}, {Cella}, {Cepeda}, {Cerd{\'a}-Dur{\'a}n},
  {Cerretani}, {Cesarini}, {Chamberlin}, {Chan}, {Chao}, {Charlton}, {Chase},
  {Chassande-Mottin}, {Chatterjee}, {Chatziioannou}, {Cheeseboro}, {Chen},
  {Chen}, {Chen}, {Cheng}, {Chia}, {Chincarini}, {Chiummo}, {Chmiel}, {Cho},
  {Cho}, {Chow}, {Christensen}, {Chu}, {Chua}, {Chua}, {Chung}, {Chung},
  {Ciani}, {Ciolfi}, {Cirelli}, {Cirone}, {Clara}, {Clark}, {Clearwater},
  {Cleva}, {Cocchieri}, {Coccia}, {Cohadon}, {Cohen}, {Colla}, {Collette},
  {Cominsky}, {Constancio}, {Conti}, {Cooper}, {Corban}, {Corbitt},
  {Cordero-Carri{\'o}n}, {Corley}, {Cornish}, {Corsi}, {Cortese}, {Costa},
  {Coughlin}, {Coughlin}, {Coulon}, {Countryman}, {Couvares}, {Covas}, {Cowan},
  {Coward}, {Cowart}, {Coyne}, {Coyne}, {Creighton}, {Creighton}, {Cripe},
  {Crowder}, {Cullen}, {Cumming}, {Cunningham}, {Cuoco}, {Dal Canton},
  {D{\'a}lya}, {Danilishin}, {D'Antonio}, {Danzmann}, {Dasgupta}, {Da Silva
  Costa}, {Dattilo}, {Dave}, {Davier}, {Davis}, {Daw}, {Day}, {De}, {DeBra},
  {Degallaix}, {De Laurentis}, {Del{\'e}glise}, {Del Pozzo}, {Demos}, {Denker},
  {Dent}, {De Pietri}, {Dergachev}, {De Rosa}, {DeRosa}, {De Rossi}, {DeSalvo},
  {de Varona}, {Devenson}, {Dhurandhar}, {D{\'\i}az}, {Dietrich}, {Di Fiore},
  {Di Giovanni}, {Di Girolamo}, {Di Lieto}, {Di Pace}, {Di Palma}, {Di Renzo},
  {Doctor}, {Dolique}, {Donovan}, {Dooley}, {Doravari}, {Dorrington},
  {Douglas}, {Dovale {\'A}lvarez}, {Downes}, {Drago}, {Dreissigacker},
  {Driggers}, {Du}, {Ducrot}, {Dudi}, {Dupej}, {Dwyer}, {Edo}, {Edwards},
  {Effler}, {Eggenstein}, {Ehrens}, {Eichholz}, {Eikenberry}, {Eisenstein},
  {Essick}, {Estevez}, {Etienne}, {Etzel}, {Evans}, {Evans}, {Factourovich},
  {Fafone}, {Fair}, {Fairhurst}, {Fan}, {Farinon}, {Farr}, {Farr},
  {Fauchon-Jones}, {Favata}, {Fays}, {Fee}, {Fehrmann}, {Feicht}, {Fejer},
  {Fernandez-Galiana}, {Ferrante}, {Ferreira}, {Ferrini}, {Fidecaro},
  {Finstad}, {Fiori}, {Fiorucci}, {Fishbach}, {Fisher}, {Fitz-Axen},
  {Flaminio}, {Fletcher}, {Fong}, {Font}, {Forsyth}, {Forsyth}, {Fournier},
  {Frasca}, {Frasconi}, {Frei}, {Freise}, {Frey}, {Frey}, {Fries}, {Fritschel},
  {Frolov}, {Fulda}, {Fyffe}, {Gabbard}, {Gadre}, {Gaebel}, {Gair},
  {Gammaitoni}, {Ganija}, {Gaonkar}, {Garcia-Quiros}, {Garufi}, {Gateley},
  {Gaudio}, {Gaur}, {Gayathri}, {Gehrels}, {Gemme}, {Genin}, {Gennai},
  {George}, {George}, {Gergely}, {Germain}, {Ghonge}, {Ghosh}, {Ghosh},
  {Ghosh}, {Giaime}, {Giardina}, {Giazotto}, {Gill}, {Glover}, {Goetz},
  {Goetz}, {Gomes}, {Goncharov}, {Gonz{\'a}lez}, {Gonzalez Castro},
  {Gopakumar}, {Gorodetsky}, {Gossan}, {Gosselin}, {Gouaty}, {Grado}, {Graef},
  {Granata}, {Grant}, {Gras}, {Gray}, {Greco}, {Green}, {Gretarsson}, {Groot},
  {Grote}, {Grunewald}, {Gruning}, {Guidi}, {Guo}, {Gupta}, {Gupta}, {Gushwa},
  {Gustafson}, {Gustafson}, {Halim}, {Hall}, {Hall}, {Hamilton}, {Hammond},
  {Haney}, {Hanke}, {Hanks}, {Hanna}, {Hannam}, {Hannuksela}, {Hanson},
  {Hardwick}, {Harms}, {Harry}, {Harry}, {Hart}, {Haster}, {Haughian}, {Healy},
  {Heidmann}, {Heintze}, {Heitmann}, {Hello}, {Hemming}, {Hendry}, {Heng},
  {Hennig}, {Heptonstall}, {Heurs}, {Hild}, {Hinderer}, {Ho}, {Hoak}, {Hofman},
  {Holt}, {Holz}, {Hopkins}, {Horst}, {Hough}, {Houston}, {Howell}, {Hreibi},
  {Hu}, {Huerta}, {Huet}, {Hughey}, {Husa}, {Huttner}, {Huynh-Dinh}, {Indik},
  {Inta}, {Intini}, {Isa}, {Isac}, {Isi}, {Iyer}, {Izumi}, {Jacqmin}, {Jani},
  {Jaranowski}, {Jawahar}, {Jim{\'e}nez-Forteza}, {Johnson},
  {Johnson-McDaniel}, {Jones}, {Jones}, {Jonker}, {Ju}, {Junker}, {Kalaghatgi},
  {Kalogera}, {Kamai}, {Kandhasamy}, {Kang}, {Kanner}, {Kapadia}, {Karki},
  {Karvinen}, {Kasprzack}, {Kastaun}, {Katolik}, {Katsavounidis}, {Katzman},
  {Kaufer}, {Kawabe}, {K{\'e}f{\'e}lian}, {Keitel}, {Kemball}, {Kennedy},
  {Kent}, {Key}, {Khalili}, {Khan}, {Khan}, {Khan}, {Khazanov}, {Kijbunchoo},
  {Kim}, {Kim}, {Kim}, {Kim}, {Kim}, {Kim}, {Kimbrell}, {King}, {King},
  {Kinley-Hanlon}, {Kirchhoff}, {Kissel}, {Kleybolte}, {Klimenko}, {Knowles},
  {Koch}, {Koehlenbeck}, {Koley}, {Kondrashov}, {Kontos}, {Korobko}, {Korth},
  {Kowalska}, {Kozak}, {Kr{\"a}mer}, {Kringel}, {Krishnan}, {Kr{\'o}lak},
  {Kuehn}, {Kumar}, {Kumar}, {Kumar}, {Kuo}, {Kutynia}, {Kwang}, {Lackey},
  {Lai}, {Landry}, {Lang}, {Lange}, {Lantz}, {Lanza}, {Larson},
  {Lartaux-Vollard}, {Lasky}, {Laxen}, {Lazzarini}, {Lazzaro}, {Leaci},
  {Leavey}, {Lee}, {Lee}, {Lee}, {Lee}, {Lee}, {Lehmann}, {Lenon}, {Leon},
  {Leonardi}, {Leroy}, {Letendre}, {Levin}, {Li}, {Linker}, {Littenberg},
  {Liu}, {Liu}, {Lo}, {Lockerbie}, {London}, {Lord}, {Lorenzini}, {Loriette},
  {Lormand}, {Losurdo}, {Lough}, {Lousto}, {Lovelace}, {L{\"u}ck}, {Lumaca},
  {Lundgren}, {Lynch}, {Ma}, {Macas}, {Macfoy}, {Machenschalk}, {MacInnis},
  {Macleod}, {Maga{\~n}a Hernandez}, {Maga{\~n}a-Sandoval}, {Maga{\~n}a
  Zertuche}, {Magee}, {Majorana}, {Maksimovic}, {Man}, {Mandic}, {Mangano},
  {Mansell}, {Manske}, {Mantovani}, {Marchesoni}, {Marion}, {M{\'a}rka},
  {M{\'a}rka}, {Markakis}, {Markosyan}, {Markowitz}, {Maros}, {Marquina},
  {Marsh}, {Martelli}, {Martellini}, {Martin}, {Martin}, {Martynov}, {Marx},
  {Mason}, {Massera}, {Masserot}, {Massinger}, {Masso-Reid}, {Mastrogiovanni},
  {Matas}, {Matichard}, {Matone}, {Mavalvala}, {Mazumder}, {McCarthy},
  {McClelland}, {McCormick}, {McCuller}, {McGuire}, {McIntyre}, {McIver},
  {McManus}, {McNeill}, {McRae}, {McWilliams}, {Meacher}, {Meadors}, {Mehmet},
  {Meidam}, {Mejuto-Villa}, {Melatos}, {Mendell}, {Mercer}, {Merilh},
  {Merzougui}, {Meshkov}, {Messenger}, {Messick}, {Metzdorff}, {Meyers},
  {Miao}, {Michel}, {Middleton}, {Mikhailov}, {Milano}, {Miller}, {Miller},
  {Miller}, {Millhouse}, {Milovich-Goff}, {Minazzoli}, {Minenkov}, {Ming},
  {Mishra}, {Mitra}, {Mitrofanov}, {Mitselmakher}, {Mittleman}, {Moffa},
  {Moggi}, {Mogushi}, {Mohan}, {Mohapatra}, {Molina}, {Montani}, {Moore},
  {Moraru}, {Moreno}, {Morisaki}, {Morriss}, {Mours}, {Mow-Lowry}, {Mueller},
  {Muir}, {Mukherjee}, {Mukherjee}, {Mukherjee}, {Mukund}, {Mullavey}, {Munch},
  {Mu{\~n}iz}, {Muratore}, {Murray}, {Nagar}, {Napier}, {Nardecchia},
  {Naticchioni}, {Nayak}, {Neilson}, {Nelemans}, {Nelson}, {Nery}, {Neunzert},
  {Nevin}, {Newport}, {Newton}, {Ng}, {Nguyen}, {Nguyen}, {Nichols}, {Nielsen},
  {Nissanke}, {Nitz}, {Noack}, {Nocera}, {Nolting}, {North}, {Nuttall},
  {Oberling}, {O'Dea}, {Ogin}, {Oh}, {Oh}, {Ohme}, {Okada}, {Oliver},
  {Oppermann}, {Oram}, {O'Reilly}, {Ormiston}, {Ortega}, {O'Shaughnessy},
  {Ossokine}, {Ottaway}, {Overmier}, {Owen}, {Pace}, {Page}, {Page}, {Pai},
  {Pai}, {Palamos}, {Palashov}, {Palomba}, {Pal-Singh}, {Pan}, {Pan}, {Pang},
  {Pang}, {Pankow}, {Pannarale}, {Pant}, {Paoletti}, {Paoli}, {Papa}, {Parida},
  {Parker}, {Pascucci}, {Pasqualetti}, {Passaquieti}, {Passuello}, {Patil},
  {Patricelli}, {Pearlstone}, {Pedraza}, {Pedurand}, {Pekowsky}, {Pele},
  {Penn}, {Perez}, {Perreca}, {Perri}, {Pfeiffer}, {Phelps}, {Piccinni},
  {Pichot}, {Piergiovanni}, {Pierro}, {Pillant}, {Pinard}, {Pinto}, {Pirello},
  {Pitkin}, {Poe}, {Poggiani}, {Popolizio}, {Porter}, {Post}, {Powell},
  {Prasad}, {Pratt}, {Pratten}, {Predoi}, {Prestegard}, {Prijatelj},
  {Principe}, {Privitera}, {Prix}, {Prodi}, {Prokhorov}, {Puncken}, {Punturo},
  {Puppo}, {P{\"u}rrer}, {Qi}, {Quetschke}, {Quintero}, {Quitzow-James},
  {Raab}, {Rabeling}, {Radkins}, {Raffai}, {Raja}, {Rajan}, {Rajbhandari},
  {Rakhmanov}, {Ramirez}, {Ramos-Buades}, {Rapagnani}, {Raymond}, {Razzano},
  {Read}, {Regimbau}, {Rei}, {Reid}, {Reitze}, {Ren}, {Reyes}, {Ricci},
  {Ricker}, {Rieger}, {Riles}, {Rizzo}, {Robertson}, {Robie}, {Robinet},
  {Rocchi}, {Rolland}, {Rollins}, {Roma}, {Romano}, {Romano}, {Romel}, {Romie},
  {Rosi{\'n}ska}, {Ross}, {Rowan}, {R{\"u}diger}, {Ruggi}, {Rutins}, {Ryan},
  {Sachdev}, {Sadecki}, {Sadeghian}, {Sakellariadou}, {Salconi}, {Saleem},
  {Salemi}, {Samajdar}, {Sammut}, {Sampson}, {Sanchez}, {Sanchez},
  {Sanchis-Gual}, {Sandberg}, {Sanders}, {Sassolas}, {Sathyaprakash},
  {Saulson}, {Sauter}, {Savage}, {Sawadsky}, {Schale}, {Scheel}, {Scheuer},
  {Schmidt}, {Schmidt}, {Schnabel}, {Schofield}, {Sch{\"o}nbeck}, {Schreiber},
  {Schuette}, {Schulte}, {Schutz}, {Schwalbe}, {Scott}, {Scott}, {Seidel},
  {Sellers}, {Sengupta}, {Sentenac}, {Sequino}, {Sergeev}, {Shaddock},
  {Shaffer}, {Shah}, {Shahriar}, {Shaner}, {Shao}, {Shapiro}, {Shawhan},
  {Sheperd}, {Shoemaker}, {Shoemaker}, {Siellez}, {Siemens}, {Sieniawska},
  {Sigg}, {Silva}, {Singer}, {Singh}, {Singhal}, {Sintes}, {Slagmolen},
  {Smith}, {Smith}, {Smith}, {Somala}, {Son}, {Sonnenberg}, {Sorazu},
  {Sorrentino}, {Souradeep}, {Spencer}, {Srivastava}, {Staats}, {Staley},
  {Steinke}, {Steinlechner}, {Steinlechner}, {Steinmeyer}, {Stevenson},
  {Stone}, {Stops}, {Strain}, {Stratta}, {Strigin}, {Strunk}, {Sturani},
  {Stuver}, {Summerscales}, {Sun}, {Sunil}, {Suresh}, {Sutton}, {Swinkels},
  {Szczepa{\'n}czyk}, {Tacca}, {Tait}, {Talbot}, {Talukder}, {Tanner},
  {T{\'a}pai}, {Taracchini}, {Tasson}, {Taylor}, {Taylor}, {Tewari}, {Theeg},
  {Thies}, {Thomas}, {Thomas}, {Thomas}, {Thorne}, {Thorne}, {Thrane},
  {Tiwari}, {Tiwari}, {Tokmakov}, {Toland}, {Tonelli}, {Tornasi},
  {Torres-Forn{\'e}}, {Torrie}, {T{\"o}yr{\"a}}, {Travasso}, {Traylor},
  {Trinastic}, {Tringali}, {Trozzo}, {Tsang}, {Tse}, {Tso}, {Tsukada}, {Tsuna},
  {Tuyenbayev}, {Ueno}, {Ugolini}, {Unnikrishnan}, {Urban}, {Usman},
  {Vahlbruch}, {Vajente}, {Valdes}, {Vallisneri}, {van Bakel}, {van Beuzekom},
  {van den Brand}, {Van Den Broeck}, {Vander-Hyde}, {van der Schaaf}, {van
  Heijningen}, {van Veggel}, {Vardaro}, {Varma}, {Vass}, {Vas{\'u}th},
  {Vecchio}, {Vedovato}, {Veitch}, {Veitch}, {Venkateswara}, {Venugopalan},
  {Verkindt}, {Vetrano}, {Vicer{\'e}}, {Viets}, {Vinciguerra}, {Vine}, {Vinet},
  {Vitale}, {Vo}, {Vocca}, {Vorvick}, {Vyatchanin}, {Wade}, {Wade}, {Wade},
  {Walet}, {Walker}, {Wallace}, {Walsh}, {Wang}, {Wang}, {Wang}, {Wang},
  {Wang}, {Ward}, {Warner}, {Was}, {Watchi}, {Weaver}, {Wei}, {Weinert},
  {Weinstein}, {Weiss}, {Wen}, {Wessel}, {We{\ss}els}, {Westerweck},
  {Westphal}, {Wette}, {Whelan}, {Whitcomb}, {Whiting}, {Whittle}, {Wilken},
  {Williams}, {Williams}, {Williamson}, {Willis}, {Willke}, {Wimmer},
  {Winkler}, {Wipf}, {Wittel}, {Woan}, {Woehler}, {Wofford}, {Wong}, {Worden},
  {Wright}, {Wu}, {Wysocki}, {Xiao}, {Yamamoto}, {Yancey}, {Yang}, {Yap},
  {Yazback}, {Yu}, {Yu}, {Yvert}, {Zadro{\.Z}ny}, {Zanolin}, {Zelenova},
  {Zendri}, {Zevin}, {Zhang}, {Zhang}, {Zhang}, {Zhang}, {Zhao}, {Zhou},
  {Zhou}, {Zhu}, {Zhu}, {Zimmerman}, {Zucker}, {Zweizig}, {LIGO Scientific
  Collaboration}, \& {Virgo Collaboration}}]{2017PhRvL.119p1101A}
{Abbott}, B.~P., {Abbott}, R., {Abbott}, T.~D., {et~al.} 2017, \prl, 119,
  161101

\bibitem[{{Akeson} {et~al.}(2021){Akeson}, {Beichman}, {Kervella}, {Fomalont},
  \& {Benedict}}]{2021AJ....162...14A}
{Akeson}, R., {Beichman}, C., {Kervella}, P., {Fomalont}, E., \& {Benedict},
  G.~F. 2021, \aj, 162, 14

\bibitem[{{Anderson} \& {Francis}(2012)}]{2012AstL...38..331A}
{Anderson}, E. \& {Francis}, C. 2012, Astronomy Letters, 38, 331

\bibitem[{{Anglada-Escud{\'e}} {et~al.}(2016){Anglada-Escud{\'e}}, {Amado},
  {Barnes}, {Berdi{\~n}as}, {Butler}, {Coleman}, {de La Cueva}, {Dreizler},
  {Endl}, {Giesers}, {Jeffers}, {Jenkins}, {Jones}, {Kiraga}, {K{\"u}rster},
  {L{\'o}pez-Gonz{\'a}lez}, {Marvin}, {Morales}, {Morin}, {Nelson}, {Ortiz},
  {Ofir}, {Paardekooper}, {Reiners}, {Rodr{\'{\i}}guez},
  {Rodr{\'{\i}}guez-L{\'o}pez}, {Sarmiento}, {Strachan}, {Tsapras}, {Tuomi}, \&
  {Zechmeister}}]{2016Natur.536..437A}
{Anglada-Escud{\'e}}, G., {Amado}, P.~J., {Barnes}, J., {et~al.} 2016, \nat,
  536, 437

\bibitem[{{Astropy Collaboration} {et~al.}(2018){Astropy Collaboration},
  {Price-Whelan}, {Sip{\H o}cz}, {G{\"u}nther}, {Lim}, {Crawford}, {Conseil},
  {Shupe}, {Craig}, {Dencheva}, {Ginsburg}, {VanderPlas}, {Bradley},
  {P{\'e}rez-Su{\'a}rez}, {de Val-Borro}, {Aldcroft}, {Cruz}, {Robitaille},
  {Tollerud}, {Ardelean}, {Babej}, {Bach}, {Bachetti}, {Bakanov}, {Bamford},
  {Barentsen}, {Barmby}, {Baumbach}, {Berry}, {Biscani}, {Boquien}, {Bostroem},
  {Bouma}, {Brammer}, {Bray}, {Breytenbach}, {Buddelmeijer}, {Burke},
  {Calderone}, {Cano Rodr{\'{\i}}guez}, {Cara}, {Cardoso}, {Cheedella},
  {Copin}, {Corrales}, {Crichton}, {D'Avella}, {Deil}, {Depagne}, {Dietrich},
  {Donath}, {Droettboom}, {Earl}, {Erben}, {Fabbro}, {Ferreira}, {Finethy},
  {Fox}, {Garrison}, {Gibbons}, {Goldstein}, {Gommers}, {Greco}, {Greenfield},
  {Groener}, {Grollier}, {Hagen}, {Hirst}, {Homeier}, {Horton}, {Hosseinzadeh},
  {Hu}, {Hunkeler}, {Ivezi{\'c}}, {Jain}, {Jenness}, {Kanarek}, {Kendrew},
  {Kern}, {Kerzendorf}, {Khvalko}, {King}, {Kirkby}, {Kulkarni}, {Kumar},
  {Lee}, {Lenz}, {Littlefair}, {Ma}, {Macleod}, {Mastropietro}, {McCully},
  {Montagnac}, {Morris}, {Mueller}, {Mumford}, {Muna}, {Murphy}, {Nelson},
  {Nguyen}, {Ninan}, {N{\"o}the}, {Ogaz}, {Oh}, {Parejko}, {Parley}, {Pascual},
  {Patil}, {Patil}, {Plunkett}, {Prochaska}, {Rastogi}, {Reddy Janga},
  {Sabater}, {Sakurikar}, {Seifert}, {Sherbert}, {Sherwood-Taylor}, {Shih},
  {Sick}, {Silbiger}, {Singanamalla}, {Singer}, {Sladen}, {Sooley},
  {Sornarajah}, {Streicher}, {Teuben}, {Thomas}, {Tremblay}, {Turner},
  {Terr{\'o}n}, {van Kerkwijk}, {de la Vega}, {Watkins}, {Weaver}, {Whitmore},
  {Woillez}, {Zabalza}, \& {Astropy Contributors}}]{2018AJ....156..123A}
{Astropy Collaboration}, {Price-Whelan}, A.~M., {Sip{\H o}cz}, B.~M., {et~al.}
  2018, \aj, 156, 123

\bibitem[{{Astropy Collaboration} {et~al.}(2013){Astropy Collaboration},
  {Robitaille}, {Tollerud}, {Greenfield}, {Droettboom}, {Bray}, {Aldcroft},
  {Davis}, {Ginsburg}, {Price-Whelan}, {Kerzendorf}, {Conley}, {Crighton},
  {Barbary}, {Muna}, {Ferguson}, {Grollier}, {Parikh}, {Nair}, {Unther},
  {Deil}, {Woillez}, {Conseil}, {Kramer}, {Turner}, {Singer}, {Fox}, {Weaver},
  {Zabalza}, {Edwards}, {Azalee Bostroem}, {Burke}, {Casey}, {Crawford},
  {Dencheva}, {Ely}, {Jenness}, {Labrie}, {Lim}, {Pierfederici}, {Pontzen},
  {Ptak}, {Refsdal}, {Servillat}, \& {Streicher}}]{2013A&A...558A..33A}
{Astropy Collaboration}, {Robitaille}, T.~P., {Tollerud}, E.~J., {et~al.} 2013,
  \aap, 558, A33

\bibitem[{{Barnes} {et~al.}(2017){Barnes}, {Jeffers}, {Haswell}, {Jones},
  {Shulyak}, {Pavlenko}, \& {Jenkins}}]{2017MNRAS.471..811B}
{Barnes}, J.~R., {Jeffers}, S.~V., {Haswell}, C.~A., {et~al.} 2017, \mnras,
  471, 811

\bibitem[{{Bedding} {et~al.}(2002){Bedding}, {Zijlstra}, {Jones}, {Marang},
  {Matsuura}, {Retter}, {Whitelock}, \& {Yamamura}}]{2002MNRAS.337...79B}
{Bedding}, T.~R., {Zijlstra}, A.~A., {Jones}, A., {et~al.} 2002, \mnras, 337,
  79

\bibitem[{{Belokurov} {et~al.}(2020){Belokurov}, {Penoyre}, {Oh}, {Iorio},
  {Hodgkin}, {Evans}, {Everall}, {Koposov}, {Tout}, {Izzard}, {Clarke}, \&
  {Brown}}]{2020MNRAS.496.1922B}
{Belokurov}, V., {Penoyre}, Z., {Oh}, S., {et~al.} 2020, \mnras, 496, 1922

\bibitem[{{Benedict} \& {McArthur}(2020)}]{2020RNAAS...4...46B}
{Benedict}, G.~F. \& {McArthur}, B.~E. 2020, Research Notes of the American
  Astronomical Society, 4, 46

\bibitem[{{Benisty} {et~al.}(2021){Benisty}, {Bae}, {Facchini}, {Keppler},
  {Teague}, {Isella}, {Kurtovic}, {P{\'e}rez}, {Sierra}, {Andrews},
  {Carpenter}, {Czekala}, {Dominik}, {Henning}, {Menard}, {Pinilla}, \&
  {Zurlo}}]{2021ApJ...916L...2B}
{Benisty}, M., {Bae}, J., {Facchini}, S., {et~al.} 2021, \apjl, 916, L2

\bibitem[{{Bessel}(1844)}]{1844MNRAS...6R.136B}
{Bessel}, F.~W. 1844, \mnras, 6, 136

\bibitem[{{Beuzit} {et~al.}(2019){Beuzit}, {Vigan}, {Mouillet}, {Dohlen},
  {Gratton}, {Boccaletti}, {Sauvage}, {Schmid}, {Langlois}, {Petit},
  {Baruffolo}, {Feldt}, {Milli}, {Wahhaj}, {Abe}, {Anselmi}, {Antichi},
  {Barette}, {Baudrand}, {Baudoz}, {Bazzon}, {Bernardi}, {Blanchard}, {Brast},
  {Bruno}, {Buey}, {Carbillet}, {Carle}, {Cascone}, {Chapron}, {Charton},
  {Chauvin}, {Claudi}, {Costille}, {De Caprio}, {de Boer}, {Delboulb{\'e}},
  {Desidera}, {Dominik}, {Downing}, {Dupuis}, {Fabron}, {Fantinel}, {Farisato},
  {Feautrier}, {Fedrigo}, {Fusco}, {Gigan}, {Ginski}, {Girard}, {Giro},
  {Gisler}, {Gluck}, {Gry}, {Henning}, {Hubin}, {Hugot}, {Incorvaia}, {Jaquet},
  {Kasper}, {Lagadec}, {Lagrange}, {Le Coroller}, {Le Mignant}, {Le Ruyet},
  {Lessio}, {Lizon}, {Llored}, {Lundin}, {Madec}, {Magnard}, {Marteaud},
  {Martinez}, {Maurel}, {M{\'e}nard}, {Mesa}, {M{\"o}ller-Nilsson}, {Moulin},
  {Moutou}, {Orign{\'e}}, {Parisot}, {Pavlov}, {Perret}, {Pragt}, {Puget},
  {Rabou}, {Ramos}, {Reess}, {Rigal}, {Rochat}, {Roelfsema}, {Rousset}, {Roux},
  {Saisse}, {Salasnich}, {Santambrogio}, {Scuderi}, {Segransan}, {Sevin},
  {Siebenmorgen}, {Soenke}, {Stadler}, {Suarez}, {Tiph{\`e}ne}, {Turatto},
  {Udry}, {Vakili}, {Waters}, {Weber}, {Wildi}, {Zins}, \&
  {Zurlo}}]{2019A&A...631A.155B}
{Beuzit}, J.~L., {Vigan}, A., {Mouillet}, D., {et~al.} 2019, \aap, 631, A155

\bibitem[{{Bortle} {et~al.}(2021){Bortle}, {Fausey}, {Ji}, {Dodson-Robinson},
  {Ramirez Delgado}, \& {Gizis}}]{2021AJ....161..230B}
{Bortle}, A., {Fausey}, H., {Ji}, J., {et~al.} 2021, \aj, 161, 230

\bibitem[{{Boyajian} {et~al.}(2012){Boyajian}, {von Braun}, {van Belle},
  {McAlister}, {ten Brummelaar}, {Kane}, {Muirhead}, {Jones}, {White},
  {Schaefer}, {Ciardi}, {Henry}, {L{\'o}pez-Morales}, {Ridgway}, {Gies}, {Jao},
  {Rojas-Ayala}, {Parks}, {Sturmann}, {Sturmann}, {Turner}, {Farrington},
  {Goldfinger}, \& {Berger}}]{2012ApJ...757..112B}
{Boyajian}, T.~S., {von Braun}, K., {van Belle}, G., {et~al.} 2012, \apj, 757,
  112

\bibitem[{{Brandenburg} {et~al.}(2017){Brandenburg}, {Mathur}, \&
  {Metcalfe}}]{2017ApJ...845...79B}
{Brandenburg}, A., {Mathur}, S., \& {Metcalfe}, T.~S. 2017, \apj, 845, 79

\bibitem[{{Brandt} {et~al.}(2021){Brandt}, {Dupuy}, {Li}, {Chen}, {Brandt},
  {Wong}, {Currie}, {Bowler}, {Liu}, {Best}, \&
  {Phillips}}]{2021arXiv210907525B}
{Brandt}, G.~M., {Dupuy}, T.~J., {Li}, Y., {et~al.} 2021, arXiv e-prints,
  arXiv:2109.07525

\bibitem[{{Brandt}(2018)}]{2018ApJS..239...31B}
{Brandt}, T.~D. 2018, The Astrophysical Journal Supplement Series, 239, 31

\bibitem[{{Brandt}(2021)}]{2021ApJS..254...42B}
{Brandt}, T.~D. 2021, \apjs, 254, 42

\bibitem[{{Brandt} {et~al.}(2019){Brandt}, {Dupuy}, \&
  {Bowler}}]{2019AJ....158..140B}
{Brandt}, T.~D., {Dupuy}, T.~J., \& {Bowler}, B.~P. 2019, \aj, 158, 140

\bibitem[{Burnham(1891)}]{10.1093/mnras/51.8.460}
Burnham, S.~W. 1891, Monthly Notices of the Royal Astronomical Society, 51, 460

\bibitem[{{Butler} {et~al.}(2017){Butler}, {Vogt}, {Laughlin}, {Burt},
  {Rivera}, {Tuomi}, {Teske}, {Arriagada}, {Diaz}, {Holden}, \&
  {Keiser}}]{2017AJ....153..208B}
{Butler}, R.~P., {Vogt}, S.~S., {Laughlin}, G., {et~al.} 2017, \aj, 153, 208

\bibitem[{{Cantat-Gaudin} \& {Brandt}(2021)}]{2021A&A...649A.124C}
{Cantat-Gaudin}, T. \& {Brandt}, T.~D. 2021, \aap, 649, A124

\bibitem[{{Capitanio} {et~al.}(2017){Capitanio}, {Lallement}, {Vergely},
  {Elyajouri}, \& {Monreal-Ibero}}]{2017A&A...606A..65C}
{Capitanio}, L., {Lallement}, R., {Vergely}, J.~L., {Elyajouri}, M., \&
  {Monreal-Ibero}, A. 2017, \aap, 606, A65

\bibitem[{{Chiavassa} {et~al.}(2011){Chiavassa}, {Pasquato}, {Jorissen},
  {Sacuto}, {Babusiaux}, {Freytag}, {Ludwig}, {Cruzal{\`e}bes}, {Rabbia},
  {Spang}, \& {Chesneau}}]{2011A&A...528A.120C}
{Chiavassa}, A., {Pasquato}, E., {Jorissen}, A., {et~al.} 2011, \aap, 528, A120

\bibitem[{{Cochetti} {et~al.}(2019){Cochetti}, {Arcos}, {Kanaan}, {Meilland},
  {Cidale}, \& {Cur{\'e}}}]{2019A&A...621A.123C}
{Cochetti}, Y.~R., {Arcos}, C., {Kanaan}, S., {et~al.} 2019, \aap, 621, A123

\bibitem[{{Cropper} {et~al.}(2018){Cropper}, {Katz}, {Sartoretti}, {Prusti},
  {de Bruijne}, {Chassat}, {Charvet}, {Boyadjian}, {Perryman}, {Sarri}, {Gare},
  {Erdmann}, {Munari}, {Zwitter}, {Wilkinson}, {Arenou}, {Vallenari},
  {G{\'o}mez}, {Panuzzo}, {Seabroke}, {Allende Prieto}, {Benson}, {Marchal},
  {Huckle}, {Smith}, {Dolding}, {Jan{\ss}en}, {Viala}, {Blomme}, {Baker},
  {Boudreault}, {Crifo}, {Soubiran}, {Fr{\'e}mat}, {Jasniewicz}, {Guerrier},
  {Guy}, {Turon}, {Jean-Antoine-Piccolo}, {Th{\'e}venin}, {David}, {Gosset}, \&
  {Damerdji}}]{2018A&A...616A...5C}
{Cropper}, M., {Katz}, D., {Sartoretti}, P., {et~al.} 2018, \aap, 616, A5

\bibitem[{{Cumming} {et~al.}(2008){Cumming}, {Butler}, {Marcy}, {Vogt},
  {Wright}, \& {Fischer}}]{2008PASP..120..531C}
{Cumming}, A., {Butler}, R.~P., {Marcy}, G.~W., {et~al.} 2008, \pasp, 120, 531

\bibitem[{{Currie} {et~al.}(2020){Currie}, {Brandt}, {Kuzuhara}, {Chilcote},
  {Guyon}, {Marois}, {Groff}, {Lozi}, {Vievard}, {Sahoo}, {Deo}, {Jovanovic},
  {Martinache}, {Wagner}, {Dupuy}, {Wahl}, {Letawsky}, {Li}, {Zeng}, {Brandt},
  {Michalik}, {Grady}, {Janson}, {Knapp}, {Kwon}, {Lawson}, {McElwain},
  {Uyama}, {Wisniewski}, \& {Tamura}}]{2020ApJ...904L..25C}
{Currie}, T., {Brandt}, T.~D., {Kuzuhara}, M., {et~al.} 2020, \apjl, 904, L25

\bibitem[{{Damasso} \& {Del Sordo}(2017)}]{2017A&A...599A.126D}
{Damasso}, M. \& {Del Sordo}, F. 2017, \aap, 599, A126

\bibitem[{{Damasso} {et~al.}(2020{\natexlab{a}}){Damasso}, {Del Sordo},
  {Anglada-Escud{\'e}}, {Giacobbe}, {Sozzetti}, {Morbidelli}, {Pojmanski},
  {Barbato}, {Butler}, {Jones}, {Hambsch}, {Jenkins}, {L{\'o}pez-Gonz{\'a}lez},
  {Morales}, {Pe{\~n}a Rojas}, {Rodr{\'\i}guez-L{\'o}pez}, {Rodr{\'\i}guez},
  {Amado}, {Anglada}, {Feng}, \& {G{\'o}mez}}]{2020SciA....6.7467D}
{Damasso}, M., {Del Sordo}, F., {Anglada-Escud{\'e}}, G., {et~al.}
  2020{\natexlab{a}}, Science Advances, 6, eaax7467

\bibitem[{{Damasso} {et~al.}(2020{\natexlab{b}}){Damasso}, {Sozzetti}, {Lovis},
  {Barros}, {Sousa}, {Demangeon}, {Faria}, {Lillo-Box}, {Cristiani}, {Pepe},
  {Rebolo}, {Santos}, {Zapatero Osorio}, {Gonz{\'a}lez Hern{\'a}ndez}, {Amate},
  {Pasquini}, {Zerbi}, {Adibekyan}, {Abreu}, {Affolter}, {Alibert}, {Aliverti},
  {Allart}, {Allende Prieto}, {{\'A}lvarez}, {Alves}, {Avila}, {Baldini},
  {Bandy}, {Benz}, {Bianco}, {Borsa}, {Bossini}, {Bourrier}, {Bouchy}, {Broeg},
  {Cabral}, {Calderone}, {Cirami}, {Coelho}, {Conconi}, {Coretti}, {Cumani},
  {Cupani}, {D'Odorico}, {Deiries}, {Dekker}, {Delabre}, {Di Marcantonio},
  {Dumusque}, {Ehrenreich}, {Figueira}, {Fragoso}, {Genolet}, {Genoni},
  {G{\'e}nova Santos}, {Hughes}, {Iwert}, {Kerber}, {Knudstrup}, {Landoni},
  {Lavie}, {Lizon}, {Lo Curto}, {Maire}, {Martins}, {M{\'e}gevand}, {Mehner},
  {Micela}, {Modigliani}, {Molaro}, {Monteiro}, {Monteiro}, {Moschetti},
  {Mueller}, {Murphy}, {Nunes}, {Oggioni}, {Oliveira}, {Oshagh}, {Pall{\'e}},
  {Pariani}, {Poretti}, {Rasilla}, {Rebord{\~a}o}, {Redaelli}, {Riva}, {Santana
  Tschudi}, {Santin}, {Santos}, {S{\'e}gransan}, {Schmidt}, {Segovia},
  {Sosnowska}, {Span{\`o}}, {Su{\'a}rez Mascare{\~n}o}, {Tabernero}, {Tenegi},
  {Udry}, \& {Zanutta}}]{2020A&A...642A..31D}
{Damasso}, M., {Sozzetti}, A., {Lovis}, C., {et~al.} 2020{\natexlab{b}}, \aap,
  642, A31

\bibitem[{{De Rosa} {et~al.}(2020){De Rosa}, {Dawson}, \&
  {Nielsen}}]{2020A&A...640A..73D}
{De Rosa}, R.~J., {Dawson}, R., \& {Nielsen}, E.~L. 2020, \aap, 640, A73

\bibitem[{{Dravins} {et~al.}(1999){Dravins}, {Lindegren}, \&
  {Madsen}}]{1999A&A...348.1040D}
{Dravins}, D., {Lindegren}, L., \& {Madsen}, S. 1999, \aap, 348, 1040

\bibitem[{{Dupuy} {et~al.}(2019){Dupuy}, {Brandt}, {Kratter}, \&
  {Bowler}}]{2019ApJ...871L...4D}
{Dupuy}, T.~J., {Brandt}, T.~D., {Kratter}, K.~M., \& {Bowler}, B.~P. 2019,
  \apj, 871, L4

\bibitem[{{El-Badry} {et~al.}(2021){El-Badry}, {Rix}, \&
  {Heintz}}]{2021MNRAS.506.2269E}
{El-Badry}, K., {Rix}, H.-W., \& {Heintz}, T.~M. 2021, \mnras, 506, 2269

\bibitem[{{ESA}(1997)}]{1997ESASP1200.....E}
{ESA}, ed. 1997, ESA Special Publication, Vol. 1200, {The HIPPARCOS and TYCHO
  catalogues. Astrometric and photometric star catalogues derived from the ESA
  HIPPARCOS Space Astrometry Mission}

\bibitem[{{Fabricius} {et~al.}(2021){Fabricius}, {Luri, X.}, {Arenou, F.},
  {Babusiaux, C.}, {Helmi, A.}, {Muraveva, T.}, {Reyl\'e, C.}, {Spoto, F.},
  {Vallenari, A.}, {Antoja, T.}, {Balbinot, E.}, {Barache, C.}, {Bauchet, N.},
  {Bragaglia, A.}, {Busonero, D.}, {Cantat-Gaudin, T.}, {Carrasco, J. M.},
  {Diakit\'e, S.}, {Fabrizio, M.}, {Figueras, F.}, {Garcia-Gutierrez, A.},
  {Garofalo, A.}, {Jordi, C.}, {Kervella, P.}, {Khanna, S.}, {Leclerc, N.},
  {Licata, E.}, {Lambert, S.}, {Marrese, P. M.}, {Masip, A.}, {Ramos, P.},
  {Robichon, N.}, {Robin, A. C.}, {Romero-G\'omez, M.}, {Rubele, S.}, \&
  {Weiler, M.}}]{GaiaEDR3validation}
{Fabricius}, C., {Luri, X.}, {Arenou, F.}, {et~al.} 2021, A\&A, 649, A5

\bibitem[{{Feng} {et~al.}(2019){Feng}, {Anglada-Escud{\'e}}, {Tuomi}, {Jones},
  {Chanam{\'e}}, {Butler}, \& {Janson}}]{2019MNRAS.490.5002F}
{Feng}, F., {Anglada-Escud{\'e}}, G., {Tuomi}, M., {et~al.} 2019, \mnras, 490,
  5002

\bibitem[{{Fischer} {et~al.}(2016){Fischer}, {Anglada-Escude}, {Arriagada},
  {Baluev}, {Bean}, {Bouchy}, {Buchhave}, {Carroll}, {Chakraborty}, {Crepp},
  {Dawson}, {Diddams}, {Dumusque}, {Eastman}, {Endl}, {Figueira}, {Ford},
  {Foreman-Mackey}, {Fournier}, {F{\H u}r{\'e}sz}, {Gaudi}, {Gregory},
  {Grundahl}, {Hatzes}, {H{\'e}brard}, {Herrero}, {Hogg}, {Howard}, {Johnson},
  {Jorden}, {Jurgenson}, {Latham}, {Laughlin}, {Loredo}, {Lovis}, {Mahadevan},
  {McCracken}, {Pepe}, {Perez}, {Phillips}, {Plavchan}, {Prato}, {Quirrenbach},
  {Reiners}, {Robertson}, {Santos}, {Sawyer}, {Segransan}, {Sozzetti},
  {Steinmetz}, {Szentgyorgyi}, {Udry}, {Valenti}, {Wang}, {Wittenmyer}, \&
  {Wright}}]{2016PASP..128f6001F}
{Fischer}, D.~A., {Anglada-Escude}, G., {Arriagada}, P., {et~al.} 2016, \pasp,
  128, 066001

\bibitem[{{Frankowski} {et~al.}(2007){Frankowski}, {Jancart}, \&
  {Jorissen}}]{2007A&A...464..377F}
{Frankowski}, A., {Jancart}, S., \& {Jorissen}, A. 2007, \aap, 464, 377

\bibitem[{{Frisch} {et~al.}(2011){Frisch}, {Redfield}, \&
  {Slavin}}]{2011ARA&A..49..237F}
{Frisch}, P.~C., {Redfield}, S., \& {Slavin}, J.~D. 2011, Annual Review of
  Astronomy and Astrophysics, 49, 237

\bibitem[{{Fuhrmann} {et~al.}(2017){Fuhrmann}, {Chini}, {Kaderhandt}, \&
  {Chen}}]{2017ApJ...836..139F}
{Fuhrmann}, K., {Chini}, R., {Kaderhandt}, L., \& {Chen}, Z. 2017, \apj, 836,
  139

\bibitem[{{Furuhjelm}(1914)}]{1914AN....197..181F}
{Furuhjelm}, R. 1914, Astronomische Nachrichten, 197, 181

\bibitem[{{Gaia Collaboration}(2020)}]{2020yCat.1350....0G}
{Gaia Collaboration}. 2020, VizieR Online Data Catalog, I/350

\bibitem[{{Gaia Collaboration} {et~al.}(2018){Gaia Collaboration}, {Brown},
  {Vallenari}, {Prusti}, {de Bruijne}, {Babusiaux}, {Bailer-Jones}, {Biermann},
  {Evans}, {Eyer}, \& et~al.}]{2018A&A...616A...1G}
{Gaia Collaboration}, {Brown}, A.~G.~A., {Vallenari}, A., {et~al.} 2018, \aap,
  616, A1

\bibitem[{{Gaia Collaboration} {et~al.}(2021{\natexlab{a}}){Gaia
  Collaboration}, {Brown}, {Vallenari}, {Prusti}, {de Bruijne}, {Babusiaux},
  {Biermann}, {Creevey}, {Evans}, {Eyer}, {Hutton}, {Jansen}, {Jordi},
  {Klioner}, {Lammers}, {Lindegren}, {Luri}, {Mignard}, {Panem}, {Pourbaix},
  {Randich}, {Sartoretti}, {Soubiran}, {Walton}, {Arenou}, {Bailer-Jones},
  {Bastian}, {Cropper}, {Drimmel}, {Katz}, {Lattanzi}, {van Leeuwen}, {Bakker},
  {Cacciari}, {Casta{\~n}eda}, {De Angeli}, {Ducourant}, {Fabricius},
  {Fouesneau}, {Fr{\'e}mat}, {Guerra}, {Guerrier}, {Guiraud}, {Jean-Antoine
  Piccolo}, {Masana}, {Messineo}, {Mowlavi}, {Nicolas}, {Nienartowicz},
  {Pailler}, {Panuzzo}, {Riclet}, {Roux}, {Seabroke}, {Sordo}, {Tanga},
  {Th{\'e}venin}, {Gracia-Abril}, {Portell}, {Teyssier}, {Altmann}, {Andrae},
  {Bellas-Velidis}, {Benson}, {Berthier}, {Blomme}, {Brugaletta}, {Burgess},
  {Busso}, {Carry}, {Cellino}, {Cheek}, {Clementini}, {Damerdji}, {Davidson},
  {Delchambre}, {Dell'Oro}, {Fern{\'a}ndez-Hern{\'a}ndez}, {Galluccio},
  {Garc{\'\i}a-Lario}, {Garcia-Reinaldos}, {Gonz{\'a}lez-N{\'u}{\~n}ez},
  {Gosset}, {Haigron}, {Halbwachs}, {Hambly}, {Harrison}, {Hatzidimitriou},
  {Heiter}, {Hern{\'a}ndez}, {Hestroffer}, {Hodgkin}, {Holl}, {Jan{\ss}en},
  {Jevardat de Fombelle}, {Jordan}, {Krone-Martins}, {Lanzafame},
  {L{\"o}ffler}, {Lorca}, {Manteiga}, {Marchal}, {Marrese}, {Moitinho}, {Mora},
  {Muinonen}, {Osborne}, {Pancino}, {Pauwels}, {Petit}, {Recio-Blanco},
  {Richards}, {Riello}, {Rimoldini}, {Robin}, {Roegiers}, {Rybizki}, {Sarro},
  {Siopis}, {Smith}, {Sozzetti}, {Ulla}, {Utrilla}, {van Leeuwen}, {van
  Reeven}, {Abbas}, {Abreu Aramburu}, {Accart}, {Aerts}, {Aguado}, {Ajaj},
  {Altavilla}, {{\'A}lvarez}, {{\'A}lvarez Cid-Fuentes}, {Alves}, {Anderson},
  {Anglada Varela}, {Antoja}, {Audard}, {Baines}, {Baker},
  {Balaguer-N{\'u}{\~n}ez}, {Balbinot}, {Balog}, {Barache}, {Barbato},
  {Barros}, {Barstow}, {Bartolom{\'e}}, {Bassilana}, {Bauchet},
  {Baudesson-Stella}, {Becciani}, {Bellazzini}, {Bernet}, {Bertone}, {Bianchi},
  {Blanco-Cuaresma}, {Boch}, {Bombrun}, {Bossini}, {Bouquillon}, {Bragaglia},
  {Bramante}, {Breedt}, {Bressan}, {Brouillet}, {Bucciarelli}, {Burlacu},
  {Busonero}, {Butkevich}, {Buzzi}, {Caffau}, {Cancelliere}, {C{\'a}novas},
  {Cantat-Gaudin}, {Carballo}, {Carlucci}, {Carnerero}, {Carrasco},
  {Casamiquela}, {Castellani}, {Castro-Ginard}, {Castro Sampol}, {Chaoul},
  {Charlot}, {Chemin}, {Chiavassa}, {Cioni}, {Comoretto}, {Cooper}, {Cornez},
  {Cowell}, {Crifo}, {Crosta}, {Crowley}, {Dafonte}, {Dapergolas}, {David},
  {David}, {de Laverny}, {De Luise}, {De March}, {De Ridder}, {de Souza}, {de
  Teodoro}, {de Torres}, {del Peloso}, {del Pozo}, {Delbo}, {Delgado},
  {Delgado}, {Delisle}, {Di Matteo}, {Diakite}, {Diener}, {Distefano},
  {Dolding}, {Eappachen}, {Edvardsson}, {Enke}, {Esquej}, {Fabre}, {Fabrizio},
  {Faigler}, {Fedorets}, {Fernique}, {Fienga}, {Figueras}, {Fouron},
  {Fragkoudi}, {Fraile}, {Franke}, {Gai}, {Garabato}, {Garcia-Gutierrez},
  {Garc{\'\i}a-Torres}, {Garofalo}, {Gavras}, {Gerlach}, {Geyer}, {Giacobbe},
  {Gilmore}, {Girona}, {Giuffrida}, {Gomel}, {Gomez}, {Gonzalez-Santamaria},
  {Gonz{\'a}lez-Vidal}, {Granvik}, {Guti{\'e}rrez-S{\'a}nchez}, {Guy},
  {Hauser}, {Haywood}, {Helmi}, {Hidalgo}, {Hilger}, {H{\l}adczuk}, {Hobbs},
  {Holland}, {Huckle}, {Jasniewicz}, {Jonker}, {Juaristi Campillo}, {Julbe},
  {Karbevska}, {Kervella}, {Khanna}, {Kochoska}, {Kontizas}, {Kordopatis},
  {Korn}, {Kostrzewa-Rutkowska}, {Kruszy{\'n}ska}, {Lambert}, {Lanza}, {Lasne},
  {Le Campion}, {Le Fustec}, {Lebreton}, {Lebzelter}, {Leccia}, {Leclerc},
  {Lecoeur-Taibi}, {Liao}, {Licata}, {Lindstr{\o}m}, {Lister}, {Livanou},
  {Lobel}, {Madrero Pardo}, {Managau}, {Mann}, {Marchant}, {Marconi}, {Marcos
  Santos}, {Marinoni}, {Marocco}, {Marshall}, {Martin Polo},
  {Mart{\'\i}n-Fleitas}, {Masip}, {Massari}, {Mastrobuono-Battisti}, {Mazeh},
  {McMillan}, {Messina}, {Michalik}, {Millar}, {Mints}, {Molina}, {Molinaro},
  {Moln{\'a}r}, {Montegriffo}, {Mor}, {Morbidelli}, {Morel}, {Morris},
  {Mulone}, {Munoz}, {Muraveva}, {Murphy}, {Musella}, {Noval}, {Ord{\'e}novic},
  {Orr{\`u}}, {Osinde}, {Pagani}, {Pagano}, {Palaversa}, {Palicio}, {Panahi},
  {Pawlak}, {Pe{\~n}alosa Esteller}, {Penttil{\"a}}, {Piersimoni}, {Pineau},
  {Plachy}, {Plum}, {Poggio}, {Poretti}, {Poujoulet}, {Pr{\v{s}}a}, {Pulone},
  {Racero}, {Ragaini}, {Rainer}, {Raiteri}, {Rambaux}, {Ramos}, {Ramos-Lerate},
  {Re Fiorentin}, {Regibo}, {Reyl{\'e}}, {Ripepi}, {Riva}, {Rixon}, {Robichon},
  {Robin}, {Roelens}, {Rohrbasser}, {Romero-G{\'o}mez}, {Rowell}, {Royer},
  {Rybicki}, {Sadowski}, {Sagrist{\`a} Sell{\'e}s}, {Sahlmann}, {Salgado},
  {Salguero}, {Samaras}, {Sanchez Gimenez}, {Sanna}, {Santove{\~n}a},
  {Sarasso}, {Schultheis}, {Sciacca}, {Segol}, {Segovia}, {S{\'e}gransan},
  {Semeux}, {Shahaf}, {Siddiqui}, {Siebert}, {Siltala}, {Slezak}, {Smart},
  {Solano}, {Solitro}, {Souami}, {Souchay}, {Spagna}, {Spoto}, {Steele},
  {Steidelm{\"u}ller}, {Stephenson}, {S{\"u}veges}, {Szabados}, {Szegedi-Elek},
  {Taris}, {Tauran}, {Taylor}, {Teixeira}, {Thuillot}, {Tonello}, {Torra},
  {Torra}, {Turon}, {Unger}, {Vaillant}, {van Dillen}, {Vanel}, {Vecchiato},
  {Viala}, {Vicente}, {Voutsinas}, {Weiler}, {Wevers}, {Wyrzykowski}, {Yoldas},
  {Yvard}, {Zhao}, {Zorec}, {Zucker}, {Zurbach}, \&
  {Zwitter}}]{2021A&A...650C...3G}
{Gaia Collaboration}, {Brown}, A.~G.~A., {Vallenari}, A., {et~al.}
  2021{\natexlab{a}}, \aap, 650, C3

\bibitem[{{Gaia Collaboration} {et~al.}(2021{\natexlab{b}}){Gaia
  Collaboration}, {Brown, A. G. A.}, {Vallenari, A.}, {Prusti, T.}, {de
  Bruijne, J. H. J.}, {Babusiaux, C.}, {Biermann, M.}, {Creevey, O. L.},
  {Evans, D. W.}, {Eyer, L.}, {Hutton, A.}, {Jansen, F.}, {Jordi, C.},
  {Klioner, S. A.}, {Lammers, U.}, {Lindegren, L.}, {Luri, X.}, {Mignard, F.},
  {Panem, C.}, {Pourbaix, D.}, {Randich, S.}, {Sartoretti, P.}, {Soubiran, C.},
  {Walton, N. A.}, {Arenou, F.}, {Bailer-Jones, C. A. L.}, {Bastian, U.},
  {Cropper, M.}, {Drimmel, R.}, {Katz, D.}, {Lattanzi, M. G.}, {van Leeuwen,
  F.}, {Bakker, J.}, {Cacciari, C.}, {Casta\~neda, J.}, {De Angeli, F.},
  {Ducourant, C.}, {Fabricius, C.}, {Fouesneau, M.}, {Fr\'emat, Y.}, {Guerra,
  R.}, {Guerrier, A.}, {Guiraud, J.}, {Jean-Antoine Piccolo, A.}, {Masana, E.},
  {Messineo, R.}, {Mowlavi, N.}, {Nicolas, C.}, {Nienartowicz, K.}, {Pailler,
  F.}, {Panuzzo, P.}, {Riclet, F.}, {Roux, W.}, {Seabroke, G. M.}, {Sordo, R.},
  {Tanga, P.}, {Th\'evenin, F.}, {Gracia-Abril, G.}, {Portell, J.}, {Teyssier,
  D.}, {Altmann, M.}, {Andrae, R.}, {Bellas-Velidis, I.}, {Benson, K.},
  {Berthier, J.}, {Blomme, R.}, {Brugaletta, E.}, {Burgess, P. W.}, {Busso,
  G.}, {Carry, B.}, {Cellino, A.}, {Cheek, N.}, {Clementini, G.}, {Damerdji,
  Y.}, {Davidson, M.}, {Delchambre, L.}, {Dell\'{}Oro, A.},
  {Fern\'andez-Hern\'andez, J.}, {Galluccio, L.}, {Garc\'{\i}a-Lario, P.},
  {Garcia-Reinaldos, M.}, {Gonz\'alez-N\'u\~nez, J.}, {Gosset, E.}, {Haigron,
  R.}, {Halbwachs, J.-L.}, {Hambly, N. C.}, {Harrison, D. L.}, {Hatzidimitriou,
  D.}, {Heiter, U.}, {Hern\'andez, J.}, {Hestroffer, D.}, {Hodgkin, S. T.},
  {Holl, B.}, {Jan\ss{}en, K.}, {Jevardat de Fombelle, G.}, {Jordan, S.},
  {Krone-Martins, A.}, {Lanzafame, A. C.}, {L\"offler, W.}, {Lorca, A.},
  {Manteiga, M.}, {Marchal, O.}, {Marrese, P. M.}, {Moitinho, A.}, {Mora, A.},
  {Muinonen, K.}, {Osborne, P.}, {Pancino, E.}, {Pauwels, T.}, {Petit, J.-M.},
  {Recio-Blanco, A.}, {Richards, P. J.}, {Riello, M.}, {Rimoldini, L.}, {Robin,
  A. C.}, {Roegiers, T.}, {Rybizki, J.}, {Sarro, L. M.}, {Siopis, C.}, {Smith,
  M.}, {Sozzetti, A.}, {Ulla, A.}, {Utrilla, E.}, {van Leeuwen, M.}, {van
  Reeven, W.}, {Abbas, U.}, {Abreu Aramburu, A.}, {Accart, S.}, {Aerts, C.},
  {Aguado, J. J.}, {Ajaj, M.}, {Altavilla, G.}, {\'Alvarez, M. A.}, {\'Alvarez
  Cid-Fuentes, J.}, {Alves, J.}, {Anderson, R. I.}, {Anglada Varela, E.},
  {Antoja, T.}, {Audard, M.}, {Baines, D.}, {Baker, S. G.},
  {Balaguer-N\'u\~nez, L.}, {Balbinot, E.}, {Balog, Z.}, {Barache, C.},
  {Barbato, D.}, {Barros, M.}, {Barstow, M. A.}, {Bartolom\'e, S.}, {Bassilana,
  J.-L.}, {Bauchet, N.}, {Baudesson-Stella, A.}, {Becciani, U.}, {Bellazzini,
  M.}, {Bernet, M.}, {Bertone, S.}, {Bianchi, L.}, {Blanco-Cuaresma, S.},
  {Boch, T.}, {Bombrun, A.}, {Bossini, D.}, {Bouquillon, S.}, {Bragaglia, A.},
  {Bramante, L.}, {Breedt, E.}, {Bressan, A.}, {Brouillet, N.}, {Bucciarelli,
  B.}, {Burlacu, A.}, {Busonero, D.}, {Butkevich, A. G.}, {Buzzi, R.}, {Caffau,
  E.}, {Cancelliere, R.}, {C\'anovas, H.}, {Cantat-Gaudin, T.}, {Carballo, R.},
  {Carlucci, T.}, {Carnerero, M. I}, {Carrasco, J. M.}, {Casamiquela, L.},
  {Castellani, M.}, {Castro-Ginard, A.}, {Castro Sampol, P.}, {Chaoul, L.},
  {Charlot, P.}, {Chemin, L.}, {Chiavassa, A.}, {Cioni, M.-R. L.}, {Comoretto,
  G.}, {Cooper, W. J.}, {Cornez, T.}, {Cowell, S.}, {Crifo, F.}, {Crosta, M.},
  {Crowley, C.}, {Dafonte, C.}, {Dapergolas, A.}, {David, M.}, {David, P.}, {de
  Laverny, P.}, {De Luise, F.}, {De March, R.}, {De Ridder, J.}, {de Souza,
  R.}, {de Teodoro, P.}, {de Torres, A.}, {del Peloso, E. F.}, {del Pozo, E.},
  {Delbo, M.}, {Delgado, A.}, {Delgado, H. E.}, {Delisle, J.-B.}, {Di Matteo,
  P.}, {Diakite, S.}, {Diener, C.}, {Distefano, E.}, {Dolding, C.}, {Eappachen,
  D.}, {Edvardsson, B.}, {Enke, H.}, {Esquej, P.}, {Fabre, C.}, {Fabrizio, M.},
  {Faigler, S.}, {Fedorets, G.}, {Fernique, P.}, {Fienga, A.}, {Figueras, F.},
  {Fouron, C.}, {Fragkoudi, F.}, {Fraile, E.}, {Franke, F.}, {Gai, M.},
  {Garabato, D.}, {Garcia-Gutierrez, A.}, {Garc\'{\i}a-Torres, M.}, {Garofalo,
  A.}, {Gavras, P.}, {Gerlach, E.}, {Geyer, R.}, {Giacobbe, P.}, {Gilmore, G.},
  {Girona, S.}, {Giuffrida, G.}, {Gomel, R.}, {Gomez, A.},
  {Gonzalez-Santamaria, I.}, {Gonz\'alez-Vidal, J. J.}, {Granvik, M.},
  {Guti\'errez-S\'anchez, R.}, {Guy, L. P.}, {Hauser, M.}, {Haywood, M.},
  {Helmi, A.}, {Hidalgo, S. L.}, {Hilger, T.}, {Hladczuk, N.}, {Hobbs, D.},
  {Holland, G.}, {Huckle, H. E.}, {Jasniewicz, G.}, {Jonker, P. G.}, {Juaristi
  Campillo, J.}, {Julbe, F.}, {Karbevska, L.}, {Kervella, P.}, {Khanna, S.},
  {Kochoska, A.}, {Kontizas, M.}, {Kordopatis, G.}, {Korn, A. J.},
  {Kostrzewa-Rutkowska, Z.}, {Kruszy\'{}nska, K.}, {Lambert, S.}, {Lanza, A.
  F.}, {Lasne, Y.}, {Le Campion, J.-F.}, {Le Fustec, Y.}, {Lebreton, Y.},
  {Lebzelter, T.}, {Leccia, S.}, {Leclerc, N.}, {Lecoeur-Taibi, I.}, {Liao,
  S.}, {Licata, E.}, {Lindstr\o{}m, E. P.}, {Lister, T. A.}, {Livanou, E.},
  {Lobel, A.}, {Madrero Pardo, P.}, {Managau, S.}, {Mann, R. G.}, {Marchant, J.
  M.}, {Marconi, M.}, {Marcos Santos, M. M. S.}, {Marinoni, S.}, {Marocco, F.},
  {Marshall, D. J.}, {Martin Polo, L.}, {Mart\'{\i}n-Fleitas, J. M.}, {Masip,
  A.}, {Massari, D.}, {Mastrobuono-Battisti, A.}, {Mazeh, T.}, {McMillan, P.
  J.}, {Messina, S.}, {Michalik, D.}, {Millar, N. R.}, {Mints, A.}, {Molina,
  D.}, {Molinaro, R.}, {Moln\'ar, L.}, {Montegriffo, P.}, {Mor, R.},
  {Morbidelli, R.}, {Morel, T.}, {Morris, D.}, {Mulone, A. F.}, {Munoz, D.},
  {Muraveva, T.}, {Murphy, C. P.}, {Musella, I.}, {Noval, L.}, {Ord\'enovic,
  C.}, {Orr\`u, G.}, {Osinde, J.}, {Pagani, C.}, {Pagano, I.}, {Palaversa, L.},
  {Palicio, P. A.}, {Panahi, A.}, {Pawlak, M.}, {Pe\~nalosa Esteller, X.},
  {Penttil\"a, A.}, {Piersimoni, A. M.}, {Pineau, F.-X.}, {Plachy, E.}, {Plum,
  G.}, {Poggio, E.}, {Poretti, E.}, {Poujoulet, E.}, {Prsa, A.}, {Pulone, L.},
  {Racero, E.}, {Ragaini, S.}, {Rainer, M.}, {Raiteri, C. M.}, {Rambaux, N.},
  {Ramos, P.}, {Ramos-Lerate, M.}, {Re Fiorentin, P.}, {Regibo, S.}, {Reyl\'e,
  C.}, {Ripepi, V.}, {Riva, A.}, {Rixon, G.}, {Robichon, N.}, {Robin, C.},
  {Roelens, M.}, {Rohrbasser, L.}, {Romero-G\'omez, M.}, {Rowell, N.}, {Royer,
  F.}, {Rybicki, K. A.}, {Sadowski, G.}, {Sagrist\`a Sell\'es, A.}, {Sahlmann,
  J.}, {Salgado, J.}, {Salguero, E.}, {Samaras, N.}, {Sanchez Gimenez, V.},
  {Sanna, N.}, {Santove\~na, R.}, {Sarasso, M.}, {Schultheis, M.}, {Sciacca,
  E.}, {Segol, M.}, {Segovia, J. C.}, {S\'egransan, D.}, {Semeux, D.}, {Shahaf,
  S.}, {Siddiqui, H. I.}, {Siebert, A.}, {Siltala, L.}, {Slezak, E.}, {Smart,
  R. L.}, {Solano, E.}, {Solitro, F.}, {Souami, D.}, {Souchay, J.}, {Spagna,
  A.}, {Spoto, F.}, {Steele, I. A.}, {Steidelm\"uller, H.}, {Stephenson, C.
  A.}, {S\"uveges, M.}, {Szabados, L.}, {Szegedi-Elek, E.}, {Taris, F.},
  {Tauran, G.}, {Taylor, M. B.}, {Teixeira, R.}, {Thuillot, W.}, {Tonello, N.},
  {Torra, F.}, {Torra, J.}, {Turon, C.}, {Unger, N.}, {Vaillant, M.}, {van
  Dillen, E.}, {Vanel, O.}, {Vecchiato, A.}, {Viala, Y.}, {Vicente, D.},
  {Voutsinas, S.}, {Weiler, M.}, {Wevers, T.}, {Wyrzykowski, L.}, {Yoldas, A.},
  {Yvard, P.}, {Zhao, H.}, {Zorec, J.}, {Zucker, S.}, {Zurbach, C.}, \&
  {Zwitter, T.}}]{GaiaEDR3content}
{Gaia Collaboration}, {Brown, A. G. A.}, {Vallenari, A.}, {et~al.}
  2021{\natexlab{b}}, A\&A, 649, A1

\bibitem[{{Gaia Collaboration} {et~al.}(2016){Gaia Collaboration}, {Prusti},
  {de Bruijne}, {Brown}, {Vallenari}, {Babusiaux}, {Bailer-Jones}, {Bastian},
  {Biermann}, {Evans}, \& et~al.}]{2016A&A...595A...1G}
{Gaia Collaboration}, {Prusti}, T., {de Bruijne}, J.~H.~J., {et~al.} 2016,
  \aap, 595, A1

\bibitem[{{Gaia Collaboration} {et~al.}(2021{\natexlab{c}}){Gaia
  Collaboration}, {Smart}, {Sarro}, {Rybizki}, {Reyl{\'e}}, {Robin}, {Hambly},
  {Abbas}, {Barstow}, {de Bruijne}, {Bucciarelli}, {Carrasco}, {Cooper},
  {Hodgkin}, {Masana}, {Michalik}, {Sahlmann}, {Sozzetti}, {Brown},
  {Vallenari}, {Prusti}, {Babusiaux}, {Biermann}, {Creevey}, {Evans}, {Eyer},
  {Hutton}, {Jansen}, {Jordi}, {Klioner}, {Lammers}, {Lindegren}, {Luri},
  {Mignard}, {Panem}, {Pourbaix}, {Randich}, {Sartoretti}, {Soubiran},
  {Walton}, {Arenou}, {Bailer-Jones}, {Bastian}, {Cropper}, {Drimmel}, {Katz},
  {Lattanzi}, {van Leeuwen}, {Bakker}, {Casta{\~n}eda}, {De Angeli},
  {Ducourant}, {Fabricius}, {Fouesneau}, {Fr{\'e}mat}, {Guerra}, {Guerrier},
  {Guiraud}, {Jean-Antoine Piccolo}, {Messineo}, {Mowlavi}, {Nicolas},
  {Nienartowicz}, {Pailler}, {Panuzzo}, {Riclet}, {Roux}, {Seabroke}, {Sordo},
  {Tanga}, {Th{\'e}venin}, {Gracia-Abril}, {Portell}, {Teyssier}, {Altmann},
  {Andrae}, {Bellas-Velidis}, {Benson}, {Berthier}, {Blomme}, {Brugaletta},
  {Burgess}, {Busso}, {Carry}, {Cellino}, {Cheek}, {Clementini}, {Damerdji},
  {Davidson}, {Delchambre}, {Dell'Oro}, {Fern{\'a}ndez-Hern{\'a}ndez},
  {Galluccio}, {Garc{\'\i}a-Lario}, {Garcia-Reinaldos},
  {Gonz{\'a}lez-N{\'u}{\~n}ez}, {Gosset}, {Haigron}, {Halbwachs}, {Harrison},
  {Hatzidimitriou}, {Heiter}, {Hern{\'a}ndez}, {Hestroffer}, {Holl},
  {Jan{\ss}en}, {Jevardat de Fombelle}, {Jordan}, {Krone-Martins}, {Lanzafame},
  {L{\"o}ffler}, {Lorca}, {Manteiga}, {Marchal}, {Marrese}, {Moitinho}, {Mora},
  {Muinonen}, {Osborne}, {Pancino}, {Pauwels}, {Recio-Blanco}, {Richards},
  {Riello}, {Rimoldini}, {Roegiers}, {Siopis}, {Smith}, {Ulla}, {Utrilla}, {van
  Leeuwen}, {van Reeven}, {Abreu Aramburu}, {Accart}, {Aerts}, {Aguado},
  {Ajaj}, {Altavilla}, {{\'A}lvarez}, {{\'A}lvarez Cid-Fuentes}, {Alves},
  {Anderson}, {Anglada Varela}, {Antoja}, {Audard}, {Baines}, {Baker},
  {Balaguer-N{\'u}{\~n}ez}, {Balbinot}, {Balog}, {Barache}, {Barbato},
  {Barros}, {Bartolom{\'e}}, {Bassilana}, {Bauchet}, {Baudesson-Stella},
  {Becciani}, {Bellazzini}, {Bernet}, {Bertone}, {Bianchi}, {Blanco-Cuaresma},
  {Boch}, {Bombrun}, {Bossini}, {Bouquillon}, {Bragaglia}, {Bramante},
  {Breedt}, {Bressan}, {Brouillet}, {Burlacu}, {Busonero}, {Butkevich},
  {Buzzi}, {Caffau}, {Cancelliere}, {C{\'a}novas}, {Cantat-Gaudin}, {Carballo},
  {Carlucci}, {Carnerero}, {Casamiquela}, {Castellani}, {Castro-Ginard},
  {Castro Sampol}, {Chaoul}, {Charlot}, {Chemin}, {Chiavassa}, {Cioni},
  {Comoretto}, {Cornez}, {Cowell}, {Crifo}, {Crosta}, {Crowley}, {Dafonte},
  {Dapergolas}, {David}, {David}, {de Laverny}, {De Luise}, {De March}, {De
  Ridder}, {de Souza}, {de Teodoro}, {de Torres}, {del Peloso}, {del Pozo},
  {Delgado}, {Delgado}, {Delisle}, {Di Matteo}, {Diakite}, {Diener},
  {Distefano}, {Dolding}, {Eappachen}, {Edvardsson}, {Enke}, {Esquej}, {Fabre},
  {Fabrizio}, {Faigler}, {Fedorets}, {Fernique}, {Fienga}, {Figueras},
  {Fouron}, {Fragkoudi}, {Fraile}, {Franke}, {Gai}, {Garabato},
  {Garcia-Gutierrez}, {Garc{\'\i}a-Torres}, {Garofalo}, {Gavras}, {Gerlach},
  {Geyer}, {Giacobbe}, {Gilmore}, {Girona}, {Giuffrida}, {Gomel}, {Gomez},
  {Gonzalez-Santamaria}, {Gonz{\'a}lez-Vidal}, {Granvik},
  {Guti{\'e}rrez-S{\'a}nchez}, {Guy}, {Hauser}, {Haywood}, {Helmi}, {Hidalgo},
  {Hilger}, {H{\l}adczuk}, {Hobbs}, {Holland}, {Huckle}, {Jasniewicz},
  {Jonker}, {Juaristi Campillo}, {Julbe}, {Karbevska}, {Kervella}, {Khanna},
  {Kochoska}, {Kontizas}, {Kordopatis}, {Korn}, {Kostrzewa-Rutkowska},
  {Kruszy{\'n}ska}, {Lambert}, {Lanza}, {Lasne}, {Le Campion}, {Le Fustec},
  {Lebreton}, {Lebzelter}, {Leccia}, {Leclerc}, {Lecoeur-Taibi}, {Liao},
  {Licata}, {Lindstr{\o}m}, {Lister}, {Livanou}, {Lobel}, {Madrero Pardo},
  {Managau}, {Mann}, {Marchant}, {Marconi}, {Marcos Santos}, {Marinoni},
  {Marocco}, {Marshall}, {Martin Polo}, {Mart{\'\i}n-Fleitas}, {Masip},
  {Massari}, {Mastrobuono-Battisti}, {Mazeh}, {McMillan}, {Messina}, {Millar},
  {Mints}, {Molina}, {Molinaro}, {Moln{\'a}r}, {Montegriffo}, {Mor},
  {Morbidelli}, {Morel}, {Morris}, {Mulone}, {Munoz}, {Muraveva}, {Murphy},
  {Musella}, {Noval}, {Ord{\'e}novic}, {Orr{\`u}}, {Osinde}, {Pagani},
  {Pagano}, {Palaversa}, {Palicio}, {Panahi}, {Pawlak}, {Pe{\~n}alosa
  Esteller}, {Penttil{\"a}}, {Piersimoni}, {Pineau}, {Plachy}, {Plum},
  {Poggio}, {Poretti}, {Poujoulet}, {Pr{\v{s}}a}, {Pulone}, {Racero},
  {Ragaini}, {Rainer}, {Raiteri}, {Rambaux}, {Ramos}, {Ramos-Lerate}, {Re
  Fiorentin}, {Regibo}, {Ripepi}, {Riva}, {Rixon}, {Robichon}, {Robin},
  {Roelens}, {Rohrbasser}, {Romero-G{\'o}mez}, {Rowell}, {Royer}, {Rybicki},
  {Sadowski}, {Sagrist{\`a} Sell{\'e}s}, {Salgado}, {Salguero}, {Samaras},
  {Sanchez Gimenez}, {Sanna}, {Santove{\~n}a}, {Sarasso}, {Schultheis},
  {Sciacca}, {Segol}, {Segovia}, {S{\'e}gransan}, {Semeux}, {Shahaf},
  {Siddiqui}, {Siebert}, {Siltala}, {Slezak}, {Solano}, {Solitro}, {Souami},
  {Souchay}, {Spagna}, {Spoto}, {Steele}, {Steidelm{\"u}ller}, {Stephenson},
  {S{\"u}veges}, {Szabados}, {Szegedi-Elek}, {Taris}, {Tauran}, {Taylor},
  {Teixeira}, {Thuillot}, {Tonello}, {Torra}, {Torra}, {Turon}, {Unger},
  {Vaillant}, {van Dillen}, {Vanel}, {Vecchiato}, {Viala}, {Vicente},
  {Voutsinas}, {Weiler}, {Wevers}, {Wyrzykowski}, {Yoldas}, {Yvard}, {Zhao},
  {Zorec}, {Zucker}, {Zurbach}, \& {Zwitter}}]{2021A&A...649A...6G}
{Gaia Collaboration}, {Smart}, R.~L., {Sarro}, L.~M., {et~al.}
  2021{\natexlab{c}}, \aap, 649, A6

\bibitem[{{Gandolfi} {et~al.}(2018){Gandolfi}, {Barrag{\'a}n}, {Livingston},
  {Fridlund}, {Justesen}, {Redfield}, {Fossati}, {Mathur}, {Grziwa}, {Cabrera},
  {Garc{\'\i}a}, {Persson}, {Van Eylen}, {Hatzes}, {Hidalgo}, {Albrecht},
  {Bugnet}, {Cochran}, {Csizmadia}, {Deeg}, {Eigm{\"u}ller}, {Endl}, {Erikson},
  {Esposito}, {Guenther}, {Korth}, {Luque}, {Monta{\~n}es Rodr{\'\i}guez},
  {Nespral}, {Nowak}, {P{\"a}tzold}, \& {Prieto-Arranz}}]{2018A&A...619L..10G}
{Gandolfi}, D., {Barrag{\'a}n}, O., {Livingston}, J.~H., {et~al.} 2018, \aap,
  619, L10

\bibitem[{{Gentile Fusillo} {et~al.}(2021){Gentile Fusillo}, {Tremblay},
  {Cukanovaite}, {Vorontseva}, {Lallement}, {Hollands}, {G{\"a}nsicke},
  {Burdge}, {McCleery}, \& {Jordan}}]{2021arXiv210607669G}
{Gentile Fusillo}, N.~P., {Tremblay}, P.~E., {Cukanovaite}, E., {et~al.} 2021,
  arXiv e-prints, arXiv:2106.07669

\bibitem[{{Gies} {et~al.}(2008){Gies}, {Dieterich}, {Richardson}, {Riedel},
  {B.~L. Team}, {McAlister}, {Bagnuolo}, {Grundstrom}, {{\v{S}}tefl},
  {Rivinius}, \& {Baade}}]{2008ApJ...682L.117G}
{Gies}, D.~R., {Dieterich}, S., {Richardson}, N.~D., {et~al.} 2008, \apjl, 682,
  L117

\bibitem[{{Gies} {et~al.}(2020){Gies}, {Lester}, {Wang}, {Couperus}, {Shepard},
  {Neiner}, {Wade}, {Dunham}, \& {Dunham}}]{2020ApJ...902...25G}
{Gies}, D.~R., {Lester}, K.~V., {Wang}, L., {et~al.} 2020, \apj, 902, 25

\bibitem[{{Ginsburg} {et~al.}(2017){Ginsburg}, {Parikh}, {Woillez}, {Groener},
  {Liedtke}, {Sipocz}, {Robitaille}, {Deil}, {Svoboda}, {Tollerud}, {Persson},
  {S{\'e}guin-Charbonneau}, {Armstrong}, {Mirocha}, {Droettboom}, {Allen},
  {Moolekamp}, {Egeland}, {Singer}, {Barbary}, {Grollier}, {Shiga}, {Moritz
  G{\"u}nther}, {Parejko}, {Booker}, {Rol}, {Edward}, {Miller}, \&
  {Willett}}]{2017ascl.soft08004G}
{Ginsburg}, A., {Parikh}, M., {Woillez}, J., {et~al.} 2017, {Astroquery: Access
  to online data resources}, Astrophysics Source Code Library

\bibitem[{{Girardi} {et~al.}(2000){Girardi}, {Bressan}, {Bertelli}, \&
  {Chiosi}}]{2000A&AS..141..371G}
{Girardi}, L., {Bressan}, A., {Bertelli}, G., \& {Chiosi}, C. 2000, Astronomy
  and Astrophysics Supplement Series, 141, 371

\bibitem[{{Gonz\'alez-Payo} {et~al.}(2021){Gonz\'alez-Payo},
  {Cort\'es-Contreras}, {Lodieu}, {Solano}, {Zhang}, \&
  {G\'alvez-Ortiz}}]{gonzalez-payo}
{Gonz\'alez-Payo}, J., {Cort\'es-Contreras}, M., {Lodieu}, N., {et~al.} 2021,
  A\&A, 650, A190

\bibitem[{{Gravity Collaboration} {et~al.}(2017){Gravity Collaboration},
  {Abuter}, {Accardo}, {Amorim}, {Anugu}, {{\'A}vila}, {Azouaoui}, {Benisty},
  {Berger}, {Blind}, {Bonnet}, {Bourget}, {Brandner}, {Brast}, {Buron},
  {Burtscher}, {Cassaing}, {Chapron}, {Choquet}, {Cl{\'e}net}, {Collin},
  {Coud{\'e} Du Foresto}, {de Wit}, {de Zeeuw}, {Deen},
  {Delplancke-Str{\"o}bele}, {Dembet}, {Derie}, {Dexter}, {Duvert}, {Ebert},
  {Eckart}, {Eisenhauer}, {Esselborn}, {F{\'e}dou}, {Finger}, {Garcia}, {Garcia
  Dabo}, {Garcia Lopez}, {Gendron}, {Genzel}, {Gillessen}, {Gonte}, {Gordo},
  {Grould}, {Gr{\"o}zinger}, {Guieu}, {Haguenauer}, {Hans}, {Haubois}, {Haug},
  {Haussmann}, {Henning}, {Hippler}, {Horrobin}, {Huber}, {Hubert}, {Hubin},
  {Hummel}, {Jakob}, {Janssen}, {Jochum}, {Jocou}, {Kaufer}, {Kellner},
  {Kendrew}, {Kern}, {Kervella}, {Kiekebusch}, {Klein}, {Kok}, {Kolb}, {Kulas},
  {Lacour}, {Lapeyr{\`e}re}, {Lazareff}, {Le Bouquin}, {L{\`e}na}, {Lenzen},
  {L{\'e}v{\^e}que}, {Lippa}, {Magnard}, {Mehrgan}, {Mellein}, {M{\'e}rand},
  {Moreno-Ventas}, {Moulin}, {M{\"u}ller}, {M{\"u}ller}, {Neumann}, {Oberti},
  {Ott}, {Pallanca}, {Panduro}, {Pasquini}, {Paumard}, {Percheron}, {Perraut},
  {Perrin}, {Pfl{\"u}ger}, {Pfuhl}, {Phan Duc}, {Plewa}, {Popovic}, {Rabien},
  {Ram{\'{\i}}rez}, {Ramos}, {Rau}, {Riquelme}, {Rohloff}, {Rousset},
  {Sanchez-Bermudez}, {Scheithauer}, {Sch{\"o}ller}, {Schuhler}, {Spyromilio},
  {Straubmeier}, {Sturm}, {Suarez}, {Tristram}, {Ventura}, {Vincent},
  {Waisberg}, {Wank}, {Weber}, {Wieprecht}, {Wiest}, {Wiezorrek}, {Wittkowski},
  {Woillez}, {Wolff}, {Yazici}, {Ziegler}, \& {Zins}}]{2017A&A...602A..94G}
{Gravity Collaboration}, {Abuter}, R., {Accardo}, M., {et~al.} 2017, \aap, 602,
  A94

\bibitem[{{Gravity Collaboration} {et~al.}(2019){Gravity Collaboration},
  {Lacour}, {Nowak}, {Wang}, {Pfuhl}, {Eisenhauer}, {Abuter}, {Amorim},
  {Anugu}, {Benisty}, {Berger}, {Beust}, {Blind}, {Bonnefoy}, {Bonnet},
  {Bourget}, {Brandner}, {Buron}, {Collin}, {Charnay}, {Chapron}, {Cl{\'e}net},
  {Coud{\'e} Du Foresto}, {de Zeeuw}, {Deen}, {Dembet}, {Dexter}, {Duvert},
  {Eckart}, {F{\"o}rster Schreiber}, {F{\'e}dou}, {Garcia}, {Garcia Lopez},
  {Gao}, {Gendron}, {Genzel}, {Gillessen}, {Gordo}, {Greenbaum}, {Habibi},
  {Haubois}, {Hau{\ss}mann}, {Henning}, {Hippler}, {Horrobin}, {Hubert},
  {Jimenez Rosales}, {Jocou}, {Kendrew}, {Kervella}, {Kolb}, {Lagrange},
  {Lapeyr{\`e}re}, {Le Bouquin}, {L{\'e}na}, {Lippa}, {Lenzen}, {Maire},
  {Molli{\`e}re}, {Ott}, {Paumard}, {Perraut}, {Perrin}, {Pueyo}, {Rabien},
  {Ram{\'\i}rez}, {Rau}, {Rodr{\'\i}guez-Coira}, {Rousset}, {Sanchez-Bermudez},
  {Scheithauer}, {Schuhler}, {Straub}, {Straubmeier}, {Sturm}, {Tacconi},
  {Vincent}, {van Dishoeck}, {von Fellenberg}, {Wank}, {Waisberg}, {Widmann},
  {Wieprecht}, {Wiest}, {Wiezorrek}, {Woillez}, {Yazici}, {Ziegler}, \&
  {Zins}}]{2019A&A...623L..11G}
{Gravity Collaboration}, {Lacour}, S., {Nowak}, M., {et~al.} 2019, \aap, 623,
  L11

\bibitem[{{Gravity Collaboration} {et~al.}(2020){Gravity Collaboration},
  {Nowak}, {Lacour}, {Molli{\`e}re}, {Wang}, {Charnay}, {van Dishoeck},
  {Abuter}, {Amorim}, {Berger}, {Beust}, {Bonnefoy}, {Bonnet}, {Brandner},
  {Buron}, {Cantalloube}, {Collin}, {Chapron}, {Cl{\'e}net}, {Coud{\'e} Du
  Foresto}, {de Zeeuw}, {Dembet}, {Dexter}, {Duvert}, {Eckart}, {Eisenhauer},
  {F{\"o}rster Schreiber}, {F{\'e}dou}, {Garcia Lopez}, {Gao}, {Gendron},
  {Genzel}, {Gillessen}, {Hau{\ss}mann}, {Henning}, {Hippler}, {Hubert},
  {Jocou}, {Kervella}, {Lagrange}, {Lapeyr{\`e}re}, {Le Bouquin}, {L{\'e}na},
  {Maire}, {Ott}, {Paumard}, {Paladini}, {Perraut}, {Perrin}, {Pueyo}, {Pfuhl},
  {Rabien}, {Rau}, {Rodr{\'\i}guez-Coira}, {Rousset}, {Scheithauer},
  {Shangguan}, {Straub}, {Straubmeier}, {Sturm}, {Tacconi}, {Vincent},
  {Widmann}, {Wieprecht}, {Wiezorrek}, {Woillez}, {Yazici}, \&
  {Ziegler}}]{2020A&A...633A.110G}
{Gravity Collaboration}, {Nowak}, M., {Lacour}, S., {et~al.} 2020, \aap, 633,
  A110

\bibitem[{{Han} {et~al.}(2014){Han}, {Wang}, {Wright}, {Feng}, {Zhao},
  {Fakhouri}, {Brown}, \& {Hancock}}]{2014PASP..126..827H}
{Han}, E., {Wang}, S.~X., {Wright}, J.~T., {et~al.} 2014, \pasp, 126, 827

\bibitem[{Harris {et~al.}(2020)Harris, Millman, van~der Walt, Gommers,
  Virtanen, Cournapeau, Wieser, Taylor, Berg, Smith, Kern, Picus, Hoyer, van
  Kerkwijk, Brett, Haldane, del R{\'\i}o, Wiebe, Peterson, G{\'e}rard-Marchant,
  Sheppard, Reddy, Weckesser, Abbasi, Gohlke, \& Oliphant}]{Harris20}
Harris, C.~R., Millman, K.~J., van~der Walt, S.~J., {et~al.} 2020, Nature, 585,
  357

\bibitem[{{Hartkopf} {et~al.}(2001){Hartkopf}, {Mason}, \&
  {Worley}}]{2001AJ....122.3472H}
{Hartkopf}, W.~I., {Mason}, B.~D., \& {Worley}, C.~E. 2001, \aj, 122, 3472

\bibitem[{Hartman \& L{\'{e}}pine(2020)}]{Hartman_2020}
Hartman, Z.~D. \& L{\'{e}}pine, S. 2020, The Astrophysical Journal Supplement
  Series, 247, 66

\bibitem[{{Hatzes} {et~al.}(2000){Hatzes}, {Cochran}, {McArthur}, {Baliunas},
  {Walker}, {Campbell}, {Irwin}, {Yang}, {K{\"u}rster}, {Endl}, {Els},
  {Butler}, \& {Marcy}}]{2000ApJ...544L.145H}
{Hatzes}, A.~P., {Cochran}, W.~D., {McArthur}, B., {et~al.} 2000, \apj, 544,
  L145

\bibitem[{{Heinze} {et~al.}(2010){Heinze}, {Hinz}, {Sivanandam}, {Kenworthy},
  {Meyer}, \& {Miller}}]{2010ApJ...714.1551H}
{Heinze}, A.~N., {Hinz}, P.~M., {Sivanandam}, S., {et~al.} 2010, \apj, 714,
  1551

\bibitem[{{Higl} \& {Weiss}(2017)}]{2017A&A...608A..62H}
{Higl}, J. \& {Weiss}, A. 2017, \aap, 608, A62

\bibitem[{{Hirsch} {et~al.}(2021){Hirsch}, {Rosenthal}, {Fulton}, {Howard},
  {Ciardi}, {Marcy}, {Nielsen}, {Petigura}, {de Rosa}, {Isaacson}, {Weiss},
  {Sinukoff}, \& {Macintosh}}]{2021AJ....161..134H}
{Hirsch}, L.~A., {Rosenthal}, L., {Fulton}, B.~J., {et~al.} 2021, \aj, 161, 134

\bibitem[{{Hobbs} {et~al.}(2019){Hobbs}, {Brown}, {H{\o}g}, {Jordi}, {Kawata},
  {Tanga}, {Klioner}, {Sozzetti}, {Wyrzykowski}, {Walton}, {Vallenari},
  {Makarov}, {Rybizki}, {Jim{\'e}nez-Esteban}, {Caballero}, {McMillan},
  {Secrest}, {Mor}, {Andrews}, {Zwitter}, {Chiappini}, {Fynbo}, {Ting},
  {Hestroffer}, {Lindegren}, {McArthur}, {Gouda}, {Moore}, {Gonzalez}, \&
  {Vaccari}}]{2019arXiv190712535H}
{Hobbs}, D., {Brown}, A., {H{\o}g}, E., {et~al.} 2019, arXiv e-prints,
  arXiv:1907.12535

\bibitem[{{Holberg} {et~al.}(2016){Holberg}, {Oswalt}, {Sion}, \&
  {McCook}}]{2016MNRAS.462.2295H}
{Holberg}, J.~B., {Oswalt}, T.~D., {Sion}, E.~M., \& {McCook}, G.~P. 2016,
  \mnras, 462, 2295

\bibitem[{{Holmberg} {et~al.}(2007){Holmberg}, {Nordstr{\"o}m}, \&
  {Andersen}}]{2007A&A...475..519H}
{Holmberg}, J., {Nordstr{\"o}m}, B., \& {Andersen}, J. 2007, \aap, 475, 519

\bibitem[{{Homan} {et~al.}(2017){Homan}, {Richards}, {Decin}, {Kervella}, {de
  Koter}, {McDonald}, \& {Ohnaka}}]{2017A&A...601A...5H}
{Homan}, W., {Richards}, A., {Decin}, L., {et~al.} 2017, \aap, 601, A5

\bibitem[{{Howard} \& {Fulton}(2016)}]{2016PASP..128k4401H}
{Howard}, A.~W. \& {Fulton}, B.~J. 2016, Publications of the Astronomical
  Society of the Pacific, 128, 114401

\bibitem[{{Huang} {et~al.}(2018){Huang}, {Burt}, {Vanderburg}, {G{\"u}nther},
  {Shporer}, {Dittmann}, {Winn}, {Wittenmyer}, {Sha}, {Kane}, {Ricker},
  {Vanderspek}, {Latham}, {Seager}, {Jenkins}, {Caldwell}, {Collins},
  {Guerrero}, {Smith}, {Quinn}, {Udry}, {Pepe}, {Bouchy}, {S{\'e}gransan},
  {Lovis}, {Ehrenreich}, {Marmier}, {Mayor}, {Wohler}, {Haworth}, {Morgan},
  {Fausnaugh}, {Ciardi}, {Christiansen}, {Charbonneau}, {Dragomir}, {Deming},
  {Glidden}, {Levine}, {McCullough}, {Yu}, {Narita}, {Nguyen}, {Morton},
  {Pepper}, {P{\'a}l}, {Rodriguez}, {Stassun}, {Torres}, {Sozzetti}, {Doty},
  {Christensen-Dalsgaard}, {Laughlin}, {Clampin}, {Bean}, {Buchhave}, {Bakos},
  {Sato}, {Ida}, {Kaltenegger}, {Palle}, {Sasselov}, {Butler}, {Lissauer},
  {Ge}, \& {Rinehart}}]{2018ApJ...868L..39H}
{Huang}, C.~X., {Burt}, J., {Vanderburg}, A., {et~al.} 2018, \apjl, 868, L39

\bibitem[{{Huby} {et~al.}(2013){Huby}, {Duch{\^e}ne}, {Marchis}, {Lacour},
  {Perrin}, {Kotani}, {Choquet}, {Gates}, {Lai}, \&
  {Allard}}]{2013A&A...560A.113H}
{Huby}, E., {Duch{\^e}ne}, G., {Marchis}, F., {et~al.} 2013, \aap, 560, A113

\bibitem[{Hunter(2007)}]{Hunter:2007}
Hunter, J.~D. 2007, Computing in Science \& Engineering, 9, 90

\bibitem[{{Janson} {et~al.}(2015){Janson}, {Quanz}, {Carson}, {Thalmann},
  {Lafreni{\`e}re}, \& {Amara}}]{2015A&A...574A.120J}
{Janson}, M., {Quanz}, S.~P., {Carson}, J.~C., {et~al.} 2015, \aap, 574, A120

\bibitem[{Jim{\'{e}}nez-Esteban {et~al.}(2019)Jim{\'{e}}nez-Esteban, Solano, \&
  Rodrigo}]{Jim_nez_Esteban_2019}
Jim{\'{e}}nez-Esteban, F.~M., Solano, E., \& Rodrigo, C. 2019, \aj, 157, 78

\bibitem[{{Jones} {et~al.}(2002){Jones}, {Paul Butler}, {Tinney}, {Marcy},
  {Penny}, {McCarthy}, {Carter}, \& {Pourbaix}}]{2002MNRAS.333..871J}
{Jones}, H. R.~A., {Paul Butler}, R., {Tinney}, C.~G., {et~al.} 2002, \mnras,
  333, 871

\bibitem[{{Jorissen} {et~al.}(2004){Jorissen}, {Jancart}, \&
  {Pourbaix}}]{2004ASPC..318..141J}
{Jorissen}, A., {Jancart}, S., \& {Pourbaix}, D. 2004, in Astronomical Society
  of the Pacific Conference Series, Vol. 318, Spectroscopically and Spatially
  Resolving the Components of the Close Binary Stars, ed. R.~W. {Hilditch},
  H.~{Hensberge}, \& K.~{Pavlovski}, 141--143

\bibitem[{{Kammerer} {et~al.}(2021{\natexlab{a}}){Kammerer}, {Lacour},
  {Stolker}, {Molli{\`e}re}, {Sing}, {Nasedkin}, {Kervella}, {Wang},
  {Ward-Duong}, {Nowak}, {Abuter}, {Amorim}, {Asensio-Torres}, {Baub{\"o}ck},
  {Benisty}, {Berger}, {Beust}, {Blunt}, {Boccaletti}, {Bohn}, {Bolzer},
  {Bonnefoy}, {Bonnet}, {Brandner}, {Cantalloube}, {Caselli}, {Charnay},
  {Chauvin}, {Choquet}, {Christiaens}, {Cl{\'e}net}, {Coud{\'e} du Foresto},
  {Cridland}, {Dembet}, {Dexter}, {de Zeeuw}, {Drescher}, {Duvert}, {Eckart},
  {Eisenhauer}, {Gao}, {Garcia}, {Garcia Lopez}, {Gendron}, {Genzel},
  {Gillessen}, {Girard}, {Haubois}, {Hei{\ss}el}, {Henning}, {Hinkley},
  {Hippler}, {Horrobin}, {Houll{\'e}}, {Hubert}, {Jocou}, {Keppler},
  {Kreidberg}, {Lagrange}, {Lapeyr{\`e}re}, {Le Bouquin}, {L{\'e}na}, {Lutz},
  {Maire}, {M{\'e}rand}, {Monnier}, {Mouillet}, {M{\"u}ller}, {Ott}, {Otten},
  {Paladini}, {Paumard}, {Perraut}, {Perrin}, {Pfuhl}, {Pueyo}, {Rameau},
  {Rodet}, {Rousset}, {Rustamkulov}, {Shangguan}, {Shimizu}, {Stadler},
  {Straub}, {Straubmeier}, {Sturm}, {Tacconi}, {van Dishoeck}, {Vigan},
  {Vincent}, {von Fellenberg}, {Widmann}, {Wieprecht}, {Wiezorrek}, {Woillez},
  {Yazici}, \& {GRAVITY Collaboration}}]{2021arXiv210608249K}
{Kammerer}, J., {Lacour}, S., {Stolker}, T., {et~al.} 2021{\natexlab{a}}, arXiv
  e-prints, arXiv:2106.08249

\bibitem[{{Kammerer} {et~al.}(2021{\natexlab{b}}){Kammerer}, {Lacour},
  {Stolker}, {Molli{\`e}re}, {Sing}, {Nasedkin}, {Kervella}, {Wang},
  {Ward-Duong}, {Nowak}, {Abuter}, {Amorim}, {Asensio-Torres}, {Baub{\"o}ck},
  {Benisty}, {Berger}, {Beust}, {Blunt}, {Boccaletti}, {Bohn}, {Bolzer},
  {Bonnefoy}, {Bonnet}, {Brandner}, {Cantalloube}, {Caselli}, {Charnay},
  {Chauvin}, {Choquet}, {Christiaens}, {Cl{\'e}net}, {Coud{\'e} du Foresto},
  {Cridland}, {Dembet}, {Dexter}, {de Zeeuw}, {Drescher}, {Duvert}, {Eckart},
  {Eisenhauer}, {Gao}, {Garcia}, {Garcia Lopez}, {Gendron}, {Genzel},
  {Gillessen}, {Girard}, {Haubois}, {Hei{\ss}el}, {Henning}, {Hinkley},
  {Hippler}, {Horrobin}, {Houll{\'e}}, {Hubert}, {Jocou}, {Keppler},
  {Kreidberg}, {Lagrange}, {Lapeyr{\`e}re}, {Le Bouquin}, {L{\'e}na}, {Lutz},
  {Maire}, {M{\'e}rand}, {Monnier}, {Mouillet}, {M{\"u}ller}, {Ott}, {Otten},
  {Paladini}, {Paumard}, {Perraut}, {Perrin}, {Pfuhl}, {Pueyo}, {Rameau},
  {Rodet}, {Rousset}, {Rustamkulov}, {Shangguan}, {Shimizu}, {Stadler},
  {Straub}, {Straubmeier}, {Sturm}, {Tacconi}, {van Dishoeck}, {Vigan},
  {Vincent}, {von Fellenberg}, {Widmann}, {Wieprecht}, {Wiezorrek}, {Woillez},
  \& {Yazici}}]{2021A&A...652A..57K}
{Kammerer}, J., {Lacour}, S., {Stolker}, T., {et~al.} 2021{\natexlab{b}}, \aap,
  652, A57

\bibitem[{{Kervella} {et~al.}(2019{\natexlab{a}}){Kervella}, {Arenou},
  {Mignard}, \& {Th{\'e}venin}}]{2019A&A...623A..72K}
{Kervella}, P., {Arenou}, F., {Mignard}, F., \& {Th{\'e}venin}, F.
  2019{\natexlab{a}}, \aap, 623, A72

\bibitem[{{Kervella} {et~al.}(2020){Kervella}, {Arenou}, \&
  {Schneider}}]{2020A&A...635L..14K}
{Kervella}, P., {Arenou}, F., \& {Schneider}, J. 2020, \aap, 635, L14

\bibitem[{{Kervella} {et~al.}(2019{\natexlab{b}}){Kervella}, {Gallenne},
  {Evans}, {Szabados}, {Arenou}, {M{\'e}rand }, {Nardetto}, {Gieren}, \&
  {Pietrzynski}}]{2019A&A...623A.117K}
{Kervella}, P., {Gallenne}, A., {Evans}, N.~R., {et~al.} 2019{\natexlab{b}},
  \aap, 623, A117

\bibitem[{{Kervella} {et~al.}(2019{\natexlab{c}}){Kervella}, {Gallenne},
  {Evans}, {Szabados}, {Arenou}, {M{\'e}rand }, {Proto}, {Karczmarek},
  {Nardetto}, {Gieren}, \& {Pietrzynski}}]{2019A&A...623A.116K}
{Kervella}, P., {Gallenne}, A., {Evans}, N.~R., {et~al.} 2019{\natexlab{c}},
  \aap, 623, A116

\bibitem[{{Kervella} {et~al.}(2016{\natexlab{a}}){Kervella}, {Homan},
  {Richards}, {Decin}, {McDonald}, {Montarg{\`e}s}, \&
  {Ohnaka}}]{2016A&A...596A..92K}
{Kervella}, P., {Homan}, W., {Richards}, A.~M.~S., {et~al.} 2016{\natexlab{a}},
  \aap, 596, A92

\bibitem[{{Kervella} {et~al.}(2016{\natexlab{b}}){Kervella}, {M{\'e}rand},
  {Ledoux}, {Demory}, \& {Le Bouquin}}]{2016A&A...593A.127K}
{Kervella}, P., {M{\'e}rand}, A., {Ledoux}, C., {Demory}, B.-O., \& {Le
  Bouquin}, J.-B. 2016{\natexlab{b}}, \aap, 593, A127

\bibitem[{{Kervella} {et~al.}(2008){Kervella}, {M{\'e}rand}, {Pichon},
  {Th{\'e}venin}, {Heiter}, {Bigot}, {ten Brummelaar}, {McAlister}, {Ridgway},
  {Turner}, {Sturmann}, {Sturmann}, {Goldfinger}, \&
  {Farrington}}]{2008A&A...488..667K}
{Kervella}, P., {M{\'e}rand}, A., {Pichon}, B., {et~al.} 2008, \aap, 488, 667

\bibitem[{{Kervella} {et~al.}(2016{\natexlab{c}}){Kervella}, {Mignard},
  {M{\'e}rand}, \& {Th{\'e}venin}}]{2016kervella}
{Kervella}, P., {Mignard}, F., {M{\'e}rand}, A., \& {Th{\'e}venin}, F.
  2016{\natexlab{c}}, \aap, 594, A107

\bibitem[{{Kervella} {et~al.}(2015){Kervella}, {Montarg{\`e}s}, {Lagadec},
  {Ridgway}, {Haubois}, {Girard}, {Ohnaka}, {Perrin}, \&
  {Gallenne}}]{2015A&A...578A..77K}
{Kervella}, P., {Montarg{\`e}s}, M., {Lagadec}, E., {et~al.} 2015, \aap, 578,
  A77

\bibitem[{{Kervella} {et~al.}(2014){Kervella}, {Montarg{\`e}s}, {Ridgway},
  {Perrin}, {Chesneau}, {Lacour}, {Chiavassa}, {Haubois}, \&
  {Gallenne}}]{2014A&A...564A..88K}
{Kervella}, P., {Montarg{\`e}s}, M., {Ridgway}, S.~T., {et~al.} 2014, \aap,
  564, A88

\bibitem[{{Kervella} {et~al.}(2004){Kervella}, {Th{\'e}venin}, {Di Folco}, \&
  {S{\'e}gransan}}]{2004A&A...426..297K}
{Kervella}, P., {Th{\'e}venin}, F., {Di Folco}, E., \& {S{\'e}gransan}, D.
  2004, \aap, 426, 297

\bibitem[{{Kervella} {et~al.}(2017){Kervella}, {Th{\'e}venin}, \&
  {Lovis}}]{2017A&A...598L...7K}
{Kervella}, P., {Th{\'e}venin}, F., \& {Lovis}, C. 2017, \aap, 598, L7

\bibitem[{{Kiefer}(2019)}]{2019A&A...632L...9K}
{Kiefer}, F. 2019, \aap, 632, L9

\bibitem[{{Kiefer} {et~al.}(2021){Kiefer}, {H{\'e}brard}, {Lecavelier des
  Etangs}, {Martioli}, {Dalal}, \& {Vidal-Madjar}}]{2021A&A...645A...7K}
{Kiefer}, F., {H{\'e}brard}, G., {Lecavelier des Etangs}, A., {et~al.} 2021,
  \aap, 645, A7

\bibitem[{{Kirk} {et~al.}(2016){Kirk}, {Conroy}, {Pr{\v{s}}a}, {Abdul-Masih},
  {Kochoska}, {Matijevi{\v{c}}}, {Hambleton}, {Barclay}, {Bloemen}, {Boyajian},
  {Doyle}, {Fulton}, {Hoekstra}, {Jek}, {Kane}, {Kostov}, {Latham}, {Mazeh},
  {Orosz}, {Pepper}, {Quarles}, {Ragozzine}, {Shporer}, {Southworth},
  {Stassun}, {Thompson}, {Welsh}, {Agol}, {Derekas}, {Devor}, {Fischer},
  {Green}, {Gropp}, {Jacobs}, {Johnston}, {LaCourse}, {Saetre}, {Schwengeler},
  {Toczyski}, {Werner}, {Garrett}, {Gore}, {Martinez}, {Spitzer}, {Stevick},
  {Thomadis}, {Vrijmoet}, {Yenawine}, {Batalha}, \&
  {Borucki}}]{2016AJ....151...68K}
{Kirk}, B., {Conroy}, K., {Pr{\v{s}}a}, A., {et~al.} 2016, \aj, 151, 68

\bibitem[{{Kraus} {et~al.}(2016){Kraus}, {Ireland}, {Huber}, {Mann}, \&
  {Dupuy}}]{2016AJ....152....8K}
{Kraus}, A.~L., {Ireland}, M.~J., {Huber}, D., {Mann}, A.~W., \& {Dupuy}, T.~J.
  2016, \aj, 152, 8

\bibitem[{{Kunovac Hod{\v{z}}i{\'c}} {et~al.}(2021){Kunovac Hod{\v{z}}i{\'c}},
  {Triaud}, {Cegla}, {Chaplin}, \& {Davies}}]{2021MNRAS.502.2893K}
{Kunovac Hod{\v{z}}i{\'c}}, V., {Triaud}, A. H.~M.~J., {Cegla}, H.~M.,
  {Chaplin}, W.~J., \& {Davies}, G.~R. 2021, \mnras, 502, 2893

\bibitem[{{Lagrange} {et~al.}(2019){Lagrange}, {Boccaletti}, {Langlois},
  {Chauvin}, {Gratton}, {Beust}, {Desidera}, {Milli}, {Bonnefoy}, {Cheetham},
  {Feldt}, {Meyer}, {Vigan}, {Biller}, {Bonavita}, {Baudino}, {Cantalloube},
  {Cudel}, {Daemgen}, {Delorme}, {D'Orazi}, {Girard}, {Fontanive}, {Hagelberg},
  {Janson}, {Keppler}, {Koypitova}, {Galicher}, {Lannier}, {Le Coroller},
  {Ligi}, {Maire}, {Mesa}, {Messina}, {M{\"u}eller}, {Peretti}, {Perrot},
  {Rouan}, {Salter}, {Samland}, {Schmidt}, {Sissa}, {Zurlo}, {Beuzit},
  {Mouillet}, {Dominik}, {Henning}, {Lagadec}, {M{\'e}nard}, {Schmid},
  {Turatto}, {Udry}, {Bohn}, {Charnay}, {Gomez Gonzales}, {Gry}, {Kenworthy},
  {Kral}, {Mordasini}, {Moutou}, {van der Plas}, {Schlieder}, {Abe}, {Antichi},
  {Baruffolo}, {Baudoz}, {Baudrand}, {Blanchard}, {Bazzon}, {Buey},
  {Carbillet}, {Carle}, {Charton}, {Cascone}, {Claudi}, {Costille}, {Deboulbe},
  {De Caprio}, {Dohlen}, {Fantinel}, {Feautrier}, {Fusco}, {Gigan}, {Giro},
  {Gisler}, {Gluck}, {Hubin}, {Hugot}, {Jaquet}, {Kasper}, {Madec}, {Magnard},
  {Martinez}, {Maurel}, {Le Mignant}, {M{\"o}ller-Nilsson}, {Llored}, {Moulin},
  {Orign{\'e}}, {Pavlov}, {Perret}, {Petit}, {Pragt}, {Szulagyi}, \&
  {Wildi}}]{2019A&A...621L...8L}
{Lagrange}, A.~M., {Boccaletti}, A., {Langlois}, M., {et~al.} 2019, \aap, 621,
  L8

\bibitem[{{Lagrange} {et~al.}(2020){Lagrange}, {Rubini}, {Nowak}, {Lacour},
  {Grandjean}, {Boccaletti}, {Langlois}, {Delorme}, {Gratton}, {Wang},
  {Flasseur}, {Galicher}, {Kral}, {Meunier}, {Beust}, {Babusiaux}, {Le
  Coroller}, {Thebault}, {Kervella}, {Zurlo}, {Maire}, {Wahhaj}, {Amorim},
  {Asensio-Torres}, {Benisty}, {Berger}, {Bonnefoy}, {Brandner}, {Cantalloube},
  {Charnay}, {Chauvin}, {Choquet}, {Cl{\'e}net}, {Christiaens}, {Coud{\'e} Du
  Foresto}, {de Zeeuw}, {Desidera}, {Duvert}, {Eckart}, {Eisenhauer},
  {Galland}, {Gao}, {Garcia}, {Garcia Lopez}, {Gendron}, {Genzel}, {Gillessen},
  {Girard}, {Hagelberg}, {Haubois}, {Henning}, {Heissel}, {Hippler},
  {Horrobin}, {Janson}, {Kammerer}, {Kenworthy}, {Keppler}, {Kreidberg},
  {Lapeyr{\`e}re}, {Le Bouquin}, {L{\'e}na}, {M{\'e}rand}, {Messina},
  {Molli{\`e}re}, {Monnier}, {Ott}, {Otten}, {Paumard}, {Paladini}, {Perraut},
  {Perrin}, {Pueyo}, {Pfuhl}, {Rodet}, {Rodriguez-Coira}, {Rousset}, {Samland},
  {Shangguan}, {Schmidt}, {Straub}, {Straubmeier}, {Stolker}, {Vigan},
  {Vincent}, {Widmann}, {Woillez}, \& {Gravity
  Collaboration}}]{2020A&A...642A..18L}
{Lagrange}, A.~M., {Rubini}, P., {Nowak}, M., {et~al.} 2020, \aap, 642, A18

\bibitem[{{Lallement} {et~al.}(2014){Lallement}, {Vergely}, {Valette},
  {Puspitarini}, {Eyer}, \& {Casagrande}}]{2014A&A...561A..91L}
{Lallement}, R., {Vergely}, J.-L., {Valette}, B., {et~al.} 2014, \aap, 561, A91

\bibitem[{{Lindegren}(1997)}]{1997ESASP.402...13L}
{Lindegren}, L. 1997, in ESA Special Publication, Vol. 402, Hipparcos - Venice
  '97, ed. R.~M. {Bonnet}, E.~{H{\o}g}, P.~L. {Bernacca}, L.~{Emiliani},
  A.~{Blaauw}, C.~{Turon}, J.~{Kovalevsky}, L.~{Lindegren}, H.~{Hassan},
  M.~{Bouffard}, B.~{Strim}, D.~{Heger}, M.~A.~C. {Perryman}, \& L.~{Woltjer},
  13--18

\bibitem[{{Lindegren} {et~al.}(2021{\natexlab{a}}){Lindegren}, {Bastian},
  {Biermann}, {Bombrun}, {de Torres}, {Gerlach}, {Geyer}, {Hern{\'a}ndez},
  {Hilger}, {Hobbs}, {Klioner}, {Lammers}, {McMillan}, {Ramos-Lerate},
  {Steidelm{\"u}ller}, {Stephenson}, \& {van Leeuwen}}]{2021A&A...649A...4L}
{Lindegren}, L., {Bastian}, U., {Biermann}, M., {et~al.} 2021{\natexlab{a}},
  \aap, 649, A4

\bibitem[{{Lindegren} \& {Dravins}(2021)}]{2021A&A...652A..45L}
{Lindegren}, L. \& {Dravins}, D. 2021, \aap, 652, A45

\bibitem[{{Lindegren} {et~al.}(2021{\natexlab{b}}){Lindegren}, {Klioner, S.
  A.}, {Hern\'andez, J.}, {Bombrun, A.}, {Ramos-Lerate, M.}, {Steidelm\"uller,
  H.}, {Bastian, U.}, {Biermann, M.}, {de Torres, A.}, {Gerlach, E.}, {Geyer,
  R.}, {Hilger, T.}, {Hobbs, D.}, {Lammers, U.}, {McMillan, P. J.},
  {Stephenson, C. A.}, {Casta\~neda, J.}, {Davidson, M.}, {Fabricius, C.},
  {Gracia-Abril, G.}, {Portell, J.}, {Rowell, N.}, {Teyssier, D.}, {Torra, F.},
  {Bartolom\'e, S.}, {Clotet, M.}, {Garralda, N.}, {Gonz\'alez-Vidal, J. J.},
  {Torra, J.}, {Abbas, U.}, {Altmann, M.}, {Anglada Varela, E.},
  {Balaguer-N\'u\~nez, L.}, {Balog, Z.}, {Barache, C.}, {Becciani, U.},
  {Bernet, M.}, {Bertone, S.}, {Bianchi, L.}, {Bouquillon, S.}, {Brown, A. G.
  A.}, {Bucciarelli, B.}, {Busonero, D.}, {Butkevich, A. G.}, {Buzzi, R.},
  {Cancelliere, R.}, {Carlucci, T.}, {Charlot, P.}, {Cioni, M.-R. L.}, {Crosta,
  M.}, {Crowley, C.}, {del Peloso, E. F.}, {del Pozo, E.}, {Drimmel, R.},
  {Esquej, P.}, {Fienga, A.}, {Fraile, E.}, {Gai, M.}, {Garcia-Reinaldos, M.},
  {Guerra, R.}, {Hambly, N. C.}, {Hauser, M.}, {Jan\ss{}en, K.}, {Jordan, S.},
  {Kostrzewa-Rutkowska, Z.}, {Lattanzi, M. G.}, {Liao, S.}, {Licata, E.},
  {Lister, T. A.}, {L\"offler, W.}, {Marchant, J. M.}, {Masip, A.}, {Mignard,
  F.}, {Mints, A.}, {Molina, D.}, {Mora, A.}, {Morbidelli, R.}, {Murphy, C.
  P.}, {Pagani, C.}, {Panuzzo, P.}, {Pe\~nalosa Esteller, X.}, {Poggio, E.},
  {Re Fiorentin, P.}, {Riva, A.}, {Sagrist\`a Sell\'es, A.}, {Sanchez Gimenez,
  V.}, {Sarasso, M.}, {Sciacca, E.}, {Siddiqui, H. I.}, {Smart, R. L.},
  {Souami, D.}, {Spagna, A.}, {Steele, I. A.}, {Taris, F.}, {Utrilla, E.}, {van
  Reeven, W.}, \& {Vecchiato, A.}}]{GaiaEDR3astrometric}
{Lindegren}, L., {Klioner, S. A.}, {Hern\'andez, J.}, {et~al.}
  2021{\natexlab{b}}, A\&A, 649, A2

\bibitem[{{Lindegren} {et~al.}(1997){Lindegren}, {Mignard}, {S{\"o}derhjelm},
  {Badiali}, {Bernstein}, {Lampens}, {Pannunzio}, {Arenou}, {Bernacca},
  {Falin}, {Froeschl{\'e}}, {Kovalevsky}, {Martin}, {Perryman}, \&
  {Wielen}}]{1997A&A...323L..53L}
{Lindegren}, L., {Mignard}, F., {S{\"o}derhjelm}, S., {et~al.} 1997, \aap, 323,
  L53

\bibitem[{{Lykou} {et~al.}(2015{\natexlab{a}}){Lykou}, {Klotz}, {Paladini},
  {Hron}, {Zijlstra}, {Kluska}, {Norris}, {Tuthill}, {Ramstedt}, {Lagadec},
  {Wittkowski}, {Maercker}, \& {Mayer}}]{2015A&A...576A..46L}
{Lykou}, F., {Klotz}, D., {Paladini}, C., {et~al.} 2015{\natexlab{a}}, \aap,
  576, A46

\bibitem[{{Lykou} {et~al.}(2015{\natexlab{b}}){Lykou}, {Klotz}, {Paladini},
  {Hron}, {Zijlstra}, {Kluska}, {Norris}, {Tuthill}, {Ramstedt}, {Lagadec},
  {Wittkowski}, {Maercker}, \& {Mayer}}]{2015A&A...581C...2L}
{Lykou}, F., {Klotz}, D., {Paladini}, C., {et~al.} 2015{\natexlab{b}}, \aap,
  581, C2

\bibitem[{{Ma{\'\i}z Apell{\'a}niz} {et~al.}(2021){Ma{\'\i}z Apell{\'a}niz},
  {Pantaleoni Gonz{\'a}lez}, \& {Barb{\'a}}}]{2021A&A...649A..13M}
{Ma{\'\i}z Apell{\'a}niz}, J., {Pantaleoni Gonz{\'a}lez}, M., \& {Barb{\'a}},
  R.~H. 2021, \aap, 649, A13

\bibitem[{{Makarov} \& {Kaplan}(2005)}]{2005AJ....129.2420M}
{Makarov}, V.~V. \& {Kaplan}, G.~H. 2005, \aj, 129, 2420

\bibitem[{{Makarov} {et~al.}(2021){Makarov}, {Zacharias}, \&
  {Finch}}]{2021arXiv210701090M}
{Makarov}, V.~V., {Zacharias}, N., \& {Finch}, C.~T. 2021, arXiv e-prints,
  arXiv:2107.01090

\bibitem[{{Makarov} {et~al.}(2008){Makarov}, {Zacharias}, \&
  {Hennessy}}]{2008ApJ...687..566M}
{Makarov}, V.~V., {Zacharias}, N., \& {Hennessy}, G.~S. 2008, \apj, 687, 566

\bibitem[{{Malkov} {et~al.}(2012){Malkov}, {Tamazian}, {Docobo}, \&
  {Chulkov}}]{2012A&A...546A..69M}
{Malkov}, O.~Y., {Tamazian}, V.~S., {Docobo}, J.~A., \& {Chulkov}, D.~A. 2012,
  \aap, 546, A69

\bibitem[{Mann {et~al.}(2015)Mann, Feiden, Gaidos, Boyajian, \& von
  Braun}]{0004-637X-804-1-64}
Mann, A.~W., Feiden, G.~A., Gaidos, E., Boyajian, T., \& von Braun, K. 2015,
  \apj, 804, 64

\bibitem[{{Mann} {et~al.}(2015){Mann}, {Feiden}, {Gaidos}, {Boyajian}, \& {von
  Braun}}]{2015ApJ...804...64M}
{Mann}, A.~W., {Feiden}, G.~A., {Gaidos}, E., {Boyajian}, T., \& {von Braun},
  K. 2015, \apj, 804, 64

\bibitem[{{Marois} {et~al.}(2010){Marois}, {Zuckerman}, {Konopacky},
  {Macintosh}, \& {Barman}}]{2010Natur.468.1080M}
{Marois}, C., {Zuckerman}, B., {Konopacky}, Q.~M., {Macintosh}, B., \&
  {Barman}, T. 2010, \nat, 468, 1080

\bibitem[{{Marrese} {et~al.}(2019){Marrese}, {Marinoni}, {Fabrizio}, \&
  {Altavilla}}]{2019A&A...621A.144M}
{Marrese}, P.~M., {Marinoni}, S., {Fabrizio}, M., \& {Altavilla}, G. 2019,
  \aap, 621, A144

\bibitem[{{Marrese} {et~al.}(2017){Marrese}, {Marinoni}, {Fabrizio}, \&
  {Giuffrida}}]{2017A&A...607A.105M}
{Marrese}, P.~M., {Marinoni}, S., {Fabrizio}, M., \& {Giuffrida}, G. 2017,
  \aap, 607, A105

\bibitem[{{Martin} {et~al.}(1997){Martin}, {Mignard}, \&
  {Froeschle}}]{1997A&AS..122..571M}
{Martin}, C., {Mignard}, F., \& {Froeschle}, M. 1997, \aaps, 122, 571

\bibitem[{{Mason} {et~al.}(2001){Mason}, {Wycoff}, {Hartkopf}, {Douglass}, \&
  {Worley}}]{2001AJ....122.3466M}
{Mason}, B.~D., {Wycoff}, G.~L., {Hartkopf}, W.~I., {Douglass}, G.~G., \&
  {Worley}, C.~E. 2001, \aj, 122, 3466

\bibitem[{Mawet {et~al.}(2019)Mawet, Hirsch, Lee, Ruffio, Bottom, Fulton,
  Absil, Beichman, Bowler, Bryan, Choquet, Ciardi, Christiaens, Defr{\`e}re,
  Gonzalez, Howard, Huby, Isaacson, Jensen-Clem, Kosiarek, Marcy, Meshkat,
  Petigura, Reggiani, Ruane, Serabyn, Sinukoff, Wang, Weiss, \&
  Ygouf}]{1538-3881-157-1-33}
Mawet, D., Hirsch, L., Lee, E.~J., {et~al.} 2019, \aj, 157, 33

\bibitem[{{McAlister} {et~al.}(2005){McAlister}, {ten Brummelaar}, {Gies},
  {Huang}, {Bagnuolo}, {Shure}, {Sturmann}, {Sturmann}, {Turner}, {Taylor},
  {Berger}, {Baines}, {Grundstrom}, {Ogden}, {Ridgway}, \& {van
  Belle}}]{2005ApJ...628..439M}
{McAlister}, H.~A., {ten Brummelaar}, T.~A., {Gies}, D.~R., {et~al.} 2005,
  \apj, 628, 439

\bibitem[{{Morel}(1997)}]{1997A&AS..124..597M}
{Morel}, P. 1997, \aaps, 124, 597

\bibitem[{{Morel} \& {Lebreton}(2008)}]{2008Ap&SS.316...61M}
{Morel}, P. \& {Lebreton}, Y. 2008, \apss, 316, 61

\bibitem[{{Morel} \& {Lebreton}(2010)}]{2010ascl.soft10059M}
{Morel}, P. \& {Lebreton}, Y. 2010, {CESAM: A Free Code for Stellar Evolution
  Calculations}

\bibitem[{{Morzinski} {et~al.}(2015){Morzinski}, {Males}, {Skemer}, {Close},
  {Hinz}, {Rodigas}, {Puglisi}, {Esposito}, {Riccardi}, {Pinna}, {Xompero},
  {Briguglio}, {Bailey}, {Follette}, {Kopon}, {Weinberger}, \&
  {Wu}}]{2015ApJ...815..108M}
{Morzinski}, K.~M., {Males}, J.~R., {Skemer}, A.~J., {et~al.} 2015, \apj, 815,
  108

\bibitem[{{Musielak} {et~al.}(2005){Musielak}, {Cuntz}, {Marshall}, \&
  {Stuit}}]{2005A&A...434..355M}
{Musielak}, Z.~E., {Cuntz}, M., {Marshall}, E.~A., \& {Stuit}, T.~D. 2005,
  \aap, 434, 355

\bibitem[{{Nidever} {et~al.}(2002){Nidever}, {Marcy}, {Butler}, {Fischer}, \&
  {Vogt}}]{2002ApJS..141..503N}
{Nidever}, D.~L., {Marcy}, G.~W., {Butler}, R.~P., {Fischer}, D.~A., \& {Vogt},
  S.~S. 2002, \apjs, 141, 503

\bibitem[{{Nielsen} {et~al.}(2019){Nielsen}, {De Rosa}, {Macintosh}, {Wang},
  {Ruffio}, {Chiang}, {Marley}, {Saumon}, {Savransky}, {Ammons}, {Bailey},
  {Barman}, {Blain}, {Bulger}, {Burrows}, {Chilcote}, {Cotten}, {Czekala},
  {Doyon}, {Duch{\^e}ne}, {Esposito}, {Fabrycky}, {Fitzgerald}, {Follette},
  {Fortney}, {Gerard}, {Goodsell}, {Graham}, {Greenbaum}, {Hibon}, {Hinkley},
  {Hirsch}, {Hom}, {Hung}, {Dawson}, {Ingraham}, {Kalas}, {Konopacky},
  {Larkin}, {Lee}, {Lin}, {Maire}, {Marchis}, {Marois}, {Metchev},
  {Millar-Blanchaer}, {Morzinski}, {Oppenheimer}, {Palmer}, {Patience},
  {Perrin}, {Poyneer}, {Pueyo}, {Rafikov}, {Rajan}, {Rameau}, {Rantakyr{\"o}},
  {Ren}, {Schneider}, {Sivaramakrishnan}, {Song}, {Soummer}, {Tallis},
  {Thomas}, {Ward-Duong}, \& {Wolff}}]{2019AJ....158...13N}
{Nielsen}, E.~L., {De Rosa}, R.~J., {Macintosh}, B., {et~al.} 2019, \aj, 158,
  13

\bibitem[{{Nowak} {et~al.}(2020){Nowak}, {Lacour}, {Lagrange}, {Rubini},
  {Wang}, {Stolker}, {Abuter}, {Amorim}, {Asensio-Torres}, {Baub{\"o}ck},
  {Benisty}, {Berger}, {Beust}, {Blunt}, {Boccaletti}, {Bonnefoy}, {Bonnet},
  {Brandner}, {Cantalloube}, {Charnay}, {Choquet}, {Christiaens}, {Cl{\'e}net},
  {Coud{\'e} Du Foresto}, {Cridland}, {de Zeeuw}, {Dembet}, {Dexter},
  {Drescher}, {Duvert}, {Eckart}, {Eisenhauer}, {Gao}, {Garcia}, {Garcia
  Lopez}, {Gardner}, {Gendron}, {Genzel}, {Gillessen}, {Girard}, {Grandjean},
  {Haubois}, {Hei{\ss}el}, {Henning}, {Hinkley}, {Hippler}, {Horrobin},
  {Houll{\'e}}, {Hubert}, {Jim{\'e}nez-Rosales}, {Jocou}, {Kammerer},
  {Kervella}, {Keppler}, {Kreidberg}, {Kulikauskas}, {Lapeyr{\`e}re}, {Le
  Bouquin}, {L{\'e}na}, {M{\'e}rand}, {Maire}, {Molli{\`e}re}, {Monnier},
  {Mouillet}, {M{\"u}ller}, {Nasedkin}, {Ott}, {Otten}, {Paumard}, {Paladini},
  {Perraut}, {Perrin}, {Pueyo}, {Pfuhl}, {Rameau}, {Rodet},
  {Rodr{\'\i}guez-Coira}, {Rousset}, {Scheithauer}, {Shangguan}, {Stadler},
  {Straub}, {Straubmeier}, {Sturm}, {Tacconi}, {van Dishoeck}, {Vigan},
  {Vincent}, {von Fellenberg}, {Ward-Duong}, {Widmann}, {Wieprecht},
  {Wiezorrek}, {Woillez}, \& {Gravity Collaboration}}]{2020A&A...642L...2N}
{Nowak}, M., {Lacour}, S., {Lagrange}, A.~M., {et~al.} 2020, \aap, 642, L2

\bibitem[{{Nuth} {et~al.}(2020){Nuth}, {Ferguson}, {Homan}, {Decin}, \&
  {Paquette}}]{2020ApJ...901..144N}
{Nuth}, J.~A., {Ferguson}, F.~T., {Homan}, W., {Decin}, L., \& {Paquette},
  J.~A. 2020, \apj, 901, 144

\bibitem[{{Ochsenbein} {et~al.}(2000){Ochsenbein}, {Bauer}, \&
  {Marcout}}]{2000A&AS..143...23O}
{Ochsenbein}, F., {Bauer}, P., \& {Marcout}, J. 2000, \aaps, 143, 23

\bibitem[{{Pathak} {et~al.}(2021){Pathak}, {Petit dit de la Roche}, {Kasper},
  {Sterzik}, {Absil}, {Boehle}, {Feng}, {Ivanov}, {Janson}, {Jones, H. R. A.},
  {Kaufer, A.}, {K\"aufl, H.-U.}, {Maire, A.-L.}, {Meyer, M.}, {Pantin, E.},
  {Siebenmorgen, R.}, {van den Ancker, M. E.}, \& {Viswanath, G.}}]{pathak2021}
{Pathak}, P., {Petit dit de la Roche}, D. J.~M., {Kasper}, M., {et~al.} 2021,
  A\&A, 652, A121

\bibitem[{Pearce {et~al.}(2020)Pearce, Kraus, Dupuy, Mann, Newton, Tofflemire,
  \& Vanderburg}]{Pearce_2020}
Pearce, L.~A., Kraus, A.~L., Dupuy, T.~J., {et~al.} 2020, \apj, 894, 115

\bibitem[{{Pecaut} \& {Mamajek}(2013)}]{2013ApJS..208....9P}
{Pecaut}, M.~J. \& {Mamajek}, E.~E. 2013, \apjs, 208, 9

\bibitem[{{Pecaut} {et~al.}(2012){Pecaut}, {Mamajek}, \&
  {Bubar}}]{2012ApJ...746..154P}
{Pecaut}, M.~J., {Mamajek}, E.~E., \& {Bubar}, E.~J. 2012, \apj, 746, 154

\bibitem[{{Perryman} {et~al.}(1997){Perryman}, {Lindegren}, {Kovalevsky},
  {Hoeg}, {Bastian}, {Bernacca}, {Cr{\'e}z{\'e}}, {Donati}, {Grenon},
  {Grewing}, {van Leeuwen}, {van der Marel}, {Mignard}, {Murray}, {Le Poole},
  {Schrijver}, {Turon}, {Arenou}, {Froeschl{\'e}}, \&
  {Petersen}}]{1997A&A...323L..49P}
{Perryman}, M.~A.~C., {Lindegren}, L., {Kovalevsky}, J., {et~al.} 1997, \aap,
  323, L49

\bibitem[{Pietrzy{\'n}ski {et~al.}(2019)Pietrzy{\'n}ski, Graczyk, Gallenne,
  Gieren, Thompson, Pilecki, Karczmarek, G{\'o}rski, Suchomska, Taormina,
  Zgirski, Wielg{\'o}rski, Ko{\l}aczkowski, Konorski, Villanova, Nardetto,
  Kervella, Bresolin, Kudritzki, Storm, Smolec, \&
  Narloch}]{Pietrzynski:2019aa}
Pietrzy{\'n}ski, G., Graczyk, D., Gallenne, A., {et~al.} 2019, Nature, 567, 200

\bibitem[{{Pittordis} \& {Sutherland}(2019)}]{2019MNRAS.488.4740P}
{Pittordis}, C. \& {Sutherland}, W. 2019, \mnras, 488, 4740

\bibitem[{{Pourbaix}(2019)}]{2019MmSAI..90..318P}
{Pourbaix}, D. 2019, \memsai, 90, 318

\bibitem[{{Rebassa-Mansergas} {et~al.}(2021){Rebassa-Mansergas}, {Solano},
  {Jim{\'e}nez-Esteban}, {Torres}, {Rodrigo}, {Ferrer-Burjachs}, {Calcaferro},
  {Althaus}, \& {C{\'o}rsico}}]{2021MNRAS.506.5201R}
{Rebassa-Mansergas}, A., {Solano}, E., {Jim{\'e}nez-Esteban}, F.~M., {et~al.}
  2021, \mnras, 506, 5201

\bibitem[{{Reffert} \& {Quirrenbach}(2011)}]{2011A&A...527A.140R}
{Reffert}, S. \& {Quirrenbach}, A. 2011, \aap, 527, A140

\bibitem[{{Reyl{\'e}}(2018)}]{2018A&A...619L...8R}
{Reyl{\'e}}, C. 2018, \aap, 619, L8

\bibitem[{{Reyl{\'e}} {et~al.}(2021){Reyl{\'e}}, {Jardine}, {Fouqu{\'e}},
  {Caballero}, {Smart}, \& {Sozzetti}}]{2021A&A...650A.201R}
{Reyl{\'e}}, C., {Jardine}, K., {Fouqu{\'e}}, P., {et~al.} 2021, \aap, 650,
  A201

\bibitem[{{Salmon} {et~al.}(2021){Salmon}, {Van Grootel}, {Buldgen}, {Dupret},
  \& {Eggenberger}}]{2021A&A...646A...7S}
{Salmon}, S.~J.~A.~J., {Van Grootel}, V., {Buldgen}, G., {Dupret}, M.~A., \&
  {Eggenberger}, P. 2021, \aap, 646, A7

\bibitem[{{Sapozhnikov} {et~al.}(2020){Sapozhnikov}, {Kovaleva}, {Malkov}, \&
  {Sytov}}]{2020ARep...64..756S}
{Sapozhnikov}, S.~A., {Kovaleva}, D.~A., {Malkov}, O.~Y., \& {Sytov}, A.~Y.
  2020, Astronomy Reports, 64, 756

\bibitem[{{Scholz} {et~al.}(2003){Scholz}, {McCaughrean}, {Lodieu}, \&
  {Kuhlbrodt}}]{2003A&A...398L..29S}
{Scholz}, R.~D., {McCaughrean}, M.~J., {Lodieu}, N., \& {Kuhlbrodt}, B. 2003,
  \aap, 398, L29

\bibitem[{{Serenelli} {et~al.}(2021){Serenelli}, {Weiss}, {Aerts}, {Angelou},
  {Baroch}, {Bastian}, {Beck}, {Bergemann}, {Bestenlehner}, {Czekala},
  {Elias-Rosa}, {Escorza}, {Van Eylen}, {Feuillet}, {Gandolfi}, {Gieles},
  {Girardi}, {Lebreton}, {Lodieu}, {Martig}, {Miller Bertolami}, {Mombarg},
  {Morales}, {Moya}, {Nsamba}, {Pavlovski}, {Pedersen}, {Ribas}, {Schneider},
  {Silva Aguirre}, {Stassun}, {Tolstoy}, {Tremblay}, \&
  {Zwintz}}]{2021A&ARv..29....4S}
{Serenelli}, A., {Weiss}, A., {Aerts}, C., {et~al.} 2021, \aapr, 29, 4

\bibitem[{{Shakht} {et~al.}(2018){Shakht}, {Gorshanov}, \&
  {Vasilkova}}]{2018RAA....18...94S}
{Shakht}, N.~A., {Gorshanov}, D.~L., \& {Vasilkova}, O.~O. 2018, Research in
  Astronomy and Astrophysics, 18, 094

\bibitem[{{Shaya} \& {Olling}(2011)}]{2011ApJS..192....2S}
{Shaya}, E.~J. \& {Olling}, R.~P. 2011, \apjs, 192, 2

\bibitem[{{Skrutskie} {et~al.}(2006){Skrutskie}, {Cutri}, {Stiening},
  {Weinberg}, {Schneider}, {Carpenter}, {Beichman}, {Capps}, {Chester},
  {Elias}, {Huchra}, {Liebert}, {Lonsdale}, {Monet}, {Price}, {Seitzer},
  {Jarrett}, {Kirkpatrick}, {Gizis}, {Howard}, {Evans}, {Fowler}, {Fullmer},
  {Hurt}, {Light}, {Kopan}, {Marsh}, {McCallon}, {Tam}, {Van Dyk}, \&
  {Wheelock}}]{2006AJ....131.1163S}
{Skrutskie}, M.~F., {Cutri}, R.~M., {Stiening}, R., {et~al.} 2006, \aj, 131,
  1163

\bibitem[{{Snellen} \& {Brown}(2018)}]{2018NatAs...2..883S}
{Snellen}, I.~A.~G. \& {Brown}, A.~G.~A. 2018, Nature Astronomy, 2, 883

\bibitem[{{Soubiran} {et~al.}(2018){Soubiran}, {Jasniewicz}, {Chemin},
  {Zurbach}, {Brouillet}, {Panuzzo}, {Sartoretti}, {Katz}, {Le Campion},
  {Marchal}, {Hestroffer}, {Th{\'e}venin}, {Crifo}, {Udry}, {Cropper},
  {Seabroke}, {Viala}, {Benson}, {Blomme}, {Jean-Antoine}, {Huckle}, {Smith},
  {Baker}, {Damerdji}, {Dolding}, {Fr{\'e}mat}, {Gosset}, {Guerrier}, {Guy},
  {Haigron}, {Jan{\ss}en}, {Plum}, {Fabre}, {Lasne}, {Pailler}, {Panem},
  {Riclet}, {Royer}, {Tauran}, {Zwitter}, {Gueguen}, \&
  {Turon}}]{2018A&A...616A...7S}
{Soubiran}, C., {Jasniewicz}, G., {Chemin}, L., {et~al.} 2018, \aap, 616, A7

\bibitem[{{Stassun} \& {Torres}(2021)}]{2021ApJ...907L..33S}
{Stassun}, K.~G. \& {Torres}, G. 2021, \apjl, 907, L33

\bibitem[{{Strand}(1943)}]{1943PASP...55...29S}
{Strand}, K.~A. 1943, \pasp, 55, 29

\bibitem[{{Strand}(1957)}]{1957AJ.....62Q..35S}
{Strand}, K.~A. 1957, \aj, 62, 35

\bibitem[{{Su{\'a}rez Mascare{\~n}o} {et~al.}(2020){Su{\'a}rez Mascare{\~n}o},
  {Faria}, {Figueira}, {Lovis}, {Damasso}, {Gonz{\'a}lez Hern{\'a}ndez},
  {Rebolo}, {Cristiani}, {Pepe}, {Santos}, {Zapatero Osorio}, {Adibekyan},
  {Hojjatpanah}, {Sozzetti}, {Murgas}, {Abreu}, {Affolter}, {Alibert},
  {Aliverti}, {Allart}, {Allende Prieto}, {Alves}, {Amate}, {Avila}, {Baldini},
  {Bandi}, {Barros}, {Bianco}, {Benz}, {Bouchy}, {Broeng}, {Cabral},
  {Calderone}, {Cirami}, {Coelho}, {Conconi}, {Coretti}, {Cumani}, {Cupani},
  {D'Odorico}, {Deiries}, {Delabre}, {Di Marcantonio}, {Dumusque},
  {Ehrenreich}, {Fragoso}, {Genolet}, {Genoni}, {G{\'e}nova Santos}, {Hughes},
  {Iwert}, {Kerber}, {Knusdstrup}, {Landoni}, {Lavie}, {Lillo-Box}, {Lizon},
  {Lo Curto}, {Maire}, {Manescau}, {Martins}, {M{\'e}gevand}, {Mehner},
  {Micela}, {Modigliani}, {Molaro}, {Monteiro}, {Monteiro}, {Moschetti},
  {Mueller}, {Nunes}, {Oggioni}, {Oliveira}, {Pall{\'e}}, {Pariani},
  {Pasquini}, {Poretti}, {Rasilla}, {Redaelli}, {Riva}, {Santana Tschudi},
  {Santin}, {Santos}, {Segovia}, {Sosnowska}, {Sousa}, {Span{\`o}}, {Tenegi},
  {Udry}, {Zanutta}, \& {Zerbi}}]{2020A&A...639A..77S}
{Su{\'a}rez Mascare{\~n}o}, A., {Faria}, J.~P., {Figueira}, P., {et~al.} 2020,
  \aap, 639, A77

\bibitem[{{van Belle} \& {von Braun}(2009)}]{2009ApJ...694.1085V}
{van Belle}, G.~T. \& {von Braun}, K. 2009, \apj, 694, 1085

\bibitem[{{van Leeuwen}(2007)}]{2007ASSL..350.....V}
{van Leeuwen}, F., ed. 2007, Astrophysics and Space Science Library, Vol. 350,
  {Hipparcos, the New Reduction of the Raw Data}

\bibitem[{{Viswanath} {et~al.}(2021){Viswanath}, {Janson}, {Dahlqvist}, {Petit
  dit de la Roche}, {Samland}, {Girard}, {Pathak}, {Kasper}, {Feng}, {Meyer},
  {Boehle}, {Quanz}, {Jones}, {Absil}, {Brandner}, {Maire}, {Siebenmorgen},
  {Sterzik}, \& {Pantin}}]{2021A&A...651A..89V}
{Viswanath}, G., {Janson}, M., {Dahlqvist}, C.-H., {et~al.} 2021, \aap, 651,
  A89

\bibitem[{{Wagner} {et~al.}(2021{\natexlab{a}}){Wagner}, {Boehle}, {Pathak},
  {Kasper}, {Arsenault}, {Jakob}, {K{\"a}ufl}, {Leveratto}, {Maire}, {Pantin},
  {Siebenmorgen}, {Zins}, {Absil}, {Ageorges}, {Apai}, {Carlotti}, {Choquet},
  {Delacroix}, {Dohlen}, {Duhoux}, {Forsberg}, {Fuenteseca}, {Gutruf}, {Guyon},
  {Huby}, {Kampf}, {Karlsson}, {Kervella}, {Kirchbauer}, {Klupar}, {Kolb},
  {Mawet}, {N'Diaye}, {de Xivry}, {Quanz}, {Reutlinger}, {Ruane}, {Riquelme},
  {Soenke}, {Sterzik}, {Vigan}, \& {de Zeeuw}}]{2021NatCo..12.2651W}
{Wagner}, K., {Boehle}, A., {Pathak}, P., {et~al.} 2021{\natexlab{a}}, Nature
  Communications, 12, 2651

\bibitem[{{Wagner} {et~al.}(2021{\natexlab{b}}){Wagner}, {Boehle}, {Pathak},
  {Kasper}, {Arsenault}, {Jakob}, {K{\"a}ufl}, {Leveratto}, {Maire}, {Pantin},
  {Siebenmorgen}, {Zins}, {Absil}, {Ageorges}, {Apai}, {Carlotti}, {Choquet},
  {Delacroix}, {Dohlen}, {Duhoux}, {Forsberg}, {Fuenteseca}, {Gutruf}, {Guyon},
  {Huby}, {Kampf}, {Karlsson}, {Kervella}, {Kirchbauer}, {Klupar}, {Kolb},
  {Mawet}, {N'Diaye}, {Orban de Xivry}, {Quanz}, {Reutlinger}, {Ruane},
  {Riquelme}, {Soenke}, {Sterzik}, {Vigan}, \& {de
  Zeeuw}}]{2021NatCo..12..922W}
{Wagner}, K., {Boehle}, A., {Pathak}, P., {et~al.} 2021{\natexlab{b}}, Nature
  Communications, 12, 922

\bibitem[{{Walker} {et~al.}(1995){Walker}, {Walker}, {Irwin}, {Larson}, {Yang},
  \& {Richardson}}]{1995Icar..116..359W}
{Walker}, G. A.~H., {Walker}, A.~R., {Irwin}, A.~W., {et~al.} 1995, \icarus,
  116, 359

\bibitem[{{Wang} {et~al.}(2014){Wang}, {Fischer}, {Xie}, \&
  {Ciardi}}]{2014ApJ...791..111W}
{Wang}, J., {Fischer}, D.~A., {Xie}, J.-W., \& {Ciardi}, D.~R. 2014, \apj, 791,
  111

\bibitem[{{Weber} \& {Strassmeier}(2011)}]{2011A&A...531A..89W}
{Weber}, M. \& {Strassmeier}, K.~G. 2011, \aap, 531, A89

\bibitem[{{Wielen} {et~al.}(1999){Wielen}, {Dettbarn}, {Jahrei{\ss}},
  {Lenhardt}, \& {Schwan}}]{1999A&A...346..675W}
{Wielen}, R., {Dettbarn}, C., {Jahrei{\ss}}, H., {Lenhardt}, H., \& {Schwan},
  H. 1999, \aap, 346, 675

\bibitem[{{Winn} \& {Fabrycky}(2015)}]{2015ARA&A..53..409W}
{Winn}, J.~N. \& {Fabrycky}, D.~C. 2015, \araa, 53, 409

\bibitem[{{Xuan} \& {Wyatt}(2020)}]{2020MNRAS.497.2096X}
{Xuan}, J.~W. \& {Wyatt}, M.~C. 2020, \mnras, 497, 2096

\bibitem[{{Zacharias} {et~al.}(2015){Zacharias}, {Finch}, {Subasavage},
  {Bredthauer}, {Crockett}, {Divittorio}, {Ferguson}, {Harris}, {Harris},
  {Henden}, {Kilian}, {Munn}, {Rafferty}, {Rhodes}, {Schultheiss}, {Tilleman},
  \& {Wieder}}]{2015AJ....150..101Z}
{Zacharias}, N., {Finch}, C., {Subasavage}, J., {et~al.} 2015, \aj, 150, 101

\bibitem[{{Zacharias} {et~al.}(2004){Zacharias}, {Monet}, {Levine}, {Urban},
  {Gaume}, \& {Wycoff}}]{2004AAS...205.4815Z}
{Zacharias}, N., {Monet}, D.~G., {Levine}, S.~E., {et~al.} 2004, in American
  Astronomical Society Meeting Abstracts, Vol. 205, 48.15

\bibitem[{Zavada \& P{\'{\i}}{\v{s}}ka(2020)}]{Zavada_2020}
Zavada, P. \& P{\'{\i}}{\v{s}}ka, K. 2020, \aj, 159, 33

\end{thebibliography}

\begin{appendix}

\section{Additional tables}
\begin{sidewaystable*}
\caption{First records of the proper motion anomaly catalog for the Hipparcos stars.}
\label{PMa-sample}
\centering
\tiny
\renewcommand{\arraystretch}{1}
\setlength\tabcolsep{4.5pt}
\begin{tabular}{ccccccccccccccccccccc}
  \hline
  \hline
HIP & EDR3 Source & $\varpi_\mathrm{G3}$ & $\sigma(\varpi_\mathrm{G3})$ & R & $\mu_{\mathrm{HG},\alpha}$ & $\sigma(\mu_{\mathrm{HG},\alpha})$ & $\mu_{\mathrm{HG},\delta}$ & $\sigma(\mu_{\mathrm{HG},\delta})$  & $\Delta \mu_\alpha$ & $\sigma(\Delta \mu_\alpha)$ & $\Delta \mu_\delta$ & $\sigma(\Delta \mu_\delta)$ & PMa & Bin. & $d\varv_\mathrm{tan}$ & $\sigma$ & $m_2$ & $\sigma^+$ & $\sigma^-$ & CPM \\
 &  & $\mathrm{mas}$ & $\mathrm{mas}$ & & $\mathrm{mas\,a^{-1}}$ & $\mathrm{mas\,a^{-1}}$ & $\mathrm{mas\,a^{-1}}$ & $\mathrm{mas\,a^{-1}}$ & $\mathrm{mas\,a^{-1}}$ & $\mathrm{mas\,a^{-1}}$ & $\mathrm{mas\,a^{-1}}$ & $\mathrm{mas\,a^{-1}}$ & S/N  & flag  & $\mathrm{m\,s^{-1}}$ & $\mathrm{m\,s^{-1}}$ & $\mathrm{M_{\rm J}}$ & $\mathrm{M_{\rm J}}$ & $\mathrm{M_{\rm J}}$ & bnd \\
  \hline  \noalign{\smallskip}
1 & 2738327528519591936 & 5.449 & 0.037 & 1.7 & -5.803 & 0.052 & -5.087 & 0.027 & 5.454 & 0.078 & 0.068 & 0.035 & 63.94 & 1 & 4745.53 & 74.22 & 594.4 & 190.7 & 72.6 & 0 \\
2 & 2341871673090078592 & 26.829 & 0.571 & 29.4 & 181.481 & 0.036 & -0.472 & 0.029 & -1.671 & 0.573 & -0.535 & 0.424 & 2.46 & 0 & 310.05 & 125.97 & 28.7 & 13.0 & 9.8 & 0 \\
3 & 2881742980523997824 & 3.029 & 0.049 & 1.0 & 5.780 & 0.013 & -2.468 & 0.009 & 0.002 & 0.030 & 0.095 & 0.025 & 2.47 & 0 & 149.38 & 60.39 & 52.9 & 19.9 & 12.2 & 1 \\
4 & 4973386040722654336 & 7.331 & 0.024 & 1.0 & 61.983 & 0.016 & 1.296 & 0.019 & -0.025 & 0.021 & 0.040 & 0.027 & 1.38 & 0 & 30.59 & 22.21 & 4.2 & 2.6 & 2.3 & 0 \\
5 & 2305974989264598272 & 2.628 & 0.025 & 1.0 & 1.001 & 0.022 & 8.756 & 0.020 & 0.018 & 0.031 & 0.011 & 0.028 & 0.50 & 0 & 37.21 & 74.47 & 7.4 & 10.1 & 9.8 & 0 \\
6 & 2740326852975975040 & 15.422 & 0.035 & 1.9 & 223.204 & 0.183 & -11.526 & 0.098 & -0.007 & 0.185 & 0.092 & 0.100 & 0.44 & 0 & 28.36 & 64.73 & 2.0 & 4.1 & 4.1 & 0 \\
7 & 2846308881856186240 & 17.525 & 0.027 & 1.1 & -211.005 & 0.038 & -196.975 & 0.029 & 4.519 & 0.047 & 0.908 & 0.031 & 81.62 & 1 & 1246.78 & 15.27 & 130.3 & 41.8 & 15.9 & 0 \\
8 & 2853169937491828608 & 1.571 & 0.068 & 1.0 & 18.804 & 0.070 & -6.584 & 0.033 & -0.040 & 0.098 & 0.145 & 0.048 & 1.38 & 0 & 455.00 & 329.37 & 29.5 & 21.1 & 19.2 & 0 \\
9 & 2880160886370458368 & 3.132 & 0.027 & 1.0 & -6.048 & 0.034 & 9.260 & 0.021 & -0.003 & 0.040 & 0.070 & 0.025 & 1.49 & 0 & 106.32 & 71.47 & 20.0 & 10.9 & 9.1 & 0 \\
10 & 4976500987226833024 & 10.721 & 0.019 & 1.0 & 42.350 & 0.028 & 40.819 & 0.027 & -0.059 & 0.030 & 0.056 & 0.030 & 1.90 & 0 & 35.96 & 18.89 & 4.2 & 2.2 & 1.8 & 0 \\
11 & 387133547311154432 & 3.395 & 0.030 & 0.9 & 11.002 & 0.015 & -2.095 & 0.014 & 0.027 & 0.029 & 0.047 & 0.025 & 1.44 & 0 & 75.76 & 52.48 & 17.6 & 9.3 & 7.7 & 0 \\
12 & 2308086876223750656 & 1.839 & 0.026 & 1.2 & -5.485 & 0.026 & 1.419 & 0.019 & -0.033 & 0.033 & 0.040 & 0.027 & 1.21 & 0 & 134.53 & 110.81 & 18.0 & 12.6 & 11.4 & 0 \\
13 & 2340148424835415552 & 2.438 & 0.027 & 1.0 & 5.951 & 0.043 & -10.491 & 0.024 & -0.108 & 0.048 & 0.032 & 0.030 & 1.99 & 0 & 219.01 & 109.90 & 43.6 & 20.1 & 15.4 & 0 \\
14 & 2449930576356314880 & 6.032 & 0.032 & 1.1 & 59.999 & 0.023 & -10.334 & 0.016 & -1.262 & 0.037 & 0.157 & 0.025 & 28.47 & 1 & 999.19 & 35.10 & 165.1 & 53.0 & 20.4 & 0 \\
15 & 394029134492039424 & 2.553 & 0.025 & 1.0 & 12.936 & 0.025 & 5.289 & 0.027 & 0.051 & 0.030 & 0.057 & 0.032 & 1.74 & 0 & 141.54 & 81.26 & 27.6 & 13.7 & 11.0 & 0 \\
16 & 4923847578691749120 & 3.301 & 0.018 & 1.1 & 259.367 & 0.057 & -97.506 & 0.065 & 0.260 & 0.059 & 0.268 & 0.066 & 4.21 & 1 & 535.93 & 127.35 & 61.4 & 22.4 & 13.0 & 0 \\
17 & 2306077724882156928 & 6.725 & 0.026 & 1.0 & -34.541 & 0.018 & -27.709 & 0.017 & -0.069 & 0.026 & -0.021 & 0.027 & 1.93 & 0 & 51.11 & 26.44 & 6.8 & 3.4 & 2.8 & 0 \\
18 & 2447815287783063040 & 22.820 & 0.023 & 1.1 & -119.293 & 0.059 & 24.096 & 0.045 & 0.014 & 0.062 & 0.062 & 0.048 & 0.81 & 0 & 13.22 & 16.28 & 1.0 & 1.1 & 1.1 & 0 \\
19 & 2880594231390895104 & 4.489 & 0.028 & 1.2 & -1.692 & 0.012 & -14.321 & 0.008 & 1.276 & 0.029 & 1.239 & 0.019 & 51.83 & 1 & 1878.65 & 36.25 & 473.9 & 151.9 & 57.5 & 0 \\
20 & 2848390257367536384 & 10.100 & 0.026 & 1.0 & 36.120 & 0.027 & -23.015 & 0.019 & 0.070 & 0.034 & 0.054 & 0.022 & 2.19 & 0 & 41.45 & 18.91 & 5.0 & 2.4 & 1.9 & 0 \\
21 & 2746745664420425344 & 3.546 & 0.031 & 1.1 & 61.284 & 0.031 & 0.026 & 0.018 & 0.022 & 0.043 & -0.042 & 0.023 & 0.98 & 0 & 63.59 & 64.83 & 12.2 & 9.1 & 8.3 & 0 \\
22 & 4977625894998735872 & 4.871 & 0.020 & 1.1 & -7.159 & 0.024 & 1.695 & 0.029 & -0.089 & 0.026 & 0.122 & 0.032 & 3.68 & 1 & 147.15 & 39.98 & 23.1 & 8.6 & 5.2 & 0 \\
23 & 2767134252131256448 & 10.767 & 0.029 & 0.9 & 53.355 & 0.025 & 9.419 & 0.015 & 0.080 & 0.034 & 0.031 & 0.022 & 2.10 & 0 & 37.51 & 17.89 & 4.7 & 2.3 & 1.8 & 0 \\
24 & 2339755869120492800 & 10.931 & 0.017 & 0.9 & 127.985 & 0.034 & 21.265 & 0.022 & -0.006 & 0.037 & -0.009 & 0.025 & 0.24 & 0 & 4.63 & 19.38 & 0.5 & 1.7 & 1.7 & 0 \\
25 & 4994581292009978112 &  &  &  &  &  &  &  &  &  &  &  &  &  &  &  &  &  &  & 0 \\
26 & 2420820593694180992 & 9.654 & 0.021 & 1.1 & -100.299 & 0.032 & -31.587 & 0.029 & 0.006 & 0.040 & 0.044 & 0.032 & 0.86 & 0 & 21.90 & 25.33 & 2.5 & 2.3 & 2.2 & 0 \\
27 & 2305871596516150656 & 8.865 & 0.019 & 1.1 & 135.285 & 0.028 & -113.910 & 0.026 & -0.836 & 0.032 & 1.684 & 0.030 & 43.25 & 1 & 1005.54 & 23.25 & 115.8 & 37.2 & 14.2 & 0 \\
28 & 4994867439910791552 & 4.416 & 0.020 & 1.0 & -10.890 & 0.024 & -7.860 & 0.019 & 0.012 & 0.030 & 0.056 & 0.024 & 1.51 & 0 & 61.95 & 41.06 & 8.8 & 5.1 & 4.4 & 0 \\
29 & 4977679594971842688 & 2.100 & 0.016 & 1.0 & 26.949 & 0.028 & 4.590 & 0.034 & -0.012 & 0.029 & 0.027 & 0.036 & 0.63 & 0 & 66.78 & 105.56 & 13.0 & 14.2 & 13.7 & 0 \\
30 & 384317728096888576 & 2.682 & 0.034 & 1.2 & -8.272 & 0.015 & -10.264 & 0.015 & 0.019 & 0.028 & -0.156 & 0.026 & 4.13 & 1 & 277.18 & 67.14 & 58.9 & 20.9 & 11.6 & 0 \\
31 & 2739891068414238464 & 2.021 & 0.029 & 1.0 & -1.347 & 0.036 & 0.207 & 0.022 & 0.097 & 0.043 & -0.015 & 0.026 & 1.94 & 0 & 231.11 & 119.41 & 43.0 & 20.0 & 15.4 & 0 \\
32 & 395612534315211520 & 1.008 & 0.023 & 0.9 & -1.589 & 0.029 & -2.271 & 0.027 & 0.015 & 0.034 & 0.089 & 0.031 & 1.96 & 0 & 422.56 & 215.26 & 172.8 & 68.0 & 44.7 & 0 \\
33 & 2422810915898660096 & 8.908 & 0.056 & 1.0 & -3.203 & 0.026 & 28.795 & 0.013 & -0.062 & 0.044 & 0.119 & 0.034 & 2.42 & 0 & 71.56 & 29.52 & 8.8 & 4.0 & 3.0 & 1 \\
34 & 2853571774632882560 & 16.087 & 0.178 & 7.0 & 43.443 & 0.020 & -53.072 & 0.012 & 1.030 & 0.147 & 0.304 & 0.081 & 6.39 & 1 & 316.35 & 49.54 & 41.1 & 14.1 & 7.0 & 0 \\
35 & 2417681762874391936 & 5.367 & 0.021 & 0.9 & 164.398 & 0.042 & -1.220 & 0.025 & 0.098 & 0.046 & 0.024 & 0.029 & 1.88 & 0 & 89.50 & 47.62 & 11.9 & 6.1 & 5.0 & 0 \\
36 & 2766909298924361600 & 5.977 & 0.040 & 1.2 & 51.358 & 0.021 & 16.571 & 0.014 & 2.017 & 0.045 & -2.754 & 0.025 & 65.69 & 1 & 2707.33 & 41.22 & 426.9 & 136.9 & 52.0 & 0 \\
38 & 4635422588982399616 & 24.949 & 0.015 & 0.9 & 162.463 & 0.018 & -62.086 & 0.019 & -0.054 & 0.024 & 0.029 & 0.024 & 1.80 & 0 & 11.74 & 6.53 & 1.2 & 0.6 & 0.5 & 0 \\
39 & 2415631586005559680 & 11.545 & 0.028 & 1.0 & 167.661 & 0.024 & -31.113 & 0.017 & -0.044 & 0.033 & 0.854 & 0.025 & 20.60 & 1 & 351.24 & 17.05 & 44.6 & 14.4 & 5.7 & 0 \\
40 & 528563384392653312 & 0.975 & 0.014 & 1.0 & -1.581 & 0.102 & -2.465 & 0.110 & -0.206 & 0.103 & 0.075 & 0.111 & 1.45 & 0 & 1065.13 & 734.50 & 533.1 & 224.3 & 159.1 & 1 \\
41 & 420335603020825856 & 1.685 & 0.044 & 1.8 & 2.713 & 0.024 & -0.374 & 0.024 & 0.493 & 0.043 & -0.132 & 0.042 & 8.50 & 1 & 1436.46 & 168.99 & 465.4 & 151.7 & 62.8 & 0 \\
42 & 2853260230588911488 & 8.556 & 0.460 & 18.6 & 20.910 & 0.027 & -9.155 & 0.018 & 3.483 & 0.430 & -3.340 & 0.238 & 9.82 & 1 & 2673.45 & 272.12 & 339.3 & 112.0 & 49.1 & 0 \\
43 & 423196566637503232 & 7.942 & 0.023 & 0.9 & -81.475 & 0.011 & -23.459 & 0.010 & 0.024 & 0.020 & 0.038 & 0.022 & 1.51 & 0 & 26.65 & 17.62 & 5.3 & 2.9 & 2.4 & 0 \\
44 & 2448142358132819456 & 3.880 & 0.031 & 1.0 & 13.054 & 0.037 & 17.012 & 0.015 & 0.045 & 0.045 & 0.027 & 0.022 & 1.04 & 0 & 64.02 & 61.45 & 12.0 & 8.5 & 7.7 & 0 \\
45 & 4701763714691371264 & 15.251 & 0.010 & 0.8 & -37.290 & 0.036 & -2.275 & 0.034 & -0.123 & 0.037 & -0.015 & 0.036 & 2.40 & 0 & 38.50 & 16.03 & 4.0 & 1.8 & 1.3 & 1 \\
46 & 2334989520573815040 & 2.955 & 0.031 & 1.0 & 16.258 & 0.036 & -13.746 & 0.021 & 0.100 & 0.043 & 0.025 & 0.030 & 1.96 & 0 & 165.47 & 84.38 & 31.6 & 14.7 & 11.3 & 0 \\
47 & 4922510057156118400 & 22.162 & 0.015 & 1.1 & -44.943 & 0.057 & -145.309 & 0.051 & 0.030 & 0.059 & -0.095 & 0.052 & 1.27 & 0 & 21.35 & 16.85 & 1.6 & 1.2 & 1.1 & 0 \\
48 & 2305958908907032832 & 2.781 & 0.024 & 1.0 & 3.535 & 0.018 & -12.472 & 0.011 & -0.027 & 0.025 & -0.004 & 0.021 & 0.83 & 0 & 46.90 & 56.34 & 11.8 & 9.1 & 8.4 & 0 \\
49 & 2772507531095941760 & 3.528 & 0.022 & 0.9 & 25.294 & 0.039 & 6.115 & 0.026 & 0.019 & 0.043 & 0.051 & 0.029 & 1.05 & 0 & 72.57 & 69.33 & 10.3 & 7.9 & 7.3 & 0 \\
50 & 4972326695628963584 & 16.550 & 0.023 & 1.1 & 53.126 & 0.011 & -19.755 & 0.012 & -0.336 & 0.017 & 0.681 & 0.019 & 29.75 & 1 & 217.56 & 7.31 & 27.8 & 8.9 & 3.5 & 1 \\
51 & 2738323714588656512 & 3.058 & 0.026 & 1.1 & 10.360 & 0.050 & 7.684 & 0.027 & 0.078 & 0.059 & 0.097 & 0.031 & 1.88 & 0 & 193.03 & 102.81 & 36.3 & 17.2 & 13.4 & 0 \\
  \noalign{\smallskip}
  \hline
\end{tabular}
\tablefoot{Part of the columns of the full catalog are not shown, but they are available in electronic form at the CDS. The `R' column lists the Gaia EDR3 RUWE value, and the companion masses $m_2$ are given for an orbital radius of 5\,au. The `CPM' column lists the number of resolved, gravitationally bound candidate companions found for each target.}
\end{sidewaystable*}

\begin{sidewaystable*}
\caption{First records of the CPM candidate catalog for the Hipparcos catalog stars.}
\label{CPMHIP-sample}
\centering
\tiny
\renewcommand{\arraystretch}{1}
\setlength\tabcolsep{4.5pt}
\begin{tabular}{cccccccccccccccccccccc}
  \hline
  \hline
T & HIP & EDR3 Source & G & K & $\varpi$ & $\sigma(\varpi)$ & R & $\mu_\alpha$ & $\sigma(\mu_\alpha)$ & $\mu_\delta$ & $\sigma(\mu_\delta)$ & LinSep & $d\varv_\mathrm{tan}$ & $\sigma(d\varv_\mathrm{tan})$ & $\varv_\mathrm{esc}$ & $P_\varpi$ & $P_\varv$ & $P_\mathrm{tot}$ & \texttt{LowV} & \texttt{Bnd} & PMa \\
 & Parent &  & mag & mag & $\mathrm{mas}$ & $\mathrm{mas}$ & & $\mathrm{mas\,a^{-1}}$ & $\mathrm{mas\,a^{-1}}$ & $\mathrm{mas\,a^{-1}}$ & $\mathrm{mas\,a^{-1}}$ & $\mathrm{au}$ & $\mathrm{km\,s^{-1}}$ & $\mathrm{km\,s^{-1}}$ & $\mathrm{km\,s^{-1}}$ &  &  &  &  & & flag \\
  \hline  \noalign{\smallskip}
1 & 1 & 2738327528519591936 & 9.010 & 7.95 & 4.550 & 1.330 & 1.7 & -5.803 & 0.052 & -5.087 & 0.027 &  &  &  &  &  &  &  &  &  & 1 \\
0 & 2 & 2389872193230266624 & 17.338 & 12.21 & 20.821 & 0.192 & 1.3 & 186.505 & 0.188 & -14.195 & 0.118 & 1.12E+05 & 3.32 & 0.37 & 0.11 & 1.00 & 0.00 & 0.00 & 1 & 0 &  \\
1 & 2 & 2341871673090078592 & 8.954 & 6.81 & 20.850 & 1.130 & 29.4 & 181.481 & 0.036 & -0.472 & 0.029 &  &  &  &  &  &  &  &  &  & 0 \\
0 & 3 & 2880990742771590144 & 12.988 & 11.26 & 3.020 & 0.025 & 0.9 & 5.510 & 0.013 & -2.236 & 0.011 & 1.68E+05 & 0.56 & 0.04 & 0.23 & 1.00 & 0.00 & 0.00 & 1 & 0 &  \\
0 & 3 & 2881742976228918912 & 11.428 & 9.66 & 3.001 & 0.032 & 1.2 & 5.053 & 0.017 & -2.174 & 0.016 & 2.18E+03 & 1.23 & 0.05 & 1.99 & 0.99 & 1.00 & 0.99 & 1 & 1 &  \\
1 & 3 & 2881742980523997824 & 6.603 & 6.51 & 3.029 & 0.049 & 1.0 & 5.780 & 0.013 & -2.468 & 0.009 &  &  &  &  &  &  &  &  &  & 0 \\
1 & 4 & 4973386040722654336 & 7.996 & 7.18 & 7.331 & 0.024 & 1.0 & 61.983 & 0.016 & 1.296 & 0.019 &  &  &  &  &  &  &  &  &  & 0 \\
1 & 5 & 2305974989264598272 & 8.319 & 6.31 & 2.628 & 0.025 & 1.0 & 1.001 & 0.022 & 8.756 & 0.020 &  &  &  &  &  &  &  &  &  & 0 \\
1 & 6 & 2740326852975975040 & 11.751 & 8.97 & 18.170 & 5.810 & 1.9 & 223.204 & 0.183 & -11.526 & 0.098 &  &  &  &  &  &  &  &  &  & 0 \\
1 & 7 & 2846308881856186240 & 9.334 & 7.58 & 17.525 & 0.027 & 1.1 & -211.005 & 0.038 & -196.975 & 0.029 &  &  &  &  &  &  &  &  &  & 1 \\
0 & 8 & 2853263597843662976 & 18.512 &  & 1.658 & 0.163 & 1.0 & 18.936 & 0.171 & -5.905 & 0.104 & 1.61E+05 & 2.09 & 0.65 & 0.08 & 0.99 & 0.00 & 0.00 & 1 & 0 &  \\
1 & 8 & 2853169937491828608 & 7.499 & 0.92 & 1.571 & 0.068 & 1.0 & 18.804 & 0.070 & -6.584 & 0.033 &  &  &  &  &  &  &  &  &  & 0 \\
1 & 9 & 2880160886370458368 & 8.260 & 5.89 & 3.132 & 0.027 & 1.0 & -6.048 & 0.034 & 9.260 & 0.021 &  &  &  &  &  &  &  &  &  & 0 \\
1 & 10 & 4976500987226833024 & 8.494 & 7.42 & 10.721 & 0.019 & 1.0 & 42.350 & 0.028 & 40.819 & 0.027 &  &  &  &  &  &  &  &  &  & 0 \\
1 & 11 & 387133547311154432 & 7.335 & 7.02 & 3.395 & 0.030 & 0.9 & 11.002 & 0.015 & -2.095 & 0.014 &  &  &  &  &  &  &  &  &  & 0 \\
1 & 12 & 2308086876223750656 & 7.906 & 4.94 & 1.839 & 0.026 & 1.2 & -5.485 & 0.026 & 1.419 & 0.019 &  &  &  &  &  &  &  &  &  & 0 \\
1 & 13 & 2340148424835415552 & 8.548 & 6.38 & 2.438 & 0.027 & 1.0 & 5.951 & 0.043 & -10.491 & 0.024 &  &  &  &  &  &  &  &  &  & 0 \\
1 & 14 & 2449930576356314880 & 6.922 & 4.51 & 6.032 & 0.032 & 1.1 & 59.999 & 0.023 & -10.334 & 0.016 &  &  &  &  &  &  &  &  &  & 1 \\
1 & 15 & 394029134492039424 & 8.267 & 5.81 & 2.553 & 0.025 & 1.0 & 12.936 & 0.025 & 5.289 & 0.027 &  &  &  &  &  &  &  &  &  & 0 \\
1 & 16 & 4923847578691749120 & 11.604 & 10.45 & 3.301 & 0.018 & 1.1 & 259.367 & 0.057 & -97.506 & 0.065 &  &  &  &  &  &  &  &  &  & 1 \\
1 & 17 & 2306077724882156928 & 8.050 & 7.04 & 6.725 & 0.026 & 1.0 & -34.541 & 0.018 & -27.709 & 0.017 &  &  &  &  &  &  &  &  &  & 0 \\
1 & 18 & 2447815287783063040 & 10.456 & 7.85 & 22.820 & 0.023 & 1.1 & -119.293 & 0.059 & 24.096 & 0.045 &  &  &  &  &  &  &  &  &  & 0 \\
1 & 19 & 2880594231390895104 & 6.295 & 4.33 & 4.489 & 0.028 & 1.2 & -1.692 & 0.012 & -14.321 & 0.008 &  &  &  &  &  &  &  &  &  & 1 \\
1 & 20 & 2848390257367536384 & 8.406 & 7.26 & 10.100 & 0.026 & 1.0 & 36.120 & 0.027 & -23.015 & 0.019 &  &  &  &  &  &  &  &  &  & 0 \\
1 & 21 & 2746745664420425344 & 7.094 & 4.51 & 3.546 & 0.031 & 1.1 & 61.284 & 0.031 & 0.026 & 0.018 &  &  &  &  &  &  &  &  &  & 0 \\
1 & 22 & 4977625894998735872 & 8.456 & 6.42 & 4.871 & 0.020 & 1.1 & -7.159 & 0.024 & 1.695 & 0.029 &  &  &  &  &  &  &  &  &  & 1 \\
1 & 23 & 2767134252131256448 & 7.474 & 6.46 & 10.767 & 0.029 & 0.9 & 53.355 & 0.025 & 9.419 & 0.015 &  &  &  &  &  &  &  &  &  & 0 \\
1 & 24 & 2339755869120492800 & 8.928 & 7.66 & 10.931 & 0.017 & 0.9 & 127.985 & 0.034 & 21.265 & 0.022 &  &  &  &  &  &  &  &  &  & 0 \\
1 & 25 & 4994581292009978112 & 6.092 & 4.37 & 12.290 & 0.770 &  &  &  &  &  &  &  &  &  &  &  &  &  &  &  \\
1 & 26 & 2420820593694180992 & 9.012 & 7.80 & 9.654 & 0.021 & 1.1 & -100.299 & 0.032 & -31.587 & 0.029 &  &  &  &  &  &  &  &  &  & 0 \\
1 & 27 & 2305871596516150656 & 9.170 & 7.67 & 8.865 & 0.019 & 1.1 & 135.285 & 0.028 & -113.910 & 0.026 &  &  &  &  &  &  &  &  &  & 1 \\
1 & 28 & 4994867439910791552 & 8.733 & 7.71 & 4.416 & 0.020 & 1.0 & -10.890 & 0.024 & -7.860 & 0.019 &  &  &  &  &  &  &  &  &  & 0 \\
1 & 29 & 4977679594971842688 & 8.894 & 6.76 & 2.100 & 0.016 & 1.0 & 26.949 & 0.028 & 4.590 & 0.034 &  &  &  &  &  &  &  &  &  & 0 \\
1 & 30 & 384317728096888576 & 8.237 & 7.83 & 2.682 & 0.034 & 1.2 & -8.272 & 0.015 & -10.264 & 0.015 &  &  &  &  &  &  &  &  &  & 1 \\
1 & 31 & 2739891068414238464 & 7.116 & 4.19 & 2.021 & 0.029 & 1.0 & -1.347 & 0.036 & 0.207 & 0.022 &  &  &  &  &  &  &  &  &  & 0 \\
1 & 32 & 395612534315211520 & 9.063 & 8.89 & 1.008 & 0.023 & 0.9 & -1.589 & 0.029 & -2.271 & 0.027 &  &  &  &  &  &  &  &  &  & 0 \\
0 & 33 & 2422810915898660352 & 15.670 & 11.99 & 9.042 & 0.067 & 1.1 & -2.400 & 0.068 & 31.465 & 0.039 & 1.74E+03 & 1.48 & 0.05 & 1.24 & 0.88 & 1.00 & 0.88 & 1 & 1 &  \\
1 & 33 & 2422810915898660096 & 8.015 & 7.00 & 8.908 & 0.056 & 1.0 & -3.203 & 0.026 & 28.795 & 0.013 &  &  &  &  &  &  &  &  &  & 0 \\
0 & 34 & 2853456772588480256 & 12.369 & 9.18 & 13.051 & 0.280 & 16.5 & 50.652 & 0.222 & -52.646 & 0.129 & 2.06E+05 & 2.55 & 0.25 & 0.13 & 0.98 & 0.00 & 0.00 & 1 & 0 &  \\
1 & 34 & 2853571774632882560 & 6.326 & 5.18 & 13.400 & 0.570 & 7.0 & 43.443 & 0.020 & -53.072 & 0.012 &  &  &  &  &  &  &  &  &  & 1 \\
1 & 35 & 2417681762874391936 & 8.844 & 6.82 & 5.367 & 0.021 & 0.9 & 164.398 & 0.042 & -1.220 & 0.025 &  &  &  &  &  &  &  &  &  & 0 \\
0 & 36 & 2766909432067530240 & 17.198 & 13.40 & 6.024 & 0.096 & 1.0 & 53.141 & 0.107 & 17.811 & 0.066 & 1.63E+04 & 1.72 & 0.10 & 0.46 & 0.99 & 0.00 & 0.00 & 1 & 0 &  \\
1 & 36 & 2766909298924361600 & 7.375 & 5.02 & 5.977 & 0.040 & 1.2 & 51.358 & 0.021 & 16.571 & 0.014 &  &  &  &  &  &  &  &  &  & 1 \\
1 & 37 &  & 10.602 & 9.35 & 2.620 & 2.550 &  & -8.070 & 2.160 & 4.930 & 1.930 &  &  &  &  &  &  &  &  &  &  \\
1 & 38 & 4635422588982399616 & 8.485 & 6.86 & 24.949 & 0.015 & 0.9 & 162.463 & 0.018 & -62.086 & 0.019 &  &  &  &  &  &  &  &  &  & 0 \\
0 & 39 & 2415630417774463872 & 15.791 & 12.09 & 11.573 & 0.052 & 1.2 & 166.164 & 0.056 & -33.168 & 0.039 & 2.85E+04 & 1.04 & 0.03 & 0.31 & 0.99 & 0.00 & 0.00 & 1 & 0 &  \\
0 & 39 & 2415630417774463744 & 15.633 & 11.96 & 11.643 & 0.043 & 1.1 & 165.782 & 0.046 & -32.884 & 0.031 & 2.83E+04 & 1.06 & 0.03 & 0.31 & 0.84 & 0.00 & 0.00 & 1 & 0 &  \\
1 & 39 & 2415631586005559680 & 7.377 & 6.38 & 11.545 & 0.028 & 1.0 & 167.661 & 0.024 & -31.113 & 0.017 &  &  &  &  &  &  &  &  &  & 1 \\
0 & 40 & 528563350037254656 & 10.338 & 7.17 & 0.983 & 0.021 & 1.6 & -1.681 & 0.019 & -2.221 & 0.020 & 8.41E+03 & 1.28 & 0.74 & 1.22 & 0.99 & 0.94 & 0.94 & 1 & 1 &  \\
1 & 40 & 528563384392653312 & 10.206 & 7.52 & 0.975 & 0.014 & 1.0 & -1.581 & 0.102 & -2.465 & 0.110 &  &  &  &  &  &  &  &  &  & 0 \\
  \noalign{\smallskip}
  \hline
\end{tabular}
\tablefoot{Part of the columns of the full catalog are not shown, but they are available in electronic form at the CDS. The `R' column lists the Gaia EDR3 RUWE value, and the `PMa' column flag indicates if a proper motion anomaly has been detected with S/N>3.}
\end{sidewaystable*}

\begin{sidewaystable*}
\caption{First records of the CPM candidate catalog for the Gaia EDR3 stars within 100\,pc.}
\label{CPMGaia-sample}
\centering
\tiny
\renewcommand{\arraystretch}{1}
\setlength\tabcolsep{4.5pt}
\begin{tabular}{cccccccccccccccccccccc}
  \hline
  \hline
T & EDR3 Parent & EDR3 Source & Gmag & Kmag & $\varpi$ & $\sigma(\varpi)$ & R & $\mu_\alpha$ & $\sigma(\mu_\alpha)$ & $\mu_\delta$ & $\sigma(\mu_\delta)$ & LinSep & $d\varv_\mathrm{tan}$ & $\sigma(d\varv_\mathrm{tan})$ & $\varv_\mathrm{esc}$ & $P_\varpi$ & $P_\varv$ & $P_\mathrm{tot}$ & \texttt{LowV} & \texttt{Bnd} \\
 &  &  &  &  & $\mathrm{mas}$ & $\mathrm{mas}$ & & $\mathrm{mas\,a^{-1}}$ & $\mathrm{mas\,a^{-1}}$ & $\mathrm{mas\,a^{-1}}$ & $\mathrm{mas\,a^{-1}}$ & $\mathrm{au}$ & $\mathrm{km\,s^{-1}}$ & $\mathrm{km\,s^{-1}}$ & $\mathrm{km\,s^{-1}}$ &  &  &  &  &  \\
  \hline  \noalign{\smallskip}
1 & 123424475948672 & 123424475948672 & 14.9007 & 10.863 & 22.499 & 0.036 & 1.2 & 176.567 & 0.038 & -42.432 & 0.033 &  &  &  &  &  &  &  &  &  \\
1 & 452212812313984 & 452212812313984 & 13.4570 & 9.925 & 24.114 & 0.018 & 1.3 & 110.206 & 0.019 & 10.807 & 0.017 &  &  &  &  &  &  &  &  &  \\
1 & 630505494719360 & 630505494719360 & 8.4576 & 5.646 & 67.933 & 0.027 & 1.1 & 389.309 & 0.044 & -925.041 & 0.040 &  &  &  &  &  &  &  &  &  \\
1 & 756571374829440 & 756571374829440 & 16.6422 & 12.058 & 26.104 & 0.094 & 0.9 & 500.694 & 0.109 & 101.946 & 0.091 &  &  &  &  &  &  &  &  &  \\
1 & 769456276704128 & 769456276704128 & 14.5992 & 10.526 & 53.084 & 0.043 & 1.3 & 180.201 & 0.041 & -320.720 & 0.033 &  &  &  &  &  &  &  &  &  \\
1 & 1013169900890624 & 1013169900890624 & 13.8886 & 10.106 & 20.550 & 0.033 & 1.8 & 200.014 & 0.036 & -134.160 & 0.033 &  &  &  &  &  &  &  &  &  \\
1 & 1155453577587072 & 1155453577587072 & 13.0448 & 9.430 & 20.864 & 0.130 & 7.7 & 119.219 & 0.132 & 9.275 & 0.130 &  &  &  &  &  &  &  &  &  \\
1 & 1227712107314688 & 1227712107314688 & 13.9193 & 10.995 & 22.195 & 0.020 & 1.2 & 1400.292 & 0.022 & -515.645 & 0.020 &  &  &  &  &  &  &  &  &  \\
1 & 1268321022907264 & 1268321022907264 & 17.2582 &  & 20.692 & 0.119 & 1.0 & 100.382 & 0.109 & -99.310 & 0.103 &  &  &  &  &  &  &  &  &  \\
0 & 1268321022907264 & 1268325318000128 & 8.0111 & 6.071 & 20.917 & 0.102 & 1.7 & 107.822 & 0.132 & -95.039 & 0.100 & 1.36E+03 & 1.97 & 0.05 & 1.23 & 0.91 & 1.00 & 0.91 & 1 & 1 \\
1 & 1268325318000128 & 1268325318000128 & 8.0111 & 6.071 & 20.885 & 0.090 & 1.7 & 107.809 & 0.132 & -95.050 & 0.100 &  &  &  &  &  &  &  &  &  \\
0 & 1268325318000128 & 1268321022907264 & 17.2582 &  & 20.723 & 0.122 & 1.0 & 100.382 & 0.109 & -99.310 & 0.103 & 1.35E+03 & 1.94 & 0.05 & 1.23 & 0.95 & 1.00 & 0.95 & 1 & 1 \\
1 & 2673638617215104 & 2673638617215104 & 13.9734 & 10.138 & 24.694 & 0.027 & 1.3 & 24.410 & 0.030 & 61.953 & 0.027 &  &  &  &  &  &  &  &  &  \\
1 & 2729232673961216 & 2729232673961216 & 13.9130 & 10.464 & 27.225 & 0.023 & 1.2 & -65.583 & 0.031 & -10.422 & 0.023 &  &  &  &  &  &  &  &  &  \\
1 & 2781562554898432 & 2781562554898432 & 18.5576 & 13.161 & 23.506 & 0.224 & 1.0 & 399.469 & 0.293 & -33.676 & 0.249 &  &  &  &  &  &  &  &  &  \\
1 & 3124060427493120 & 3124060427493120 & 17.9569 & 12.591 & 26.891 & 0.158 & 0.9 & -56.456 & 0.174 & -23.817 & 0.156 &  &  &  &  &  &  &  &  &  \\
1 & 3179036008830848 & 3179036008830848 & 12.1014 & 7.833 & 116.268 & 0.043 & 1.4 & 1741.875 & 0.047 & 86.494 & 0.044 &  &  &  &  &  &  &  &  &  \\
1 & 3499474928879104 & 3499474928879104 & 14.5659 & 10.267 & 25.773 & 0.027 & 1.3 & -60.619 & 0.031 & -34.159 & 0.026 &  &  &  &  &  &  &  &  &  \\
1 & 3499474928879232 & 3499474928879232 & 15.4037 &  & 25.201 & 0.395 & 3.1 & -73.794 & 0.516 & -30.225 & 0.382 &  &  &  &  &  &  &  &  &  \\
0 & 3499474928879232 & 3499474928879104 & 14.5659 & 10.267 & 25.830 & 0.028 & 1.3 & -60.619 & 0.031 & -34.159 & 0.026 & 9.80E+01 & 2.59 & 0.13 & 2.61 & 0.88 & 1.00 & 0.88 & 1 & 1 \\
1 & 3897223260297728 & 3897223260297728 & 11.9129 & 8.969 & 27.223 & 0.022 & 1.2 & 457.416 & 0.019 & -38.237 & 0.018 &  &  &  &  &  &  &  &  &  \\
1 & 4061462809542400 & 4061462809542400 & 14.3658 & 10.693 & 22.917 & 0.038 & 1.2 & 146.420 & 0.031 & -208.567 & 0.027 &  &  &  &  &  &  &  &  &  \\
1 & 4267552520347776 & 4267552520347776 & 10.8856 & 7.532 & 55.241 & 0.033 & 1.4 & -122.154 & 0.114 & -563.222 & 0.118 &  &  &  &  &  &  &  &  &  \\
1 & 4307169298670464 & 4307169298670464 & 16.7811 & 12.362 & 23.644 & 0.091 & 0.8 & 43.227 & 0.108 & -95.985 & 0.098 &  &  &  &  &  &  &  &  &  \\
1 & 4656161161454464 & 4656161161454464 & 11.2865 & 8.128 & 26.179 & 0.024 & 1.3 & 94.676 & 0.114 & -12.432 & 0.135 &  &  &  &  &  &  &  &  &  \\
1 & 5837071009282816 & 5837071009282816 & 15.6198 & 11.513 & 32.610 & 0.060 & 1.2 & -170.186 & 0.064 & -665.861 & 0.054 &  &  &  &  &  &  &  &  &  \\
1 & 5850333868501248 & 5850333868501248 & 5.9800 & 5.297 & 20.372 & 0.063 & 1.2 & 69.830 & 0.009 & -40.718 & 0.011 &  &  &  &  &  &  &  &  &  \\
1 & 6260619209062272 & 6260619209062272 & 18.8828 & 13.543 & 20.593 & 0.316 & 0.9 & 86.469 & 0.443 & -111.127 & 0.491 &  &  &  &  &  &  &  &  &  \\
1 & 6316591222362496 & 6316591222362496 & 13.1439 & 9.162 & 21.449 & 0.021 & 1.4 & 74.488 & 0.024 & -53.621 & 0.021 &  &  &  &  &  &  &  &  &  \\
1 & 6346041813693696 & 6346041813693696 & 14.4636 & 8.286 & 21.645 & 0.036 & 1.5 & -85.906 & 0.038 & -301.169 & 0.033 &  &  &  &  &  &  &  &  &  \\
0 & 6346041813693696 & 6346041813693824 & 9.9620 & 7.718 & 21.787 & 0.017 & 1.0 & -77.611 & 0.016 & -299.502 & 0.014 & 1.93E+02 & 1.85 & 0.01 & 3.22 & 0.86 & 1.00 & 0.86 & 1 & 1 \\
1 & 6346041813693824 & 6346041813693824 & 9.9620 & 7.718 & 21.753 & 0.015 & 1.0 & -77.706 & 0.045 & -299.487 & 0.051 &  &  &  &  &  &  &  &  &  \\
0 & 6346041813693824 & 6346041813693696 & 14.4636 & 8.286 & 21.696 & 0.038 & 1.5 & -85.906 & 0.038 & -301.169 & 0.033 & 1.92E+02 & 1.82 & 0.02 & 3.22 & 0.98 & 1.00 & 0.98 & 1 & 1 \\
1 & 6348240836955520 & 6348240836955520 & 14.3046 & 10.738 & 21.472 & 0.023 & 1.1 & 486.485 & 0.030 & -126.405 & 0.024 &  &  &  &  &  &  &  &  &  \\
1 & 6397856299125120 & 6397856299125120 & 15.6955 & 11.511 & 22.178 & 0.049 & 1.1 & 193.064 & 0.053 & -298.907 & 0.045 &  &  &  &  &  &  &  &  &  \\
1 & 6452209110182016 & 6452209110182016 & 12.0893 & 8.539 & 40.821 & 0.025 & 1.4 & 469.045 & 0.025 & -122.976 & 0.024 &  &  &  &  &  &  &  &  &  \\
1 & 6963383233077632 & 6963383233077632 & 14.9644 &  & 31.933 & 0.045 & 1.1 & 227.794 & 0.045 & 54.000 & 0.040 &  &  &  &  &  &  &  &  &  \\
0 & 6963383233077632 & 6963761190199424 & 6.7926 & 5.570 & 32.014 & 0.032 & 1.3 & 232.162 & 0.030 & 49.848 & 0.026 & 3.67E+02 & 0.89 & 0.01 & 2.33 & 0.99 & 1.00 & 0.99 & 1 & 1 \\
1 & 6963761190199424 & 6963761190199424 & 6.7926 & 5.570 & 31.979 & 0.029 & 1.3 & 232.099 & 0.014 & 49.833 & 0.011 &  &  &  &  &  &  &  &  &  \\
0 & 6963761190199424 & 6963383233077632 & 14.9644 &  & 31.968 & 0.048 & 1.1 & 227.794 & 0.045 & 54.000 & 0.040 & 3.66E+02 & 0.89 & 0.01 & 2.33 & 1.00 & 1.00 & 1.00 & 1 & 1 \\
1 & 6989290475763456 & 6989290475763456 & 7.8996 & 6.192 & 30.322 & 0.026 & 1.2 & 698.204 & 0.028 & -158.185 & 0.030 &  &  &  &  &  &  &  &  &  \\
1 & 6989977670532096 & 6989977670532096 & 13.1090 & 9.498 & 32.159 & 0.020 & 1.3 & 78.657 & 0.020 & -74.622 & 0.018 &  &  &  &  &  &  &  &  &  \\
1 & 8011385317379968 & 8011385317379968 & 12.5541 & 8.925 & 22.949 & 0.154 & 7.0 & 112.114 & 0.167 & -83.471 & 0.172 &  &  &  &  &  &  &  &  &  \\
1 & 8193113974019072 & 8193113974019072 & 16.0723 & 11.836 & 22.283 & 0.078 & 1.3 & 9.094 & 0.074 & -129.972 & 0.067 &  &  &  &  &  &  &  &  &  \\
1 & 8367803179074816 & 8367803179074816 & 9.6229 & 7.531 & 21.351 & 0.014 & 0.9 & 186.196 & 0.063 & -43.933 & 0.046 &  &  &  &  &  &  &  &  &  \\
1 & 8479094371605632 & 8479094371605632 & 7.7199 & 5.841 & 42.078 & 0.028 & 1.0 & 328.040 & 0.034 & 21.148 & 0.036 &  &  &  &  &  &  &  &  &  \\
1 & 8578256576520320 & 8578256576520320 & 15.7810 & 14.847 & 33.937 & 0.043 & 1.0 & 479.722 & 0.048 & -75.686 & 0.045 &  &  &  &  &  &  &  &  &  \\
1 & 8684153290176000 & 8684153290176000 & 14.5669 & 10.734 & 32.805 & 0.032 & 1.3 & 312.205 & 0.034 & -162.257 & 0.029 &  &  &  &  &  &  &  &  &  \\
1 & 8777268181161216 & 8777268181161216 & 16.7696 & 12.023 & 25.913 & 0.102 & 1.1 & 158.392 & 0.105 & 13.262 & 0.092 &  &  &  &  &  &  &  &  &  \\
1 & 8810116091022208 & 8810116091022208 & 5.8609 & 4.845 & 23.148 & 0.087 & 1.4 & 66.207 & 0.009 & -89.190 & 0.009 &  &  &  &  &  &  &  &  &  \\
  \noalign{\smallskip}
  \hline
\end{tabular}
\tablefoot{Part of the columns of the full catalog are not shown, but they are available in electronic form at the CDS. The `R' column lists the Gaia EDR3 RUWE value.}
\end{sidewaystable*}

\begin{table*}
\caption{Very bright stars ($m < 3$ in the $V$, $H_P$ or $G$ bands) with gravitationally bound candidate companions.}
\label{superbright}
\tiny
\renewcommand{\arraystretch}{0.8}
\begin{tabular}{ccrrrcrrrrrrrrrr}
\hline
\hline  \noalign{\smallskip}
HIP & EDR3 Source & Name & $\varpi$ & $\sigma_\varpi$ & $\varpi$ & $m_V$ & $H_P$ & $G$ & Mass & Radius & Sep. & Sep. & Sc.\tablefootmark{b} \\
 &  &  &  $\mathrm{mas}$ & $\mathrm{mas}$ & Ref.\tablefootmark{a} &  &  &  & $\mathrm{M_{\odot}}$ & $\mathrm{R_{\odot}}$ & $\mathrm{{}^{\prime\prime}}$ & $\mathrm{kau}$ &  \\
 \noalign{\smallskip}  \hline  \noalign{\smallskip}
3092 & 2858629802998456576 & $\delta$ And & 30.910 & 0.150 & H & 3.27 & 3.43 & 2.81 & 2.7 & 15.4 &  &  &  \\
 & 2858629802997575936 &  & 30.946 & 0.036 & G &  &  & 11.43 & 0.5 & 0.5 & 29.0 & 0.94 & 1.00 \\
11767 &  & $\alpha$ UMi & 7.540 & 0.110 & H & 1.98 &  & 10.46 & 8.4 & 47.8 &  &  &  \\
 & 576402619921510144 &  & 7.305 & 0.020 & G &  &  & 8.63 & 1.5 &  & 18.3 & 2.43 & 0.78 \\
13847 & 5044368071870093312 & $\theta$ Eri & 20.230 & 0.550 & H & 3.18 & 2.94 & 3.16 &  &  &  &  &  \\
 & 5044368071868204160 &  & 19.544 & 0.157 & G & 4.11 &  & 4.32 & 2.8 &  & 8.4 & 0.41 & 0.92 \\
15474 & 5099455184968674560 & $\tau^4$ Eri & 10.710 & 0.540 & H & 3.74 & 3.73 & 2.50 & 4.4 & 130.2 &  &  &  \\
 & 5099455184967336832 &  & 10.668 & 0.120 & G$ ^\star$ &  &  & 9.64 & 0.9 &  & 5.8 & 0.54 & 1.00 \\
18543 & 5111187420714898304 & $\gamma$ Eri & 16.040 & 0.580 & H & 2.96 & 3.07 & 2.19 & 4.3 & 66.3 &  &  &  \\
 & 5111187420713057280 &  & 16.172 & 0.090 & G &  &  & 16.14 & 0.1 &  & 16.6 & 1.04 & 1.00 \\
19780 & 4676528209144882304 & $\alpha$ Ret & 20.180 & 0.100 & H & 3.33 & 3.50 & 2.99 & 3.3 & 13.5 &  &  &  \\
 & 4676528243502157696 &  & 20.309 & 0.014 & G & 12.96 &  & 11.18 & 0.3 & 1.1 & 48.4 & 2.40 & 0.94 \\
21421 &  & $\alpha$ Tau & 48.940 & 0.770 & H & 0.86 & 1.00 &  & 3.9 & 57.9 &  &  &  \\
 & 3313069881590622848 &  & 47.253 & 0.117 & G$ ^\star$ &  &  & 11.95 & 0.2 &  & 31.3 & 0.64 & 0.81 \\
24608 &  & $\alpha$ Aur & 76.200 & 0.460 & H & 0.08 & 0.24 &  & 3.7 & 14.6 &  &  &  \\
 & 211810233512673920 &  & 75.184 & 0.062 & G &  &  & 11.64 & 0.5 & 0.5 & 725.2 & 9.52 & 0.95 \\
 & 211810233512673792 &  & 74.993 & 0.021 & G & 14.72 &  & 9.32 & 0.6 & 0.6 & 722.3 & 9.48 & 0.93 \\
26634 & 2900546759663847168 & $\alpha$ Col & 12.480 & 0.360 & H & 2.62 & 2.61 & 8.60 & 5.3 & 7.5 &  &  &  \\
 & 2900579779371768832 &  & 12.566 & 0.016 & G & 14.96 &  & 13.96 & 0.5 & 0.4 & 1213.4 & 97.22 & 0.97 \\
30438 &  & $\alpha$ Car & 10.550 & 0.560 & H & -0.74 & -0.55 &  & 14.2 & 69.5 &  &  &  \\
 & 5500822971164705792 &  & 11.764 & 0.092 & G &  &  & 17.14 & 0.2 & 0.2 & 220.1 & 20.87 & 0.78 \\
32349 &  & $\alpha$ CMa & 379.210 & 1.580 & H & -1.46 & -1.09 &  & 2.2 & 1.7 &  &  &  \\
 & 2947050466531873024 &  & 374.511 & 0.261 & G$ ^\star$ &  &  & 8.52 &  &  & 7.3 & 0.02 & 1.00 \\
35264 & 5589311357728452608 & $\pi$ Pup & 4.040 & 0.330 & H & 2.69 & 2.83 & 2.01 & 10.0 & 275.1 &  &  &  \\
 & 5589305482213195648 &  & 3.357 & 0.042 & G & 7.93 &  & 8.10 & 2.9 & 2.4 & 69.0 & 17.08 & 0.79 \\
 & 5589305580993769856 &  & 3.506 & 0.157 & G &  &  & 18.93 & 0.3 & 0.3 & 240.6 & 59.55 & 0.72 \\
 & 5589310807971543808 &  & 3.285 & 0.064 & G & 17.74 &  & 16.69 & 0.6 & 0.5 & 244.7 & 60.57 & 0.71 \\
36377 & 5512070906394195968 & $\sigma$ Pup & 16.840 & 0.480 & H & 3.25 & 3.38 & 2.53 & 3.8 & 47.9 &  &  &  \\
 & 5512071009471894912 &  & 17.658 & 0.013 & G &  &  & 8.67 & 0.9 &  & 22.1 & 1.31 & 0.85 \\
37819 & 5538814190283894656 & c Pup & 2.880 & 0.190 & H & 3.62 & 3.72 & 2.78 & 10.2 & 301.0 &  &  &  \\
 & 5538814568238884864 &  & 2.695 & 0.064 & G &  &  & 16.89 & 0.3 &  & 58.7 & 20.39 & 0.88 \\
40526 & 3098404220680931968 & $\beta$ Cnc & 10.750 & 0.190 & H & 3.52 & 3.67 & 2.97 & 4.6 & 55.5 &  &  &  \\
 & 3098404216385467392 &  & 10.279 & 0.026 & G &  &  & 12.93 & 0.7 & 0.6 & 29.5 & 2.74 & 0.72 \\
49669 &  & $\alpha$ Leo & 41.130 & 0.350 & H & 1.40 & 1.32 &  & 3.7 & 3.9 &  &  &  \\
 & 3880785530720066176 &  & 41.310 & 0.031 & G & 8.15 &  & 7.89 & 0.6 & 0.9 & 176.2 & 4.28 & 0.99 \\
50801 & 804753180515722624 & $\mu$ UMa & 14.160 & 0.540 & H & 3.04 & 3.15 & 2.24 & 4.4 & 73.2 &  &  &  \\
 & 804753008712665472 &  & 13.768 & 0.275 & G &  &  & 18.83 & 0.1 &  & 55.2 & 3.90 & 0.98 \\
52419 &  & $\theta$ Car & 7.160 & 0.210 & H & 2.76 & 2.65 & 2.73 & 7.2 & 8.9 &  &  &  \\
 & 5239823134370969088 &  & 6.764 & 0.097 & G & 4.73 &  & 4.76 & 4.4 & 4.6 & 381.7 & 53.30 & 0.84 \\
 & 5239825642632266624 &  & 6.758 & 0.031 & G & 16.24 &  & 15.54 & 0.5 & 0.5 & 421.5 & 58.87 & 0.79 \\
 & 5239829800128060672 &  & 6.716 & 0.059 & G & 17.42 &  & 16.62 & 0.4 & 0.4 & 230.1 & 32.14 & 0.79 \\
 & 5239824435712298752 &  & 6.689 & 0.033 & G & 16.64 &  & 15.69 & 0.5 & 0.5 & 665.6 & 92.96 & 0.75 \\
 & 5239829937599494016 &  & 6.628 & 0.036 & G & 17.13 &  & 15.82 & 0.4 & 0.4 & 115.7 & 16.16 & 0.71 \\
 & 5239823787213813888 &  & 6.629 & 0.037 & G & 7.24 &  & 7.25 & 2.3 & 1.9 & 209.6 & 29.27 & 0.70 \\
54061 &  & $\alpha$ UMa & 26.540 & 0.480 & H & 1.79 & 1.95 &  & 4.3 & 27.3 &  &  &  \\
 & 862234033499968640 &  & 25.223 & 0.131 & G &  &  & 15.92 & 0.1 &  & 14.5 & 0.55 & 0.60 \\
59747 & 6071060144089351808 & $\delta$ Cru & 9.450 & 0.150 & H & 2.74 & 2.71 & 2.74 & 6.1 & 6.4 &  &  &  \\
 & 6071112989362835584 &  & 9.224 & 0.042 & G & 16.94 &  & 16.07 & 0.4 & 0.4 & 678.8 & 71.83 & 0.89 \\
60718 &  & $\alpha$ Cru & 10.130 & 0.500 & H &  & 0.67 &  &  &  &  &  &  \\
 & 6053807844583576064 &  & 9.396 & 0.148 & G & 4.81 &  & 4.81 & 3.6 & 3.2 & 90.0 & 8.88 & 0.90 \\
60965 & 3520586071217872896 & $\delta$ Crv & 37.550 & 0.160 & H & 2.93 & 2.94 & 3.02 & 2.6 & 2.3 &  &  &  \\
 & 3520585968137789184 &  & 37.384 & 0.027 & G &  &  & 8.18 & 0.7 & 0.8 & 24.0 & 0.64 & 0.98 \\
62434 &  & $\beta$ Cru & 11.710 & 0.980 & H & 1.25 & 1.15 &  & 8.0 & 10.6 &  &  &  \\
 & 6056717736475418368 &  & 10.631 & 0.037 & G & 16.32 &  & 15.67 & 0.4 & 0.4 & 388.9 & 33.21 & 0.85 \\
 & 6056695059045701120 &  & 10.791 & 0.110 & G$ ^\star$ & 11.69 &  & 10.47 & 0.5 & 2.2 & 421.0 & 35.95 & 0.85 \\
 & 6056690729718614272 &  & 10.659 & 0.018 & G & 13.82 &  & 12.73 & 0.3 & 1.0 & 742.6 & 63.41 & 0.74 \\
63125 & 1517698716348324992 & $\alpha^2$ CVn & 28.410 & 0.900 & H & 2.85 & 2.85 & 5.56 & 3.1 & 2.8 &  &  &  \\
 & 1517698613271954304 &  & 30.612 & 0.076 & G &  &  & 5.54 & 1.5 & 1.5 & 19.2 & 0.68 & 0.72 \\
65378 & 1563590579347125632 & $\zeta$ UMa & 38.010 & 1.710 & H & 2.22 & 2.25 & 2.28 &  &  &  &  &  \\
 & 1563590510627624064 &  & 40.280 & 0.327 & G$ ^\star$ & 3.86 &  & 3.91 & 2.0 & 2.9 & 14.4 & 0.38 & 0.91 \\
 & 1565090003967879424 &  & 40.465 & 0.158 & G$ ^\star$ & 3.99 &  & 3.97 & 1.9 & 2.4 & 708.5 & 18.64 & 0.89 \\
72105 &  & $\epsilon$ Boo & 16.100 & 0.660 & H & 2.39 & 2.52 & 2.18 & 4.9 &  &  &  &  \\
 & 1279752168030730496 &  & 15.730 & 0.597 & G &  &  & 19.53 & 0.1 &  & 78.4 & 4.87 & 0.99 \\
73714 & 6227443304915069056 & $\sigma$ Lib & 11.310 & 0.250 & H & 3.28 & 3.31 & 2.18 & 4.9 & 134.8 &  &  &  \\
 & 6227443098756005504 &  & 11.874 & 0.026 & G &  &  & 14.48 & 0.2 &  & 102.8 & 9.09 & 0.76 \\
80331 & 1625209684868707328 & $\eta$ Dra & 35.420 & 0.090 & H & 2.71 & 2.87 & 2.52 & 2.8 & 10.9 &  &  &  \\
 & 1625209684868172672 &  & 35.820 & 0.026 & G &  &  & 8.70 & 0.6 &  & 4.7 & 0.13 & 0.84 \\
80704 & 1381119031215320576 & g Her & 9.210 & 0.180 & H & 4.91 & 4.48 & 2.69 & 3.4 & 334.9 &  &  &  \\
 & 1381119031214099968 &  & 9.248 & 0.095 & G$ ^\star$ &  &  & 13.40 & 0.4 &  & 9.8 & 1.06 & 1.00 \\
83081 & 5917537534527580160 & $\zeta$ Ara & 6.710 & 0.190 & H & 3.10 & 3.24 & 2.42 & 6.9 & 133.2 &  &  &  \\
 & 5917537912480048384 &  & 6.339 & 0.018 & G &  &  & 13.15 & 0.5 &  & 138.4 & 20.63 & 0.81 \\
85670 & 1415230383034813824 & $\beta$ Dra & 8.580 & 0.100 & H & 2.77 & 2.95 & 2.42 & 6.4 & 42.3 &  &  &  \\
 & 1415230383034347264 &  & 8.362 & 0.128 & G$ ^\star$ &  &  & 11.76 & 0.6 &  & 4.6 & 0.53 & 0.91 \\
93747 & 4314399312979641728 & $\zeta$ Aql & 39.280 & 0.160 & H & 2.95 & 2.99 & 2.90 & 2.5 & 2.4 &  &  &  \\
 & 4314399312966563968 &  & 39.113 & 0.078 & G &  &  & 10.85 & 0.3 &  & 7.4 & 0.19 & 0.98 \\
100751 &  & $\alpha$ Pav & 18.240 & 0.520 & H & 1.92 & 1.86 &  & 5.2 & 5.6 &  &  &  \\
 & 6468761643074086912 &  & 18.513 & 0.093 & G$ ^\star$ & 17.39 &  & 15.38 & 0.3 & 0.3 & 1409.2 & 77.26 & 0.74 \\
104060 & 2162316545207257600 & $\xi$ Cyg & 3.870 & 0.160 & H & 3.70 & 3.83 & 2.98 & 8.0 & 181.0 &  &  &  \\
 & 2162314792856443904 &  & 3.518 & 0.060 & G & 17.75 &  & 17.14 & 0.4 & 0.4 & 256.0 & 66.15 & 0.63 \\
108085 & 6586825380598949632 & $\gamma$ Gru & 15.450 & 0.670 & H & 2.98 & 2.98 & 2.93 & 4.3 & 4.3 &  &  &  \\
 & 6586813320330106624 &  & 15.922 & 0.033 & G &  &  & 14.39 & 0.3 & 0.3 & 162.1 & 10.49 & 0.97 \\
113368 &  & $\alpha$ PsA & 129.810 & 0.470 & H & 1.16 & 1.18 &  & 2.1 & 1.7 &  &  &  \\
 & 6604147121141267712 &  & 131.580 & 0.031 & G & 6.48 &  & 6.09 & 0.5 & 0.7 & 7062.5 & 54.41 & 0.98 \\
 \noalign{\smallskip}
  \hline
  \end{tabular}
\tablefoot{
\tablefoottext{a}{G = Gaia EDR3, G$^\star$ = Gaia EDR3 with RUWE>1.4, H = Hipparcos.}
\tablefoottext{b}{Total score of the candidate companion (Sect.~\ref{totalscore}).}
}
\end{table*}

\begin{table*}[ht]
 \caption{Properties and observed tangential velocity anomaly $\Delta v_\mathrm{T,G3}$ for the 17 white dwarfs of our sample.
 \label{wd-PMa}}
 \centering
 \renewcommand{\arraystretch}{1.2}
 \small
  \begin{tabular}{lrcrrrcc}
  \hline
  \hline
         Name    & HIP & $m_1$ & $\Delta_\mathrm{G2}$ & $\Delta v_\mathrm{T,G3}$ & $\Delta_\mathrm{G3}$ & $\Delta$ & $m_2^\dag$  \\ 
             & & ($M_\odot)$ & & (m\,s$^{-1}$) & & & ($M_J$ at 5\,au)  \\ 
  \hline  \noalign{\smallskip}
\object{Wolf 28} & 3829 & $0.68_{0.02}$ & 2.1 & $9.4_{4.6}$ & 2.1 & $\circ$ & $0.70^{+0.37}_{-0.31}$ \\
\object{GD 279} & 8709 & $0.64_{0.03}$ & 2.7 & $12.2_{9.2}$ & 1.3 &  & $0.87^{+0.65}_{-0.59}$ \\
\object{Feige 22} & 11650 & $0.59_{0.02}$ & 1.0 & $16.1_{27.1}$ & 0.6 &  & $1.09^{+1.67}_{-1.64}$ \\
\object{CPD-69 177} & 14754 & $0.68_{0.02}$ & 0.7 & $5.2_{4.6}$ & 1.1 &  & $0.39^{+0.32}_{-0.30}$ \\
\object{LAWD 23} & 32560 & $0.69_{0.03}$ & 0.8 & $6.8_{11.4}$ & 0.6 &  & $0.52^{+0.76}_{-0.75}$ \\
\object{GD 140} & 56662 & $0.97_{0.03}$ & 3.4 & $35.5_{10.0}$ & 3.5 & $\bullet$ & $3.74^{+1.43}_{-0.90}$ \\
\object{LAWD 37} & 57367 & $0.61_{0.01}$ & 5.1 & $8.7_{2.1}$ & 4.1 & $\bullet$ & $0.60^{+0.23}_{-0.15}$ \\
\object{BD-07 3632} & 65877 & $0.53_{0.08}$ & 1.7 & $18.3_{13.2}$ & 1.4 &  & $1.19^{+0.85}_{-0.77}$ \\
\object{LAWD 52} & 66578 & $0.58_{0.01}$ & 1.3 & $19.9_{15.9}$ & 1.2 &  & $1.34^{+1.04}_{-0.97}$ \\
\object{CD-38 10980} & 80300 & $0.68_{0.02}$ & 1.1 & $7.8_{7.6}$ & 1.0 &  & $0.59^{+0.52}_{-0.50}$ \\
\object{DN Dra} & 82257 & $0.75_{0.03}$ & 1.3 & $7.1_{7.1}$ & 1.0 &  & $0.58^{+0.52}_{-0.49}$ \\
\object{LAWD 74} & 95071 & $0.62_{0.02}$ & 0.6 & $8.8_{7.9}$ & 1.1 &  & $0.62^{+0.53}_{-0.49}$ \\
\object{CD-30 17706} & 99438 & $0.61_{0.02}$ & 0.4 & $5.8_{13.8}$ & 0.4 &  & $0.40^{+0.86}_{-0.85}$ \\
\object{HD 340611} & 101516 & $0.64_{0.03}$ & 2.2 & $10.1_{7.2}$ & 1.4 &  & $0.72^{+0.51}_{-0.46}$ \\
\object{EGGR 141} & 102207 & $0.62_{0.02}$ & 1.0 & $2.0_{14.1}$ & 0.1 &  & $0.14^{+0.87}_{-0.87}$ \\
\object{EGGR 150} & 107968 & $0.63_{0.02}$ & 1.0 & $8.1_{17.9}$ & 0.5 &  & $0.57^{+1.13}_{-1.12}$ \\
\object{LAWD 93} & 117059 & $0.56_{0.05}$ & 0.4 & $36.4_{66.5}$ & 0.6 &  & $2.40^{+3.99}_{-3.93}$ \\
  \noalign{\smallskip}
  \hline
\end{tabular}
\tablefoot{The estimated mass of the WD is listed in the $m_1$ column, $\Delta_\mathrm{G2}$ is the signal-to-noise ratio of the Hip-DR2 PMa, $\Delta_\mathrm{G3}$ is the signal-to-noise ratio of the Hip-EDR3 PMa and $m_2^\dag$ is the normalized mass of the companion at an orbital radius of 5\,au (or an upper limit).
The column $\Delta$ is set to $\bullet$ for 3<$\Delta_\mathrm{G3}$<5 and $\circ$ for 2<$\Delta_\mathrm{G3}$<3.}
\end{table*}

\end{appendix}

\end{document}